\documentclass[12pt]{article}
\usepackage[margin=2.5cm]{geometry}
\usepackage{amsthm}
\usepackage[round]{natbib}
\usepackage{tabularx}
\newtheorem{theorem}{Theorem}
\newtheorem{lemma}{Lemma}
\newtheorem{definition}{Definition}
\theoremstyle{remark}
\newtheorem{remark}{Remark}
\usepackage[ruled,vlined]{algorithm2e}
\usepackage{algpseudocode}
\usepackage{caption}
\usepackage{subcaption}
\usepackage{booktabs}
\usepackage{amssymb}
\usepackage{url}
\usepackage{makeidx}         
\usepackage{graphicx}        
\usepackage{tikz}
\usetikzlibrary{positioning, arrows.meta}
\usepackage{float}
\usepackage{xcolor}
\usepackage{multicol}        
\usepackage[bottom]{footmisc}

\usepackage{newtxtext}       %
\usepackage{newtxmath}       
\usepackage{setspace}
\usepackage{titling}
\usepackage{authblk}
\makeatletter
\DeclareRobustCommand{\change}{%
  \@bsphack
  \leavevmode
  \color{red}%
  \@esphack
}
\DeclareRobustCommand{\stopchange}{%
  \@bsphack
  \normalcolor
  \@esphack
}
\makeatother

\title{\textbf{Causal Portfolio Optimization:\\ Principles and Sensitivity-Based Solutions}}

\author{Alejandro Rodriguez Dominguez\thanks{Head of Quantitative Analysis and Quantitative Research at Miralta Finance Bank S.A., and Quantitative Advisor at Inspiration Q. The views expressed in this monograph are solely those of the author and do not necessarily reflect the opinions or positions of the institutions with which he is affiliated. The author would like to express his sincere gratitude to Miquel Noguer i Alonso for the insightful discussions on this topic and for his continued support in promoting this framework through the AIFI Bootcamps in 2024 and 2025, as well as for including it as a chapter in his latest book on Quantitative Portfolio Management published by Wiley. Special thanks are also extended to Igor Halperin and JD Opdyke for their engaging conversations and valuable feedback over the past two years. The author is equally grateful to his colleague Hugo Valle Varcarcel for his assistance with several sensitivity analysis computations and for posing challenging and thought-provoking questions. This work has also benefited from stimulating conversations with Brian Brice, Daniel Ung, Arun Varma, Peter Cotton, Florian Campuzan, Frank Fabozzi, Cristian Homescu, Mishel Qyrana, and Giuseppe Paleologo—thank you all for your time and insight. Finally, a heartfelt thank you to José Rodríguez Pérez, Managing Director at Miralta Finance Bank, for his unwavering support throughout this project.} \\
    \small Head of Quantitative Analysis \\
    \small Miralta Finance Bank, S.A. \\
    \small \texttt{arodriguez@miraltabank.com} \\
}

\makeindex

\begin{document}
\maketitle

\vspace{2cm}

\begin{center}
    \textit{This monograph was prepared within the Quantitative Analysis Team of Miralta Finance Bank, S.A.}\\
    First version: 07/04/2025
\end{center}

\vfill

\clearpage
\begin{abstract}
Fundamental and necessary principles for achieving efficient portfolio optimization based on asset and diversification dynamics are presented. The Commonality Principle is a necessary and sufficient condition for identifying optimal drivers of a portfolio in terms of its diversification dynamics. The proof relies on the Reichenbach Common Cause Principle, along with the fact that the sensitivities of portfolio constituents with respect to the common causal drivers are themselves causal. A conformal map preserves idiosyncratic diversification from the unconditional setting while optimizing systematic diversification on an embedded space of these sensitivities. Causal methodologies for combinatorial driver selection are presented, such as the use of Bayesian networks and correlation-based algorithms from Reichenbach's principle. Limitations of linear models in capturing causality are discussed, and included for completeness alongside more advanced models such as neural networks. Portfolio optimization methods are presented that map risk from the sensitivity space to other risk measures of interest. Finally, the work introduces a novel risk management framework based on Common Causal Manifolds, including both theoretical development and experimental validation. The sensitivity space is predicted along the common causal manifold, which is modeled as a causal time system. Sensitivities are forecasted using SDEs calibrated to data previously extracted from neural networks to move along the manifold via its tangent bundles. An optimization method is then proposed that accumulates information across future predicted tangent bundles on the common causal time system manifold. It aggregates sensitivity-based distance metrics along the trajectory to build a comprehensive sensitivity distance matrix. This matrix enables trajectory-wide optimal diversification, taking into account future dynamics.
\end{abstract}

\clearpage
\tableofcontents
\newpage


\section{Introduction}
\label{sec:sample1}

This monograph focuses on a framework for portfolio optimization that addresses some of the drawbacks of existing approaches. It builds on several long-standing assumptions about investor behavior toward risk and return, the risk-return trade-off, and the crucial concept of portfolio risk diversification introduced by H. Markowitz in Modern Portfolio Theory (MPT) \citep{10.2307/2975974}. These are fundamental financial concepts within the realm of social science that have been empirically validated for more than half a century and take precedence over any mathematical or optimization tool. This distinction is important because some researchers conflate financial concepts with mathematical models. While mathematical models are user-dependent, financial concepts are not.

\subsection{Liquidity, Technological, and Operational Constraints Create a Gap Between Market and External World Dynamics}

\begin{figure}[H]
    \centering
    \includegraphics[width=0.5\linewidth]{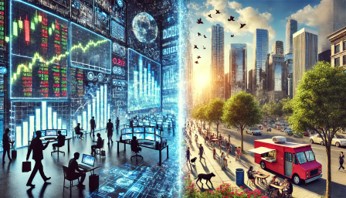}
    \caption{Financial Markets vs Outside World}
    \label{fig:enter-label1}
\end{figure}

In this monograph’s framework, markets are assumed to be complex dynamical systems with a multitude of interacting variables that operate at different frequencies and in various ways. In reality, two distinct worlds can be identified: the financial markets and the external world. An illustrative representation can be seen in Figure \ref{fig:enter-label1}.

The financial market world comprises all information embedded in market prices, originating from orders and executions across various systems and platforms. This includes different types of orders and transactions across multiple exchanges, provided they exert some influence on the overall market \citep{harris2003trading}. In contrast, the external world refers to the entirety of what humans perceive or comprehend over the course of their lives.

While the market world can be represented as tabular data at different frequencies, the external world is far more complex and cannot be captured realistically 100\% in full by any model. However, one thing must be certain: most market participants base their decisions on dynamic events (ie time dependent) occurring in the external world or based on market information (ie coming from both worlds) \citep{harris2003trading,Froot_Teo_2008}. There is a spectrum of traders and investors from pure systematic which are based solely on market information and pure fundamental investors which are based only on outside markets information. They process this information, make decisions, and execute trades or orders accordingly. Once trades are executed, their profit and loss (P\&L) becomes a dynamic system of its own, evolving based on price fluctuations and closing quotes over time \citep{harris2003trading,Froot_Teo_2008,abergel2012market,bouchaud2018trades}.

There exists a widespread misconception among market participants—that trading or investment strategies can be implemented with perfect precision. In reality, most fundamental decisions are influenced by real-world events and are subject to a range of operational, liquidity, technological, and market-related constraints \citep{kirilenko2011microstructure}. Although blockchain technologies offer a promising avenue for addressing some of these inefficiencies, their implementation remains limited in scale and scope \citep{lipton2021blockchain}. Trade execution is rarely instantaneous; rather, it occurs sequentially and is often shaped by the convergence of decision-making processes among market participants, resulting in overlapping behaviors and implementation bottlenecks. These challenges extend to post-trade operations and banking transactions, where similar inefficiencies persist.

The following illustration highlights well-established facts regarding financial market operations, as extensively documented in the literature on market microstructure \citep{harris2003trading,Froot_Teo_2008,abergel2012market,bouchaud2018trades}. These observations underscore a persistent misconception: that the process of translating investment ideas into executable trades is straightforward. In reality, a significant disconnect often exists between the intentions of investors—or the decision logic of algorithmic systems—and the eventual behavior of the market. This gap reflects the complexity of implementation and the influence of numerous frictions within the trading environment.

\begin{figure}[H]
    \centering
    \includegraphics[width=0.65\linewidth]{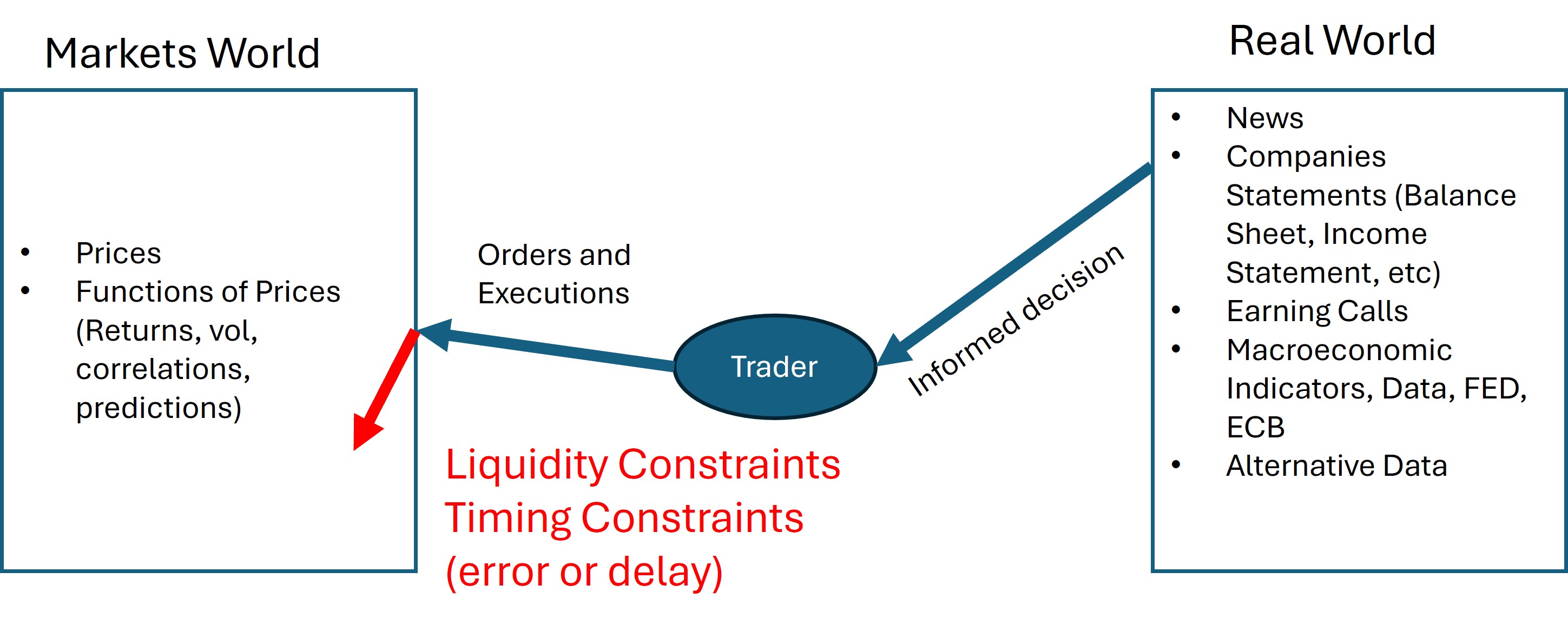}
    \caption{A trader making a decision based on external market information (e.g., news, corporate accounting data, earnings calls), yet introducing inefficiencies such as delays or missmatch errors between both worlds.}
    \label{fig:enter-label2}
\end{figure}

\begin{figure}[H]
    \centering
    \includegraphics[width=0.65\linewidth]{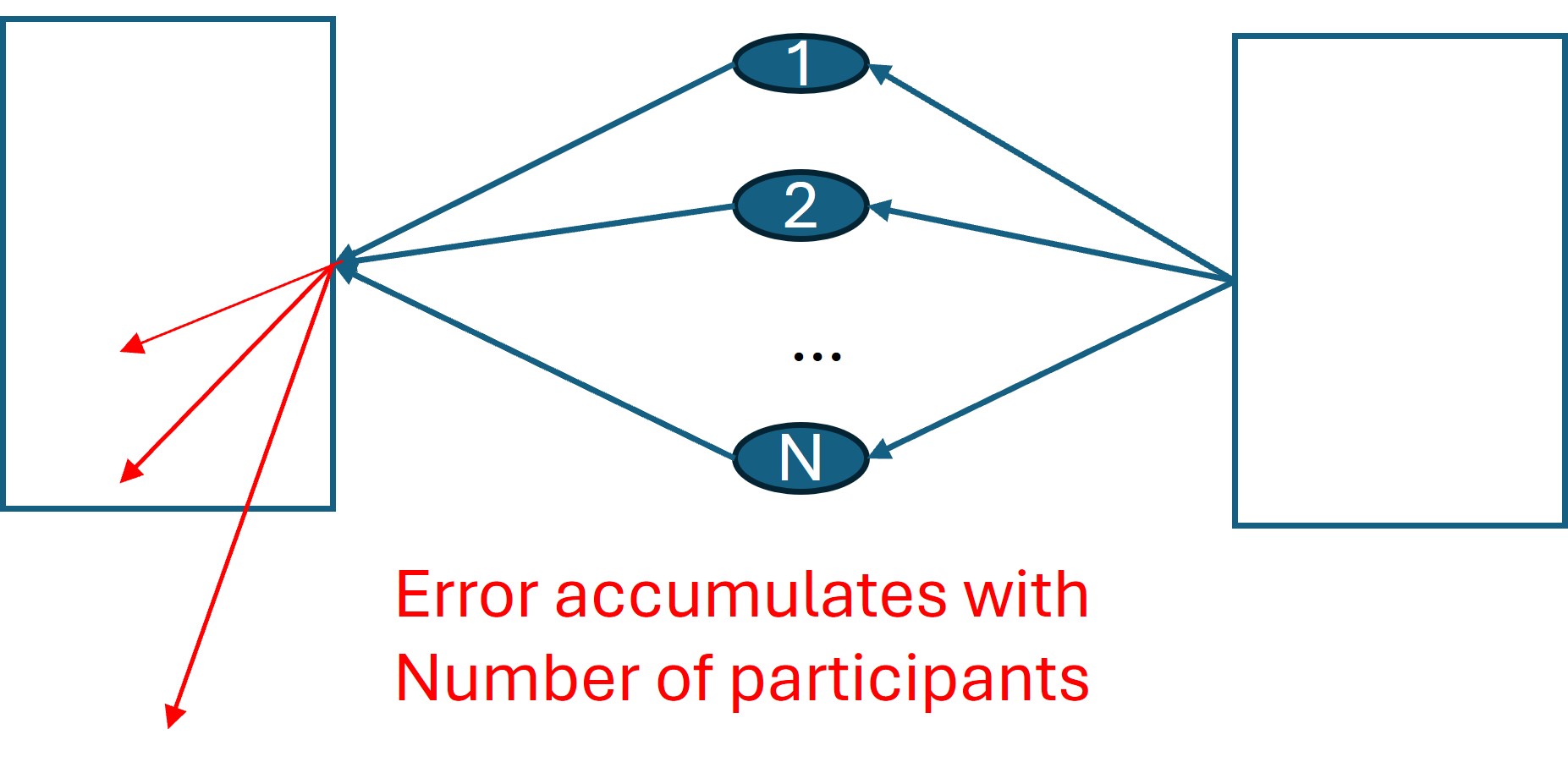}
    \caption{The aggregate error increases with the number of market participants, as operational and liquidity constraints become more pronounced.}

    \label{fig:enter-label3}
\end{figure}

If a participant—let’s call them \textbf{Trader A}—is unable to execute their trade efficiently due to timing or liquidity constraints, they may receive worse execution prices than others who acted first. This inefficiency directly impacts how Trader A translates their outside-world decision into the financial market world (Figure \ref{fig:enter-label2}). More importantly, this execution error compounds over time in two ways:  
\begin{enumerate}
    \item As the number of participants increases (Figure \ref{fig:enter-label3}).  
    \item As this pattern persists over longer time horizons.  
\end{enumerate}

Thus, daily price quotes for financial products reflect decisions made at various prior points in time rather than in real time. The accumulated timing error in execution implies that prices do not precisely capture the intentions of the participants who executed the trades. Furthermore, this phenomenon has not only occurred at time $t$, but has persisted over an extended period, resulting in a compounded lag that is often more significant than market participants might anticipate. Notably, this lag may extend both backward and forward in time when investment decisions are based on anticipated expectations regarding outside-market information about the future producing a negative lag in time between the market impact dynamics and the dynamics of the ideas in the real world that decided the trading, as illustrated in Figure \ref{fig:enter-label6}.

On the other hand, there exists the systematic spectrum of investors, ranging in size from large institutional players such as systematic hedge funds and asset managers \citep{quantblueprint2025largest}, to retail traders relying on technical analysis and chart trading \citep{Ponsi2016TechnicalAA}. This type of investor operates within a market feedback loop, where decisions are based on past market price data and, in turn, influence the behavior of that data at future points in time. This short-term feedback loop is largely driven by the size distribution of orders in the market.

\subsection{Consequences}

\begin{figure}[H]
    \centering
    \includegraphics[width=0.65\linewidth]{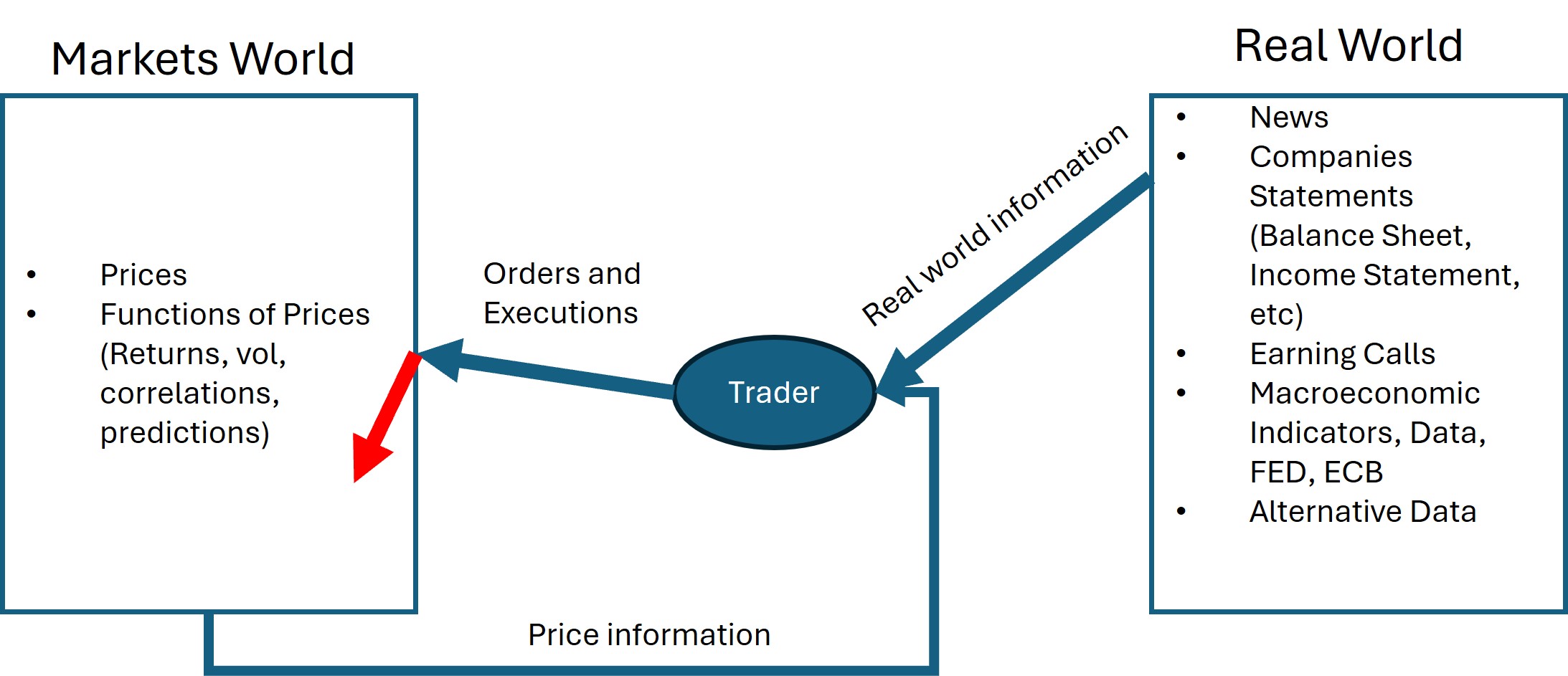}
    \caption{In this case the Trader A is influenced by market information and outside information so the error is feedback-loop in the market information and continue to operate in the outside market driven decisions.}
    \label{fig:enter-label5}
\end{figure}

\begin{figure}[H]
    \centering
    \includegraphics[width=0.5\linewidth]{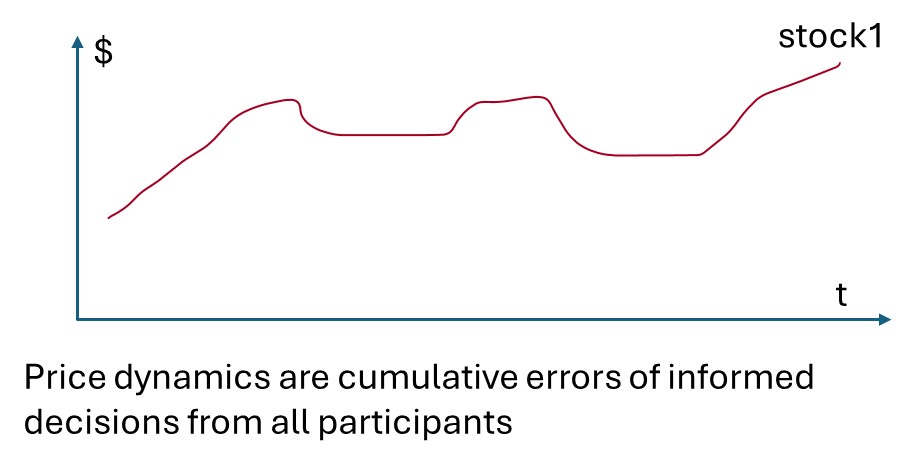}
    \caption{Asset, derivative, index or any financial product prices quotes over time contain these accumulated error $\forall t$}
    \label{fig:enter-label4}
\end{figure}

\begin{figure}
    \centering
    \includegraphics[width=0.65\linewidth]{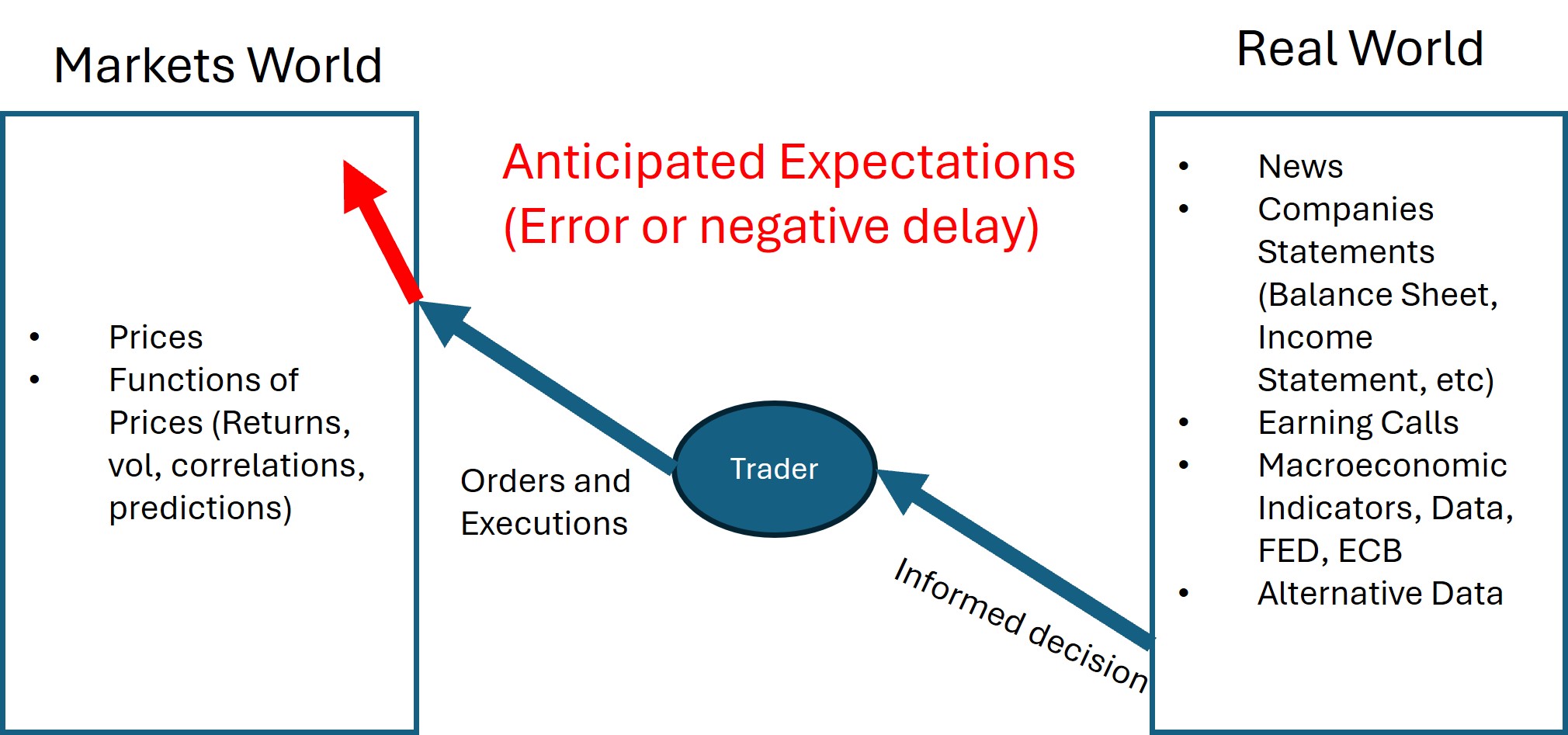}
    \caption{A negative lag in time arises from the discrepancy between the dynamics of market impact resulting from executed trades and the dynamics that originally initiated the decision-making process in the outside-market world }
    \label{fig:enter-label6}
\end{figure}

\begin{itemize}
    \item \textbf{Prices and market dynamics are not an accurate reflection of the outside-market world} in which participants make their decisions. Instead, they incorporate a significant error due to the inherent technical limitations of financial markets.  
    \item \textbf{Some participants recognize this inefficiency} and account for both financial market data and external-world information in their decision-making. These traders leverage market data—including price movements, order flow, and execution patterns at various frequencies—to gain an advantage over those who rely solely on external information.  
    \item \textbf{This feedback loop further isolates financial markets from the external world}, as sophisticated participants increasingly rely on market-generated signals rather than real-world events (illustrated in Figure \ref{fig:enter-label5}, which depicts the feedback mechanism with the market). Consequently, market dynamics become more self-referential, intensifying the disconnection between asset prices and the external realities they are intended to represent. Asset and index prices reflect not only investor expectations and beliefs but also their conditional responses to information derived from both the external world and the market itself (see Figure \ref{fig:enter-label4}).
\end{itemize}

Another reality is that market participants operate at different investment horizons, which leads to another common misconception: the belief that the longer the investment horizon, the less one should care about the short term. While this might hold true for alpha-seeking strategies, this monograph focuses on optimal portfolio diversification from the perspective of assets and portfolio dynamics. As common knowledge quote, "the most important aspect of solving a problem is understanding the boundaries of the problem".  

If your primary focus is diversification and hedging portfolio risks, the compounding effect of all short-term movements—and hence, short-term market dynamics—becomes critically important. Therefore, even for long-term investors, diversification dynamics should be analyzed from an almost continuous-time perspective—or, at a minimum, in discrete time. This implies that risk management rules must be designed to remain applicable across all short-term scenarios.

Some strategies focus on black swan events or market anomalies, but this monograph does not cover those approaches. Instead, the focus lies on optimal diversification, as originally conceptualized by Nobel laureate and pioneer Harry Markowitz, and on developing a framework to understand it from a short-term—potentially continuous-time—perspective, while ensuring its applicability across all investment horizons.

Several researchers have attempted to formalize this misconception through various scholarly frameworks, including the debate between behavioral and rational finance paradigms \citep{fama1970efficient,shiller2003efficient,kahneman2011thinking}, as well as the development and refinement of the Efficient Market Hypothesis and its extensions \citep{malkiel2003efficient,Gabaix2020,Bouchaud2022}. Furthermore, other studies have employed a dynamical systems perspective to model financial markets, offering a more realistic and systemic representation \citep{Bouchaud2023farmer,Bouchaud2024SOC,Halperin2025}. In contrast, purely idiosyncratic, bottom-up approaches often fall into this same misconception, as they fail to address the structural nature of the market and lack a coherent framework for its practical implementation \citep{Lopez_de_Prado_2023}.

This phenomenon manifests in practice when large hedge funds systematically employ algorithmic trading strategies based solely on price movements. Such activity can significantly influence market dynamics, thereby impeding other participants' ability to execute trades based on real-world events, including news releases, unforeseen incidents, or major corporate announcements. As this lag increases, the market increasingly behaves as a closed probability space—a self-referential system governed more by its own internal mechanisms than by exogenous information.

The characterization of the market as a closed probability space permits the application of analytical tools from dynamical systems theory, particularly those developed in the context of probabilistic causality. Relevant concepts include common cause closeness \citep{Gyenis2004} and the completeability of probability spaces \citep{Hofer1999}.

This rationale extends to specific segments of financial markets, such as individual asset classes or trading venues, and with appropriate approximations, may be further applied to subsets of exchanges or markets within defined geographical or asset-type boundaries. Such observations are valuable, as they provide a basis for modeling the market using established frameworks for dynamic behavior.

In implementing the approach proposed in this monograph, a pragmatic perspective is adopted. Rather than modeling the entire market system, certain assumptions or bounded errors are introduced to restrict the analysis to specific segments. Within these localized segments, it becomes feasible to construct closed probability spaces in which the assets comprising the portfolio are embedded. Although this approach introduces approximation errors and limits universality, it offers a practical and operational methodology consistent with standard modeling trade-offs. These are modeling choices aimed at reducing data-related errors; however, the overall framework is defined in general terms both in this work and in \citep{RODRIGUEZDOMINGUEZ2023100447}.

Another aspect of this monograph's framework is that, as part of a complex dynamical system, asset and portfolio dynamics can be represented through time-dependent differential equations of various types. These include linear and nonlinear ordinary differential equations (ODEs), stochastic differential equations (SDEs), partial differential equations (PDEs), and stochastic partial differential equations (SPDEs).  

These equations may incorporate both endogenous and exogenous variables and are often partially or completely unknown. However, they can be approximated using data, albeit with some degree of error.

The framework employs the concept of factors, similar to those introduced in the Capital Asset Pricing Model (CAPM) \citep{sharpe1964,lintner1965} and later expanded in the Arbitrage Pricing Theory (APT) \citep{ross1976,ross1977}, along with contributions from Fama, French, and others \citep{fama1970efficient,Fama1992}. However, in this context, they are referred to as drivers of asset and portfolio dynamics. These drivers act as the exogenous variables in the differential equations and must satisfy certain properties \citep{RODRIGUEZDOMINGUEZ2023100447}.


\begin{definition}[\textbf{Drivers Optimality}]
A driver is optimal for an asset if it is:
\begin{itemize}
    \item \textbf{Optimal in persistence:} the amount of time it remains a driver.
    \item \textbf{Optimal in the selection, based on the probability of causality:} since causality cannot be guaranteed, it is considered in terms of probabilities. An optimal driver should maximize the likelihood of influencing asset dynamics.
\end{itemize}    
\end{definition} 

Until now, no fundamental novelty has been introduced, as these conditions naturally emerge from the study of dynamical systems and differential equations. A review of the literature in these fields reveals that these conditions are necessary and sufficient for a variable to act as a driver of another variable’s dynamics. This holds regardless of the model used—whether it is a linear regression, a neural network, or any other framework.



However, when the problem is framed specifically within the context of portfolio optimization, the relevant set of drivers is naturally restricted. Since the primary objective is to maximize diversification, the drivers of interest are those that govern portfolio dynamics rather than those affecting individual portfolio constituents or assets. This leads to the introduction of the Commonality Principle for Optimal Portfolio Drivers \citep{RODRIGUEZDOMINGUEZ2023100447}, a concept that is both highly intuitive and deeply significant, forming the foundation of the thesis behind this framework.

\begin{definition}[\textbf{Specific Drivers}]
Specific drivers are the optimal drivers for individual assets (portfolio constituents).
\end{definition}

The principle is formalized in the following theorem:

\begin{theorem}[\textbf{The Commonality Principle for Optimal Portfolio Drivers}]
The optimal drivers for a portfolio are the specific drivers that are most frequently selected across portfolio constituents, both in terms of persistence and probability of causality.
\end{theorem}

The proof was first introduced by the same author in the referenced work~\citep{RODRIGUEZDOMINGUEZ2023100447}, relying on \textit{Reichenbach’s Common Cause Principle}~\citep{Reichenbach1956-REITDO-2} and \textit{Modern Portfolio Theory}~\citep{10.2307/2975974} to formally establish that the optimal drivers of a portfolio must be \textbf{common, causal, and persistent}. Additionally, the paper includes a \textbf{complementary geometric proof} using \textit{projective geometry}, inspired by the original construction introduced by Markowitz in 1952. 

In this paper, the author demonstrated that to maintain \textit{investment efficiency} — defined as the combination of:
\begin{itemize}
    \item Idiosyncratic diversification (arising from unconditional probabilities, as in the classical Markowitz framework), and
    \item Optimal systematic diversification through exogenous causal drivers (via conditional probabilities),
\end{itemize}
A conformal map must exist between three spaces:
\begin{enumerate}
    \item The \textit{unconditional probability space} embedded in time;
    \item The \textit{conditional probability space} embedded in time;
    \item The \textit{sensitivity space} (also referred to as the \textit{beta space}) of the portfolio’s assets.
\end{enumerate}

The necessity of using a conformal map in the proposed framework is grounded in the mathematical definition of conformality—namely, the preservation of angles between vectors under transformation. For readers unfamiliar with this notion, consider that the expected returns of portfolio constituents can be embedded in a time-indexed vector space, where axes represent time stamps and vectors encode expected returns. In such a space, the cosine of the angle between any pair of vectors corresponds to their correlation, which underpins idiosyncratic diversification in the classical, unconditional Markowitz framework.

If the angle-preserving structure of the unconditional embedding space is maintained through the transitions into the conditional and sensitivity spaces, then the relative geometric relationships among portfolio constituents—including pairwise correlations and clustering—remain preserved. This preservation allows idiosyncratic diversification (stemming from unconditional return correlations) and systematic diversification (arising from sensitivities to causal drivers) to coexist within a unified geometric framework. Consequently, conformality is not simply an illustrative or aesthetic feature; it constitutes a necessary condition in the theoretical foundation of this diversification approach.

Furthermore, as established over five decades ago, the degree of idiosyncratic diversification achievable in the unconditional case depends on the number of assets in the portfolio \citep{RePEc:bla:jfinan:v:19:y:1964:i:3:p:425-442}. While this level of diversification can either be maximized or held constant, the only way to preserve it in the transition to a conditional and ultimately causal framework—while simultaneously maximizing systematic diversification through causal drivers—is through the existence of a conformal mapping between these spaces.

In this context, the Commonality Principle—used to identify optimal causal drivers—serves as both a necessary and sufficient condition for ensuring the conformal structure required for optimal diversification. Therefore, it provides not only a practical selection criterion for drivers but also a rigorous theoretical guarantee that the resulting sensitivity space supports both idiosyncratic and systematic diversification to their fullest, under causally consistent portfolio dynamics.

Moreover, the use of conformal maps in this context resonates with foundational concepts in the geometry of causal spaces, tracing back over a century to the work of Einstein and Minkowski in relativistic space-time, where conformality characterizes causal structures. This historical parallel is not coincidental but indicative of a deeper alignment between causal inference in physics and portfolio dynamics under causal drivers \citep{minkowski1909space,einstein1916foundation}.

Lastly, the sensitivity space is favored over the time-embedded conditional space because, from the perspective of dynamical systems theory, sensitivities with respect to causal drivers are themselves causal. This property renders the sensitivity space not only causally interpretable but also dynamically informative, facilitating more robust and directionally-aware portfolio optimization. These implications are direct and well-established in the literature on dynamical systems \citep{windeknecht1967causality}.

In essence, for the conformal mapping to hold and ensure investment efficiency, the drivers must satisfy causal, persistent, and commonality properties, aligning with the geometric and probabilistic structure outlined in~\citep{RODRIGUEZDOMINGUEZ2023100447}.

This monograph revisits the core principle and its corresponding proof, reformulating the argument to accommodate a wider audience. For readers unfamiliar with Reichenbach’s Common Cause Principle \citep{Reichenbach1956-REITDO-2}, the relationship to the Markov Common Cause Principle and its treatment in causal inference literature may not be immediately apparent. The original formulation in \citep{RODRIGUEZDOMINGUEZ2023100447} employs Pearl’s structural causal model (SCM) framework \citep{neuberg_2003} and adopts the Markov condition for causal sufficiency \citep{gyenis2004reichenbach}. In this monograph, the proof is reconstructed through both frameworks—Reichenbach’s probabilistic screening-off condition and the Markovian causal sufficiency assumption—while preserving the original proof in the appendix for completeness.

\subsection{Research Context}

The investigation into the integration of asset and portfolio dynamics into portfolio optimization was initiated at Miraltabank with the objective of addressing a longstanding limitation in financial theory: \textbf{the inherent unpredictability of portfolio risk}, an issue originating with the foundational work of Markowitz and Modern Portfolio Theory (MPT).At the time of publication, this represented the state-of-the-art. Regardless of subsequent developments, promising claims, or heightened expectations, the fundamental limitations identified in the existing methodologies remain unresolved. Several established approaches have sought to address this challenge, including:

\begin{itemize}
    \item \textbf{Time series forecasting}, involving econometric models, machine learning (ML), and deep learning (DL), has generally failed to deliver reliable predictions beyond a one-day forecast horizon, thus limiting practical utility in portfolio applications.
    
    \item \textbf{Dynamical systems, partial differential equations (PDEs), and stochastic PDEs (SPDEs)} have been employed to capture the evolution of asset prices and portfolios. Nonetheless, their adoption within the financial industry has remained limited due to \textbf{restrictive assumptions and high computational complexity}.
    
    \item \textbf{Distance metric approaches}, which leverage parametric distance functions based on preselected features, inherit many of the same limitations as traditional forecasting techniques, particularly in terms of robustness and scalability.
    
    \item \textbf{Causal inference methodologies} present several fundamental obstacles:
    \begin{itemize}
        \item Extracting genuine causal relationships from time-stamped financial data is inherently difficult and often inconclusive.
        \item Controlled experiments, while theoretically sound, are costly and frequently too specific to generalize.
        \item Knowledge graph-based representations are often constrained by \textbf{limited user knowledge}, restricting their efficacy in broad market settings.
        \item \textbf{Econometric techniques such as Granger causality} rely on temporal precedence and statistical associations, failing to establish true structural causality.
        \item All of the aforementioned methods are limited by the presence of \textbf{latent or unobserved confounding variables}.
    \end{itemize}
    
    \item Emerging approaches in deep learning have begun to explore \textbf{autoencoder architectures and manifold learning} in the context of causal inference. However, the central issue remains unresolved: \textbf{portfolio risk remains fundamentally difficult to predict}.
\end{itemize}

\subsection{Classical Settings in Mean-Variance Portfolio Optimization}

\begin{figure}[h]
    \centering
    \includegraphics[width=0.65\linewidth]{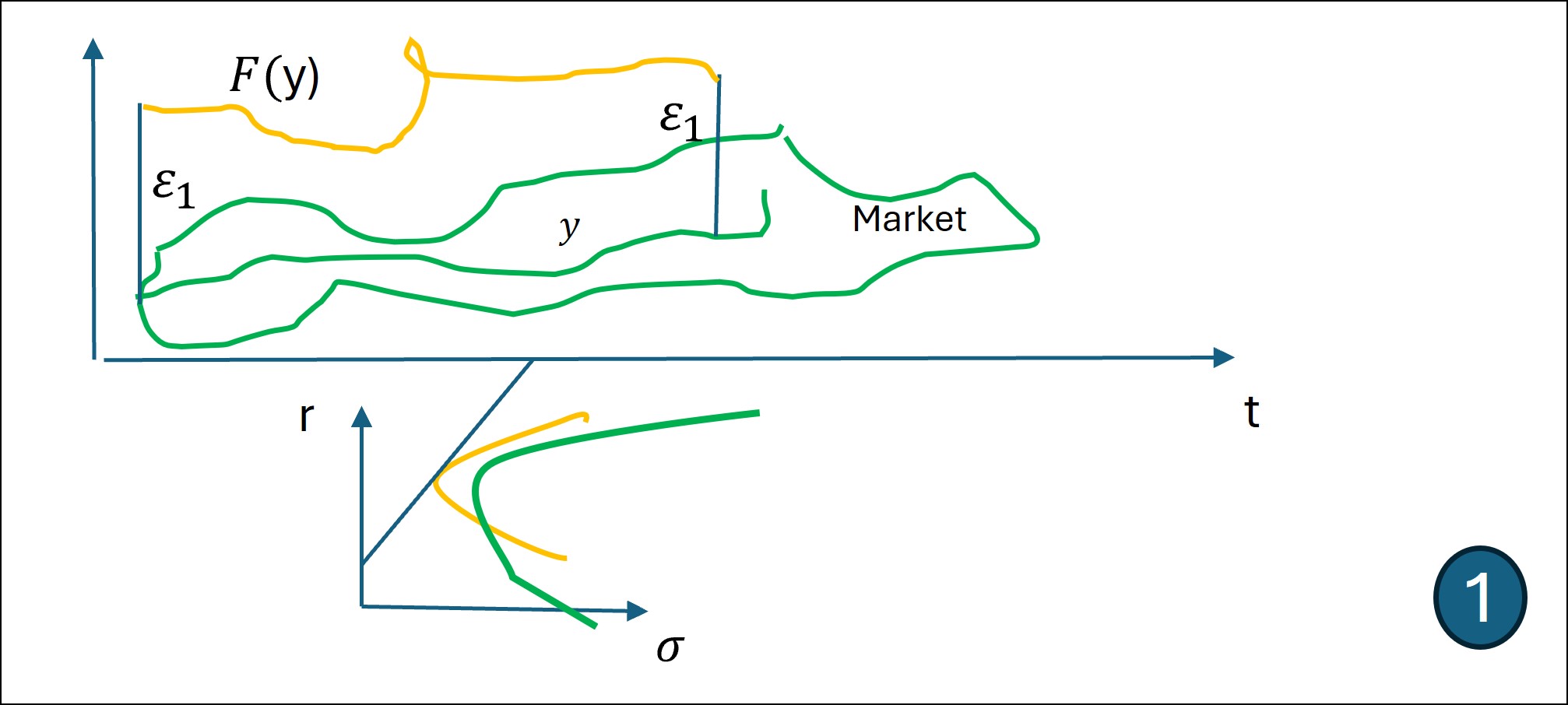}
    \caption{Mean-variance optimization with statistical factors (unconditional case). The top figure shows error $\sum\varepsilon_1^2$ in predicting asset trajectory $y$ using a statistical factor model $F$. The lower part illustrates estimation error in the efficient frontier. Image from \citep{rodriguez2024aifi}}
    \label{fig:Im1}
\end{figure}

\begin{figure}[h]
    \centering
    \includegraphics[width=0.65\linewidth]{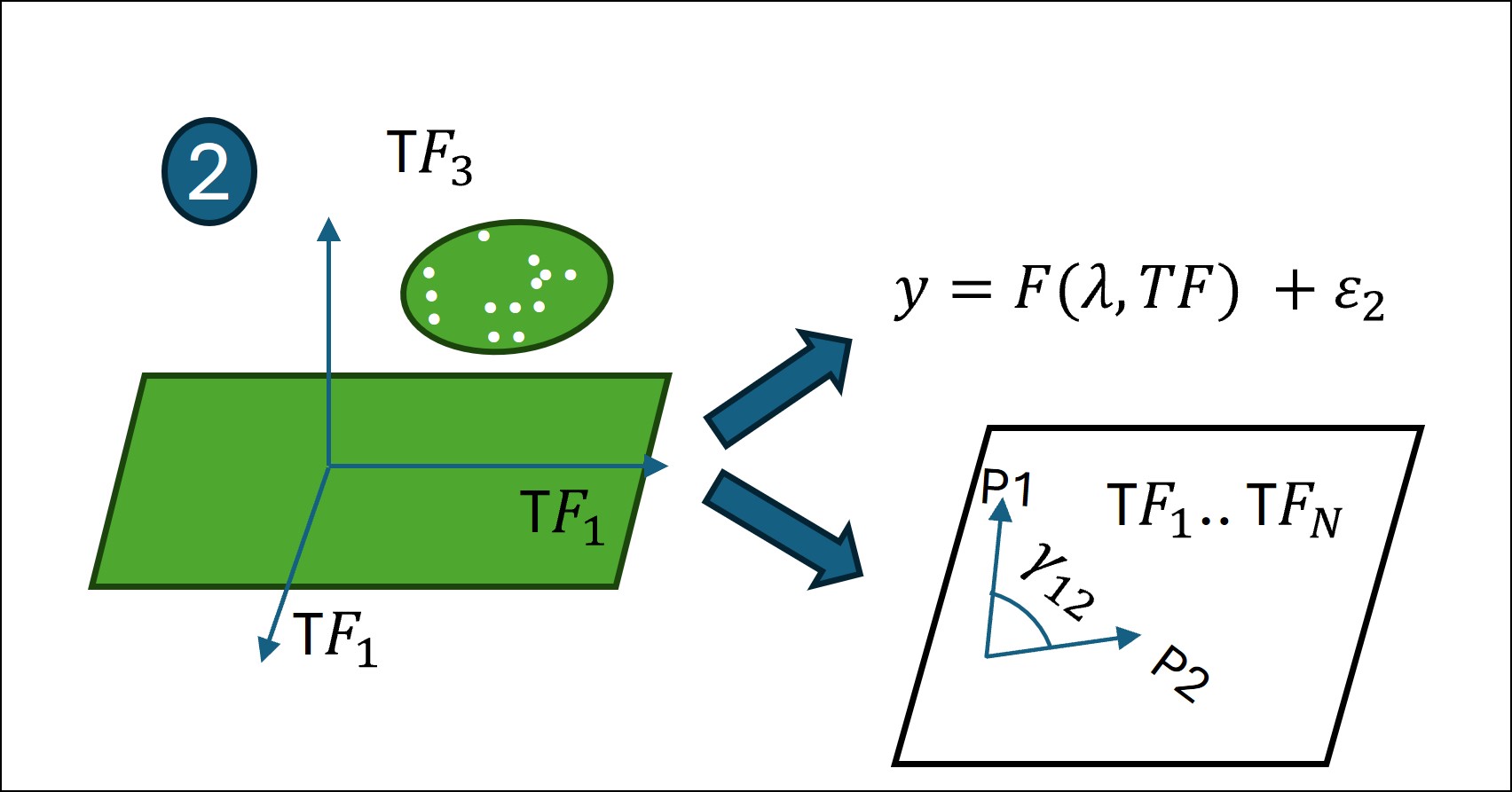}
    \caption{Mean-variance setting with a thematic factor model. Prediction model and hedging are based on a hyperplane (linear case) or hypersurface (nonlinear case). Image from \citep{rodriguez2024aifi}}
    \label{fig:Im2}
\end{figure}

A review of the prevailing approaches to portfolio optimization within the mean-variance framework reveals two predominant settings. Figure~\ref{fig:Im1} presents a simplified schematic of a statistical factor model, typically employed for forecasting or as an extension of the traditional mean-variance setup. In contrast, Figure~\ref{fig:Im2} illustrates a cross-sectional factor model based on thematic factors or smart beta indices. These models use exogenous variables from which portfolio exposures can be inferred via regression. Such models find applications in forecasting, portfolio optimization, and hedging strategies—particularly in settings where portfolio weights are constructed to hedge specific exposures, as shown in the bottom-right corner of Figure~\ref{fig:Im2}.

In the statistical factor model case (Figure~\ref{fig:Im1}), historical returns are used as proxies for future behavior. Two scenarios typically arise: either the forecasting horizon is known, or it is not. In both cases, model selection error and trajectory prediction error contribute significantly to the overall uncertainty. The top-right part of the figure visualizes the evolution of a financial asset $y$ (green) across a data manifold formed by public historical information. The statistical forecast (yellow), however, lies outside this manifold, and its deviation from reality introduces a persistent error $\varepsilon_1$. When translated into a mean-variance context, this leads to a predicted efficient frontier (yellow) that deviates from the true, yet unknown, efficient frontier (green), due to estimation and model errors \citep{rodriguez2024aifi}.

\begin{figure}[t]
    \centering
    \includegraphics[width=0.65\linewidth]{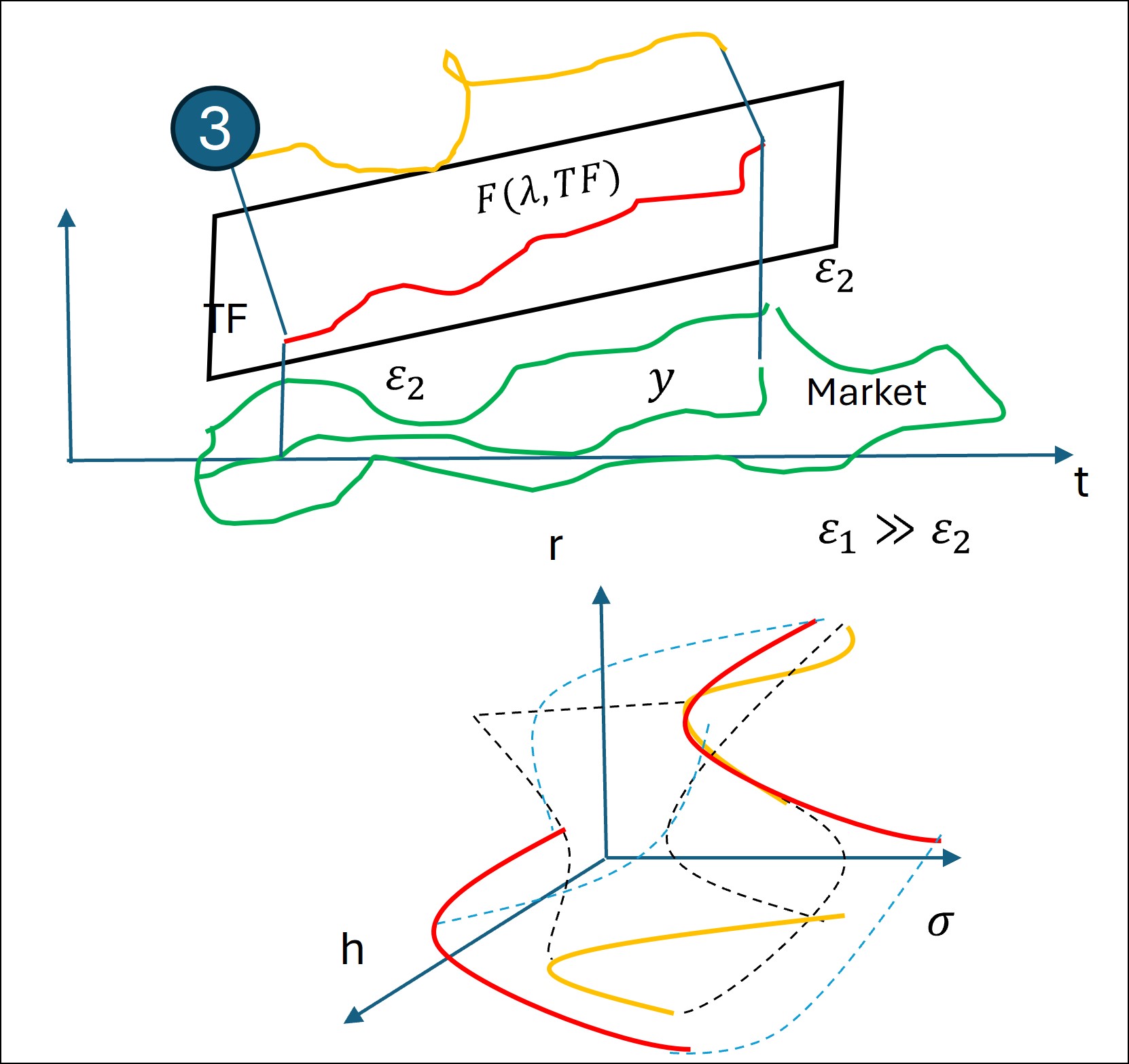}
    \caption{Hyperplane(surface) formed by the thematic factors in which now the prediction is projected will always minimize the error $\sum\varepsilon_2^2$. Image from \citep{rodriguez2024aifi}}
    \label{fig:Im3}
\end{figure}

Figure~\ref{fig:Im2} corresponds to thematic factor models where the explanatory variables are exogenous and selected a priori, as opposed to being statistical properties derived from the asset universe. While similar issues regarding forecasting horizon, trajectory uncertainty, and model selection persist, certain advantages arise if the selected factors are representative of market dynamics. In such cases, the factor space forms a hyperplane (for linear models) or a hypersurface (for nonlinear models) into which asset behaviors can be projected (shown in red in Figure \ref{fig:Im3}). This projection yields reduced forecasting error relative to direct modeling (i.e., comparing red to green rather than yellow to green trajectories). These benefits arise because factor dynamics—when properly chosen—are often more stable and predictable than the underlying asset dynamics themselves. As a result, the projected manifold enables a more accurate forecast and efficient frontier approximation \citep{rodriguez2024aifi}.

\subsection{Limited Framework or Wrong Use}
Focusing on the portfolio optimization problem allows us to use projective geometry as a valuable tool for demonstrating many of the framework’s hypotheses, just as it has been successfully applied in previous cases, such as in Modern Portfolio Theory (MPT) with Markowitz \citep{10.2307/2975974}.

In this work, a similar representation is developed, but from a dynamic perspective of risk and return at both the asset and portfolio levels:
\begin{itemize}
    \item A vast body of literature is dedicated to addressing, one by one, the numerous drawbacks associated with the leading portfolio optimization and factor investing frameworks. However, listing their main limitations is a much simpler task.

    \item It is also essential to understand the chronological evolution of portfolio theory, which follows a logical progression: starting with Markowitz and MPT \citep{10.2307/2975974}, then moving to CAPM \citep{sharpe1964,lintner1965}, followed by APT \citep{ross1976,ross1977}, and eventually leading to the rise of factor investing.
\end{itemize}

\subsubsection{Chronological Order of the Background is not a Coincidence}

It is evident that Markowitz laid the foundation for defining the concept of risk premia through his framework, which describes the relationship between risk and return both at the asset and portfolio levels. This, in turn, introduced the concepts of diversification and hedging within a portfolio. One of the key tools used to demonstrate his thesis was projective geometry, which remains unfamiliar to many because most business schools focus on teaching the results of his work rather than its mathematical proof. Given this, the next natural step is to consider exogenous variables beyond the hyperplane defined by the portfolio and the geometric set formed by the investment universe \citep{10.2307/2975974}.

The most immediate consideration is the market factor, which leads to the development of CAPM \citep{sharpe1964,lintner1965}. However, CAPM does not deviate from the fundamental concept of a purely projective geometry-based thesis, where diversification is merely a geometric hedge, defined by a projection onto a common factor. Business schools often present an idealized version of this theory, suggesting the possible existence of these market factors, but their empirical validation remains an ongoing challenge.

The next layer of complexity follows naturally and arises from the introduction of additional factors, all of which are rooted in the projective geometry framework initiated by Markowitz, known as risk premia. However, this time, the projective spaces are defined based on these new factors. Initially, these factors were linear, but over time, more complex models emerged, incorporating non-linear models, machine learning \citep{RePEc:pra:mprapa:30534,coqueret2020machine}, and even causal inference \citep{Lopez_de_Prado_2023,Howard2025}.

This historical progression explains why the sequence of developments unfolded as it did, rather than in reverse order. It also clarifies why, in any attempt to redefine the foundations of a new framework, one must start with assets and the portfolio, before advancing to factor models. Some approaches in the literature attempt to build new frameworks based on factors, redefining diversification and risk formulas, without first establishing a foundational structure based on portfolios and their underlying assets, as Markowitz did. This is a fundamental mathematical error.

While specific ad hoc methods may prove effective in particular cases, they will never achieve the general applicability of frameworks like MPT, CAPM, and APT, regardless of criticism. Their effectiveness can be debated, but their structural robustness and interconnections as frameworks remain unquestionable. The best evidence of their enduring relevance is that, nearly a century later, they continue to serve as the preferred theoretical starting points in finance.

\subsection{The Issue with Modern Portfolio Theory is about Projective Spaces}

Many researchers criticize MPT (and, by extension, CAPM and APT) by merely highlighting some of their individual drawbacks without proposing an alternative framework. This occurs for two main reasons: first, because identifying flaws is easier than developing a new framework, and second, and more importantly, because the framework itself is not flawed—rather, its assumptions and methods of application are. If MPT could be perfectly adapted to any environment, which is not always feasible, it would be an ideal framework. Therefore, the issue is not with the framework itself, and it makes no sense to criticize or replace it. This reflects a common confusion between framework and method; critiques are often directed at specific methods as though they represent the entire framework. However, the objective is to retain the core principles of the framework while improving upon the methods used to implement it. Examples include \citep{Lopez_de_Prado_2023}.

This monograph argues that MPT (and, consequently, CAPM and APT) depends on the projective space in which it is applied. This has two key consequences. First, since MPT is dependent on projective space, CAPM and APT are special cases of MPT in which exogenous variables are introduced into the projective plane. Second, this dependency extends to the type of geometry used, whether Euclidean, Riemannian, or any other kind. MPT is inherently linked to the chosen geometry, which, for instance, could allow causality concepts to be more easily incorporated if Riemannian geometries or those that admit time curvature are used.

This perspective opens the door to the field of information geometry, which has significant potential applications in portfolio optimization \citep{Marti2021,armstrong2024optimal}. MPT should be understood as a financial concept that is applicable to any geometric framework, and as users, by selecting a specific geometry, we inherently restrict the problem space. For example, causality cannot be meaningfully addressed within a Euclidean geometry, just as once a specific geometry is chosen, the statistical methods available are constrained by that selection. Some geometries work better in linear spaces, while others are more suited to nonlinear spaces, meaning that the initial choice of geometry restricts the subsequent statistical approach \citep{rodriguez2024aifi}.

One example of this limitation is correlation, a linear measure that may be inappropriate for certain financial applications. Financial markets, for instance, often exhibit nonlinear and non-Euclidean inference properties, making the use of correlation potentially misleading. However, above all, the core concepts of risk premia, diversification, risk, return, the rational investor, utility functions, etc., as defined in MPT, are independent of the choice of geometry. This confirms that the framework itself is fundamentally sound, while misapplications resulting from incorrect geometric and statistical assumptions are the actual sources of failure.

\begin{figure}
    \centering
    \includegraphics[width=1\linewidth]{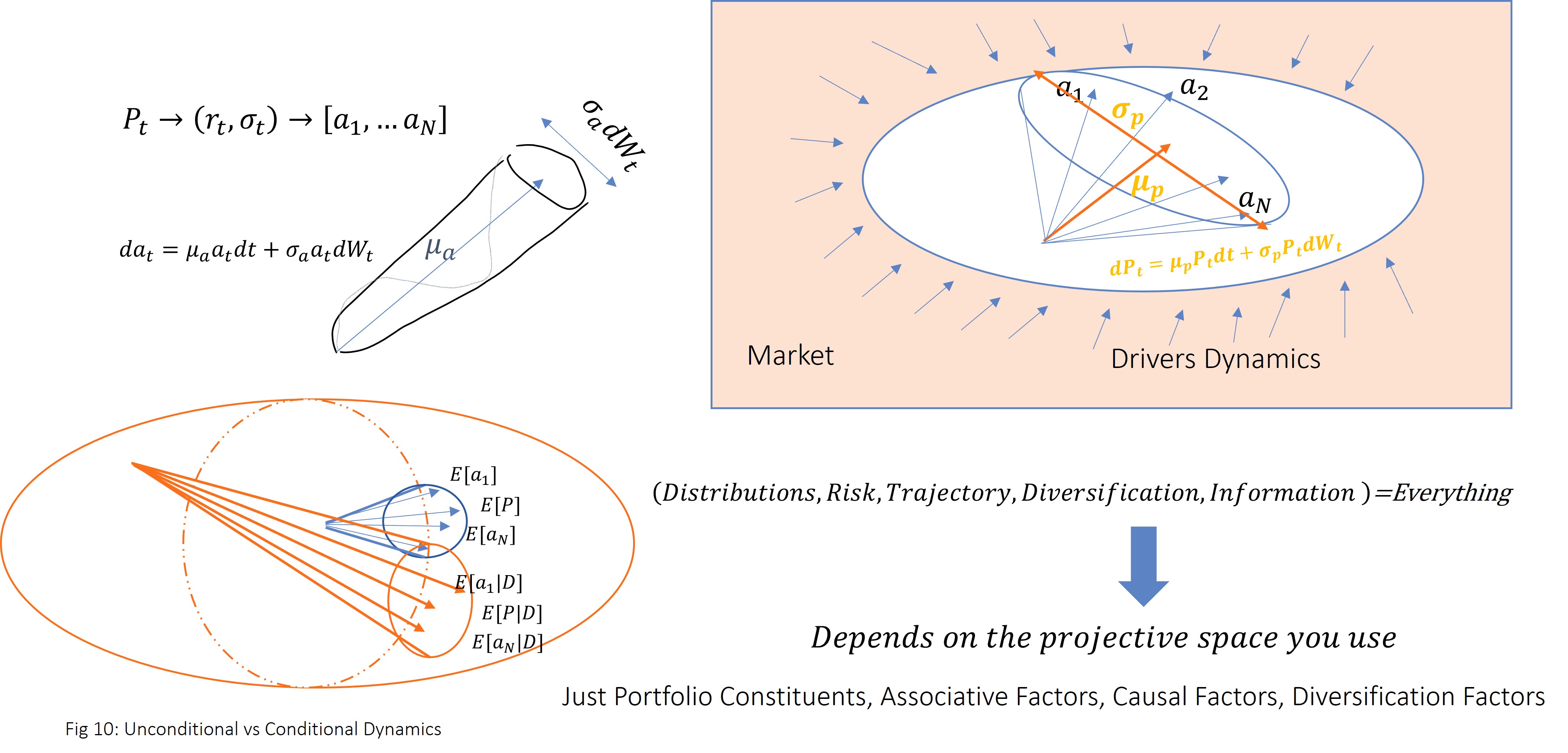}
    \caption{Conic or parabolic expansions can be used to represent the dynamics of assets or portfolios through drift and diffusion parameters. Notably, the conditional and unconditional cases yield fundamentally different geometric expansions—differences that are further amplified when the selection of drivers is suboptimal. Figure adapted from \citep{rodriguez2024aifi}.}
    \label{Conic}
\end{figure}

In Figure \ref{Conic}, a schematic representation of the claim that MPT depends on the chosen geometric space is shown. Portfolio return and risk are functions of the constituents' returns, which can be represented by stochastic differential equations (SDEs) and graphically depicted as a conic expansion, where the drift term defines the direction and the dispersion parameter determines the diameter of the cone.

On the right side, a financial market is illustrated, with an ellipse enclosing all available information about the portfolio constituents. Within the ellipse, vectors indicate their respective directions, each with its own conic representation. The portfolio itself also has a conic structure, shown in blue, formed by the individual cones of its constituents. These are geometric representations of risk and return dynamics.

The key issue is that the market interacts with this elliptical information space through a set of drivers’ dynamics. On the bottom left, an orange ellipsoid represents both the portfolio constituents and their drivers’ dynamics. Here, the portfolio cone is also in orange, fundamentally different from the blue one obtained in a mean-variance framework as seen above. This distinction arises because, in this case, returns are modeled as conditional expectations rather than unconditional expected returns.

From this, we can conclude that all aspects of portfolio risk management—including distributions, risk trajectory, diversification, and information—depend on the projective space in which they are represented \citep{rodriguez2024aifi}.

\subsection{What Do We Need to Overcome?}
There is a clear lack of dynamism in current asset optimization and risk management methods. This can be addressed in various ways — through predictive models, factor models ranging from one to N factors, which themselves can be linear or nonlinear, or based on partial differential equations (PDEs) or stochastic partial differential equations (SPDEs). Alternatively, one can leverage the mathematical and physical properties of the market’s causal dynamics, as outlined in previous works \citep{RODRIGUEZDOMINGUEZ2023100447,dominguez2024portfolios,noguer2025quantitative}, to simplify the problem. For example, by identifying a system that models the dynamics of diversification over time and its trajectory.

Incorporating exogenous and endogenous variables in portfolio optimization (PO) introduces both idiosyncratic and systematic diversification. The factor-based approach enhances exogeneity and enables systematic risk hedging by integrating factor information into the portfolio optimization process. However, to preserve the highest level of idiosyncratic risk diversification while adding systematic diversification, the exogenous variables and the risk diversification dynamics must meet specific properties, which we will outline in the following sections.  

Causality is inherently challenging to measure and prove. However, in the context of diversification, effective tools exist for estimating the probability of causality, making it a more practical approach for our problem.

\subsection{Proposed Solution and Monograph Additional Content}
In the original paper \citep{RODRIGUEZDOMINGUEZ2023100447}, the proposed framework was implemented using PDEs, SPDEs, and neural networks, with sensitivities extracted via automatic adjoint differentiation and constituents embedded in a sensitivity space. A hierarchical mapping was then applied from this space to the desired risk measure. These modeling choices represented one of the central contributions of that work.

In this monograph, the core framework remains unchanged—centered on the Commonality Principle and the causal sensitivity space—but different modeling approaches are adopted. Specifically, linear models are employed instead of PDEs or neural networks, and the mapping from the sensitivity space to the risk measure is varied depending on the specific metric in use. For example, mappings differ when targeting volatility versus CVaR.

Thus, while the theoretical structure based on causal inference and sensitivity embeddings is preserved, the implementation methods are adapted to provide a broader range of practical solutions depending on the risk objective.

\newpage
 
\section{Theoretical Framework}
For the framework to function, the following components are required:

\begin{itemize}
    \item \textbf{A theoretical model for asset and portfolio dynamics.} This can be any well-defined mathematical model. However, the choice of model will significantly affect the quality, robustness, and interpretability of the framework’s outputs.

    \item \textbf{Input data for the chosen model.} The data must correspond to the optimally selected portfolio drivers, as these are fundamental for accurately modeling the dynamic behavior of assets.

    \item \textbf{A model for approximating the theoretical dynamics.} This refers to the empirical or computational method used to approximate the theoretical model using observed data. The chosen approximation method (e.g., linear regression, neural networks, kernel methods) will directly impact the estimation of sensitivities and, consequently, the quality of the optimization.

    \item \textbf{An objective function.} The framework seeks to determine an optimal diversification strategy grounded in the notion of \emph{investment efficiency}, originally introduced by Markowitz in Modern Portfolio Theory~\citep{10.2307/2975974} and later extended by works such as \citep{sharpe1964,lintner1965} to include exogenous diversification and the broader risk premia landscape \citep{fama1970efficient}. Within this framework, optimality is redefined in terms of selecting the optimal drivers of portfolio dynamics according to the \emph{Commonality Principle}, which ensures the preservation of idiosyncratic diversification from the unconditional case while incorporating systematic diversification based on causal drivers~\citep{RODRIGUEZDOMINGUEZ2023100447}. Alternative definitions of efficiency in portfolio optimization are demonstrated to be suboptimal under this framework, as formally shown in~\citep{RODRIGUEZDOMINGUEZ2023100447}.

    \item \textbf{An algorithm for solving the portfolio optimization problem.} The appropriate algorithm depends on the model used to describe asset and portfolio dynamics. This will be addressed in Chapter 7. Depending on the chosen risk measure and the method used to estimate sensitivities, various solution methods can be employed—several of which will be discussed in that chapter.
\end{itemize}

\subsection{A Theoretical Model for Asset and Portfolio Dynamics}
This framework considers asset dynamics as a function of sensitivities with respect to a set of explanatory drivers. These models can take a variety of forms—ranging from simple linear structures, where sensitivities correspond to factor betas, to more complex, non-linear functional relationships. The goal is to estimate asset behavior through a function that captures the dependence of asset returns on selected drivers and their dynamic properties.

In the general case, asset behavior is modeled as an unknown function \( F \) of both the drivers and their sensitivities. Although this function may exhibit considerable complexity and lack a closed-form analytical expression, it can be approximated using a variety of function approximation techniques that support sensitivity analysis. These include kernel methods, splines, and machine learning models such as decision trees and neural networks. 

\begin{remark}
Among these, neural networks have been demonstrated to yield superior performance, justifying their use in the original study~\citep{RODRIGUEZDOMINGUEZ2023100447}. It is important to clarify that this framework does not pose a prediction problem in the traditional sense. Rather, the objective is to accurately fit the sensitivity function, enabling effective generalization for subsequent use in the portfolio optimization stage. In the case of neural networks, the approach qualifies as a white-box methodology, since the selection of optimal causal drivers precedes model fitting. Consequently, the interpretability of the model is enhanced, and its causal structure is preserved. For completeness, the methodology is presented using various model classes.   
\end{remark}

Let $y$ denote the asset value, and $x = (x_1, \dots, x_N)$ the vector of portfolio drivers. A general formulation of the dynamic relationship can be written as:

\begin{equation}
y(t) = F\left(\frac{\partial y(t)}{\partial x_1(t)}, \dots, \frac{\partial y(t)}{\partial x_N(t)}, \frac{\partial x_1}{\partial t}, \dots, \frac{\partial x_N}{\partial t}, \frac{\partial y}{\partial t}, x_1(t), \dots, x_N(t)\right)
\label{FUNC1}
\end{equation}

If $F$ is linear, the model reduces to a traditional linear factor structure:

\begin{equation}
y(t) = x_1(t) \frac{\partial y(t)}{\partial x_1(t)} + x_2(t) \frac{\partial y(t)}{\partial x_2(t)} + \dots + x_N(t) \frac{\partial y(t)}{\partial x_N(t)} + \frac{\partial y}{\partial t} + \frac{\partial x_1}{\partial t} + \dots + \frac{\partial x_N}{\partial t}
\label{FUNC2}
\end{equation}

To approximate the functional form $F$, any universal or sufficiently flexible function approximator may be used, depending on the desired level of expressiveness and interpretability. A generic form of such a dynamical system can be expressed as:

\begin{equation}
d[k] = g\left(d^{\{k-1\}}, u^{\{k\}}, \varepsilon^{\{k-1\}}\right) + \varepsilon[k]
\label{GENSYS}
\end{equation}

Where:
\begin{itemize}
  \item $g$ is a generic (possibly non-linear) approximating function.
  \item $d[k]$ is the predicted asset return at time $k$.
  \item $d^{\{k-1\}} = [d[k-1], d[k-2], \dots]^T$ is the vector of lagged asset returns.
  \item $u^{\{k\}} = [u[k-1], u[k-2], \dots]^T$ contains the historical values of external drivers.
  \item $\varepsilon^{\{k-1\}} = [\varepsilon[k-1], \varepsilon[k-2], \dots]^T$ is the vector of past noise terms.
  \item $\varepsilon[k]$ is the innovation or noise at time $k$.
\end{itemize}

The system in Equation~(\ref{FUNC1}) or~(\ref{GENSYS}) can be implemented using any suitable approximation strategy. The choice of model will influence both the fidelity of the sensitivity estimates and the tractability of the optimization problem.

Once the functional relationship has been approximated, sensitivities of asset values with respect to the drivers can be extracted using analytical methods, numerical differentiation, or automatic differentiation, depending on the modeling choice.

\begin{figure}[h!]
  \centering
  \includegraphics[width=65mm]{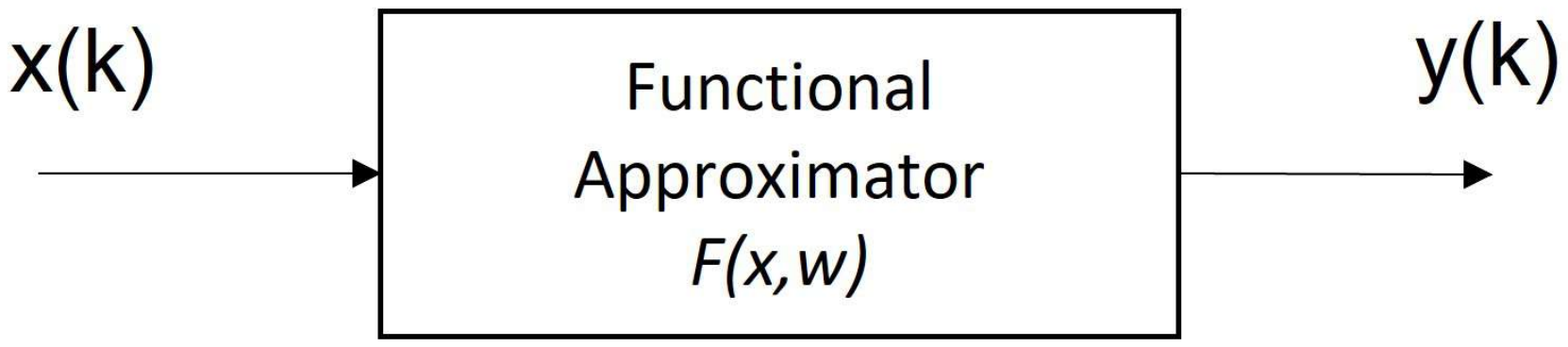}
  \caption{Generic Functional Approximator Scheme}
  \label{figure1}
\end{figure}

\section{Optimal Drivers Selection: Causality and Persistence}
\label{sectionCausality}

Drivers are selected based on two primary objectives: (1) maximizing the accuracy in approximating asset and portfolio dynamics, and (2) enabling the use of sensitivity information to project portfolio constituents into a sensitivity space, which allows for diversification grounded in dynamic risk and return characteristics. 

Candidate drivers are defined as any exogenous variables—that is, variables external to the portfolio—that may influence asset or portfolio behavior. In practice, the set of available driver candidates is constrained by the user's data access. However, this limitation is common across all frameworks and methodologies and does not undermine the generality or applicability of the proposed approach.

\begin{definition}
	Drivers Optimality.\\
	A driver is optimal for an asset if it is:
	\begin{itemize}
		\item Optimal in persistence: the amount of time it remains a driver.
		\item Optimal in selection based on the probability of causality: since causality cannot be guaranteed, it is considered in terms of probabilities. An optimal driver should maximize the probability of influencing asset dynamics.
	\end{itemize}    
\end{definition} 

\begin{definition}
	Specific drivers are the optimal drivers for individual assets (portfolio constituents).
\end{definition}

The selection of optimal drivers plays a fundamental role in enhancing the approximation of asset dynamics and sensitivities. For effective portfolio optimization, it is essential to identify and utilize the most commonly shared specific drivers across all portfolio constituents. This ensures that the resulting sensitivity space supports optimal diversification. The shared set of drivers identified in this manner also constitutes the optimal portfolio drivers, as determined by their statistical persistence and relevance.

The following theorem and its proof, as introduced in~\citep{RODRIGUEZDOMINGUEZ2023100447}, are foundational to the framework. Although intuitive, they are necessary both for improving the approximation of portfolio dynamics and for attaining optimal diversification.

\begin{theorem}[\textbf{Commonality Principle for Portfolio Drivers, \citep{RODRIGUEZDOMINGUEZ2023100447}}]
	The Commonality Principle for Portfolio Drivers states that the optimal drivers for a portfolio are those specific drivers that are most frequently selected across portfolio constituents, according to both their persistence and their estimated probability of causality.
\end{theorem}

The proof of the Commonality Principle begins by demonstrating that the drivers most frequently selected across constituents—i.e., the common drivers—are the most persistent and therefore optimal for representing the collective dynamics of the portfolio. The argument draws upon foundational results from Modern Portfolio Theory (MPT)~\citep{10.2307/2975974}.

\begin{proof}
	\textbf{Modern portfolio theory and portfolio drivers' persistence:}\\
    Drivers influence asset risks by determining their dynamics. Risks are categorized as idiosyncratic or     systematic, with specific drivers contributing to both types. Most specific drivers of a particular         portfolio constituent primarily contribute to its idiosyncratic risk, affecting only that asset.            However, some drivers may also contribute to the systematic risk of both this individual asset and other    constituents.
	
    In the context of Modern Portfolio Theory (MPT), the total systematic risk explained by focusing on all specific drivers from all constituents is not maximal, as most specific drivers focus solely on idiosyncratic risks. By identifying common drivers across the portfolio, the focus shifts to those specific drivers that maximize systematic risk explanation across all constituents. These common drivers ensure optimal driver persistence as they contribute the maximum amount of systematic risk explanation for the portfolio.
\end{proof}

To further support the Commonality Principle, it must be proved that the common drivers are the specific drivers for all constituents that exhibit the highest probability of causality for portfolio dynamics. This can be shown by proving that the maximum probability of causality for a portfolio, given any possible selection of drivers, is achieved by selecting drivers according to the Commonality Principle. Since causality cannot be guaranteed, the probability of causality is utilized in this analysis.

\begin{definition}[\textbf{Probability of Causality}]
Let $a_i$ be the return of asset $i$ at time $t+k$, and let $\boldsymbol{SD}_i = \{SD_{i1}, \dots, SD_{iM_i}\}_t$ be its set of specific drivers at time $t$. The \emph{probability of causality} for $a_i$ with respect to $\boldsymbol{SD}_i$ is:
\[
P(a_{i,t+k} \mid do(\boldsymbol{SD}_i)) > P(a_{i,t+k} \mid do(\sim \boldsymbol{SD}_i))
\]
This expresses that $\boldsymbol{SD}_i$ is a direct causal driver of $a_i$.
\end{definition}

\begin{definition}[\textbf{Common Drivers}]
Let $p = \{a_1, \dots, a_N\}$ be a portfolio of $N$ assets. A set $\boldsymbol{D}_t = \{D_1, \dots, D_M\}$ represents \emph{common drivers} if:
\[
P(p_{t+k} \mid do(\boldsymbol{D}_t)) > P(p_{t+k} \mid do(\sim \boldsymbol{D}_t))
\]
and if these drivers causally influence all $a_i \in p$ under the commonality principle.
\end{definition}

\begin{lemma}[Reichenbach Causal Screening for Portfolio Constituents]
Let $a_i$, $a_j$ be two correlated asset returns such that:
\[
P(a_i \cap a_j) > P(a_i)P(a_j)
\]
Then, a common cause $\boldsymbol{D}$ exists such that the following hold:
\begin{align}
P(a_i \cap a_j \mid \boldsymbol{D}) &= P(a_i \mid \boldsymbol{D}) P(a_j \mid \boldsymbol{D}) \\
P(a_i \cap a_j \mid \overline{\boldsymbol{D}}) &= P(a_i \mid \overline{\boldsymbol{D}}) P(a_j \mid \overline{\boldsymbol{D}}) \\
P(a_i \mid \boldsymbol{D}) &> P(a_i \mid \overline{\boldsymbol{D}}) \\
P(a_j \mid \boldsymbol{D}) &> P(a_j \mid \overline{\boldsymbol{D}})
\end{align}
\end{lemma}

\begin{proof}
Follows directly from Reichenbach’s Common Cause Principle \citep{Reichenbach1956-REITDO-2}, where $\boldsymbol{D}$ is a set of common drivers satisfying the four independent screening-off and influence conditions for each pair $(a_i, a_j)$.
\end{proof}

\begin{theorem}[Optimality of Common Drivers under the Commonality Principle]
Let $p = \{a_1, \dots, a_N\}$ be a portfolio, and let each $a_i$ have specific drivers $\boldsymbol{SD}_i$. Suppose there exists a set $\boldsymbol{D}$ such that:
\[
\forall i, \quad P(a_i \mid do(\boldsymbol{D})) > P(a_i \mid do(\sim \boldsymbol{D}))
\]
and $\boldsymbol{D} \equiv \boldsymbol{SD}_1 \equiv \dots \equiv \boldsymbol{SD}_N$.

Then, $\boldsymbol{D}$ is the set of portfolio-level drivers with:
\[
P(p_{t+k} \mid \boldsymbol{D}) > \prod_{i=1}^N P(a_{i,t+k})
\]
and satisfies the generalization of the Common Cause Principle for all asset pairs in $p$.
\end{theorem}

\begin{proof}
By Lemma 1, each pair $(a_i, a_j)$ satisfies the Reichenbach conditions given $\boldsymbol{D}$. Assuming causal sufficiency and equivalence of driver sets $\boldsymbol{SD}_i$, $\boldsymbol{D}$ screens off all correlations and explains the joint probability structure. Thus, $\boldsymbol{D}$ maximizes the joint causal effect over the portfolio.
\end{proof}

\begin{remark}[MCCP as a Structural Basis for RCCP in the Proof]
Although the previous theorem is stated in terms of Reichenbach’s Common Cause Principle (RCCP), the mathematical structure of the proof implicitly relies on assumptions from the Markov Common Cause Principle (MCCP). Specifically, we assume that all portfolio constituents $a_i$ are generated by a shared set of common drivers $\boldsymbol{D}$, such that:
\[
a_i \perp a_j \mid \boldsymbol{D} \quad \text{and} \quad P(a_1, \dots, a_N \mid \boldsymbol{D}) = \prod_{i=1}^N P(a_i \mid \boldsymbol{D}),
\]
which is the Causal Markov Condition applied to a star-shaped DAG where $\boldsymbol{D} \rightarrow a_i$ for all $i$.

From this assumption, we then derive:
\[
P(a_i \cap a_j \mid \boldsymbol{D}) = P(a_i \mid \boldsymbol{D}) P(a_j \mid \boldsymbol{D}),
\]
and similar inequalities for the absence of $\boldsymbol{D}$, along with monotonicity conditions (i.e., $P(a_i \mid \boldsymbol{D}) > P(a_i \mid \overline{\boldsymbol{D}})$). These are exactly the four conditions that define RCCP \citep{Reichenbach1956-REITDO-2}.

Therefore, the proof uses the structure and independence assumptions of MCCP to validate RCCP probabilistically. This shows that the structural model not only satisfies the Reichenbach conditions but also explains why they hold in a system governed by shared causal drivers under the commonality principle.
\end{remark}

\newpage

Finally, a third component is missing in the proof which has to do with the concept of optimal diversification, which needs to be connected to the prior two sections of the proof and optimal portfolio drivers selection as is shown in the next Section, taken from the original work \citep{RODRIGUEZDOMINGUEZ2023100447}. But first, optimal diversification is defined.

Idiosyncratic diversification refers to the diversification obtained through unconditional portfolio optimization methods—those that exclude exogenous information—such as in the classical Markowitz framework and Modern Portfolio Theory (MPT). This type of diversification can be either suboptimal or optimal, depending on the number and characteristics of the assets included in the portfolio.

Systematic diversification, on the other hand, arises from conditional portfolio optimization approaches, which incorporate exogenous information. Traditional factor models, as well as more advanced and robust methodologies \citep{10.1016/j.eswa.2021.116308, 4744738}, fall into this category.

\begin{lemma}[\textbf{Optimal Diversification under the Optimal Selection of Drivers of Portfolio Dynamics}~\citep{RODRIGUEZDOMINGUEZ2023100447}]
    Optimal diversification under the Commonality Principle is defined as the ability to preserve the level of idiosyncratic diversification achieved in the unconditional case while simultaneously incorporating optimal systematic diversification through the inclusion of common causal and persistent drivers.
\end{lemma}

This distinction is critical, as the term "investment efficiency" is often used without a formal definition. In the unconditional setting, optimal diversification is obtained by including a sufficient number of assets, as supported by established results on diversification limits. In the conditional setting, however, achieving true optimal diversification is only possible if the Commonality Principle is satisfied. Without it, the addition of systematic diversification via exogenous drivers does not guarantee optimality—since the drivers may not be causally responsible for the portfolio’s diversification dynamics.

Moreover, there exists a well-documented trade-off between idiosyncratic and systematic diversification: increasing exposure to systematic factors may reduce idiosyncratic diversification. This phenomenon is well-recognized in the literature on factor models from a rigorous mathematical perspective~\citep{10.2307/3533246}, as well as in research on factor investing and smart beta strategies~\citep{fama1970efficient,Fama1992}, with the foundational example being the Capital Asset Pricing Model (CAPM)~\citep{RePEc:bla:jfinan:v:19:y:1964:i:3:p:425-442}. Therefore, any claim of investment efficiency through factor-based methods that do not satisfy the Commonality Principle cannot be substantiated for portfolios with more than two assets. The only condition under which the trade-off can be circumvented—thus allowing the preservation of idiosyncratic diversification while adding systematic diversification—is when the selected drivers are both common across portfolio constituents and causal and persistent in nature.

It is in Lemma~1 and the subsequent proof that one finds a fundamental critique of investment efficiency claims based on traditional factor models, as commonly used in the factor investing literature~\citep{RePEc:bla:jfinan:v:19:y:1964:i:3:p:425-442,ross1976,Fama1992}. These approaches implicitly assume that investment efficiency can be achieved by accurately predicting asset returns. However, such predictive accuracy is unrealistic in practice and, more importantly, does not guarantee optimal diversification.

Achieving optimal diversification requires satisfying specific geometric properties that enable the portfolio optimization problem to be solved effectively. This distinction is central to the argument presented in the following lemma and its proof, which demonstrate that only under these geometric conditions—particularly those satisfied when the Commonality Principle holds—can true investment efficiency be attained.

\subsection{Proof of Lemma 1: Optimal Diversification under the Commonality Principle}
\label{sectionGeometry}

In \citep{10.2307/3533246}, the authors develop a representation theory for dynamic factor models. Before proceeding, let us clarify a potential source of confusion, particularly for some smart beta aficionados.

\begin{definition}[Factors vs. Drivers in terms of causing asset and diversification dynamics]
    In this work, a \emph{Factor} and a \emph{Driver} are considered equivalent. Both denote variables that cause or explain the dynamics of asset returns and portfolio diversification. The distinction between the two is not theoretical, but rather methodological, depending on whether the variable is directly observable (e.g., from public datasets) or constructed (e.g., a smart beta factor).
\end{definition}

From a theoretical standpoint, both terms are interchangeable within the context of optimality defined by the Commonality Principle. For the purpose of optimal diversification, what matters is whether these variables are responsible for the dynamics of asset returns and diversification. From a methodological perspective, one could argue about the preferable representation of these variables—whether using raw data from public or private sources, or transformations such as smart beta factors. However, such distinctions do not alter the core theoretical results presented here.
   
\begin{figure}[ht]
    \centering
    \begin{subfigure}[b]{0.45\textwidth}
        \includegraphics[width=\linewidth]{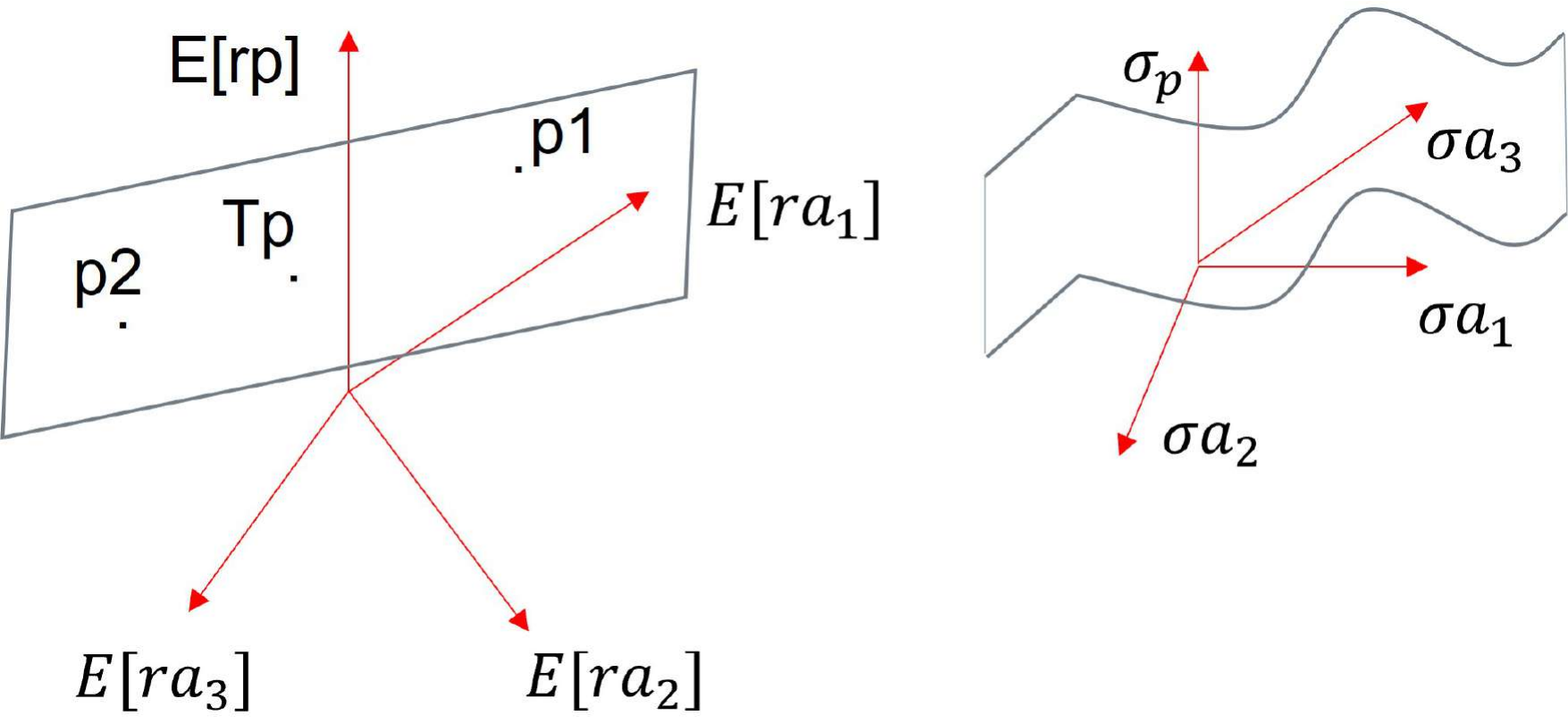}
        \caption{Unconditional case, Mean Variance, portfolio expected return hyperplane and risk hypersurface}
        \label{figure2}
    \end{subfigure}
    \hfill
    \begin{subfigure}[b]{0.45\textwidth}
        \includegraphics[width=\linewidth]{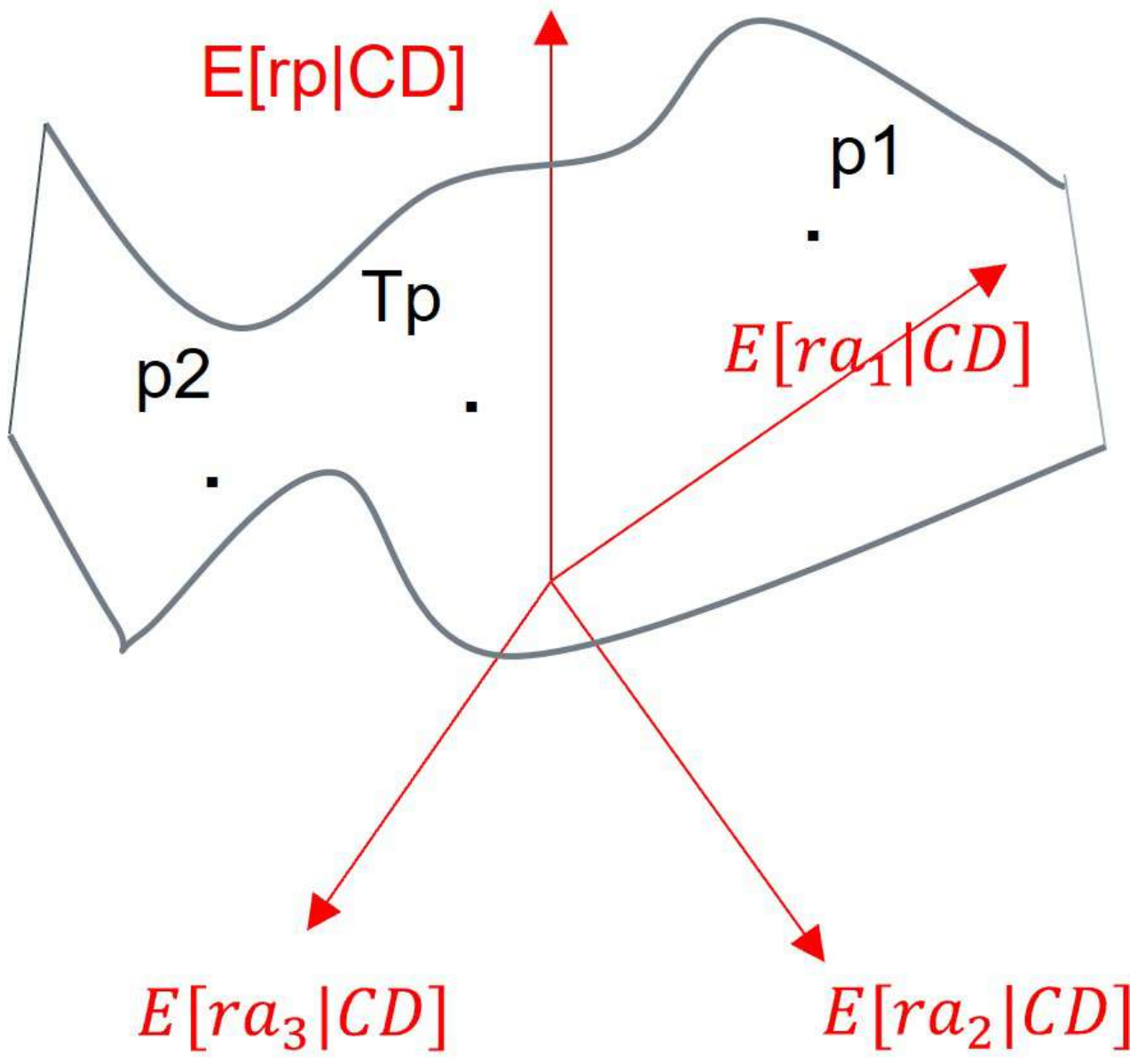}
        \caption{Conditional portfolio expected returns hypersurface}
        \label{figure4}
    \end{subfigure}
    \caption{Comparison between unconditional and conditional portfolio expected returns.}
    \label{fig:combined}
\end{figure}

\begin{figure}
    \centering
    \includegraphics[width=0.5\linewidth]{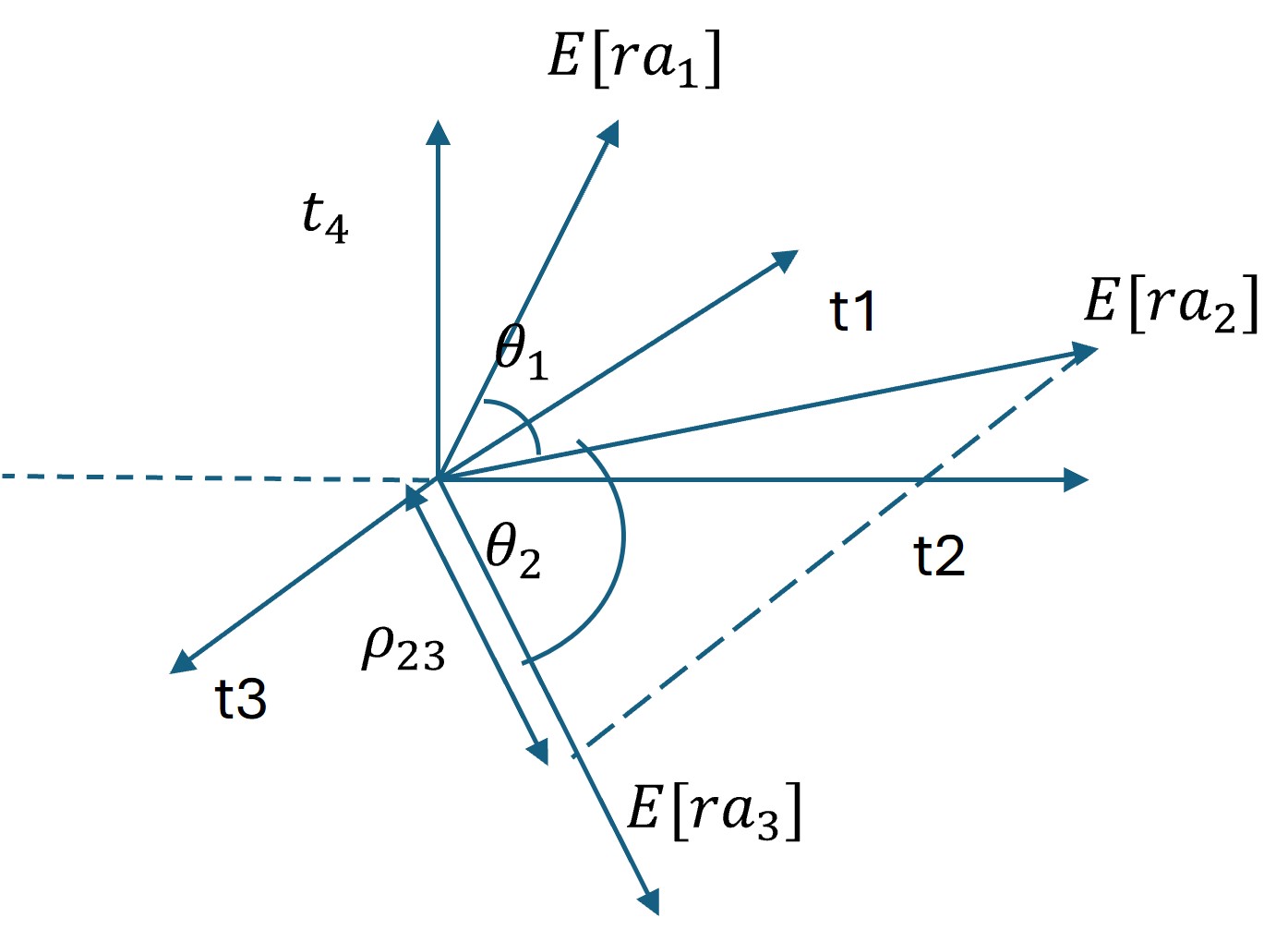}
    \caption{Mean-Variance Time Embedding}
    \label{fig:enter-Markowitztimeembedding}
\end{figure}

Optimal common drivers selected by the commonality principle can refer to common factors in \citep{10.2307/3533246}, and share the same canonical decomposition:
\begin{equation}
    x_{it}=\ proj(x_{it}|\mathcal{G}(x))+\delta_{it}\ 
\end{equation}
Implying $\chi_{it}\ \in\ \mathcal{G}(x)$ and $\xi _{it}\ \bot\ \mathcal{G}(x)$, so that $\chi_{it}\ =\ proj(x_{it}|\mathcal{G}(x))$ and $\xi_{it}\ =\ \delta_{it}$. With Common Drivers(CD), $\boldsymbol{CD}=\mathcal{G}(x)$, $\xi _{it}\ \bot\ \boldsymbol{CD}$, $\chi_{it}\ =\ proj(x_{it}|\boldsymbol{CD}) = E[x_{it}|\boldsymbol{CD}]$. 
It is shown now with geometry how the canonical decomposition with common drivers preserves the idiosyncratic risk representation in the embedded spaces of sensitivities (or betas) while adding systematic risk representation. In the unconditional case, for the mean-variance framework from MPT, the portfolio’s expected returns lie on a hyperplane of the constituents' expected returns, and portfolio risk lies on a hypersurface, as seen in Figure \ref{figure2}. The hyperplane is given by: 
\begin{equation}
E\left[r_p\right]=\sum_{i=1}^{n}w_iE[r_{a_i}]   
\label{equationww}
\end{equation} 

where, $E\left[r_p\right]$ are the portfolio’s expected returns, Tp is the Tangency Portfolio, $E\left[r_{a_i}\right]=\mu_i$ are the constituents' expected returns, as in Figure \ref{figure2}. $E\left[r_p\right]$ is linear in $E[r_{a_i}]$, portfolio risk $\sigma_p$ is non-linear in constituents' risk ${\sigma_{a_i}}$, and w are the weights, solution to the quadratic optimization in: 
\begin{equation}
\label{equation27}
w = \min_w{w^{T}\mathrm{\Sigma}\ w}
\end{equation}
with the tangency portfolio as the optimal solution. In Figure \ref{fig:enter-Markowitztimeembedding}, it is shown the representation of portfolio constituents' expected returns for a period, in the time-dimensional space or what is called a time-embedding space. Axis are points in time, $\theta$ are angles between expected returns, and the cosines are the correlations, ie, $\rho_{23}=\cos{\theta_2}$ is the correlation between $r_{a_2}$ and $r_{a_3}$ and $\theta_2=\hat{E[r_{a_2}]E[r_{a_3}]}$.
The expected portfolio returns conditional on the common drivers is a linear combination of constituents’ expected returns conditional on the same drivers: 
\begin{equation}
E\left[r_p|\boldsymbol{CD}\right]=\sum_{i=1}^{n}w_iE[r_{a_i}|\boldsymbol{CD}]    
\end{equation}
is a hypersurface. Figure~\ref{figure4} illustrates the nonlinear approximation case—such as one modeled using neural networks—of the conditional expectations with the hypersurface or manifold. The tangency portfolio corresponds to the optimal solution obtained through the portfolio optimization procedure.

\begin{figure}
    \centering
    \includegraphics[width=1\linewidth]{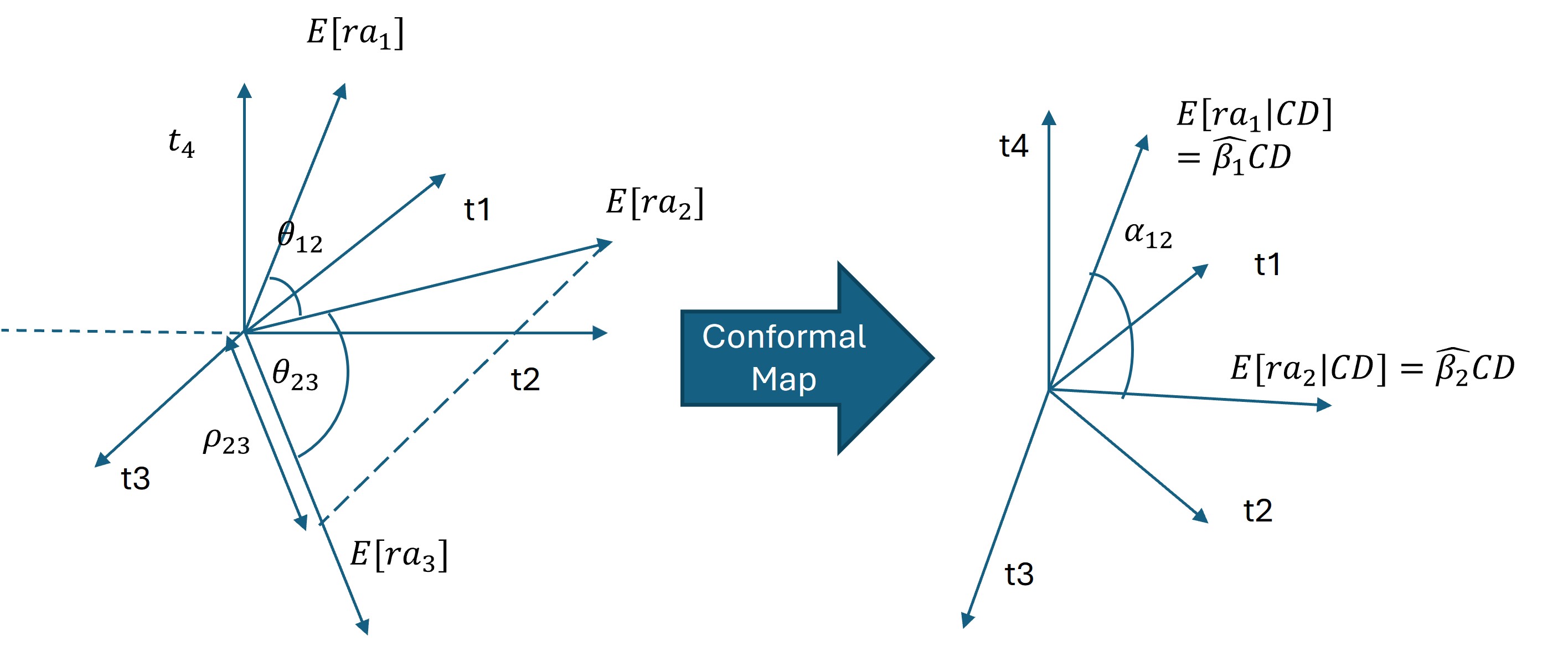}
    \caption{Conformal map between the unconditional Mean-Variance (MV) framework and the conditional framework. Time embeddings preserve proportional angles for all $\theta$ and $\alpha$.}
    \label{fig:enter-1stconformalmap.jpg}
\end{figure}

In Figure~\ref{fig:enter-1stconformalmap.jpg} (right graph), the expected returns of portfolio constituents, conditional on common drivers, are shown embedded in a time-dimensional space where timestamps are the axes. As demonstrated in \citep{RODRIGUEZDOMINGUEZ2023100447}, the key step in the proof is to establish the existence of a conformal map between this time-embedding space (for the conditional case with common drivers) and the space defined by the sensitivities of the portfolio constituents with respect to those drivers, which can be seen in Figure \ref{fig:enter-SecondConformalMap}.

A conformal map preserves angles between vectors in both spaces, which is crucial because the cosines of those angles encode relative projective heading information. In the unconditional case, this is directly represented by the cosine being equal to the correlation coefficient (see Figure~\ref{fig:enter-Markowitztimeembedding}). In the conditional case, the time-embedding space reflects the correlation of conditional expected returns (also illustrated in Figure~\ref{fig:enter-1stconformalmap.jpg}, right graph). 

In the sensitivity space (or beta embedding space) from Figure \ref{fig:enter-SecondConformalMap}, the structure depends on the choice of metric, but what remains certain is that the proportions encoded in the projective structure are preserved when mapping from the conditional space to the sensitivity space. For any given time $t$, the variation across these spaces is solely due to the change in embedding—driven by sensitivity (beta) information or by the influence of common drivers—while the relative angular and cosine relationships (from the unconditional to the conditional, and from the conditional to the sensitivity space) are maintained.

As shown in \citep{RODRIGUEZDOMINGUEZ2023100447}, the conformal map between the conditional embedding and the unconditional embedding becomes trivial once the conformal relationship between the conditional and sensitivity spaces has been established. For completeness, both mappings are demonstrated in this chapter.

In the conditional expectations with respect to common drivers time embedded space, angles between conditional expected returns are a sum of two components, a systematic component, and an idiosyncratic component from the unconditional expectation case from MPT: 
\begin{equation}
\alpha_{ij}={\gamma_1\theta}_{ij}+\gamma_2\omega_{ij}
\label{angles}
\end{equation}
where ${\theta}_{ij}$ represent the angles of the unconditional time-embedding case in Figure \ref{fig:enter-1stconformalmap.jpg} left plot before the conformal map and $\alpha_{ij}$ are the angles of the right plot for the conditional case, after the conformal map. $\cos{\theta_{ij}}$\ is $\rho_{ij}$ from $\mathrm{\Sigma}$\ \ in (\ref{equationww}) and are the angles of the unconditional time-embedding case in Figure \ref{fig:enter-1stconformalmap.jpg} left graph, with ${\ \theta}_{ij}$ the idiosyncratic and $\omega_{ij}$  the systematic component in (\ref{angles}). It is clear that the systematic component has been added after the conformal map and that it will be proved next. 

For diversification optimality, meaning that maximum idiosyncratic diversification is achieved while incorporating systematic diversification, the map between the time embedding of the unconditional Markowitz case in Figure \ref{fig:enter-Markowitztimeembedding} and the sensitivity space embedding (Figure \ref{fig:enter-SecondConformalMap} right subfigure) must be conformal, i.e., preserve proportional angles. We demonstrate that this conformality is only possible when the drivers are common and causal—like is was proven in \citep{RODRIGUEZDOMINGUEZ2023100447}.

\subsubsection{Justification for Using the Sensitivity Space in Causal Frameworks When Seeking Optimality}

\begin{figure}
    \centering
    \includegraphics[width=1\linewidth]{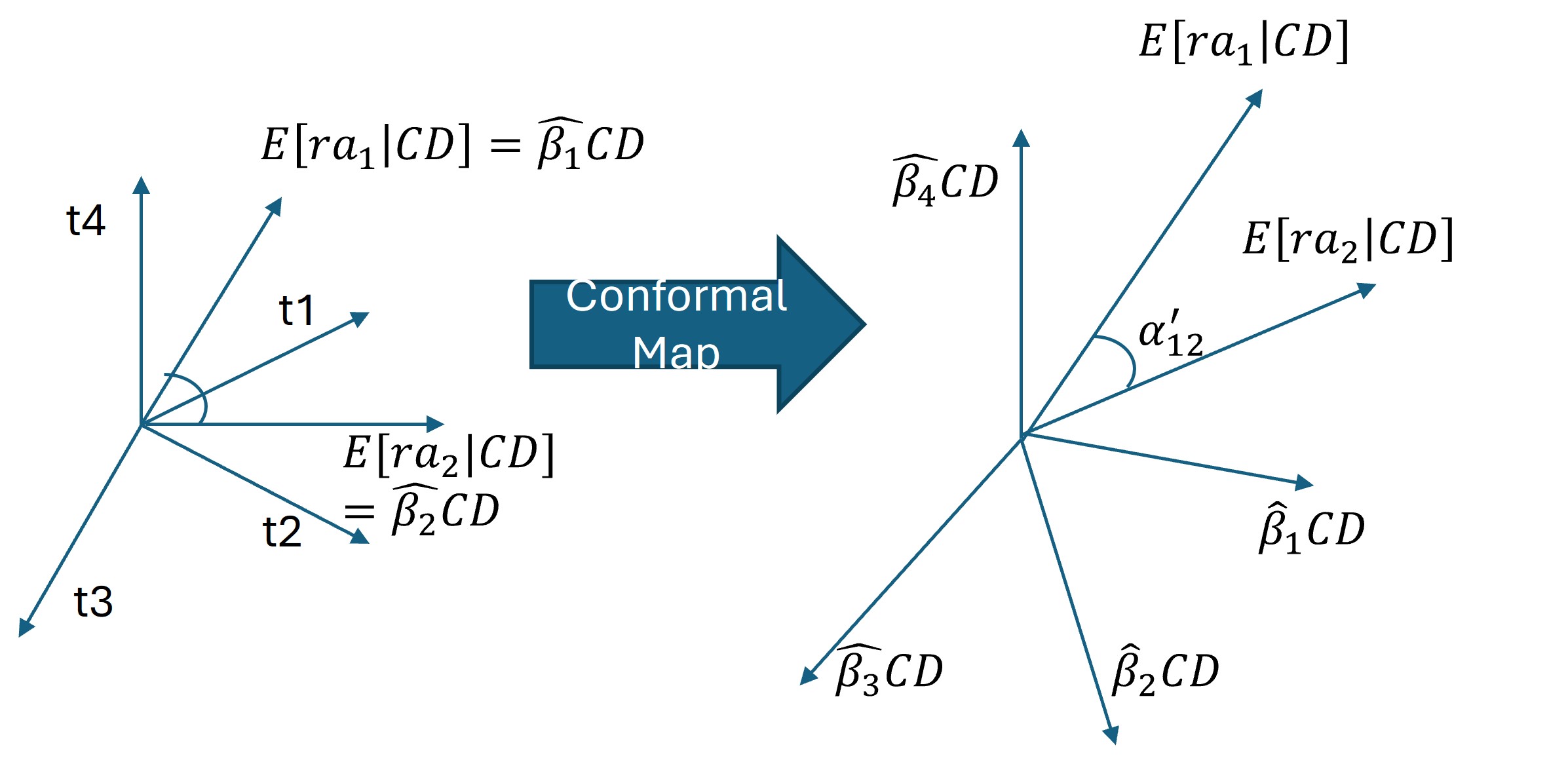}
    \caption{Conformal map between the time embedding framework and the sensitivity space: $\forall \alpha \text{ and } \alpha'$}
    \label{fig:enter-SecondConformalMap}
\end{figure}

It is important to clarify why a second conformal map—leading to the sensitivity space—is necessary (Figure \ref{fig:enter-SecondConformalMap}), rather than stopping at the initial time-embedded conditional space composed of common drivers. Given that the embedding space is already formed by causal and persistent drivers, and the conformality condition has been satisfied, one might question the need for further mapping.

However, although optimization could be performed directly in the conditional embedding space using a factor model, this approach is inherently suboptimal compared to operating in the sensitivity space. The reason is twofold: first, using correlations or covariances in the conditional space undermines the causal structure that has been established; second, regardless of the optimization methodology, sensitivities derived from common causal drivers inherently contain more information about the trajectory of asset dynamics. 

In particular, while conditional expectations reflect average behavior, sensitivities capture the local directional responses, offering a richer geometric and dynamic representation. Therefore, even though the use of the conditional space is technically valid, optimality—in the sense of preserving both idiosyncratic and systematic diversification as defined in the framework—is only achieved within the causal sensitivity space. The choice ultimately lies with the practitioner, but the theoretical justification strongly favors the use of the sensitivity-based embedding.

What’s the point of finding the best common causal drivers for a portfolio if the method used to capture the portfolio’s behavior is poorly chosen? Causal dynamics in financial markets are often nonlinear and complex. Identifying the correct drivers is only the first step—what really matters next is how those dynamics are modeled. If, after selecting the right drivers, one uses simple linear models or correlation-based methods to track how the portfolio reacts, a significant amount of valuable information is lost. In fact, this second step can introduce more error than the initial selection phase.

To truly benefit from causal modeling, both steps must be treated with the same level of care: the drivers must be optimally selected, and the portfolio’s behavior with respect to those drivers must be captured in a way that reflects its true dynamics. This is why sensitivity analysis—how much assets respond to changes in drivers—is crucial, and why partial differential equations (PDEs) and neural networks outperform linear models or standard machine learning tools in this context.

For the Proof, two conformal maps are needed. First the one between the unconditional case and the conditional case, which is proved in the next slide. Then, you need a second conformal map between the conditional time embedding case and the sensitivity embedding. Which has the same proof as the previous map.

\subsubsection{Sensitivity Space for Optimal Diversification Based on Drivers of Portfolio Dynamics}

\begin{proof}
To prove that the map is conformal, the construction can be divided into two sequential conformal mappings. These mappings connect the unconditional and conditional time-embedding spaces to the sensitivity space. For the overall transformation to preserve angles (i.e., to be conformal), the intermediate mappings must each satisfy the properties of conformality. In particular, this requires the sensitivity space to retain the geometric relationships—such as inner products and angle structures—found in the unconditional and conditional embedding representations. This layered structure enables a rigorous link between classical mean-variance optimization and dynamic, causality-driven diversification.

\textbf{First conformal map}

Lets assume n assets with returns $ra_i$, $i=1\dots,n$, and m drivers candidates $CD_p$, $p=1,\dots,m$.

\paragraph{1)} If CD (causal drivers) are causal, then each pair of conditional expectations (in Figure \ref{fig:enter-1stconformalmap.jpg}, right plot)  is a function of time and of the specific driver that is the causal candidate for each constituent at that point in time:

\begin{equation}
    \begin{split}
    \alpha_{ij} = f\left(t, \mathbb{E}[ra_i \mid CD_p], \mathbb{E}[ra_j \mid CD_q]\right) = g(t, CD_k),\\
    \quad i \ne j,\quad i,j= 1, \dots, n\quad p \ne q,\quad p,q,k= 1, \dots, m    
    \end{split}
\end{equation}

\paragraph{2)} However, to represent these relations in a sensitivity space in which the canonical basis is meet as stated by projective spaces and algebra, the drivers must be common. That is:

\[
CD_p = CD_q
\]

If the causal drivers are common and truly causal, the pair of conditional expectations becomes a function of time alone:

\[
\alpha_{ij} = f\left(t, \mathbb{E}[ra_i \mid \boldsymbol{CD}], \mathbb{E}[ra_j \mid \boldsymbol{CD}]\right) = g(t), \quad i \ne j,\quad i,j = 1, \dots, n
\]

Hence, the ratio between the unconditional and conditional cases is time-proportional $\forall t$, and the transformation map is conformal:

\[
\frac{f(t)}{g(t)} = \frac{\theta_{ij}}{\alpha_{ij}} = \text{constant}
\]

If they are not common, the embedding sensitivity space does not work mathematically, and if they are not causal, they are neither a function of time nor a constant ratio, as driver candidates will change for each particular time $t$. Same reasoning applies in the second conformal map to the sensitivity space, the sensitivity functions will change with the candidate change. Hence, drivers must be common and causal, making the sensitivities causal.

\textbf{Second conformal map}

\[
\alpha_{ij}^\prime = f\left(\boldsymbol{\beta}, \mathbb{E}[ra_i \mid \boldsymbol{CD}], \mathbb{E}[ra_j \mid \boldsymbol{CD}]\right) = g(t;\boldsymbol{\beta},\mathbb{E}[\boldsymbol{ra} \mid \boldsymbol{CD}), \quad i \ne j,\quad i,j = 1, \dots, n
\]

but firstly, the drivers $\boldsymbol{CD}$ must be common as stated by the rules of projective spaces. Also, sensitivities $\boldsymbol{\hat{\beta}}$ are a function of time $t$ and the conditional expectation of asset return $ra_i$ with respect to the common drivers, $\mathbb{E}[ra_i \mid \boldsymbol{CD}]$:
\[
\boldsymbol{\hat{\beta}}_i = \left[
\frac{\partial \mathbb{E}[ra_i \mid \boldsymbol{CD}]}{\partial CD_1},
\frac{\partial \mathbb{E}[ra_i \mid \boldsymbol{CD}]}{\partial CD_2},
\dots,
\frac{\partial \mathbb{E}[ra_i \mid \boldsymbol{CD}]}{\partial CD_m}
\right]
\]

For $i = 1, \dots, n$, and based on the first conformal map, the conditional expectation $\mathbb{E}[ra_i \mid \boldsymbol{CD}]$ is a function of time alone if and only if the drivers are causal. Given this, and the requirement that the drivers must be common to satisfy the projective space conditions in the sensitivity space, it follows that $\boldsymbol{CD}$ must be both common and causal, as stated by the Commonality Principle and established in the proof of probabilistic causality.

To conclude, $\alpha_{ij}^\prime$ is a function of time alone for all $i, j = 1, \dots, n$, and together with $\alpha_{ij}$, their ratio remains constant over time. This condition guarantees the conformality of the map.

\[
 \frac{\alpha_{ij}}{\alpha_{ij}^\prime} = \text{constant}
 \]

\end{proof} 

This proof is intentionally presented in a shorter and less mathematically intensive form to enhance accessibility for a broader audience. For readers seeking the complete formal derivation and more rigorous treatment of the result, reference is made to the original open-access article \citep{RODRIGUEZDOMINGUEZ2023100447}.

\subsection{Why Causal Factor Investing May Not Achieve Investment Efficiency in Terms of Diversification}

\begin{theorem}[Necessary and Sufficient Conditions for Causal Factors to Enable Investment Efficiency]
A necessary and sufficient condition for achieving investment efficiency is that the factors are causal, common, and that their dynamics are accurately captured across all time points $t$.
\end{theorem}

\begin{proof}
Based on the proofs presented in this monograph—namely, the Proof of Persistence, the Proof of Probabilistic Causality, and the Proof of Optimality in Diversification—as well as those in \citep{RODRIGUEZDOMINGUEZ2023100447}, the following conclusions hold:

\begin{itemize}
    \item Any causal factor approach claiming to enable optimal diversification in portfolio optimization must ensure that the factors used are both common and causal, as established by the Commonality Principle.
    
    \item Any method that claims to attain investment efficiency must clearly define its meaning. If investment efficiency refers to the notion developed in this monograph and in \citep{RODRIGUEZDOMINGUEZ2023100447}, then it is mathematically impossible to achieve without satisfying the Commonality Principle.
    
    \item Some claims in the literature suggesting that investment efficiency has been attained are therefore mathematically inconsistent, unless they explicitly meet the above conditions. Such assertions may mislead practitioners and researchers.
    
    \item In practice, causal factor models that do not adhere to the Commonality Principle may reduce systematic risk but inevitably increase idiosyncratic risk. Due to the geometric properties of projective spaces and causality proven herein and in \citep{RODRIGUEZDOMINGUEZ2023100447}, these methods are unable to maintain the conformal structure required to preserve idiosyncratic diversification from the unconditional case unless the factors are common and causal.
    
    \item Sensitivities derived from causal drivers contain additional information regarding the trajectory of diversification as has been proved to be connected via conformal map with the unconditional case. Neglecting this information, as is often the case in standard causal factor models \citep{Lopez_de_Prado_2023,lopez2024case}, renders such models suboptimal solutions.
    
    \item Finally, when models are incapable of adequately representing causal mechanisms—due to, for example, their inability to handle nonlinearity or dynamic complexity—any hypothesis tested using them will be subject to model error. This limitation applies broadly to many frameworks found in the existing literature.
\end{itemize}
\end{proof}

For instance, examples of methods that rely on suboptimal assumptions, frameworks, or models while claiming optimal solutions can be found in \citep{Lopez_de_Prado_2023,lopez2024case} and the reader is advised to take carefully consideration of this aspects in practical applications.

\newpage

\section{Methodology}
\subsection{Summary of the Portfolio Optimization Methodology}

\subsection{Unified Methodological Summary}

The methods presented throughout this monograph are modular implementations of a single coherent framework introduced in the previous section. This framework is designed around the principle of \textbf{optimal diversification based on causal drivers of portfolio dynamics}. The flexibility of the framework allows for various methodological choices at each stage, depending on the modeling assumptions, data availability, and computational constraints.

\vspace{1em}
\noindent The full procedure consists of the following four core modules:

\begin{enumerate}
    \item \textbf{Driver Selection Methods} \\
    These methods are described in detail in Chapter~\ref{Subsection41}:
    \begin{itemize}
        \item \textbf{RCCP Reverse Engineering}: Ensures that the correlation structure satisfies the Reichenbach screen-off condition, formulated inversely.
        \item \textbf{Bayesian Networks}: Learns probabilistic graphical models to infer causal dependencies between drivers and constituents.
        \item \textbf{Maximum Likelihood Correlation-Based Method (ML-Corr)}: Ranks drivers based on their joint likelihood contributions to observed asset behavior.
    \end{itemize}

    \item \textbf{Predictive Modeling of Dynamics} \\
    These models approximate portfolio constituent dynamics using data and common drivers. Causal sensitivities are then extracted from these models to support diversification decisions. The methods are discussed in Chapter~\ref{subsection421}:
    \begin{itemize}
        \item \textbf{Feed-forward Neural Networks (FNNs)}: Nonlinear function approximators used to model complex dynamics between assets and drivers.
        \item \textbf{Linear Regression Models}: Serve as benchmarks or proxies for classical factor models.
    \end{itemize}

    \item \textbf{Sensitivity Extraction Methods}
    \begin{itemize}
        \item \textbf{Automatic Adjoint Differentiation (AAD)}: Used with neural networks to compute sensitivities of asset returns with respect to driver inputs~\citep{huge2020differential}.
        \item \textbf{Analytical Derivatives}: In linear models, sensitivities are equivalent to regression coefficients (betas)~\citep{pizarroso2021neuralsens}.
        \item \textbf{Statistical Calibration and Modeling}: Stochastic differential equations (SDEs) are fitted to sensitivity estimates and calibrated to simulate future trajectories of the sensitivity matrix and diversification dynamics. Full details are provided in Chapter~\ref{HSP_sens_fut}.
    \end{itemize}

    \item \textbf{Risk Mapping-Based Portfolio Optimization Strategies} \\
    This module maps information from the sensitivity space to the risk measure of interest. Detailed implementations are discussed in Chapter~\ref{RiskMap}:
    \begin{itemize}
        \item \textbf{Hierarchical Methods}: Hierarchical clustering of assets using sensitivity distances, followed by numerical optimization procedures from the literature~\citep{Prado2016BuildingDP,raffinot2018hierarchical,cotton2024schurcomplementaryallocationunification}. When applied to the sensitivity distance matrix using HRP-type solutions, the approach is referred to as \textit{Hierarchical Sensitivity Parity (HSP)}, originally introduced in~\citep{RODRIGUEZDOMINGUEZ2023100447} and revisited in Chapter~\ref{HSP_Section}.
        \item \textbf{Mean-Variance Optimization (MVO)}: Includes traditional approaches such as Maximum Sharpe Ratio, Minimum Volatility, Quadratic Utility, and Target Return strategies.
        \item \textbf{CVaR Optimization}: Uses the Rockafellar-Uryasev formulation to cast Conditional Value-at-Risk minimization as a linear programming problem~\citep{ROCKAFELLAR20021443}.
        \item \textbf{Hierarchical Methods with Simulated Sensitivity Paths}: An extension of the hierarchical approach, using simulated future sensitivity paths to account for diversification trajectory over time. Full details are provided in Chapter~\ref{HSP_sens_fut}.
    \end{itemize}
\end{enumerate}

\noindent Each module contributes to the end goal of \textit{optimal portfolio diversification informed by causal dynamics}, and the interplay between them defines multiple variants of the same theoretical framework. While different combinations may be selected based on performance or data constraints, all configurations maintain the core structure: \emph{causal driver identification} $\rightarrow$ \emph{dynamic modeling} $\rightarrow$ \emph{sensitivity embedding} $\rightarrow$ \emph{risk-mapping based optimization}.

This modular approach not only allows for interpretability and benchmarking against classical approaches, but also ensures generality, extensibility, and practical applicability to real-world investment workflows.

\section{Optimal Portfolio Common-Cause Drivers Identification}
\label{Subsection41}

The concept of Statistical Common Cause Systems (SCCS) was first introduced as a formalization of Hans Reichenbach’s original concept of a common cause (\citep{Reichenbach1956-REITDO-2}). It defined a system comprising multiple events that collectively explain correlations between two phenomena. This generalization provided a robust framework for analyzing probabilistic dependencies and has since influenced subsequent research on probabilistic causality and its applications in diverse fields (\citep{Hofer-Szabo2004-HOFRCC}).

\subsection{\textbf{Reichembach Common Cause Principle (RCCP)}}

1. \textbf{Correlation Between \( A \) and \( B \):}
\[
P(A \cap B) \neq P(A) \cdot P(B)
\]
This means \( A \) and \( B \) are statistically correlated.

2. \textbf{Existence of a Common Cause \( C \):}
There is a third event \( C \) that influences both \( A \) and \( B \), accounting for their correlation.

3. \textbf{Screening Off by \( C \):}
\[
P(A \cap B \mid C) = P(A \mid C) \cdot P(B \mid C)
\]
Once \( C \) is known, \( A \) and \( B \) become independent.

4. \textbf{Independence of \( C \):}
\[
P(A \mid C) \neq P(A \mid \neg C) \quad \text{and} \quad P(B \mid C) \neq P(B \mid \neg C)
\]
The probabilities of \( A \) and \( B \) must depend on \( C \), ensuring \( C \) genuinely explains the correlation. These conditions ensure that the correlation between \( A \) and \( B \) has a non-coincidental explanation rooted in the common cause \( C \) (\citep{Reichenbach1956-REITDO-2}).

\begin{figure}[h!]
\centering
\begin{tikzpicture}[
    node distance=2.5cm and 3cm,
    every node/.style={circle, draw=black, thick, minimum size=1.2cm},
    every path/.style={->, thick}
]

\node (C)  {\(C\)};
\node (A) [below left=of C] {\(A\)};
\node (B) [below right=of C] {\(B\)};

\draw (C) -- (A);
\draw (C) -- (B);

\end{tikzpicture}
\caption{Causal graph representing Reichenbach's Common Cause Principle: \( C \) is a common cause of \( A \) and \( B \), accounting for their correlation.}
\label{fig:reichembach_dag}
\end{figure}
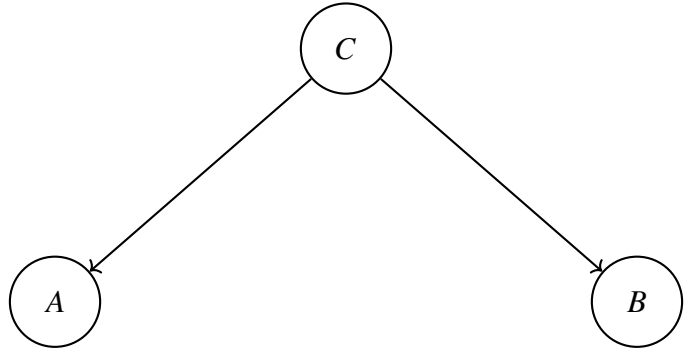

\begin{definition}[Screening Off]
Given a probability space $(\Omega, \mathcal{F}, P)$, let $A, B \in \mathcal{F}$.  
We say that an event $C$ is a \textit{screening off} event for the pair $\{A, B\}$ if:
\[
P(A \cap B \mid C) = P(A \mid C) P(B \mid C)
\]
In cases where $A$ and $B$ are correlated, we also say that $C$ \textit{screens off} the correlation.
\end{definition}

\subsection{$\varepsilon$-Common Cause Systems}

Consider $Y_1,\ldots,Y_n$ as a set of constituents in a portfolio. Suppose that every pair $Y_i,Y_j$ belongs to some market $\mathcal{F}_r$, for $r=1,\ldots,m$. A partition of unity of $\mathcal{F}_k$ is said to be an $\epsilon$-SCCS for $Y_i$ and $Y_j$ if it satisfies the statistical relevance condition with respect to $Y_i$ and $Y_j$, all its members are different from $Y_i$ and $Y_j$, and each portfolio driver $X_k$, $k=1,\dots,M$, and $M<<m$, fulfills (\citep{Hofer-Szabo2004-HOFRCC}):
\[
P(Y_i Y_j \mid X_k) = P(Y_i \mid X_k) P(Y_j \mid X_k) \leq \epsilon.
\]
The cardinality of the partition is called the size of the $\epsilon$-SCCS. For any pair of constituents \(Y_i\) and \(Y_j\), the conditional independence rule with \(\varepsilon\)-threshold is defined as:
\[
P(Y_i Y_j \mid X_k) = P(Y_i \mid X_k) P(Y_j \mid X_k), \quad \text{and } P(Y_i \mid X_k) P(Y_j \mid X_k) \leq \varepsilon,
\]
for all drivers \(X_k\) in their respective partitions $\mathcal{F}_1,\dots,\mathcal{F}_M$.

This ensures:
\begin{itemize}
    \item \(Y_i\) and \(Y_j\) are conditionally independent given \(X_k\), and
    \item Their joint probability under \(X_k\) is bounded by \(\varepsilon > 0\), limiting their dependence.
\end{itemize}










\subsection{Commonality Principle for the Optimal Selection of Portfolio Drivers}

Given:
\begin{itemize}
    \item A large set of potential portfolio drivers \( X = \{X_1, X_2, \ldots, X_m\} \) with \( m \gg M \),
    \item A subset \( S = \{X_{i_1}, X_{i_2}, \ldots, X_{i_M}\} \subseteq X \) with size \( |S| = M \),
    \item \( n \) random variables \( Y_1, Y_2, \ldots, Y_n \), representing assets or outcomes in a portfolio.
\end{itemize}

The goal is to select \( S \) such that the total deviation of the joint probability \( P(Y_i Y_j \mid S) \) from the product form \( P(Y_i \mid S) P(Y_j \mid S) \) is minimized across all pairs \( (Y_i, Y_j) \). The deviation is controlled by the threshold \( M \varepsilon \), ensuring approximate independence under the conditioning events in \( S \). Expressing the optimization problem as:
\begin{equation}
 \text{minimize:} \quad \mathcal{G} = \left| \sum_{i=1}^{n} \sum_{\substack{j=1 \\ j \neq i}}^{n} \left( P(Y_i Y_j \mid S) - P(Y_i \mid S) P(Y_j \mid S) \right) \right|,   
 \label{eq13}
\end{equation}

subject to:
\[
\mathcal{G} \leq M \varepsilon,
\]
where:
\begin{itemize}
    \item \( S = \{X_{i_1}, X_{i_2}, \ldots, X_{i_M}\} \subseteq X \),
    \item \( |S| = M \),
    \item \( P(Y_i Y_j \mid S) = P(Y_i \mid S) P(Y_j \mid S) \) holds approximately for all pairs \((i, j)\),
    \item \( M \) ranges from 1 to a maximum value \( m \).
\end{itemize}

\begin{enumerate}
    \item Approximate independence is measured by the term \( P(Y_i Y_j \mid S) - P(Y_i \mid S) P(Y_j \mid S) \). Minimizing this sum ensures that, under \( S \), the random variables \( Y_i \) and \( Y_j \) are nearly independent.
    \item The threshold \( M \varepsilon \) controls the total deviation across all pairs and scales proportionally with the size of \( S \).
    \item Select \( M \) portfolio drivers from the market of candidates \( X \) (with \( m \gg M \)) to achieve approximate independence among portfolio variables. \( S \) contains the portfolio drivers that satisfy these properties optimally.
\end{enumerate}

Generalize the objective function as:
\begin{equation}
\text{minimize:} \quad \mathcal{G}(S) = \sum_{i=1}^{n} \sum_{\substack{j=1 \\ j \neq i}}^{n} \left| P(Y_i Y_j \mid S) - P(Y_i \mid S) P(Y_j \mid S) \right|,
\label{eq14}
\end{equation}
subject to:
\[
\mathcal{G}(S) \leq M \varepsilon,
\]
where:
\begin{itemize}
    \item \( |S| = M \),
    \item \( P(Y_i Y_j \mid S) = P(Y_i \mid S) P(Y_j \mid S) \) holds approximately for all pairs \((i, j)\).
\end{itemize}

Allow \( M \) to range from \( 1 \) to \( m \), scaling the selection process for varying subset sizes. Select \( S \) such that the total joint deviation is minimized while adhering to the independence approximation. This formulation applies to selecting a small set of representative drivers from a market of millions (e.g., \( X \)) to construct a portfolio where the joint behavior of assets \( Y_1, \ldots, Y_n \) is approximately independent under the chosen conditioning events. The threshold \( M \varepsilon \) ensures that the total dependency is controlled, stabilizing and simplifying the portfolio.

\begin{figure}
    \centering
    \includegraphics[width=1\linewidth]{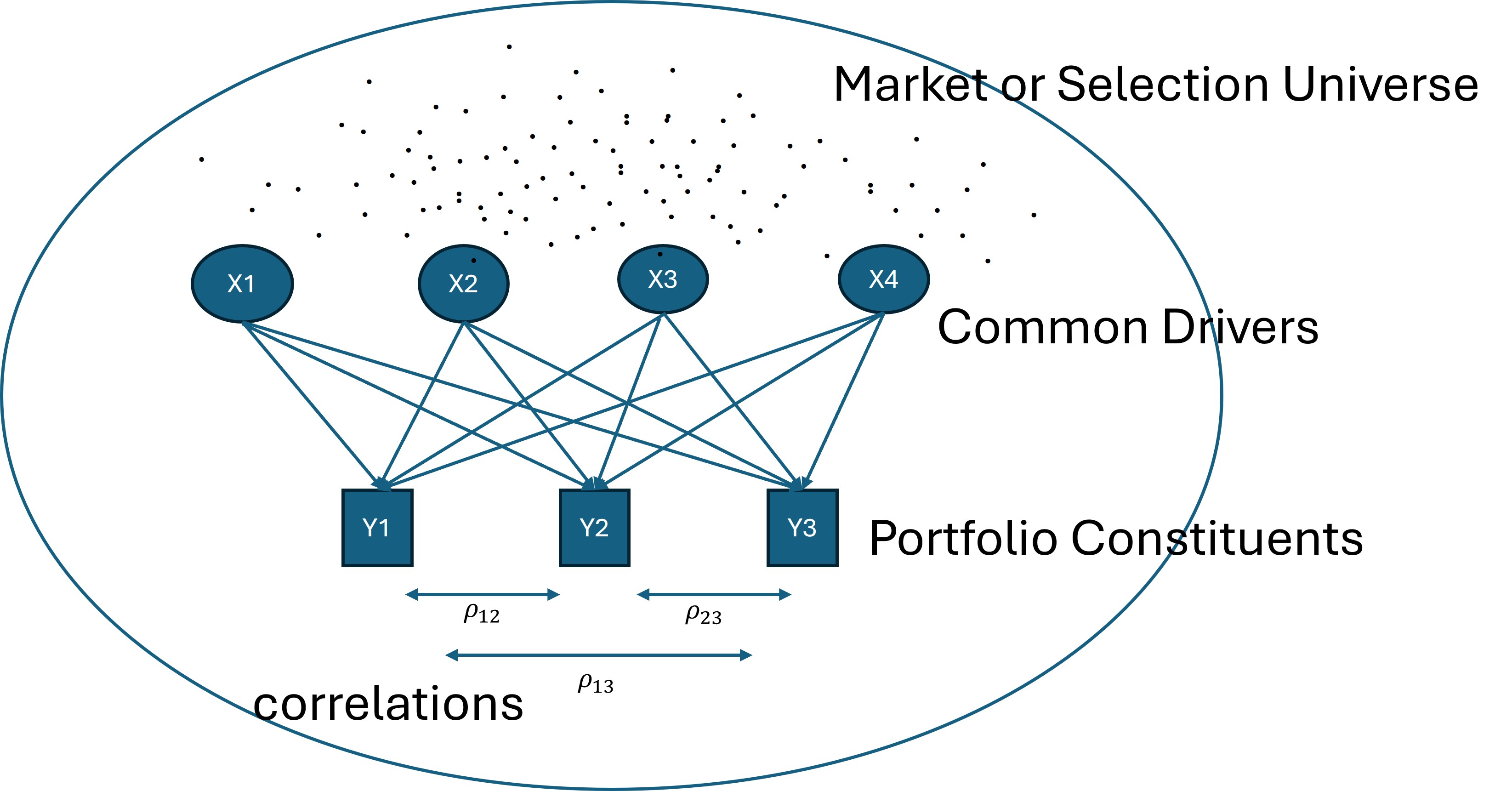}
\caption{Implementation of Common Causal Combinatorial Search Algorithms for a portfolio of 3 constituents and 4 selected Common Causal Drivers, based on the hyperparameter, from $m$ candidates in the dataset as dots.}

    \label{fig:enter-label}
\end{figure}

\newpage

\begin{remark}[\textbf{Confounder and Collider Bias Avoidance via RCCP-Based Optimization}]
Consider the objective function:
\[
\mathcal{G}(S) = \sum_{i=1}^{n} \sum_{j \neq i} \left| P(Y_i Y_j \mid S) - P(Y_i \mid S) P(Y_j \mid S) \right|
\]
where $S$ is a subset of candidate drivers $X = \{X_1, \dots, X_m\}$. Minimizing $\mathcal{G}(S)$ seeks to find a set of variables such that the joint behavior of the outcome variables $Y_1, \dots, Y_n$ is approximately conditionally independent.

\textbf{Why this avoids confounder bias:} If there exists a latent common cause $Z$ that affects both $Y_i$ and $Y_j$, then omitting $Z$ will cause:
\[
P(Y_i Y_j \mid S) \neq P(Y_i \mid S) P(Y_j \mid S)
\]
unless $Z \in S$. Therefore, to minimize $\mathcal{G}(S)$, any omitted confounder must be included in $S$, or the objective function remains large. This enforces the Reichenbach screening-off condition:
\[
P(Y_i \cap Y_j \mid Z) = P(Y_i \mid Z) P(Y_j \mid Z)
\]

\textbf{Why this avoids collider bias:} Suppose $S$ includes a collider $C$ such that $Y_i \rightarrow C \leftarrow Y_j$. Then conditioning on $C$ introduces a dependence:
\[
P(Y_i Y_j \mid C) \neq P(Y_i \mid C) P(Y_j \mid C)
\]
which violates RCCP. Thus, including a collider in $S$ will \emph{increase} $\mathcal{G}(S)$, making such a selection suboptimal.

 The optimization process inherently favors the inclusion of common causes and discourages the inclusion of colliders, as both influence $\mathcal{G}(S)$ in opposite directions. By minimizing this functional, one effectively enforces the probabilistic independence structure that underlies Reichenbach’s Common Cause Principle and respects the Causal Markov Condition.
\end{remark}

\begin{theorem}[\textbf{Equivalence Between RCCP Screen-Off Condition and Conditional Covariance Minimization Under Linear Models}]
Let \( Y_1, \dots, Y_n \) denote portfolio constituents and \( S = \{X_1, \dots, X_m\} \subset X \) a selected subset of drivers. Suppose the conditional expectations \( \mathbb{E}[Y_i \mid S] \) are modeled using linear regressions. Then, the loss function from the Commonality Principle,

\[
\mathcal{G}(S) = \sum_{i=1}^{n} \sum_{\substack{j=1 \\ j \neq i}}^{n} \left| P(Y_i Y_j \mid S) - P(Y_i \mid S) P(Y_j \mid S) \right|,
\]

under Gaussian assumptions and linear conditional models, reduces to a function of the pairwise covariances of the conditional expectations:

\[
\mathcal{G}(S) \propto \sum_{i=1}^{n} \sum_{\substack{j=1 \\ j \neq i}}^{n} \text{Cov}(\mathbb{E}[Y_i \mid S], \mathbb{E}[Y_j \mid S]).
\]

Hence, minimizing \( \mathcal{G}(S) \) becomes equivalent to minimizing the total off-diagonal covariance between the conditional expectations—i.e., a conditional covariance minimization problem.
\end{theorem}

\begin{remark}
This result highlights a fundamental limitation of linear models, both in causal portfolio optimization and in common causal factor modeling frameworks. While linear models offer analytical simplicity and interpretability, they reduce the structure of conditional dependence to a form that is mathematically indistinguishable from the unconditional case. Specifically, under linearity and Gaussian assumptions, the loss function associated with the causal screening condition collapses into a covariance minimization objective, thereby failing to retain the richness of causal interactions among variables.

This limitation extends beyond portfolio construction and applies equally to factor-based representations of financial systems. In causal factor models, if linear approximations are used, key properties such as asymmetry, nonlinear feedback, and regime-dependent sensitivities are lost. This undermines the very rationale for adopting causal models in the first place, as the resulting factor exposures do not faithfully reflect the directional or probabilistic structure of the underlying drivers.

Therefore, while linear methods may serve as diagnostic tools or first-order approximations\citep{lopez2024case}, they fall short when aiming to capture the full complexity of market dynamics. In contrast, nonlinear models—such as neural networks, kernel methods, or other flexible function approximators—allow for richer representations of the causal relationships between drivers and responses \citep{RODRIGUEZDOMINGUEZ2023100447}. These models preserve sensitivity to dynamic, heterogeneous, and nonlinear structures, making them better suited for both identifying true causal factors and achieving optimal diversification in practice.
\end{remark}

\begin{proof}
Let \( S = \{X_{i_1}, X_{i_2}, \ldots, X_{i_m}\} \subset X \) denote a subset of selected drivers. The conditional expectation of asset \( Y_i \) given \( S \) is modeled using a linear regression:
\[
Y_i = \beta_{i0} + \sum_{k \in S} \beta_{ik} X_k + \epsilon_i,
\]
where \( \epsilon_i \) is a zero-mean noise term uncorrelated with the regressors. Similarly for asset \( Y_j \):
\[
Y_j = \beta_{j0} + \sum_{k \in S} \beta_{jk} X_k + \epsilon_j.
\]

Assuming a joint linear model, the joint conditional expectation of the product is:
\[
\mathbb{E}[Y_i Y_j \mid S] = \mathbb{E}[(\hat{Y}_i + \epsilon_i)(\hat{Y}_j + \epsilon_j)] = \mathbb{E}[\hat{Y}_i \hat{Y}_j] + \mathbb{E}[\epsilon_i \epsilon_j],
\]
and given that \( \epsilon_i \) and \( \epsilon_j \) are uncorrelated, this simplifies to:
\[
\mathbb{E}[Y_i Y_j \mid S] = \mathbb{E}[\hat{Y}_i \hat{Y}_j].
\]

The marginal conditional expectations are:
\[
\mathbb{E}[Y_i \mid S] = \hat{Y}_i = \sum_{k \in S} \beta_{ik} X_k, \quad \text{and} \quad \mathbb{E}[Y_j \mid S] = \hat{Y}_j = \sum_{k \in S} \beta_{jk} X_k.
\]

Thus, the product of marginal expectations is:
\[
\mathbb{E}[Y_i \mid S] \mathbb{E}[Y_j \mid S] = \mathbb{E}[\hat{Y}_i] \mathbb{E}[\hat{Y}_j].
\]

The Commonality Principle loss function becomes:
\[
\mathcal{G}(S) = \sum_{i=1}^n \sum_{j \ne i} \left| \mathbb{E}[Y_i Y_j \mid S] - \mathbb{E}[Y_i \mid S] \mathbb{E}[Y_j \mid S] \right| = \sum_{i=1}^n \sum_{j \ne i} \left| \operatorname{Cov}(\hat{Y}_i, \hat{Y}_j) \right|.
\]

The covariance is explicitly:
\[
\operatorname{Cov}(\hat{Y}_i, \hat{Y}_j) = \sum_{k \in S} \sum_{l \in S} \beta_{ik} \beta_{jl} \operatorname{Cov}(X_k, X_l).
\]

Therefore, minimizing \(\mathcal{G}(S)\) corresponds to minimizing the total off-diagonal pairwise covariances between predicted assets, as influenced by the shared structure of \( S \). In the case where the features \( X_k \in S \) are uncorrelated (or orthogonalized), the loss reduces to:
\[
\operatorname{Cov}(\hat{Y}_i, \hat{Y}_j) = \sum_{k \in S} \beta_{ik} \beta_{jk} \operatorname{Var}(X_k),
\]
and the loss function becomes:
\[
\mathcal{G}(S) = \sum_{i=1}^n \sum_{j \ne i} \left| \sum_{k \in S} \beta_{ik} \beta_{jk} \operatorname{Var}(X_k) \right|.
\]

This result shows that in the linear setting, the loss function implied by the Commonality Principle reduces to the sum of pairwise covariances of conditional expectations. Therefore, minimizing this loss is equivalent to performing conditional covariance minimization over the portfolio constituents, where dependencies are evaluated given the selected drivers.

\end{proof}

\newpage

\subsection{Methodologies for Identifying Common Drivers Based on the Commonality Principle}

A collection of methods is introduced, combining both novel methodologies and tailored implementations of existing techniques suited to this specific application. The first approach, as proposed in the original article \citep{RODRIGUEZDOMINGUEZ2023100447}, leverages the Reichenbach Common Cause Principle (RCCP) in conjunction with correlations, providing a statistical proxy that is both practical and efficient. This approach holds particular value given that RCCP remains one of the few theoretical frameworks directly linking causality and correlation. While alternative theories of causality may offer appealing conceptual formulations, they are often computationally intractable for practical implementation.

Due to the widespread use of correlation metrics in the financial industry, the methodology becomes especially powerful: it relies on correlation, is computationally efficient, data-dependent, and grounded in a causal selection framework through RCCP. This method serves as the baseline for model selection.

However, as established in the framework and further implied by the use of the Markov Common Cause Principle (MCCP) in the proof of the Commonality Principle in \citep{RODRIGUEZDOMINGUEZ2023100447}, or in this monograph in Section \ref{sectionCausality} any methodology based on conditional independence and structural models is equally applicable. In this setting, the problem reduces to a combinatorial search over a driver map, which does not require predefined structural assumptions—contrary to some graph-based approaches in causal investing. Structural constraints may still be introduced at later modeling stages, but their absence does not hinder the causal interpretation, distinguishing this approach from others that depend heavily on prior structural assumptions. Optimization can proceed by directly minimizing the objective in (\ref{eq13}) without needing additional structure.

Additional methods presented are selected based on criteria of traceability and practical implementation. The second approach explores the use of Bayesian networks as a modeling tool for representing probabilistic causality through conditional dependencies. This method facilitates interpretable graphical structures and can capture complex causal relationships under uncertainty.

The third approach introduces the application of the Maximum Likelihood Correlation-Based Method, which aims to identify the most probable structure or set of drivers that explain the observed dependencies using likelihood-based inference.

Finally, a fourth approach is also based on the Maximum Likelihood Correlation-Based Method, adapted to the particular structure of financial datasets and tailored for scenarios where direct causal interpretability is required while maintaining computational efficiency.

\subsubsection{Inverse-Engineered RCCP-Based Driver Selection via Correlation Optimization}
\label{InverseRCCP}
In portfolio theory, identifying drivers that explain the behavior of individual assets and the portfolio as a whole is critical. Drivers are divided into two categories: \textit{specific drivers} and \textit{common drivers}. Specific drivers are relevant for individual portfolio constituents, while common drivers are shared across multiple constituents. The \textit{Principle of Commonality} ensures that the selected drivers balance individual explanatory power with collective portfolio-wide relevance.

\begin{itemize}
    \item \textbf{Specific Drivers:} For each portfolio constituent \(Y_i\), specific drivers \(S_i\) are those that maximize persistence and probability of causality for \(Y_i\). These drivers are relevant to \(Y_i\) alone and satisfy the condition of statistical independence with other constituents.
    \item \textbf{Common Drivers:} A common driver \(X_k\) is a specific driver for at least one portfolio constituent \(Y_i\) and is repeatedly selected as a specific driver for the largest possible number of portfolio constituents \( \{Y_1, \ldots, Y_n\} \). Furthermore, common drivers must maximize correlation strength across the portfolio.
\end{itemize}

The Principle of Commonality states that optimal common drivers must satisfy two conditions (\citep{RODRIGUEZDOMINGUEZ2023100447}):
\begin{enumerate}
    \item Be specific drivers for at least one portfolio constituent \(Y_i\).
    \item Be repeatedly selected as specific drivers across the portfolio, maximizing the number of correlated constituents and the correlation strength.
\end{enumerate}

The approach is based on the inverse engineering of Reichenbach’s Common Cause Principle (RCCP), where the existence of sufficiently strong and repeated correlations is used as a proxy to induce and validate potential causal structure. Rather than proving that correlation is a result of a common cause, the method ensures that correlations exist in a consistent and statistically meaningful direction, thereby increasing the posterior probability of $X_k$ acting as a plausible common cause. In addition, the selection process incorporates a multicollinearity control mechanism. The database is pre-screened to ensure that there is no overlapping information between portfolio constituents and candidate drivers that could lead to collinearity. This step ensures that each selected driver contributes distinct explanatory power, improving the robustness of the sensitivity estimation and preserving the validity of projections within the sensitivity space.

\subsubsection*{Mathematical Formulation}

The goal is to find a subset of $m$ drivers from the set of candidates $X = \{X_1, \dots, X_K\}$ such that two objectives are jointly optimized:
\begin{enumerate}
    \item The selected drivers maximize \textbf{repeated significant correlation} across portfolio constituents.
    \item The selection implicitly reduces conditional dependence between constituents, aligning with RCCP.
\end{enumerate}

The optimization problem is:
\[
\min_{S \subset X, |S| = m} \left[ \sum_{i=1}^n \sum_{j \neq i}^n \left| P(Y_i Y_j | S) - P(Y_i | S)P(Y_j | S) \right| - \lambda \sum_{X_k \in S} R(X_k) \right],
\]
where $R(X_k)$ is the count of constituents $Y_i$ for which driver $X_k$ exhibits strong correlation, and $\lambda > 0$ is a regularization parameter.

\subsubsection*{Correlation-Based Approximation}

In practical implementation, the conditional independence term is approximated using a correlation structure as follows:

\begin{itemize}
    \item Let $C[k,i] = \text{Corr}(X_k, Y_i)$, the Pearson correlation between driver $X_k$ and asset $Y_i$.
    \item Define a threshold $\epsilon$ such that:
    \[
    R[k, i] = \begin{cases}
    1 & \text{if } |C[k,i]| \geq \epsilon \\
    0 & \text{otherwise}
    \end{cases}
    \]
    \item Repeatedness:
    \[
    R(X_k) = \sum_{i=1}^n R[k,i]
    \]
    \item Cumulative correlation strength:
    \[
    S(X_k) = \sum_{i=1}^n |C[k,i]| \cdot R[k,i]
    \]
\end{itemize}

\subsubsection*{Selection Algorithm}

Drivers are ranked lexicographically using the tuple $(R(X_k), S(X_k))$ to select the top $m$ candidates:

\begin{algorithm}[H]
\SetAlgoLined
\KwIn{$Y$: asset returns, $X$: candidate drivers, $\epsilon$: correlation threshold, $m$: number of drivers}
\KwOut{$S$: selected drivers of size $m$}

Compute correlation matrix $C[k,i] = \text{Corr}(X_k, Y_i)$\;

Define relevance matrix $R[k,i] = \mathbb{1}_{\{ |C[k,i]| \geq \epsilon \}}$\;

Compute $R(X_k) = \sum_i R[k,i]$\;

Compute $S(X_k) = \sum_i |C[k,i]| \cdot R[k,i]$\;

Sort drivers by $(R(X_k), S(X_k))$ in decreasing order\;

Return top $m$ drivers
\caption{Correlation-Based Inverse RCCP Driver Selection}
\end{algorithm}

\subsubsection*{Remarks}

This method is efficient, interpretable, and aligned with causal inference principles under data constraints. By using correlations to approximate conditional dependencies, it sidesteps the need for full causal discovery algorithms while still aligning with RCCP. The method forms the basis of the baseline selection algorithm in \citep{RODRIGUEZDOMINGUEZ2023100447} and can be extended to non-linear correlations and mutual information metrics.

\textbf{Input:}
\begin{itemize}
    \item Portfolio: \(Y = \{ Y_1, Y_2, Y_3 \}\)
    \item Drivers: \(X = \{ X_1, X_2, X_3, X_4 \}\)
    \item Correlation threshold: \(\epsilon = 0.5\)
    \item Number of drivers to select: \(m = 2\)
\end{itemize}

\textbf{Correlation Matrix:}
\[
C = 
\begin{bmatrix}
0.6 & 0.4 & 0.3 \\
0.8 & 0.7 & 0.5 \\
0.2 & 0.6 & 0.4 \\
0.9 & 0.8 & 0.7
\end{bmatrix}
\]

\textbf{Relevance Matrix:}
\[
R = 
\begin{bmatrix}
1 & 0 & 0 \\
1 & 1 & 1 \\
0 & 1 & 0 \\
1 & 1 & 1
\end{bmatrix}
\]

\textbf{Repeatedness:}
\[
R(X_1) = 1, \quad R(X_2) = 3, \quad R(X_3) = 1, \quad R(X_4) = 3
\]

\textbf{Cumulative Correlation Strength:}
\[
S(X_1) = 0.6, \quad S(X_2) = 2.0, \quad S(X_3) = 0.6, \quad S(X_4) = 2.4
\]

\textbf{Ranked Drivers:}
\[
\text{Rank: } X_4, X_2, X_1, X_3
\]

\textbf{Output:}
\[
\text{Selected Drivers: } \{ X_4, X_2 \}
\]

The algorithm ensures that common drivers are selected based on their repeatedness and correlation strength across the portfolio constituents, maximizing their explanatory relevance while adhering to the Principle of Commonality.

\newpage

\subsubsection{A Bayesian Network Approach}

Given a subset \( S \), the goal is to ensure approximate conditional independence between each pair \( (Y_i, Y_j) \), such that:
\[
P(Y_i, Y_j \mid S) \approx P(Y_i \mid S) P(Y_j \mid S).
\]
The deviation from independence is measured as:
\[
\Delta_{ij}(S) = \left| P(Y_i Y_j \mid S) - P(Y_i \mid S) P(Y_j \mid S) \right|.
\]
To minimize the total deviation from independence, we define the objective function:
\[
\mathcal{G}(S) = \sum_{i=1}^{n} \sum_{\substack{j=1 \\ j \neq i}}^{n} \Delta_{ij}(S).
\]

The optimization is subject to two constraints: first, the total deviation must satisfy \( \mathcal{G}(S) \leq m \varepsilon \), where \( m \) controls the size of the selected subset \( S \) and \( \varepsilon \) is the threshold for independence. Second, the subset \( S \) must satisfy \( S \subseteq X \) and \( |S| = m \).

To solve this problem, we use observed data to estimate the conditional probabilities \( P(Y_i \mid X_k) \) and the joint probabilities \( P(Y_i, Y_j \mid S) \). Optimization techniques, such as score-based or constraint-based methods, are then employed to identify \( S \) that minimizes \( \mathcal{G}(S) \). For large \( M \), heuristic methods or greedy algorithms may be used to efficiently explore the space of subsets. Finally, the selected subset \( S \) is validated to ensure that \( \mathcal{G}(S) \leq m \varepsilon \).

\subsubsection{Approaches to Select \( S \)}

\textbf{Dynamic Programming Approach:}

The dynamic programming approach explores all possible subsets of \( S \) to identify the optimal subset that minimizes the objective function \( \mathcal{G}(S) \). Let \( F(k, m) \) represent the minimum \( \mathcal{G}(S) \) achievable using the first \( k \) variables from \( X \) to form a subset \( S \) of size \( m \). The recurrence relation is defined as:
\[
F(k, m) = \min \left\{ F(k-1, m), F(k-1, m-1) + \Delta(k) \right\},
\]
where:
\[
\Delta(k) = \sum_{i=1}^n \sum_{j \neq i} \left| \hat{P}(Y_i Y_j \mid S \cup \{X_k\}) - \hat{P}(Y_i \mid S \cup \{X_k\}) \hat{P}(Y_j \mid S \cup \{X_k\}) \right|.
\]
Here, \( \Delta(k) \) is computed using linear regression for \( \hat{P}(Y_i \mid S) \) and multivariate regression for \( \hat{P}(Y_i Y_j \mid S) \).

The procedure is as follows:
\begin{itemize}
    \item Initialize \( F(k, m) \) with \( F(k, 0) = 0 \) and \( F(0, m) = \infty \) for \( m > 0 \).
    \item Compute \( \Delta(k) \) for each variable \( X_k \) using the regression models.
    \item Populate \( F(k, m) \) iteratively and retrieve the optimal subset \( S \) from the table.
\end{itemize}

This approach guarantees the globally optimal subset but can be computationally expensive for large \( M \).

\textbf{Greedy Approach:}

The greedy approach incrementally constructs \( S \) by adding variables that most improve \( \mathcal{G}(S) \) at each step. The objective function \( \mathcal{G}(S) \) is computed as:
\[
\mathcal{G}(S) = \sum_{i=1}^n \sum_{j \neq i} \left| \hat{P}(Y_i Y_j \mid S) - \hat{P}(Y_i \mid S) \hat{P}(Y_j \mid S) \right|.
\]

The procedure is as follows:
\begin{itemize}
    \item Start with \( S = \emptyset \).
    \item At each step, evaluate all remaining variables \( X_k \in X \setminus S \).
    \item Add the variable \( X_k \) that minimizes \( \mathcal{G}(S \cup \{X_k\}) \), using linear regression for \( \hat{P}(Y_i \mid S \cup \{X_k\}) \) and multivariate regression for \( \hat{P}(Y_i Y_j \mid S \cup \{X_k\}) \).
    \item Stop when \( |S| = m \) or when adding further variables increases \( \mathcal{G}(S) \).
\end{itemize}

In score-based learning, the predictive performance of the regression models is optimized. A score for \( S \) is defined as:
\begin{equation}
   \text{Score}(S) = -\mathcal{G}(S) 
   \label{score}
\end{equation}

and subsets that improve this score are prioritized. In constraint-based learning, independence is enforced by adding variables to \( S \) only if:
\[
\Delta_{ij}(S) \leq \varepsilon,
\]
where \( \varepsilon \) is a small tolerance for independence violations. This ensures that the selected subset \( S \) maintains approximate independence between \( Y_i \) and \( Y_j \).

\subsubsection{Maximum Likelihood Correlation-based Method}

Consider a system with two target variables \( X_1 \) and \( X_2 \), and one causal variable \( Z \). The system assumes that \( X_1 \) and \( X_2 \) are conditionally independent given \( Z \), which is expressed as:
\[
P(X_1, X_2 \mid Z) = P(X_1 \mid Z) P(X_2 \mid Z).
\]
The joint probability of \( X_1 \) and \( X_2 \) is then given by:
\[
P(X_1, X_2) = \int P(X_1 \mid Z) P(X_2 \mid Z) P(Z) \, dZ.
\]

Graphically, this relationship can be represented as a directed acyclic graph (DAG):
\[
Z \to X_1, \quad Z \to X_2.
\]

This structure implies that \( Z \) acts as the common cause of \( X_1 \) and \( X_2 \), explaining their dependence. For example, if \( Z \) represents a latent factor like market sentiment, and \( X_1 \) and \( X_2 \) represent stock prices, the conditional independence ensures that the relationship between the stock prices is fully explained by \( Z \).

\textbf{Joint Probability for Two Variables with One Common Cause:}

Consider two target variables \( X_1 \) and \( X_2 \), and one common causal variable \( Z \). Assuming linear relationships:
\[
X_1 = \alpha_1 Z + \epsilon_1, \quad X_2 = \alpha_2 Z + \epsilon_2,
\]
where \( \epsilon_1 \) and \( \epsilon_2 \) are independent noise terms with zero mean and uncorrelated with \( Z \).

The joint probability of \( X_1 \) and \( X_2 \) is modeled as a bivariate normal distribution:
\[
P(X_1, X_2) = \frac{1}{2 \pi \sigma_{X_1} \sigma_{X_2} \sqrt{1 - \rho_{X_1 X_2}^2}} \exp\left(-\frac{1}{2(1 - \rho_{X_1 X_2}^2)} Q\right),
\]
where the quadratic term \( Q \) is given by:
\[
Q = \frac{(X_1 - \alpha_1 \mu_Z)^2}{\sigma_{X_1}^2} - 2 \rho_{X_1 X_2} \frac{(X_1 - \alpha_1 \mu_Z)(X_2 - \alpha_2 \mu_Z)}{\sigma_{X_1} \sigma_{X_2}} + \frac{(X_2 - \alpha_2 \mu_Z)^2}{\sigma_{X_2}^2}.
\]

Variances of \( X_1 \) and \( X_2 \):
   \[
   \sigma_{X_1}^2 = \alpha_1^2 \text{Var}(Z) + \sigma_{\epsilon_1}^2, \quad \sigma_{X_2}^2 = \alpha_2^2 \text{Var}(Z) + \sigma_{\epsilon_2}^2.
   \]

Correlation Between \( X_1 \) and \( X_2 \):
   \[
   \rho_{X_1 X_2} = \frac{\text{Cov}(X_1, X_2)}{\sqrt{\text{Var}(X_1) \text{Var}(X_2)}} = \frac{\alpha_1 \alpha_2 \text{Var}(Z)}{\sqrt{\sigma_{X_1}^2 \sigma_{X_2}^2}}.
   \]

Mean of \( Z \):
   \[
   \mu_Z = \mathbb{E}[Z].
   \]

The observed correlation \( \rho_{X_1 X_2} \) reflects the shared dependence of \( X_1 \) and \( X_2 \) on the common cause \( Z \). The joint probability \( P(X_1, X_2) \) is fully described in terms of:
\begin{itemize}
    \item The variances of \( X_1 \), \( X_2 \), and \( Z \)
    \item The strength of the causal effects (\( \alpha_1, \alpha_2 \))
    \item The noise variances (\( \sigma_{\epsilon_1}^2, \sigma_{\epsilon_2}^2 \))
\end{itemize}
The correlation structure in the system is described as:
\begin{itemize}
    \item Correlation between \( X_1 \) and \( X_2 \):
    \[
    \rho_{X_1 X_2} = \frac{\alpha_1 \alpha_2 \text{Var}(Z)}{\sqrt{\left(\alpha_1^2 \text{Var}(Z) + \sigma_{\epsilon_1}^2\right)\left(\alpha_2^2 \text{Var}(Z) + \sigma_{\epsilon_2}^2\right)}}.
    \]
    \item Correlation between \( Z \) and \( X_1 \):
    \[
    \rho_{Z X_1} = \frac{\alpha_1 \sqrt{\text{Var}(Z)}}{\sqrt{\alpha_1^2 \text{Var}(Z) + \sigma_{\epsilon_1}^2}}.
    \]
    \item Correlation between \( Z \) and \( X_2 \):
    \[
    \rho_{Z X_2} = \frac{\alpha_2 \sqrt{\text{Var}(Z)}}{\sqrt{\alpha_2^2 \text{Var}(Z) + \sigma_{\epsilon_2}^2}}.
    \]
\end{itemize}

These relationships fully describe the dependencies between the variables in terms of their correlations and variances.

\textbf{Maximum Likelihood Estimation for \( Z \):}
Now, the goal is to find the value of \( Z \) that maximizes the likelihood of observing the data \( \{(X_1^{(i)}, X_2^{(i)})\}_{i=1}^n \), given the correlation structure and a sample \( \{Z_1, Z_2, \dots, Z_m\} \) of potential values for \( Z \). The joint probability \( P(X_1, X_2) \) is modeled as a bivariate normal distribution:
\[
P(X_1, X_2) = \frac{1}{2 \pi \sigma_{X_1} \sigma_{X_2} \sqrt{1 - \rho_{X_1 X_2}^2}} \exp\left(-\frac{1}{2(1 - \rho_{X_1 X_2}^2)} Q\right),
\]
where:
\[
Q = \frac{(X_1 - \mu_{X_1})^2}{\sigma_{X_1}^2} - 2 \rho_{X_1 X_2} \frac{(X_1 - \mu_{X_1})(X_2 - \mu_{X_2})}{\sigma_{X_1} \sigma_{X_2}} + \frac{(X_2 - \mu_{X_2})^2}{\sigma_{X_2}^2}.
\]

Here, the parameters are defined as:
\begin{itemize}
    \item \( \mu_{X_1} = \alpha_1 Z, \quad \mu_{X_2} = \alpha_2 Z \),
    \item \( \sigma_{X_1}^2 = \alpha_1^2 \text{Var}(Z) + \sigma_{\epsilon_1}^2 \),
    \item \( \sigma_{X_2}^2 = \alpha_2^2 \text{Var}(Z) + \sigma_{\epsilon_2}^2 \),
    \item \( \rho_{X_1 X_2} = \frac{\alpha_1 \alpha_2 \text{Var}(Z)}{\sqrt{\sigma_{X_1}^2 \sigma_{X_2}^2}} \).
\end{itemize}

The likelihood of observing the data \( \{(X_1^{(i)}, X_2^{(i)})\}_{i=1}^n \) is:
\[
L(Z) = \prod_{i=1}^n P(X_1^{(i)}, X_2^{(i)} \mid Z).
\]
Substituting \( P(X_1, X_2) \), the likelihood becomes:
\[
L(Z) = \prod_{i=1}^n \frac{1}{2 \pi \sigma_{X_1} \sigma_{X_2} \sqrt{1 - \rho_{X_1 X_2}^2}} \exp\left(-\frac{1}{2(1 - \rho_{X_1 X_2}^2)} Q^{(i)}\right),
\]
where:
\[
Q^{(i)} = \frac{(X_1^{(i)} - \alpha_1 Z)^2}{\sigma_{X_1}^2} - 2 \rho_{X_1 X_2} \frac{(X_1^{(i)} - \alpha_1 Z)(X_2^{(i)} - \alpha_2 Z)}{\sigma_{X_1} \sigma_{X_2}} + \frac{(X_2^{(i)} - \alpha_2 Z)^2}{\sigma_{X_2}^2}.
\]

The log-likelihood for computational convenience is:
\[
\log L(Z) = -n \log(2\pi) - n \log(\sigma_{X_1}) - n \log(\sigma_{X_2}) - \frac{n}{2} \log(1 - \rho_{X_1 X_2}^2) - \frac{1}{2(1 - \rho_{X_1 X_2}^2)} \sum_{i=1}^n Q^{(i)},
\]
where \( Q^{(i)} \) is substituted as above. The goal is to maximize the log-likelihood with respect to \( Z \). Specifically:
\[
\hat{Z} = \arg\max_{Z_k \in \{Z_1, Z_2, \dots, Z_m\}} \log L(Z_k).
\]

This solution leverages the correlation structure between \( Z \), \( X_1 \), and \( X_2 \), as well as their variances.

\textbf{Maximum Likelihood Estimation for \( M \) Causes with \( N \) Targets:}

Given \( N \) target variables \( \mathcal{X} = \{X_1, X_2, \dots, X_N\} \) and \( m \) candidate causal variables \( \mathcal{Z}_{\text{candidate}} = \{Z_1, Z_2, \dots, Z_m\} \), we aim to select \( M \) optimal causal variables \( \mathcal{Z}^* \subseteq \mathcal{Z}_{\text{candidate}} \) such that \( |\mathcal{Z}^*| = M \) and the likelihood of the observed data \( \{X_1^{(i)}, X_2^{(i)}, \dots, X_N^{(i)}\}_{i=1}^n \) is maximized. The joint probability of the target variables \( \mathcal{X} \) given a subset of \( M \) causal variables \( \mathcal{Z}^* = \{Z_{j_1}, Z_{j_2}, \dots, Z_{j_M}\} \) is:
\[
P(\mathcal{X} \mid \mathcal{Z}^*) = \frac{1}{(2\pi)^{N/2} \prod_{i=1}^N \sigma_{X_i} \sqrt{\det(I - \rho_{\mathcal{X}})}} \exp\left(-\frac{1}{2(1 - \det(\rho_{\mathcal{X}}))} Q\right),
\]
where:
\[
Q = \sum_{i=1}^N \frac{(X_i - \mu_{X_i})^2}{\sigma_{X_i}^2} - \sum_{i \neq j} \rho_{X_i X_j} \frac{(X_i - \mu_{X_i})(X_j - \mu_{X_j})}{\sigma_{X_i} \sigma_{X_j}},
\]
and:
\begin{itemize}
    \item \( \mu_{X_i} = \sum_{k=1}^M \alpha_{ik} Z_k \) is the mean of \( X_i \) conditioned on the \( M \) selected causal variables.
    \item \( \sigma_{X_i}^2 = \sum_{k=1}^M \alpha_{ik}^2 \text{Var}(Z_k) + \sigma_{\epsilon_i}^2 \), where \( \sigma_{\epsilon_i}^2 \) is the variance of the noise for \( X_i \).
    \item \( \rho_{X_i X_j} \) is the correlation coefficient between \( X_i \) and \( X_j \), defined by the shared dependence on \( \mathcal{Z}^* \).
\end{itemize}

The likelihood of observing the data \( \{X_1^{(i)}, X_2^{(i)}, \dots, X_N^{(i)}\}_{i=1}^n \) is:
\[
L(\mathcal{Z}^*) = \prod_{i=1}^n P(\mathcal{X}^{(i)} \mid \mathcal{Z}^*),
\]
where \( \mathcal{X}^{(i)} = \{X_1^{(i)}, X_2^{(i)}, \dots, X_N^{(i)}\} \). Substituting the expression for \( P(\mathcal{X} \mid \mathcal{Z}^*) \), we have:
\[
L(\mathcal{Z}^*) = \prod_{i=1}^n \frac{1}{(2\pi)^{N/2} \prod_{k=1}^N \sigma_{X_k} \sqrt{\det(I - \rho_{\mathcal{X}})}} \exp\left(-\frac{1}{2(1 - \det(\rho_{\mathcal{X}}))} Q^{(i)}\right),
\]
where \( Q^{(i)} \) is the quadratic term for the \( i \)-th observation. Taking the log for computational convenience:
\[
\log L(\mathcal{Z}^*) = -\frac{nN}{2} \log(2\pi) - n \sum_{k=1}^N \log(\sigma_{X_k}) - \frac{n}{2} \log(\det(I - \rho_{\mathcal{X}})) - \frac{1}{2(1 - \det(\rho_{\mathcal{X}}))} \sum_{i=1}^n Q^{(i)}.
\]

The goal is to select the subset \( \mathcal{Z}^* \) of \( M \) variables from \( \mathcal{Z}_{\text{candidate}} \) that maximizes the log-likelihood:
\[
\mathcal{Z}^* = \arg\max_{\mathcal{Z} \subseteq \mathcal{Z}_{\text{candidate}}, |\mathcal{Z}| = M} \log L(\mathcal{Z}).
\]

\newpage

\section{Approximation Model for Asset and Portfolio Dynamics}
\label{subsection421}
A key concept is that once the drivers are causally identified and used as inputs to model asset or portfolio dynamics, the resulting sensitivities or betas derived from this model are also causal functions. It does not make sense to apply causal identification methods to select drivers and then revert to a traditional factor model that relies solely on correlations for portfolio optimization. This inconsistency defeats the purpose of causally sound modeling.

The existence of a conformal map—one that preserves angular (i.e., directional) relationships between the unconditional, conditional, and sensitivity spaces—is both a necessary and sufficient condition for maintaining optimal idiosyncratic and systematic diversification. The final step of this conformal map is the embedding into the sensitivity space, valid at every time $t$. Since the sensitivities are causally derived, they retain trajectory-level information, which is essential for optimizing diversification dynamically over time.\par

Traditional factor models typically select statistics, smart beta factors or macroeconomic indices without explicit causal validation. They estimate the factor loadings matrix based on correlations and apply it in mean-variance optimization frameworks. Identifying causal factors and then apply factor engine associational tools is contradictory at first. In contrast, this framework focuses on an emsbdeded sensitivyt space after comomn causal selection is implemented. In the case of linear models that will be shown first for completeness, it interprets the beta vectors from linear regressions as coordinates in a low-dimensional causal embedding space, where similarity is based on shared structural influences. Portfolio weights are then optimized using a distance-based diversification criterion that reflects variation in causal exposure.

Traditional factor models typically rely on statistical constructs, smart beta factors, or macroeconomic indices without explicit causal validation. These models estimate the factor loadings matrix based on correlations and subsequently apply it within a mean-variance optimization framework. However, identifying causal factors and then applying association-based tools for factor modeling introduces a conceptual inconsistency. 

In contrast, the framework proposed here emphasizes the use of an embedded sensitivity space, which is constructed after the implementation of common causal driver selection. In the case of linear models—which are presented first for completeness—beta vectors obtained from linear regressions are interpreted as coordinates in a low-dimensional causal embedding space, where similarity between assets reflects shared structural (i.e., causal) influences. Portfolio weights are then optimized using a distance-based diversification criterion, designed to capture variation in causal exposure rather than mere statistical association.


\subsection{Linear Models Selection}

Consider $N$ assets and $M$ causal drivers. Each asset return $R_n$ is modeled via a linear regression on the common drivers:
\[
R_n = \sum_{m=1}^{M} \beta_{nm} D_m + \varepsilon_n
\]
Let $\boldsymbol{\beta}_n = [\beta_{n1}, \dots, \beta_{nM}]^\top \in \mathbb{R}^M$ denote the beta vector associated with asset $n$, and define the matrix $\boldsymbol{B} \in \mathbb{R}^{N \times M}$ as the collection of all such vectors.

The pairwise Euclidean distance between the regression coefficients of assets $i$ and $j$ is computed as:
\[
D_{ij} = \|\boldsymbol{\beta}_i - \boldsymbol{\beta}_j\|_2
\]
The optimization problem aims to allocate portfolio weights $\boldsymbol{w} \in \mathbb{R}^N$ to achieve diversification in the causal embedding space.

\subsubsection{Optimization Without Non-Negativity Constraints}

The unconstrained problem is defined as:
\[
\min_{\boldsymbol{w}} \quad \boldsymbol{w}^\top \boldsymbol{D} \boldsymbol{w}
\quad \text{subject to} \quad \boldsymbol{1}^\top \boldsymbol{w} = 1
\]

Applying the method of Lagrange multipliers yields the closed-form solution:
\[
\boxed{
\boldsymbol{w}^* = \frac{\boldsymbol{D}^{-1} \boldsymbol{1}}{\boldsymbol{1}^\top \boldsymbol{D}^{-1} \boldsymbol{1}}
}
\]

\subsubsection{Optimization With Non-Negativity Constraints}

Introducing non-negativity conditions on the weights, the optimization becomes:
\[
\min_{\boldsymbol{w}} \quad \boldsymbol{w}^\top \boldsymbol{D} \boldsymbol{w}
\quad \text{subject to} \quad \boldsymbol{1}^\top \boldsymbol{w} = 1, \quad \boldsymbol{w} \ge 0
\]

This formulation constitutes a convex quadratic program (QP), for which a closed-form solution is not generally available. The necessary optimality conditions are given by the Karush–Kuhn–Tucker (KKT) conditions:

\begin{align*}
\text{Stationarity:} & \quad 2\boldsymbol{D}\boldsymbol{w} - \lambda \boldsymbol{1} - \boldsymbol{\mu} = 0 \\\\
\text{Primal feasibility:} & \quad \boldsymbol{1}^\top \boldsymbol{w} = 1, \quad \boldsymbol{w} \ge 0 \\\\
\text{Dual feasibility:} & \quad \mu_i \ge 0 \quad \forall i \\\\
\text{Complementary slackness:} & \quad \mu_i w_i = 0 \quad \forall i
\end{align*}

These conditions can be solved numerically using standard quadratic programming solvers.

\newpage

\section{Partial Differential Equations, Neural Networks, and Automatic Differentiation}
\label{PDEsAAD}
Asset dynamics can be modeled by a system of unknown partial differential equations (PDEs), where the independent variables are external drivers. Let the return of an asset at time \( t \) be denoted by \( a_t \), and let the vector of lagged driver returns at time \( \tau \) be given by \( \boldsymbol{D}_\tau = \{D_{1\tau}, D_{2\tau}, \dots, D_{M\tau}\} \), where \( t > \tau \).

Assuming that asset dynamics can be described by a first-order PDE that is analytically unsolvable and unknown, the solution can be expressed as a general nonlinear function \( F : \mathbb{R}^{3M+1} \rightarrow \mathbb{R} \):

\begin{equation}
	a_t = F\left(
	\frac{\partial a_t}{\partial D_{1\tau}}, \dots, \frac{\partial a_t}{\partial D_{M\tau}},
	\frac{\partial D_{1\tau}}{\partial t}, \dots, \frac{\partial D_{M\tau}}{\partial t},
	\frac{\partial a_t}{\partial t},
	D_{1\tau}, \dots, D_{M\tau}
	\right)
	\label{equation161}
\end{equation}

In the special case where the relationship is linear, the expression simplifies to:

\begin{equation}
	a_t = \sum_{j=1}^{M} \left( 
	\frac{\partial a_t}{\partial D_{j\tau}} + 
	\frac{\partial D_{j\tau}}{\partial t} + 
	D_{j\tau} \right) 
	+ \frac{\partial a_t}{\partial t}
	\label{equation162}
\end{equation}

This class of PDEs, despite being analytically intractable, can be effectively approximated using time series data and neural networks, due to their universal approximation capabilities~\citep{Cybenko1989ApproximationBS}.

For each asset in the portfolio, the most suitable neural network architecture is selected to model its conditional behavior based on a predefined set of common causal drivers. Sensitivities—defined as partial derivatives of the model’s output with respect to each input driver—are computed using automatic differentiation techniques. These sensitivities are treated as discrete time series, and their statistical summaries (e.g., mean) over the training horizon are used as key metrics. The choice of summarization function significantly affects performance, as highlighted in \citep{pizarroso2021neuralsens}.

For any constituent $a_i$, the conditional expectation is given by:
\begin{equation}
\begin{split}
E\left[a_{it} \middle| \boldsymbol{D_{\tau}}\right] = F\left(
\frac{\partial a_t}{\partial D_{1\tau}}, \ldots, \frac{\partial a_t}{\partial D_{M\tau}},
\frac{\partial D_{1\tau}}{\partial t}, \ldots, \frac{\partial D_{M\tau}}{\partial t},
\frac{\partial a_t}{\partial t}, D_{1\tau}, \ldots, D_{M\tau}
\right)
\end{split}
\end{equation}
This function is approximated by a neural network $({NN}_i)$, and the sensitivities \( \frac{\partial a_t}{\partial D_{j\tau}} \) are computed using automatic adjoint differentiation (AAD). Sensitivities are defined as:
\[ {s_{ij}|}_{\boldsymbol{D_{\tau}}}^{{NN}_i} = \frac{\partial a_{it}}{\partial D_{j\tau}}(\boldsymbol{D_{\tau}}) \]
indicating the sensitivity of the neural network output to the $j$-th driver at input sample $\boldsymbol{D_{\tau}}$, calculated through the chain rule across network layers.

\begin{figure}[h] \centering \includegraphics[width=0.75\linewidth]{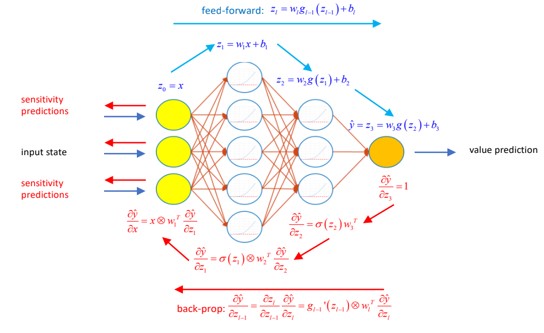} \caption{Automatic Adjoint Differentiation (AAD) for neural models, as described in \citep{huge2020differential}. While extensively utilized on the sell side, its application can be highly beneficial for the buy side, particularly in portfolio risk management.}
\label{fig:enter-AAD} \end{figure}

\begin{figure}[h] 
\centering 
\includegraphics[width=1\linewidth]{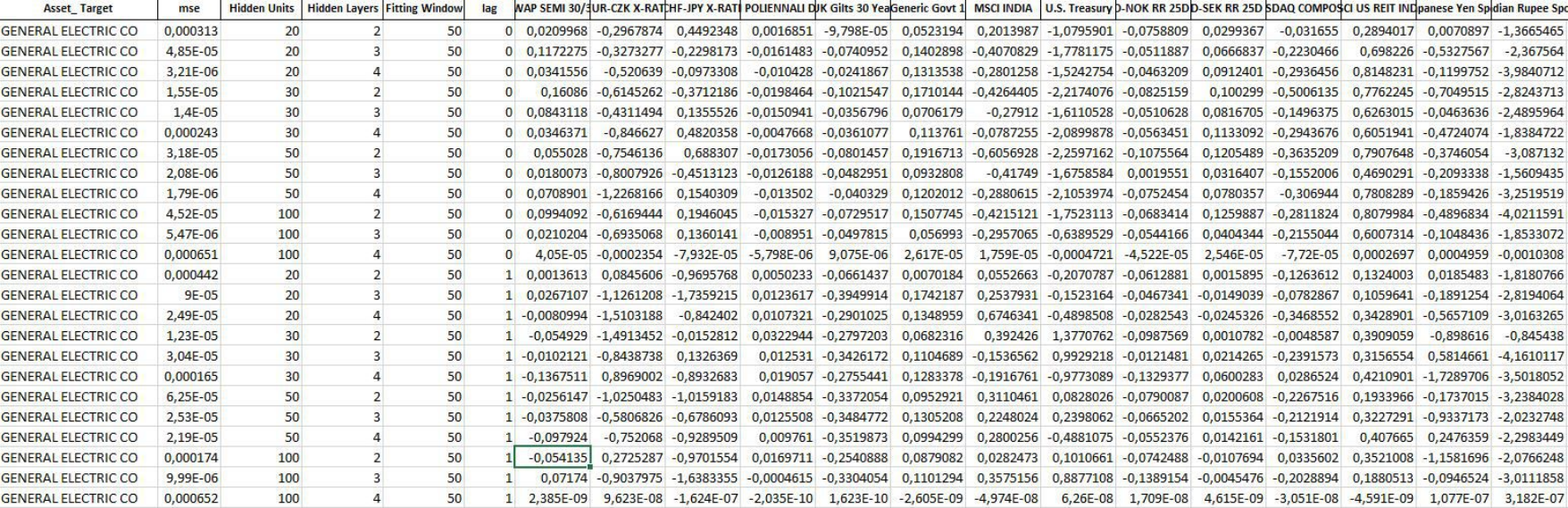} 
\caption{Sensitivity data obtained for a particular portfolio constituent. Each row corresponds to a different neural network architecture, detailing the RMSE, number of hidden units and layers, and the lag structure between inputs and outputs. The rightmost columns report the average sensitivity values of the constituent with respect to the selected set of common drivers.}

\label{fig:enter-tablesens} 
\end{figure}

The AAD methodology is implemented using TensorFlow's GradientTape, consistent with the approach used in \citep{huge2020differential} for solving pricing problems involving differential equations. Sensitivities are collected over time and aggregated into representative statistics, most commonly their average. Alternative aggregation strategies could be employed to improve robustness and stability. In Figure \ref{fig:enter-AAD}, the process of extracting sensitivities from feed-forward neural networks is shown schematically. This example illustrates the simplest case, where no specific architectural structure is imposed on the network. However, in the work by \citep{huge2020differential}, the authors apply similar techniques to hedge complex derivatives, using more sophisticated network architectures. A similar approach can be extended to portfolio optimization when structural knowledge about the portfolio constituents is available. In such cases, this structure can be incorporated at the network architecture level, and the resulting sensitivities—which are causal by construction—preserve this structure. This information becomes valuable during the optimization stage, particularly for guiding diversification decisions and constraints. An example table of the sensitivities obtained from the experiments is presented in Figure~\ref{fig:enter-tablesens}.

\subsection{Sensitivity Distance Matrix}
Following the extraction of sensitivities for all portfolio constituents with respect to common drivers, the construction of a distance-based optimization framework is facilitated. Each asset is embedded in a $M$-dimensional space, where $M$ is the number of common drivers. The coordinate of each asset in this space corresponds to the average sensitivity of that asset with respect to each driver.

The embedded form is:
\begin{equation}
\left[\begin{matrix}y_1\\\vdots\\y_N\end{matrix}\right] =
\left[\begin{matrix}
\frac{\partial y_1}{\partial x_1}(t) & \cdots & \frac{\partial y_1}{\partial x_M}(t) \\
\vdots & \ddots & \vdots \\
\frac{\partial y_N}{\partial x_1}(t) & \cdots & \frac{\partial y_N}{\partial x_M}(t)
\end{matrix}\right]
\left[\begin{matrix}x_1\\\vdots\\x_M\end{matrix}\right] 
\cong
\left[\begin{matrix}
{s_{11}|}_{x_n}^{{NN}_1} & \cdots & {s_{1M}|}_{x_n}^{{NN}_1} \\
\vdots & \ddots & \vdots \\
{s_{N1}|}_{x_n}^{{NN}_N} & \cdots & {s_{NM}|}_{x_n}^{{NN}_N}
\end{matrix}\right]
\left[\begin{matrix}x_1\\\vdots\\x_M\end{matrix}\right]
\end{equation}

In this sensitivity-based coordinate system, a distance matrix is constructed.

\begin{definition}[Sensitivity Distance Matrix]
Let $S$ be the sensitivity distance matrix whose entries represent the pairwise distance between portfolio constituents in the embedded space of average sensitivity vectors:
\begin{equation}
\begin{split}
S = \left[\begin{matrix}s_{11} & \cdots & s_{1N} \\
\vdots & \ddots & \vdots \\
s_{N1} & \cdots & s_{NN}\end{matrix}\right], \\ 
s_{ij} = d\left(\frac{\partial a_i}{\partial \boldsymbol{CD}}, \frac{\partial a_j}{\partial \boldsymbol{CD}}\right) = d\left(\vec{\beta}_i, \vec{\beta}_j\right) = d\left([\beta_i^1, \ldots, \beta_i^M], [\beta_j^1, \ldots, \beta_j^M]\right)
\end{split}
\label{sensmatrix}
\end{equation}
The coordinates correspond to average sensitivity values in the training dataset, and the resulting distance matrix is referred to as the Sensitivity Distance Matrix.
\end{definition}

\begin{figure}[h]
\label{Image7}
\centering
\includegraphics[width=95mm]{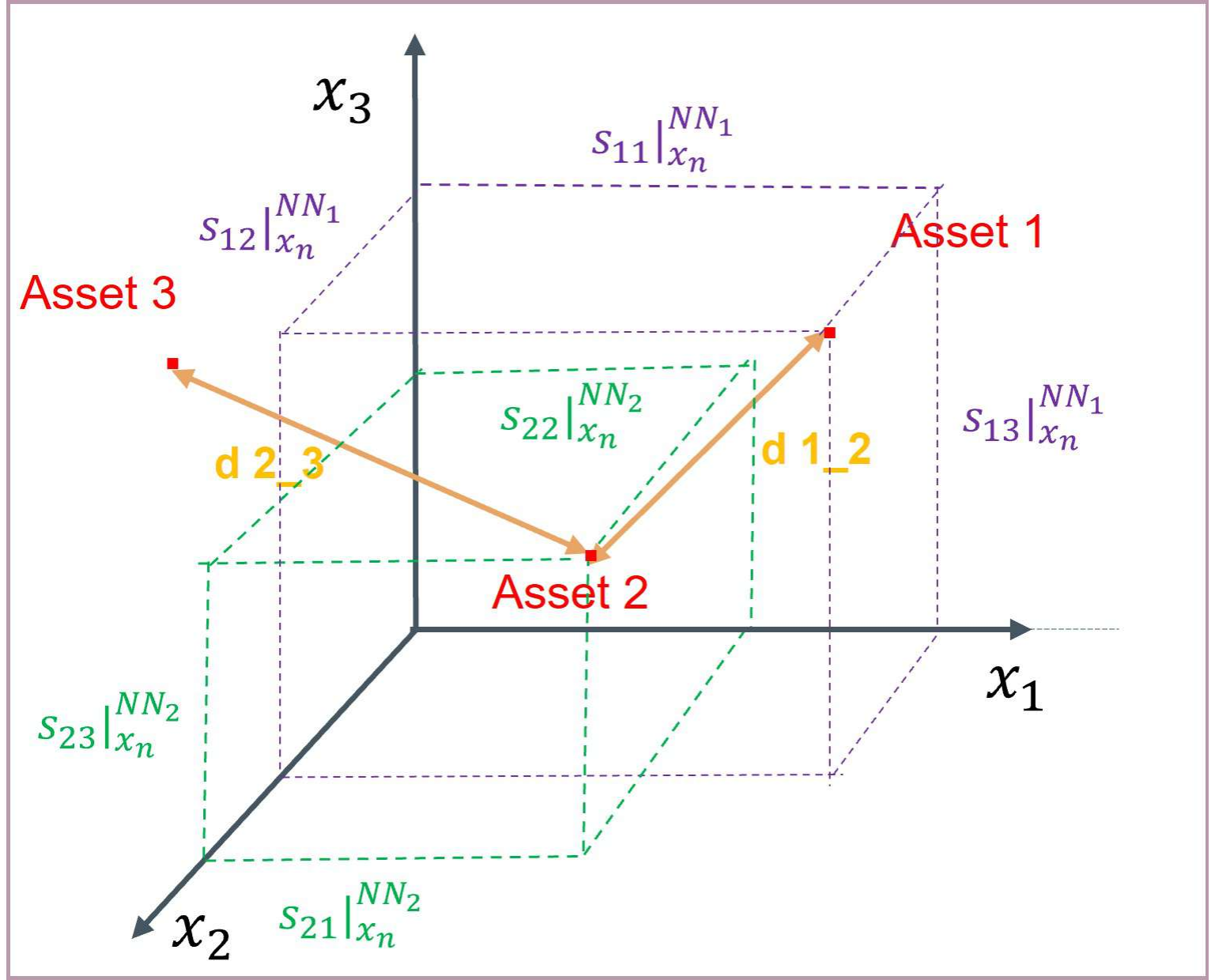}
\caption{Embedded space of sensitivities: coordinates are the average sensitivities and the orange line represents pairwise distances.}
\label{embed15}
\end{figure}

\begin{figure}
    \centering
    \includegraphics[width=1\linewidth]{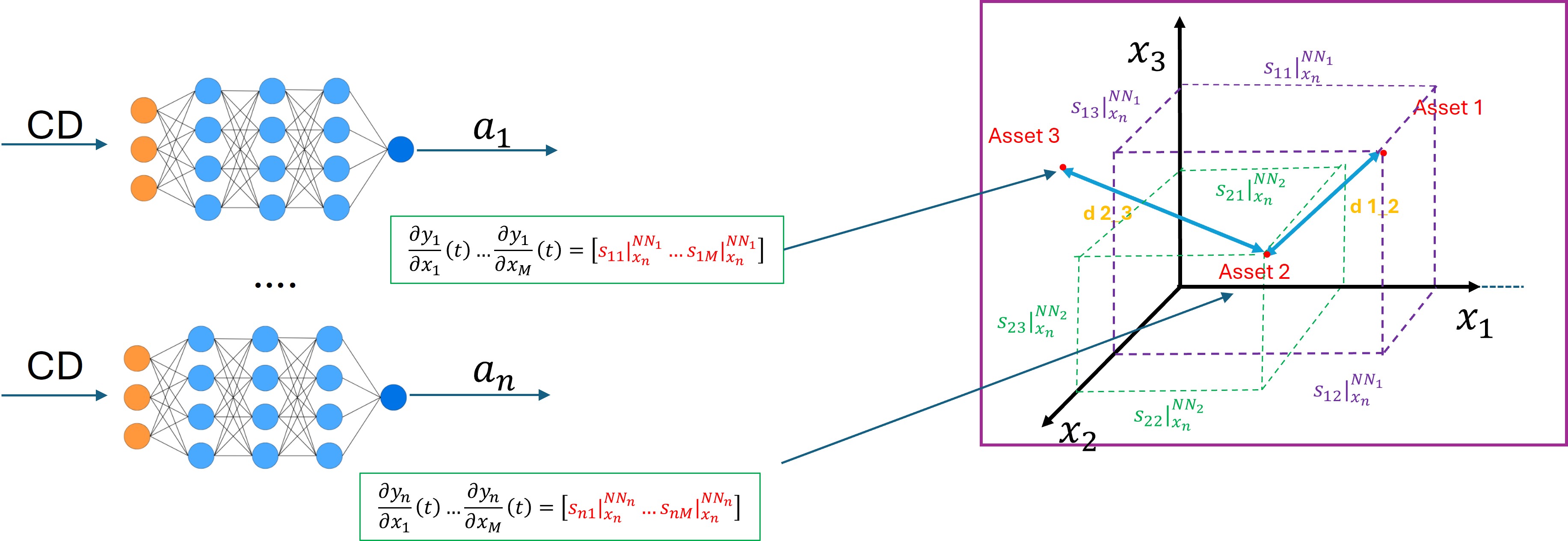}
\caption{Neural networks with the common drivers as input and each portfolio constituent as output, used to obtain sensitivities via AAD and embed the portfolio constituents into a sensitivity space.}

    \label{fig:enter-NN}
\end{figure}

The embedded representation reflects both causal and persistent characteristics of the portfolio constituents' dynamics over the training period (see Figure \ref{embed15}). It preserves idiosyncratic diversification consistent with the unconditional mean-variance case while enabling systematic diversification through exogenous common causal drivers. Furthermore, it allows for diversification across directional dynamics, capturing responses of asset returns to public drivers rather than purely statistical factors. This enhances the interpretability and robustness of the portfolio construction process by incorporating economically meaningful external structure. In Figure~\ref{fig:enter-NN}, the left part schematically illustrates how neural networks predict each portfolio constituent using the common drivers as inputs. Sensitivities are then extracted using Automatic Adjoint Differentiation (AAD). The averaged sensitivities are subsequently used as coordinates for the embedding, shown in the right part of Figure~\ref{fig:enter-NN}.

\begin{figure}
    \centering
    \includegraphics[width=1\linewidth]{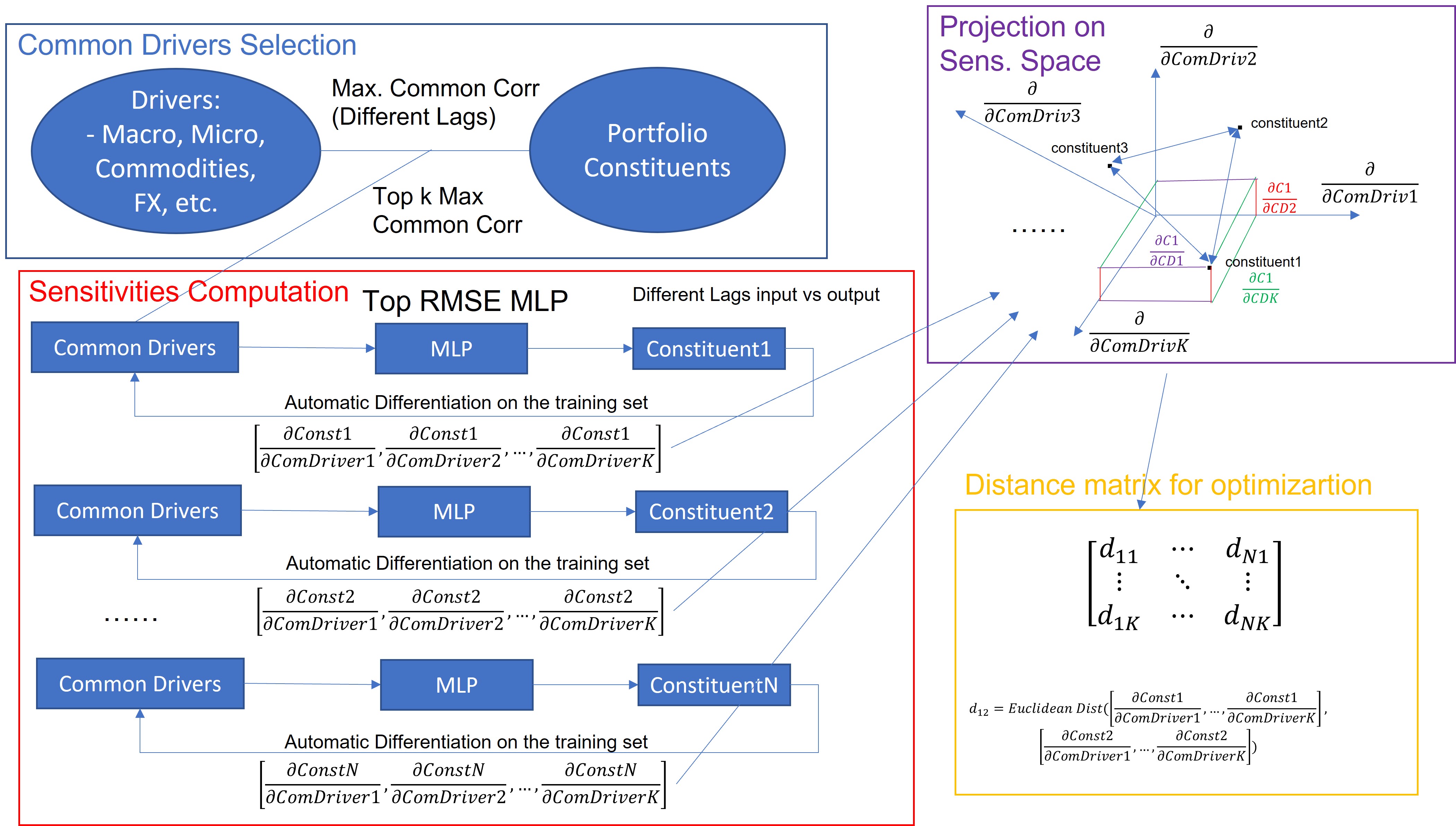}
    \caption{Methodology consisting of Common Causal Drivers Identification, Sensitivity Approximation and Embedding Space, and Sensitivity Distance Matrix Computation.}

    \label{fig:enter-NNArchitecture}
\end{figure}

Finally, for reference and guidance, the complete methodology—excluding the convex optimization step—is illustrated in Figure \ref{fig:enter-NNArchitecture}.

\newpage

\section{Mapping Sensitivity Space to Risk Measures of Interest (a.k.a) Solving the Optimization}
\label{RiskMap}

In the proposed framework, assets are projected into a space defined by their sensitivities to a set of common drivers. While this space provides meaningful causal coordinates, unless for the case of linear models, for more complex models the notion of risk within it diverges from traditional approaches based on elliptically distributed returns and moment-based measures such as variance, skewness, or kurtosis. To reconcile the geometry of the sensitivity space with conventional portfolio risk management practices, a mapping is required that connects asset positions in the sensitivity space with portfolio-level risk metrics.

\subsection{Hierarchical Risk Allocation Methods}
The original paper approach to perform this mapping through hierarchical clustering of the constituents based on their pairwise distances in the sensitivity space \citep{RODRIGUEZDOMINGUEZ2023100447} (See Figure \ref{RiskMap}). These distances can be of many different kinds, including the Euclidean norm:
\[
D_{ij} = \| \boldsymbol{s}_i - \boldsymbol{s}_j \|_2
\]
where $\boldsymbol{s}_i$ is the sensitivity vector for asset $i$. Given this distance matrix, a dendrogram is constructed using linkage methods (e.g., single, average, or Ward linkage), and a hierarchical tree is formed.

Once the hierarchical structure is established, portfolio allocation can be performed using:
\begin{itemize}
    \item \textbf{Hierarchical Risk Parity (HRP)}: Allocates capital recursively through the tree to achieve local risk parity at each node.
    \item \textbf{Hierarchical Equal Risk Contribution (HERC)}: Allocates based on equal risk contributions across hierarchical clusters.
\end{itemize}
In both HRP and HERC, cluster-level volatilities are computed from the time series of asset returns grouped according to sensitivity similarity. The volatility of each cluster informs weight assignment, typically through inverse-volatility scaling.

\subsection{Alternative Non-Hierarchical Risk Mappings}
The mapping from the sensitivity space to volatility-weighted portfolios does not necessarily require hierarchical clustering. Several alternative approaches include:
\begin{itemize}
    \item \textbf{Kernel smoothing of volatility surfaces}: Estimate the volatility of each asset by weighting nearby assets in the sensitivity space.
    \item \textbf{Distance-weighted shrinkage}: Assign weights inversely proportional to average volatility among $k$-nearest neighbors in sensitivity space.
    \item \textbf{Manifold learning}: Perform dimensionality reduction (e.g., t-SNE, UMAP) on sensitivity vectors, followed by clustering or direct volatility estimation in the embedded space.
    \item \textbf{Graph-based approaches}: Construct a graph where nodes represent assets and edges encode similarity (e.g., via sensitivity distance), and propagate volatility or CVaR estimates using message-passing schemes.
\end{itemize}
These methods allow for volatility-informed risk allocation even in cases where hierarchical clustering is unstable or inappropriate.

\begin{figure}[H]
    \centering
    \includegraphics[width=1\linewidth]{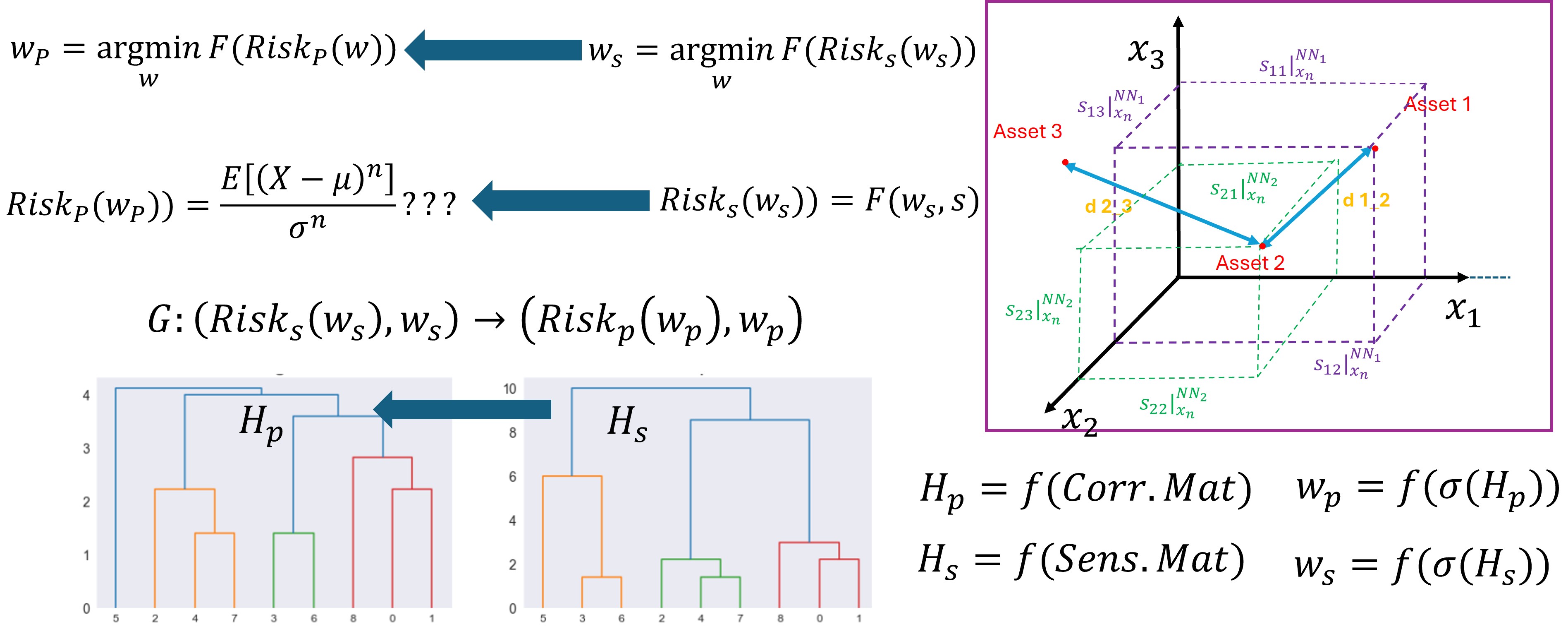}
    \caption{Risk map from the sensitivity space to the target risk measure. In this example, the mapping is performed via hierarchical clustering, which preserves the diversification dynamics by identifying hierarchical structures among assets. Clusters formed through this hierarchy are then evaluated in terms of volatility by computing the risk associated with each cluster.}

    \label{fig:enter-RiskMap}
\end{figure}

\subsection{Linear Model in Sensitivity Space and Portfolio Optimization Methods}

In the linear model setup, asset returns are regressed on a common set of causal drivers to estimate their sensitivities. These sensitivities are then used to construct a distance matrix in sensitivity space, which informs portfolio optimization under various risk objectives.

\subsubsection{Linear Sensitivity Estimation}
For each asset \( a_i \), its return at time \( t \) is modeled as:
\[
ra_i(t) = \beta_{i1} D_1(t) + \cdots + \beta_{iM} D_M(t) + \varepsilon_i(t)
\]
where \( \beta_{ij} \) is the sensitivity of asset \( i \) to driver \( D_j \), and \( \varepsilon_i(t) \) is the residual. The resulting sensitivity vector \( \boldsymbol{\beta}_i = [\beta_{i1}, \ldots, \beta_{iM}] \in \mathbb{R}^M \) defines the coordinates of asset \( i \) in sensitivity space.

A distance matrix \( \boldsymbol{S} \) is constructed using Euclidean distances between these vectors:
\[
S_{ij} = \| \boldsymbol{\beta}_i - \boldsymbol{\beta}_j \|_2
\]
This matrix captures similarity in causal structure and is used for diversification.

\subsubsection{Portfolio Optimization Methods for the Linear Models Case}
Three approaches are considered using this linear sensitivity structure:

\paragraph{1. Mean-Variance Optimization with Sensitivity Regularization}
\begin{itemize}
    \item Estimate expected returns \( \mu \) and covariance matrix \( \Sigma \).
    \item Use distance matrix \( D \) as a regularizer to promote diversification in sensitivity space.
    \item Solve:
    \[ \min_{\boldsymbol{w}} \quad \boldsymbol{w}^\top \Sigma \boldsymbol{w} + \lambda \boldsymbol{w}^\top D \boldsymbol{w} \quad \text{s.t.} \quad \sum w_i = 1, \ w_i \geq 0 \]
\end{itemize}

\paragraph{2. Volatility-Based Allocation}
\begin{itemize}
    \item Use the distance matrix \( D \) to form clusters (e.g., hierarchical clustering).
    \item Compute volatility of clusters or assets.
    \item Allocate weights inversely proportional to volatility (direct or hierarchical methods like HRP or HERC).
\end{itemize}

\paragraph{3. CVaR-Based Optimization Using Linear Approximation}
\begin{itemize}
    \item Approximate returns as \( ra_i(t) \approx \boldsymbol{\beta}_i^\top \boldsymbol{D}_t \).
    \item Simulate portfolio returns: \( ra_p(t) = \boldsymbol{w}^\top B \boldsymbol{D}_t \).
    \item Solve CVaR optimization:
    \[ \min_{\boldsymbol{w}, \zeta, \boldsymbol{z}} \zeta + \frac{1}{(1 - \alpha) T} \sum_{t=1}^T z_t \]
    \text{subject to:}
    \begin{align*}
    z_t &\geq -\boldsymbol{w}^\top B \boldsymbol{D}_t - \zeta \\
    z_t &\geq 0 \\
    \sum w_i &= 1, \quad w_i \geq 0
    \end{align*}
\end{itemize}

\subsection{Mappings from Sensitivity Space to Risk and Return Measures (Recap Methods)}

The sensitivity space represents each asset by its vector of sensitivities \( \boldsymbol{\beta}_i \in \mathbb{R}^M \) to a common set of causal drivers. Several practical mappings exist to translate this geometric representation into traditional portfolio metrics such as expected return, volatility, covariance, and tail risk.

1. Expected Return via Driver Forecasts
\[
\mathbb{E}[R_i] \approx \boldsymbol{\beta}_i^\top \mathbb{E}[\boldsymbol{D}]
\]
This allows constructing portfolios using mean-variance or utility-based objectives given driver forecasts.

2. Covariance Matrix Approximation
\[
\Sigma \approx B \cdot \Sigma_D \cdot B^\top + \Sigma_{\text{residual}}
\]
Useful for plug-in to standard risk models or optimization schemes.

3. Volatility from Sensitivity Magnitudes
\[
\sigma_i \approx \sqrt{\boldsymbol{\beta}_i^\top \Sigma_D \boldsymbol{\beta}_i}
\]
This supports inverse volatility weighting and volatility-based clustering.

4. Tail Risk via Simulation (CVaR)
Simulated portfolio return:
\[
R_p(t) = \boldsymbol{w}^\top B \boldsymbol{D}_t
\]
CVaR objective:
\[
\text{CVaR}_\alpha(R_p) = \min_{\zeta} \left\{ \zeta + \frac{1}{1 - \alpha} \mathbb{E}[( -R_p - \zeta )_+ ] \right\}
\]
Applicable in empirical or scenario-based optimization.

5. Geometric Diversification via Distance
Define a pairwise distance matrix:
\[
D_{ij} = \| \boldsymbol{\beta}_i - \boldsymbol{\beta}_j \|_2
\]
Minimize similarity or maximize geometric spread:
\[
\min_{\boldsymbol{w}} \quad \sum_{i,j} w_i w_j D_{ij}
\]

6. Cluster-Based Volatility Allocation
Cluster assets using sensitivity vectors, compute cluster-level volatility:
\[
\sigma_{\text{cluster}_k} = \text{volatility}\left( \text{PCA}(\{R_i\}_{i \in \text{cluster}_k}) \right)
\]
Allocate inversely proportional to \( \sigma_{\text{cluster}_k} \).

7. Directional Risk Attribution
Decompose portfolio risk exposure by driver:
\[
\text{Risk}_{D_j} = \sum_{i=1}^N w_i \cdot \left( \frac{\partial R_i}{\partial D_j} \cdot \sigma_{D_j} \right)
\]
Supports constraint or tilt-based control over causal factors.

\begin{table}[h!]
\centering
\begin{tabular}{|l|l|l|}
\hline
\textbf{Mapping Type} & \textbf{Output} & \textbf{Use Case} \\
\hline
Expected Return         & \( \mathbb{E}[R_i] \)           & Forecast-driven allocation \\
Covariance Approximation & \( \Sigma \approx B \Sigma_D B^\top \) & Mean-variance, CVaR \\
Volatility Estimate     & \( \sigma_i \approx \sqrt{\beta_i^\top \Sigma_D \beta_i} \) & Risk parity, inverse vol \\
Simulated CVaR          & \( \text{CVaR}_\alpha(R_p) \)      & Tail-risk minimization \\
Distance Diversification & \( D_{ij} = \| \beta_i - \beta_j \| \) & Spread-based diversification \\
Cluster Volatility      & \( \sigma_{\text{cluster}_k} \)     & HRP/HERC strategies \\
Directional Attribution & \( \sum w_i \cdot \partial R_i/\partial D_j \) & Thematic/macro control \\
\hline
\end{tabular}
\caption{Practical mappings from sensitivity space to portfolio risk and return metrics.}
\end{table}

\begin{table}[h!]
\centering
\begin{tabular}{|l|l|l|l|}
\hline
\textbf{Method} & \textbf{Inputs} & \textbf{Objective} & \textbf{Tools} \\
\hline
Mean-Variance & \( \mu, \Sigma, D \) & \( \min \boldsymbol{w}^\top \Sigma \boldsymbol{w} + \lambda \boldsymbol{w}^\top D \boldsymbol{w} \) & QP \\
\hline
Volatility-Based & Return vol, \( D \) & Inverse vol / HRP, HERC, Schur & Clustering  \\
\hline
CVaR & \( B, D_t \) simulations & Rockafellar–Uryasev CVaR & LP / Convex Solv. \\
\hline
\end{tabular}
\caption{Summary of portfolio optimization methods using linear sensitivity structure. Rockafellar–Uryasev CVaR \citep{ROCKAFELLAR20021443}, HRP \citep{Prado2016BuildingDP}, HERC \citep{raffinot2018hierarchical}, Schur \citep{cotton2024schurcomplementaryallocationunification}.}
\end{table}

\subsection{Risk Mapping from the Sensitivity Space to Risk Measures Using Copulas}
\label{sec:copula_risk_mapping}

A central objective in portfolio risk management is to map the structure of asset exposures—captured in the sensitivity space—into concrete risk metrics such as variance, Value-at-Risk (VaR), or Expected Shortfall (ES). The sensitivity space is derived from causal modeling of asset dynamics with respect to common drivers. However, to incorporate joint behavior and nonlinear dependencies, a copula-based framework provides a principled approach.

\subsubsection{Sensitivity Space and Dependency Structure}

Let \( \boldsymbol{S} \in \mathbb{R}^{n \times m} \) be the matrix of sensitivities for \( n \) assets to \( m \) common causal drivers. The row \( \boldsymbol{s}_i \in \mathbb{R}^m \) represents the sensitivities of asset \( a_i \) to all drivers. These vectors form the coordinates of portfolio constituents in the tangent space of the Common Causal Manifold at time \( t \).

\textbf{Objective:} Model the dependence between these vectors and translate that into meaningful joint risk measures.

\subsubsection{Copula-Based Modeling Framework}

By Sklar's Theorem, the joint distribution \( H(\boldsymbol{s}_1, \dots, \boldsymbol{s}_n) \) can be expressed as:
\[
H(\boldsymbol{s}_1, \dots, \boldsymbol{s}_n) = C\left(F_1(\boldsymbol{s}_1), \dots, F_n(\boldsymbol{s}_n)\right),
\]
where \( F_i \) is the marginal distribution of sensitivities for asset \( i \), and \( C \) is a copula encoding their dependency structure.

\textbf{Popular Copulas:}
\begin{itemize}
    \item \textbf{Gaussian Copula:} Captures linear correlation but no tail dependence.
    \item \textbf{t-Copula:} Models both correlation and tail dependence, suitable for financial applications.
    \item \textbf{Clayton, Gumbel:} Capture asymmetric lower/upper tail dependencies.
\end{itemize}

\subsubsection{Mapping to Risk Measures}

Assuming a Gaussian copula and normal marginal distributions of asset sensitivities:
\[
\text{Cov}(a_i, a_j) = \rho_{ij} \cdot \sigma_i \cdot \sigma_j,
\]
where:
\begin{itemize}
    \item \( \rho_{ij} \): Copula-implied correlation.
    \item \( \sigma_i \): Standard deviation of sensitivity \( \boldsymbol{s}_i \).
\end{itemize}

The portfolio variance is then:
\[
\sigma_p^2 = \boldsymbol{w}^\top \Sigma \boldsymbol{w},
\]
where \( \boldsymbol{w} \) is the vector of portfolio weights, and \( \Sigma \) is the covariance matrix derived from the copula.

\textbf{Value-at-Risk (VaR)}

Given the copula \( C \), generate \( M \) joint samples of asset sensitivities:
\[
\{ \boldsymbol{s}_1^{(k)}, \dots, \boldsymbol{s}_n^{(k)} \}_{k=1}^M,
\]
and simulate corresponding portfolio returns:
\[
r_p^{(k)} = \sum_{i=1}^n w_i \cdot \boldsymbol{s}_i^{(k)}.
\]
Estimate VaR at confidence level \( \alpha \) as:
\[
\text{VaR}_{\alpha} = - \text{Quantile}_{\alpha} \left( \{ r_p^{(1)}, \dots, r_p^{(M)} \} \right).
\]

\subsubsection{Remarks}

This approach allows for:
\begin{itemize}
    \item Capturing nonlinear dependence across assets rooted in causal sensitivities.
    \item Leveraging heavy-tailed marginals and realistic dependence via copulas.
    \item Seamlessly integrating into portfolio optimization frameworks via copula-implied risk metrics.
\end{itemize}

Future extensions may include dynamic copulas for time-evolving sensitivity structures or high-dimensional copula models for complex portfolios.

\subsection{Handling Non-Invertibility of Distance Matrices in Sensitivity Space}

When projecting portfolio constituents into a sensitivity space and defining a distance matrix \( \boldsymbol{D} \) (e.g., using Euclidean distances), the assumption that \( \boldsymbol{D} \) is invertible does not always hold. Although Euclidean distance matrices are symmetric and encode valuable structural information, they may still be singular or ill-conditioned, particularly when there are redundant or linearly dependent sensitivity vectors.

\subsubsection{Causes of Non-Invertibility}
\begin{itemize}
    \item \textbf{Redundant Sensitivities}: Identical or nearly identical sensitivity vectors yield duplicate rows and columns in \( \boldsymbol{D} \).
    \item \textbf{Low-Rank Structures}: When the dimensionality of the sensitivity space is much smaller than the number of assets (i.e., \( M \ll N \)), the resulting distance matrix may be rank-deficient.
    \item \textbf{Zero Diagonal and Lack of Diagonal Dominance}: Euclidean distance matrices are hollow (\( D_{ii} = 0 \)) and may not satisfy the conditions for positive definiteness.
\end{itemize}

\subsubsection{Strategies for Handling Non-Invertibility}

\paragraph{1. Regularization}
Introduce a small perturbation to the diagonal to ensure positive definiteness:
\[
\boldsymbol{D}_{\text{reg}} = \boldsymbol{D} + \lambda \boldsymbol{I}, \quad \lambda > 0
\]
This modification guarantees invertibility and improves numerical stability. The regularized matrix \( \boldsymbol{D}_{\text{reg}} \) can be used in analytical portfolio weight formulations.

\paragraph{2. Pseudoinverse}
Use the Moore–Penrose pseudoinverse \( \boldsymbol{D}^{\dagger} \) in place of \( \boldsymbol{D}^{-1} \):
\[
\boldsymbol{w}^* = \frac{\boldsymbol{D}^{\dagger} \mathbf{1}}{\mathbf{1}^\top \boldsymbol{D}^{\dagger} \mathbf{1}}
\]
This is suitable in unconstrained optimization settings but may lack interpretability in financial applications.

\paragraph{3. Numerical Optimization}
Avoid the need for inversion entirely by solving optimization problems directly:
\[
\min_{\boldsymbol{w}} \quad \boldsymbol{w}^\top \boldsymbol{D} \boldsymbol{w} \quad \text{s.t.} \quad \mathbf{1}^\top \boldsymbol{w} = 1, \quad \boldsymbol{w} \geq 0
\]
This approach remains valid regardless of whether \( \boldsymbol{D} \) is invertible, and is compatible with CVaR-based or regularized formulations.

\paragraph{4. Kernel Transformation}
Transform the distance matrix into a positive-definite kernel matrix, such as using the Gaussian kernel:
\[
K_{ij} = \exp\left(-\frac{\|\boldsymbol{s}_i - \boldsymbol{s}_j\|^2}{2\sigma^2}\right)
\]
This guarantees invertibility and allows application of kernel-based learning and optimization techniques.

While Euclidean distance matrices offer intuitive structure in sensitivity-based asset embeddings, their use in analytical optimization procedures must account for potential singularity. Regularization and direct optimization are practical solutions, while kernelization provides a theoretically robust alternative. The method selected should reflect the modeling assumptions, numerical properties, and interpretability requirements of the portfolio optimization task.

\newpage

\section{Hierarchical Sensitivity Parity: Hierarchical Clustering and Convex Optimization}
\label{HSP_Section}
It is assumed that the previous steps of the methodology have already been carried out, including the driver selection process if applicable (which can be performed at the same time as the portfolio rebalancing or maintained fixed over a longer period), the computation of sensitivities, and the construction of the sensitivity distance matrix. These steps precede the portfolio optimization phase, where risk is mapped from the sensitivity space to the volatility space.

To achieve this, a hierarchical clustering algorithm is applied to preserve the hierarchical relationships between clusters within the sensitivity space. Portfolio weights are then computed based on the covariances and volatilities of those clusters, using numerical methods analogous to those employed in Hierarchical Risk Parity (HRP), \citep{Prado2016BuildingDP}.

The following section summarizes the methodology and lists the relevant hyperparameters for this case \citep{RODRIGUEZDOMINGUEZ2023100447}:

\begin{itemize}
    \item \textbf{Common driver selection}: The method is RCCP Reverse Engineering from Section \ref{InverseRCCP}, with another algorithm found in the Appendix \ref{appselectdrov}.  Select the top \( K \) drivers (hyperparameter 1) that exhibit the highest cumulative correlation with all portfolio constituents. Correlation is computed from daily returns over a historical window \( W_{CD} \) (hyperparameter 2).
    
    \item \textbf{Neural network modeling}: For each asset in the portfolio, train multiple feed-forward neural networks using the selected common drivers as inputs and the asset's return series as output. Across various architectures (differing in depth and width), the best-performing model is selected based on prediction accuracy (RMSE) using a training window of length \( W_{NN} \) (hyperparameter 3).
    
    \item \textbf{Sensitivity extraction}: Compute the sensitivities of each asset with respect to the selected common drivers using Automatic Adjoint Differentiation (AAD) on the optimal network models. Sensitivities are computed over the training window.
    
    \item \textbf{Embedding and distance computation}: Average the sensitivities across the training window to obtain a representative sensitivity vector for each asset. These vectors are used to calculate a pairwise distance matrix, embedding the portfolio constituents in a sensitivity space.
    
    \item \textbf{Hierarchical optimization}: Use numerical methods to find the nearest positive semi-definite approximation of the distance matrix. Apply hierarchical clustering to this matrix to determine a hierarchy of sensitivity-based relationships. Finally, compute portfolio weights using a recursive bisection procedure based on the cluster structure and corresponding covariance matrices.
\end{itemize}

 By employing hierarchical clustering, portfolio constituents are grouped based on structures derived from their dynamic properties in the sensitivity space. Following the rationale in \citep{Prado2016BuildingDP}, which leverages hierarchical clustering on the correlation matrix to enhance diversification via hierarchical projections, analogous insights can be extracted from the hierarchical organization of sensitivity vectors with respect to common causal drivers. This approach benefits from the inherent directionality encoded in those sensitivities and further promotes diversification.

Performance improvements are observed when the sensitivity distance matrix is replaced by its nearest positive semi-definite neighbor, computed according to \citep{HIGHAM1988103}. Once this matrix is obtained, single-linkage clustering is applied using the following distance function:
\begin{equation}
\begin{split}
D(X,Y)=\min_{x\in X, y\in Y}d(x,y) \\
d(x,y)=\sqrt{\sum_{n=1}^{N}\left(S_{n,x}-S_{n,y}\right)^2} 
\end{split}
\label{singlelink}
\end{equation}
Here, \( S \) may refer to either the original sensitivity distance matrix defined in \eqref{sensmatrix}, or its nearest positive semi-definite neighbor. Using this distance metric, the clustering process organizes constituents such that similar ones are positioned nearby in the hierarchy, while dissimilar ones are placed farther apart.

The Recursive Bijection algorithm from \citep{Prado2016BuildingDP} is subsequently applied to compute portfolio weights. A sorted distance matrix is used to traverse the clustering tree from top to bottom. At each bifurcation, two clusters are evaluated. Within each, the volatility is computed using the weight vector:
\begin{equation}
    w=\frac{\text{diag}\left[V\right]^{-1}}{\text{trace}(\text{diag}\left[V\right]^{-1})}
\end{equation}
where \( V \) is the covariance matrix of assets within the cluster. The respective variances of the clusters are then given by:
\begin{equation}
\sigma_1={w_1}^TV_1w_1, \quad \sigma_2={w_2}^TV_2w_2
\end{equation}
Rescaling factors \( \alpha_1 \) and \( \alpha_2 \) are computed as:
\begin{equation}
\alpha_1=1-\frac{\sigma_1}{\sigma_1+\sigma_2}, \quad \alpha_2=1-\alpha_1
\end{equation}
and the weights are updated by \( w_1=w_1\alpha_1, \ w_2=w_2\alpha_2 \). When there are \( 2^n \) assets, the initial weight vector is uniformly distributed. Each recursion step halves the total weight sum, leading to a final normalization factor of 1 after \( n \) steps, consistent with \citep{HRPmiker}.

This methodology applies recursive bisection to a clustering tree derived from the sensitivity distance matrix or its nearest positive semi-definite approximation using single-linkage distance. Unlike Hierarchical Risk Parity (HRP), which constructs the tree from correlation distances, this approach—termed Hierarchical Sensitivity Parity (HSP)—derives the structure directly from causal sensitivity information.

Recursive bisection is implemented using a clustering tree generated from the single-linkage algorithm, where the distance metric is based on either the original sensitivity distance matrix \eqref{sensmatrix} or its nearest positive semi-definite neighbor. This differs from the methodology in Hierarchical Risk Parity (HRP) described in \citep{Prado2016BuildingDP}, where recursive bisection is applied to a tree derived from correlation distances. The current approach, referred to as Hierarchical Sensitivity Parity (HSP), instead employs the structure induced by causal sensitivities.

\IncMargin{1em}
\begin{algorithm}[H]
\SetAlgoLined
\KwData{\texttt{Constituents}: Matrix of daily returns for portfolio constituents\\
\texttt{Drivers}: Matrix of daily returns for candidate explanatory variables\\
\texttt{RebalanceDates}: List of portfolio rebalancing dates\\
\texttt{DriverUpdateDates}: List of dates for updating common driver selection}

\KwResult{Dictionary \texttt{W}, mapping each rebalancing date to a vector of optimized portfolio weights}

\texttt{CD} $\leftarrow$ \texttt{SelectCommonDrivers(Constituents, Drivers)}\;
\ForEach{$\tau \in$ 	exttt{RebalanceDates}}{
    \If{$\tau \in$ 	exttt{DriverUpdateDates}}{
        \texttt{CD} $\leftarrow$ \texttt{SelectCommonDrivers(Constituents, Drivers)}\;
    }
    \texttt{Sensitivities} $\leftarrow$ [\ ]\;
    \ForEach{asset $a \in$ 	exttt{Constituents}}{
        \texttt{NN} $\leftarrow$ \texttt{TrainNN}(asset, \texttt{CD})\;
        $\vec{s}_a \leftarrow$ \texttt{AAD}(\texttt{NN}, asset, \texttt{CD})\;
        $\bar{s}_a \leftarrow$ \texttt{AggregateSensitivity}($\vec{s}_a$)\;
        Append $\bar{s}_a$ to \texttt{Sensitivities}\;
    }
    \texttt{S} $\leftarrow$ \texttt{PairwiseDistances(Sensitivities, metric="Euclidean")}\;
    \texttt{S\_PD} $\leftarrow$ \texttt{NearestPositiveDefinite(S)}\;
    \texttt{LinkageTree} $\leftarrow$ \texttt{HierarchicalClustering(S\_PD)}\;
    \texttt{SortedIndex} $\leftarrow$ \texttt{OrderFromLinkage(LinkageTree)}\;
    \texttt{ReturnWindow} $\leftarrow$ \texttt{Constituents}[$(\tau - 3\text{months}):\tau$]\;
    \texttt{CovMatrix} $\leftarrow$ \texttt{Covariance(ReturnWindow)}\;
    \texttt{W[$\tau$]} $\leftarrow$ \texttt{RecursiveBisection(CovMatrix, SortedIndex)}\;
}
\caption{Hierarchical Sensitivity Parity (HSP)}
\label{HSP}
\end{algorithm}

Algorithm \ref{HSP} outlines the implementation of Hierarchical Sensitivity Parity (HSP). The function\\ \texttt{SelectCommonDrivers()} selects a set of common causal drivers using procedures defined in Chapter \ref{Subsection41}. The function \texttt{TrainNN()} trains a neural network for each asset, and \texttt{AAD()} computes the sensitivity of asset returns to the common drivers using automatic differentiation. The function \texttt{AggregateSensitivity()} reduces the sensitivity vector to a summary metric, typically using an average.

A pairwise Euclidean distance matrix is computed over the asset sensitivities using \\ \texttt{PairwiseDistances()}. This matrix is regularized to its nearest positive semi-definite form via \\ \texttt{NearestPositiveDefinite()}, following \citep{HIGHAM1988103}. Hierarchical clustering is then performed with \texttt{HierarchicalClustering()}, and \texttt{OrderFromLinkage()} retrieves the asset sort order.

A trailing window of asset returns is used to compute the sample covariance matrix with \texttt{Covariance()}, and portfolio weights are computed using \texttt{RecursiveBisection()}, which operates on the covariance matrix and the sorted asset order.

In contrast to Hierarchical Risk Parity (HRP) \citep{Prado2016BuildingDP}, which derives its clustering from correlation matrices, HSP constructs the hierarchy from causal sensitivity measures, thereby aligning diversification with structural causality. In the conducted experiments, \texttt{DriverUpdateDates} occur semi-annually, while \texttt{RebalanceDates} are set on a monthly basis.

\subsection{Implementation}

The data used in the implementation is sourced from Bloomberg and is organized into two primary datasets stored in a structured database:

\begin{itemize}
    \item The first dataset includes approximately 1200 time series spanning from 2012 to the end of 2021. It comprises:
    \begin{itemize}
        \item Spot and option prices (across various strikes and tenors) for key global foreign exchange (FX) crosses;
        \item Government bond yields with multiple maturities from major economies across all continents;
        \item Macroeconomic indicators (monthly and quarterly) for leading global economies;
        \item Equity indices from multiple geographic regions;
        \item Mutual fund indices covering government, corporate, investment-grade, and high-yield debt in Europe, the USA, Asia, and emerging markets;
        \item Credit market data, including Credit Default Swap (CDS) prices and mutual fund indices for investment-grade and high-yield credit in Europe and the USA;
        \item Commodity futures prices for globally traded assets;
        \item Smart beta Exchange Traded Funds (ETFs) representing widely accepted risk factors;
        \item Leading crypto assets;
        \item Option-implied volatility time series for equity indices across strikes and tenors (used as market risk indicators);
        \item Sector-specific equity ETFs for the USA and Europe.
    \end{itemize}
    This dataset represents a comprehensive collection of publicly available financial market data commonly utilized by practitioners, and serves as the pool of candidate drivers.

    \item The second dataset contains two groups of 14 daily stock price time series each, drawn respectively from the Stoxx 600 and S\&P 500 indices. The selection spans various sectors to ensure initial diversification. These subsets are summarized in Table \ref{porttable}.

    \item For all computations, including common driver selection and downstream analysis, all time series are transformed into return format by calculating the percentage change.
\end{itemize}
    
\begin{table}[!hbt]
\centering
\begin{tabularx}{\columnwidth} {@{} l*{3}{X} @{}}
\toprule
SXXP (Portfolio EU) & SPX (Portfolio USA)\\
\midrule
ASML HOLDING NV      & GENERAL ELECTRIC CO         \\
LVMH MOET HENNESSY LOUIS        & GOLDMAN SACHS GROUP INC     \\
SAP SE       & APPLE INC                   \\
SIEMENS AG-REG      & NVIDIA CORP                 \\
L'OREAL        & DOVER CORP                  \\
SANOFI       & FORD MOTOR CO               \\
ALLIANZ SE-REG       & ORACLE CORP                 \\
SCHNEIDER ELECTRIC SE        & PACKAGING CORP OF AMERICA   \\
TELEFONICA SA       & MCDONALD'S CORP             \\
BANCO SANTANDER SA       & PFIZER INC                  \\
INTL CONSOLIDATED AIRLINE-DI      & SCHLUMBERGER LTD            \\
REPSOL SA      & BLACKROCK INC               \\
INDRA SISTEMAS SA     & PHILIP MORRIS INTERNATIONAL \\
	GRIFOLS SA      & EQUINIX INC \\
\bottomrule
\end{tabularx}
\caption{Portfolios for experiments: One for Europe, other for USA markets}
\label{porttable}
\end{table}

The experiments are conducted by varying several hyperparameters. The number of selected common drivers, denoted by \( K \), is tested using values 10, 15, 20, and 30. The correlation window \( W_{CD} \), used for optimal driver selection, is tested using both 6-month and 12-month daily time series data. The neural network training window \( W_{NN} \) is evaluated with durations of 60, 90, and 125 market days. For correlation thresholds \( T_1 \) and \( T_0 \) at lags 1 and 0 respectively, a ranking-based selection rule is applied: for each candidate driver, correlations with all constituents are summed, and the top \( K \) candidates by total correlation are selected. Alternatively, thresholds \( T_1 \) and \( T_0 \) may be calibrated either to model performance or chosen such that the number of candidates passing the threshold equals \( K \). This flexibility accommodates cases where certain drivers overlap with existing portfolio constituents (e.g., an index containing a stock in the portfolio) or when the presence of spurious correlations necessitates exclusion.

Lag values of 0, 1, 2, 5, 10, and 20 days are considered between input drivers and output returns in neural networks (hyperparameter 6). Two methodological variants of the HSP algorithm are defined: SELECT, which permits manual adjustment to exclude redundant or spurious drivers or address multicollinearity; and OPT, where the algorithm selects the top \( K \) drivers purely based on correlation rankings without intervention. Driver selection may incorporate either lag 0, lag 1, or both, producing different driver sets and resulting portfolio performances. 

Portfolio rebalancing is scheduled monthly, with weights held constant throughout the following month. Experiments are run with two configurations: one where driver selection is updated at every rebalancing date, and another where common drivers are selected semiannually and kept fixed for six rebalancing cycles, with other modules (neural network modeling, sensitivity extraction, HSP execution) operating with the same driver set.

All performance metrics are computed out-of-sample. Historical data is used exclusively for training and inference, ensuring that all decisions are based only on past information. This condition is consistently applied to all benchmark methods. Portfolio constituents are fixed across all methods, and constraints are imposed on portfolio weights (between 3\% and 10\%) to avoid investment exclusion or concentration.

The benchmark methods include various mean-variance optimizations (Maximum Sharpe Ratio, Minimum Volatility, Quadratic Utility, Target Return), as well as the Hierarchical Risk Parity (HRP) algorithm \citep{Prado2016BuildingDP}. The experimental hypothesis is that the use of hierarchical clustering based on causal sensitivity dynamics improves performance compared to correlation-based approaches. Indirect comparison is also performed by substituting the common drivers with smart beta ETFs or sector/equity indices, simulating typical factor-based methods; observed performance is consistently inferior to the optimal driver selection method. This is a method similar to other methods in the literature that applies smart betas as causal driver candidates. 

\begin{remark}\textbf{Smart Beta Causal Models}:
 This example of the methodology is conceptually aligned with others in the literature that employ smart beta factors as causal driver candidates. However, by applying a selection process over this set and implementing the full methodology based on sensitivities, it was shown in \citep{RODRIGUEZDOMINGUEZ2023100447} that such approaches tend to perform poorly. The reason is that portfolio constituents often exhibit high correlation and significant multicollinearity with smart beta factors, hence they were the first to be omitted in the Reverse Engineering RCCP method together with equity indexes that contain the constituents itself (One variable cannot be the cause of itself). This introduces considerable challenges in using these factors effectively for portfolio risk management, as proposed in works such as \citep{Lopez_de_Prado_2023,Howard2025}. Not to mention in portfolio optimization, were apart from this drawback other limitations of these methods have been commented in the theoretical Section \ref{sectionGeometry} of this monograph, and validated empirically and theoretically in the original paper \citep{RODRIGUEZDOMINGUEZ2023100447}.   
\end{remark}

Regarding the portfolio optimization techniques, all methods utilize the same historical window for covariance or correlation matrix computation in portfolio optimization and recursive bisection procedures.

Due to clarity considerations, only top-performing strategies are visualized. The excluded results generally exhibit inferior performance.

The experimental results are presented for both U.S. and European portfolios. Performance across all methods is evaluated using Net Asset Value (NAV), which reflects the time series of dollar value of the strategy assuming an initial investment of 100 USD. The NAV is computed from 01/06/2020 to 01/12/2021. For the proposed method, common drivers are selected on 01/06/2020, 01/01/2021, and 01/07/2021. All methods implement rebalancing on the first day of each month, with weights held fixed for 30 days. 

Additional long-term experiments are conducted over a 6-year horizon, applying all modules as described, including portfolio selection carried out in each re-balancing date monthly. \par

\subsection{Short-term Investments}

\subsubsection{Portfolio USA}
\label{USAcase}
Experiments are conducted across all methods, including the equally weighted (1/N) strategy. Constraints are imposed to ensure practical weight distributions, such as a maximum allocation of 10\% per asset. For the 1/N strategy, with 14 assets, this equates to approximately 7.1\% per constituent. These constraints are applied to prevent portfolio concentration. Net Asset Values (NAVs) are computed assuming an initial investment of 100 USD, covering the period from 01/06/2020 to 01/12/2021.

\begin{figure}
	\centering
	\includegraphics[width=120 mm]{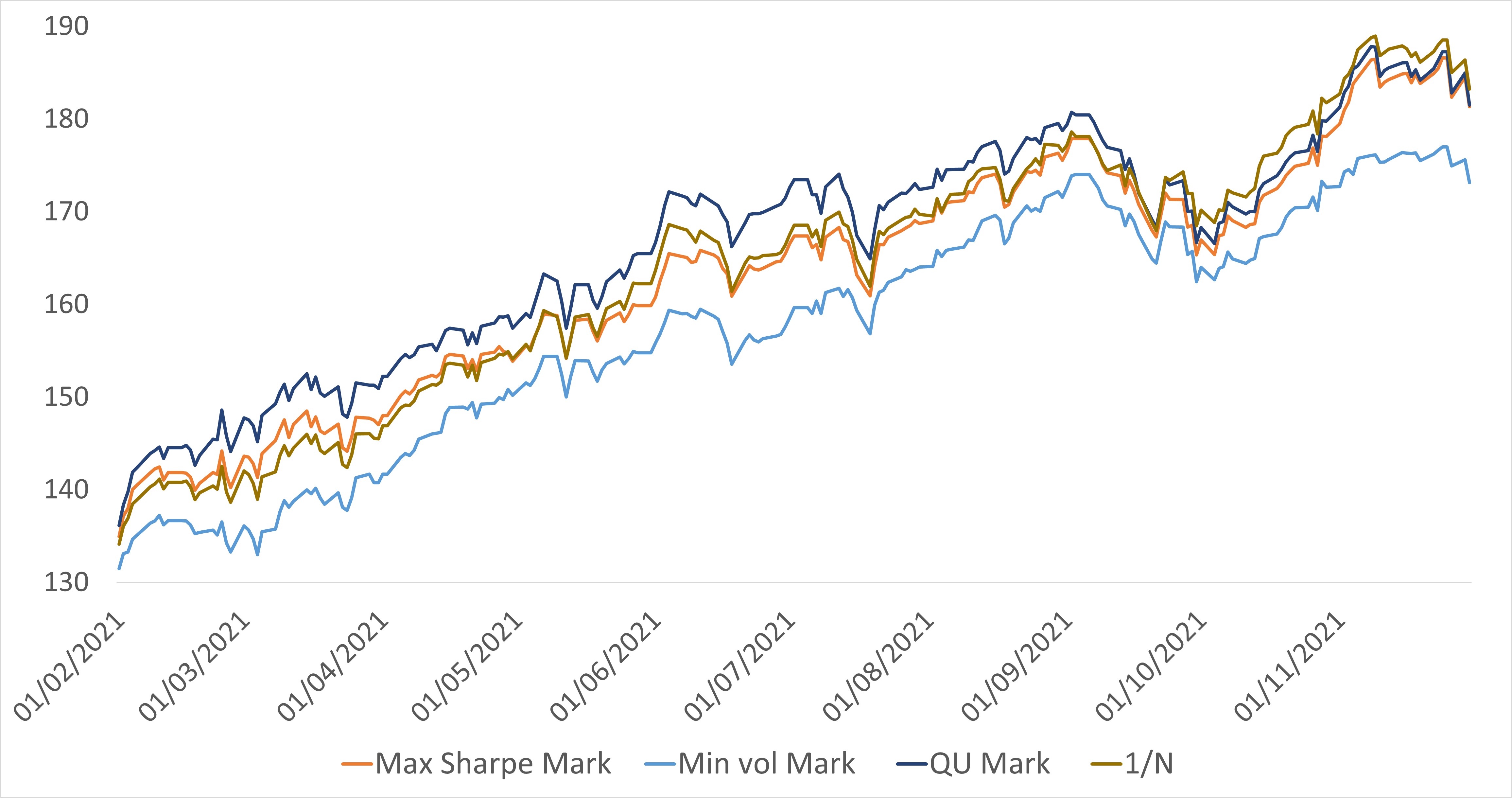}
	\caption{NAVs for the USA portfolio comparing top-performing mean-variance optimization methods with the equally weighted strategy. NAVs are tracked from 01/06/2020, highlighting the subperiod from 02/2021 to 12/2021.}
	\label{figure5}
\end{figure}

Table \ref{table2} presents the performance metrics—total return, annualized volatility, and Sharpe ratio—for the leading mean-variance optimization methods and the 1/N benchmark over the full evaluation period (01/06/2020 to 01/12/2021). Figure \ref{figure5} illustrates the NAV trajectories for these selected strategies during the subperiod from February to December 2021.

\begin{table}[!hbt]
\centering
\begin{tabularx}{\columnwidth} {@{} l*{4}{X} @{}}
\toprule
    & Max Sharpe(Mark) & Min Vol(Mark) & QU Mark & 1/N   \\
\toprule
\midrule
Return    & 49\%            & 44\%         & 49\%    & 50\%  \\
Vol (Ann) & 16\%            & 15\%         & 17\%    & 17\%  \\
Sharpe    & 3,061           & 3,002        & 2,920   & 3,006 \\
\bottomrule
\end{tabularx}
\caption{USA portfolio performance metrics for top mean-variance methods and 1/N: Returns, Risks, and Sharpes for full period: 01/06/2020 – 01/12/2021}
\label{table2}
\end{table}

The proposed method, Hierarchical Sensitivity Parity (HSP), is evaluated alongside benchmark portfolio optimization strategies. Table \ref{table3} reports comparative performance metrics for top mean-variance strategies, the Maximum Sharpe Ratio method, the equally weighted benchmark (1/N), and the Hierarchical Risk Parity (HRP) method from \citep{Prado2016BuildingDP}. The assessment also includes results for three high-performing HSP variants: HSP 6m LAG1 SELECT, HSP 6m LAG0 OPT, and HSP 6m LAG1 OPT, each of which outperforms the benchmark strategies.

Driver selection is based on correlation analysis computed over 6-month or 12-month historical windows, consistent with the methodology outlined in Section \ref{Subsection41App} (hyperparameter 2). Correlation thresholds \( T_1 \) and \( T_0 \), corresponding to lags 1 and 0 respectively (hyperparameters 4 and 5), are calibrated to yield a suitable number of candidate drivers such that the final set aligns with the number of portfolio constituents (hyperparameter 1). A discussion on the interdependencies between hyperparameters 1, 2, and 3 is presented in the subsequent section.

Neural network architectures for each portfolio asset are selected from a configuration space defined by the temporal lag structure of inputs and outputs—namely, lag 0, lag 1, or both—following the procedure described in Section \ref{subsection421} (hyperparameter 6). Model training is performed over windows of 60, 90, or 125 market days (hyperparameter 3) to identify the optimal neural network for sensitivity estimation.

The configuration HSP 6m LAG1 OPT employs a 6-month window for correlation-driven driver selection and applies lag-1 relationships in neural network modeling. In this variant, driver selection proceeds in a fully automated fashion based on correlation thresholds. Conversely, the SELECT configuration permits manual refinement of the driver set to exclude variables exhibiting spurious correlations, redundancy (such as a constituent index), or multicollinearity. Ranking in this case adheres to the commonality principle while honoring these constraints.

Due to variation in correlation structures over time, the number of selected drivers may change across rebalancing dates. As a result, hyperparameter 1 is dynamically linked to thresholds \( T_1 \) and \( T_0 \). 

\begin{figure}
	\centering
	\includegraphics[width=130 mm]{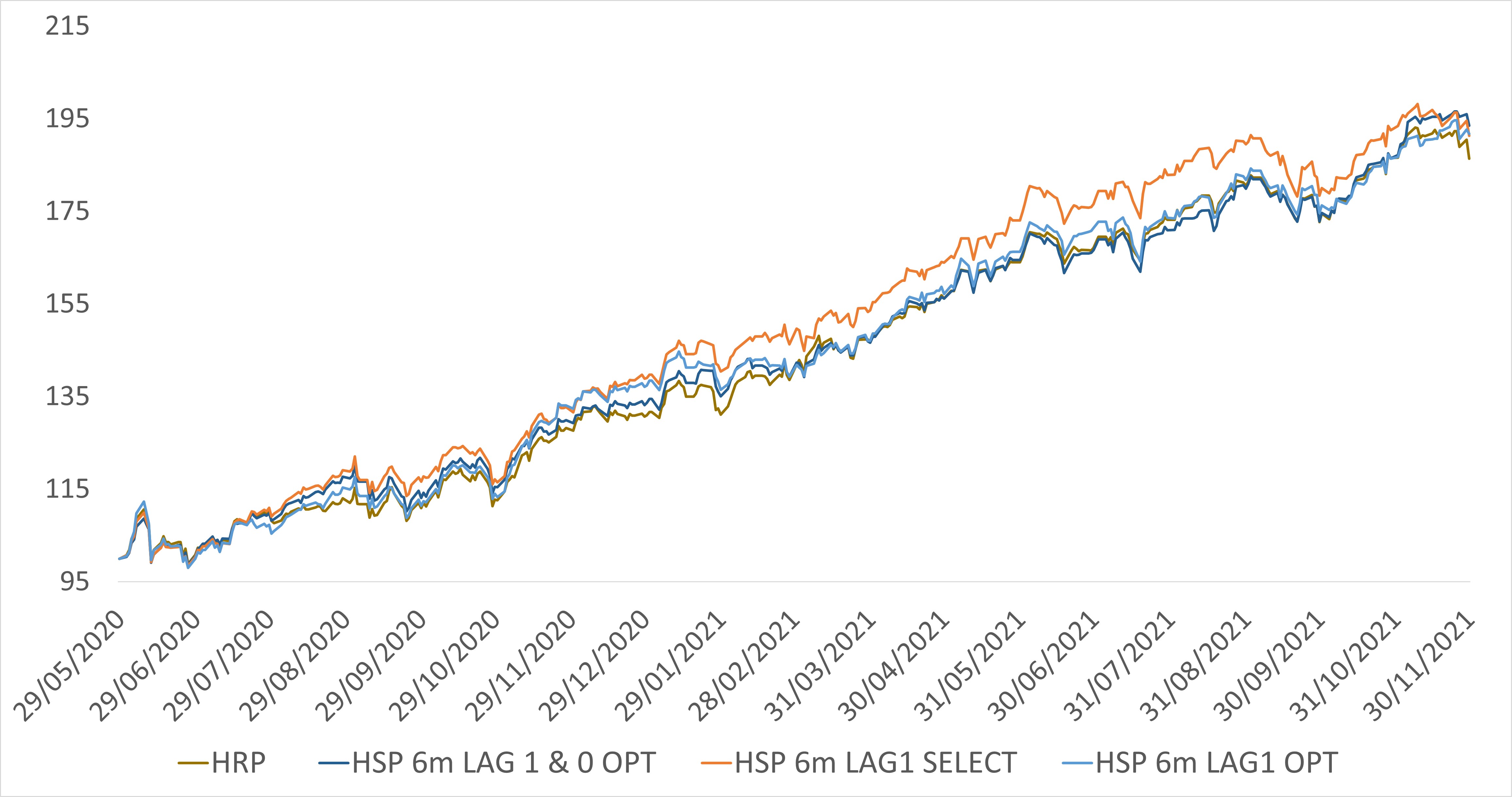}
	\caption{NAVs for USA portfolio for top mean-variance methods, 1/N, HRP, and HSP for different model hyperparameters: NAV starting from 01/06/2020. Showing the top 4 performers}
	\label{figure6}
\end{figure}

\begin{figure}
	\centering
	\includegraphics[width=130 mm]{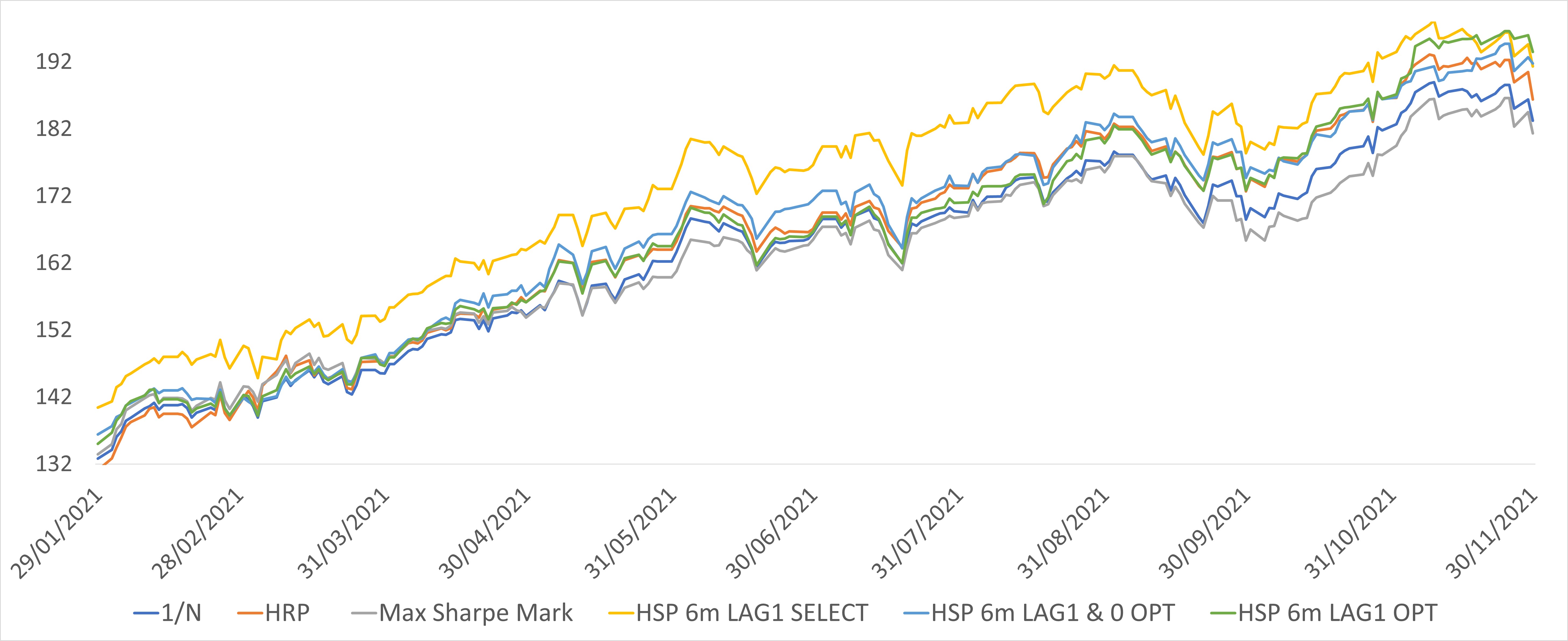}
	\caption{NAVs for USA portfolio for top mean-variance methods, 1/N, HRP, and HSP for different model hyperparameters: NAV starting from 01/06/2020, zoom of Figure \ref{figure6} showing subperiod from 01/2021}
	\label{figure7}
\end{figure}

Figure \ref{figure6} displays the Net Asset Value (NAV) trajectories for the full evaluation period, while Figure \ref{figure7} focuses on the subperiod spanning from 01/02/2021 to 01/12/2021. It is noteworthy that the final common driver selection occurred on 01/07/2021, approximately five months prior to the onset of performance deterioration observed across most models beginning in November 2021, as illustrated in Figure \ref{figure7}. This timing suggests that the previously optimal driver configuration may have become outdated. More frequent updates to the set of common drivers—rather than the fixed six-month interval used—might have mitigated the performance decline experienced by the models during this later period.

Subsection 5.3 presents evidence that increasing the frequency of driver updates enhances long-term performance. Notably, despite the general decline, the configuration HSP 6m LAG0 OPT maintained superior performance relative to other models during the subperiod in question.

\begin{table}[!hbt]
\centering
\begin{tabularx}{\columnwidth} {@{} l*{5}{X} @{}}
\toprule
          & HSP 6m LAG 1 SELECT & HSP 6m LAG 0 OPT & HSP 6m LAG 1 OPT & 1/N  & HRP \\
\toprule
\midrule
Return    & 54\%                & 55\%             & 54\%             & 50\% & 52\%               \\
Vol (Ann) & 17\%                & 17\%             & 17\%             & 17\% & 17\%               \\
Sharpe    & 3,157               & 3,340            & 3,116            & 3,0  & 2,954              \\
\bottomrule
\end{tabularx}
\caption{USA portfolio performance metrics for 1/N, HRP, HSP for different model hyperparameters: Returns, Risk and Sharpes for full period: 01/06/2020 – 01/12/2021}
\label{table3}
\end{table}

\subsubsection{Portfolio EU}
\label{EUcase}

The experimental setup for the European (EU) portfolio replicates that of the U.S. portfolio, with the only difference being the set of equity names. The proposed method continues to deliver the strongest performance in this context. Table \ref{table4} summarizes results for the leading mean-variance configuration, the equally weighted benchmark (1/N), the Hierarchical Risk Parity (HRP) method, and two selected configurations of the Hierarchical Sensitivity Parity (HSP) method.

The first configuration, HSP 6m LAG1 OPT, employs a 6-month correlation window for hyperparameter 2 and lag-1 for hyperparameter 6. The "OPT" designation indicates that the selection of common drivers is fully algorithmic, based on correlation thresholds (hyperparameters 4 and 5), which are tuned identically to the U.S. case to determine hyperparameter 1. 

The second configuration, HSP 6m LAG0\&1 SELECT, utilizes both lag-0 and lag-1 in the neural network configuration (hyperparameter 6) and allows manual refinement of the common driver set based on filtering spurious or redundant drivers. This variant modifies the selection of drivers that pass correlation thresholds (hyperparameters 4 and 5) according to the procedure previously described. Table \ref{table4} reports performance metrics for the full evaluation period, from 01/06/2020 to 01/12/2021.

\begin{figure}
	\centering
	\includegraphics[width=120 mm]{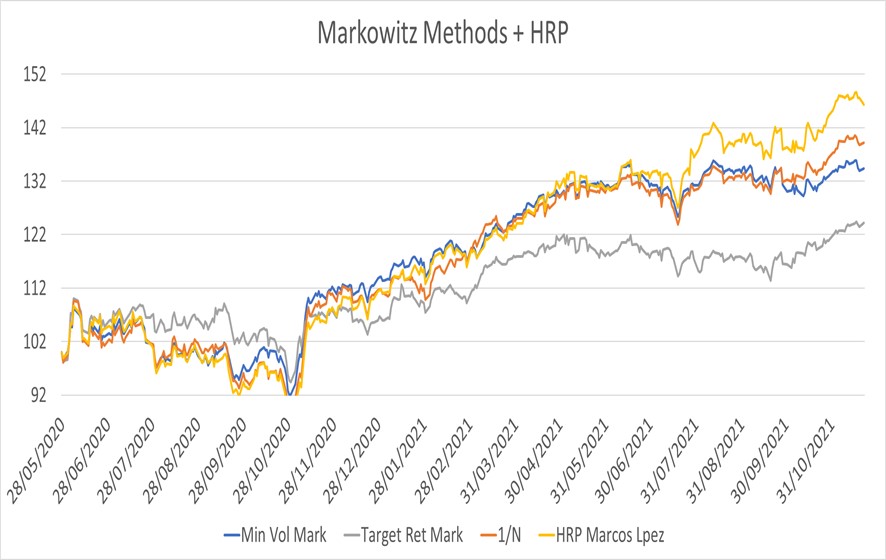}
	\caption{NAVs for the EU portfolio using top mean-variance methods, 1/N, and HRP, beginning from 01/06/2020.}
	\label{figure8}
\end{figure}

\begin{figure}
	\centering
	\includegraphics[width=120 mm]{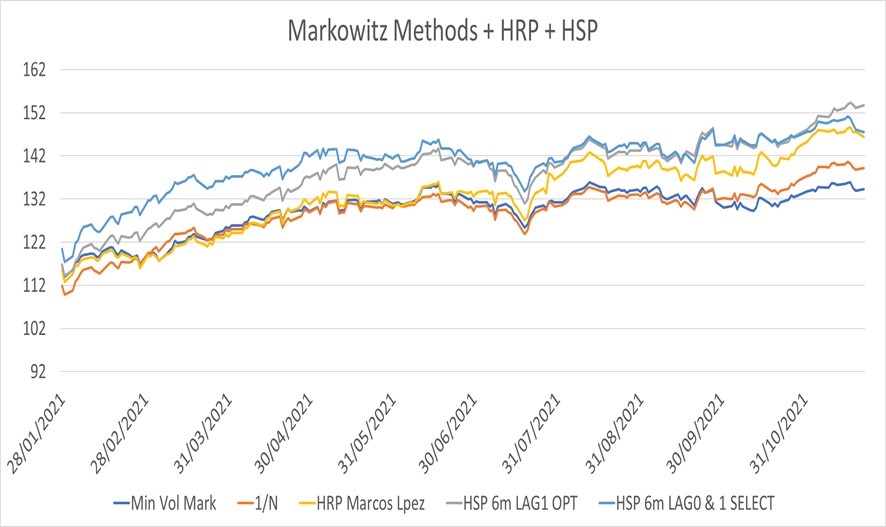}
	\caption{NAVs for the EU portfolio comparing top mean-variance methods, 1/N, HRP, and HSP variants. The plot shows only the final subperiod.}
	\label{figure9}
\end{figure}

Figure \ref{figure8} displays NAVs for benchmark methods only—mean-variance, 1/N, and HRP—while Figure \ref{figure9} includes the NAV trajectories of HSP configurations alongside the strongest-performing alternatives. The HSP method consistently outperforms the other models. Insights drawn from the U.S. portfolio analysis are also applicable here. Performance gains are likely attainable through further tuning of model parameters, such as common driver selection windows, selection rationale, update frequency, and neural network fitting periods.

\begin{table}[!hbt]
\centering
\begin{tabularx}{\columnwidth} {@{} l*{7}{X} @{}}
\toprule
          & Min vol  & Target Ret & 1/N    & HRP  & HSP 6m LAG 1 OPT & HSP 6m LAG0 \& 1 SELECT \\
\toprule
\midrule
Return    & 22\%         & 16\%            & 25\%   & 30\%               & 34\%             & 30\%                    \\
Vol (Ann) & 17\%         & 16\%            & 18\%   & 19\%               & 21\%             & 21\%                    \\
Sharpe    & 1,3014       & 0,9688          & 1,3740 & 1,5242             & 1,6494           & 1,433  
                \\
\bottomrule
\end{tabularx}
\caption{EU portfolio performance metrics for top mean-variance methods, 1/N, HRP, HSP for different model hyperparameters. Returns, Risk and Sharpes for full period: 01/06/2020 – 01/12/2021}
\label{table4}
\end{table}

\subsection{Long-term investments}

\begin{figure}
	\vspace*{-30mm}
	\centering
	\includegraphics[width=140 mm, height=10cm]{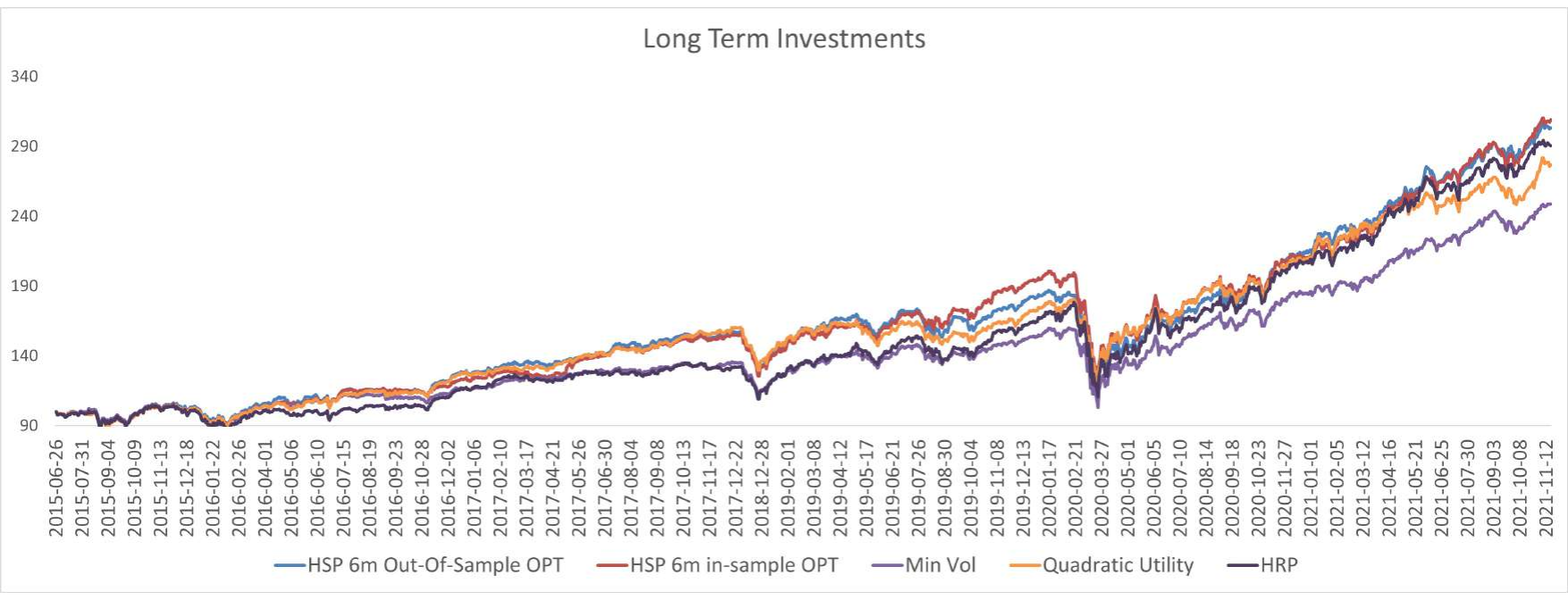}
	\vspace*{-20mm}
	\caption{NAV performance for the U.S. portfolio from 06/2015 to 12/2021.}
	\vspace*{-15mm}
	\label{figure17}
	\centering
	\includegraphics[width=140 mm,height=10cm]{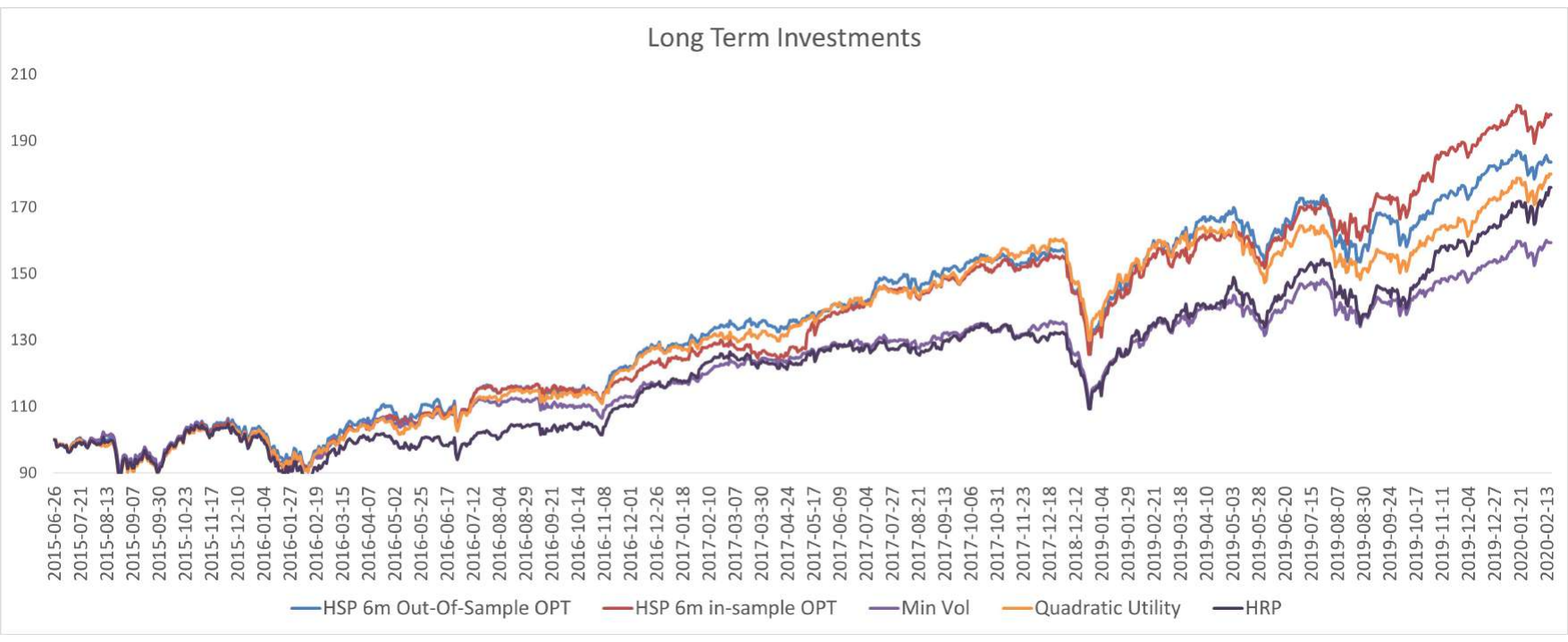}
	\vspace*{-20mm}
	\caption{Zoomed-in NAVs from Figure \ref{figure17}, highlighting the period 06/2015 to 03/2020 (pre-COVID).}
	\label{figure18}
\end{figure}

Long-term investment experiments for the U.S. portfolio, covering the period from June 2015 to December 2021, are conducted using the same methodological framework. As reported in Table \ref{table5} and visualized in Figures \ref{figure17}-\ref{figure18}, the Hierarchical Sensitivity Parity (HSP) method consistently demonstrates superior performance across return, Sharpe ratio, and Net Asset Value (NAV) metrics. Both configurations of HSP employ a 6-month window for the selection of common drivers and adopt the OPT strategy, indicating fully algorithmic selection based on correlation thresholds.

All results are out-of-sample, with the neural network sensitivities computed using test data rather than in-sample training data. Driver selection is updated monthly at each rebalancing point, which contributes to the performance enhancement compared to previous experiments involving fixed 6-month driver selection intervals. This improvement is evident in Figures \ref{figure17}-\ref{figure18} and Tables \ref{table5}-\ref{table6}.

\begin{table}[!hbt]
	\centering
	\begin{tabular}{lccccc}
		\hline\noalign{\smallskip}
		& HSP 6m Out-Of-Sa OPT & HSP 6m In-Sa OPT & Min Vol & Quadr Ut & HRP \\
		\noalign{\smallskip}\hline\noalign{\smallskip}
		Return    & 18.9\%  & 19.3\%  & 15.3\%  & 17.2\%  & 18.1\%  \\
		Vol (Ann) & 21.2\%  & 21.2\%  & 19.2\%  & 20.4\%  & 21.8\%  \\
		Sharpe    & 0.89    & 0.91    & 0.80    & 0.85    & 0.83    \\
		\noalign{\smallskip}\hline
	\end{tabular}
	\caption{Performance metrics for the U.S. portfolio over the period 06/2015–12/2021: comparison of top-performing mean-variance strategies, HRP, and HSP variants.}
	\label{table5}
\end{table}

Empirical evidence indicates that the optimal number of common drivers depends on the number of portfolio constituents. For a 14-asset portfolio, the optimal range is between 10 and 20 drivers. Selecting fewer drivers diminishes explanatory power, while using more introduces multicollinearity. Correlation thresholds are selected such that, on each driver selection date, the number of candidate drivers falls within this range. Shorter correlation windows (6 months) allow for quicker adaptation to evolving market dynamics.

In the OPT version of driver selection, correlation thresholds are applied to ensure the desired number of driver candidates. These thresholds may be calibrated using historical performance data. The SELECT configuration introduces manual filtering to address redundancy or spurious correlations, and may be used to limit the driver set to relevant instruments such as smart beta ETFs. These selections can also incorporate sector-based, geographical, or cross-asset information.

Regarding optimal neural network selection (as discussed in Section \ref{subsection421}), fitting windows generally range from 3 to 4.5 months. Longer windows degrade performance. Maintaining a consistent lag structure across all rebalancing dates yields better results than varying lags. Sensitivities can be computed using either training or test datasets. Euclidean distance is adopted for sensitivity-based distance calculations, although alternative metrics are permissible.

The HSP method applies single-linkage hierarchical clustering, but alternative clustering strategies may also be employed. Using a positive semi-definite neighbor matrix of the sensitivity distance matrix, as proposed in \citep{HIGHAM1988103}, improves stability and performance by aligning with recursive bisection-based weight computations.

\subsection{Two-Decade Investment Horizon: Application to Retirement and Pension Funds with Focus on the Credit Crisis}

\begin{figure}[t]
	\centering
	\includegraphics[width=140 mm]{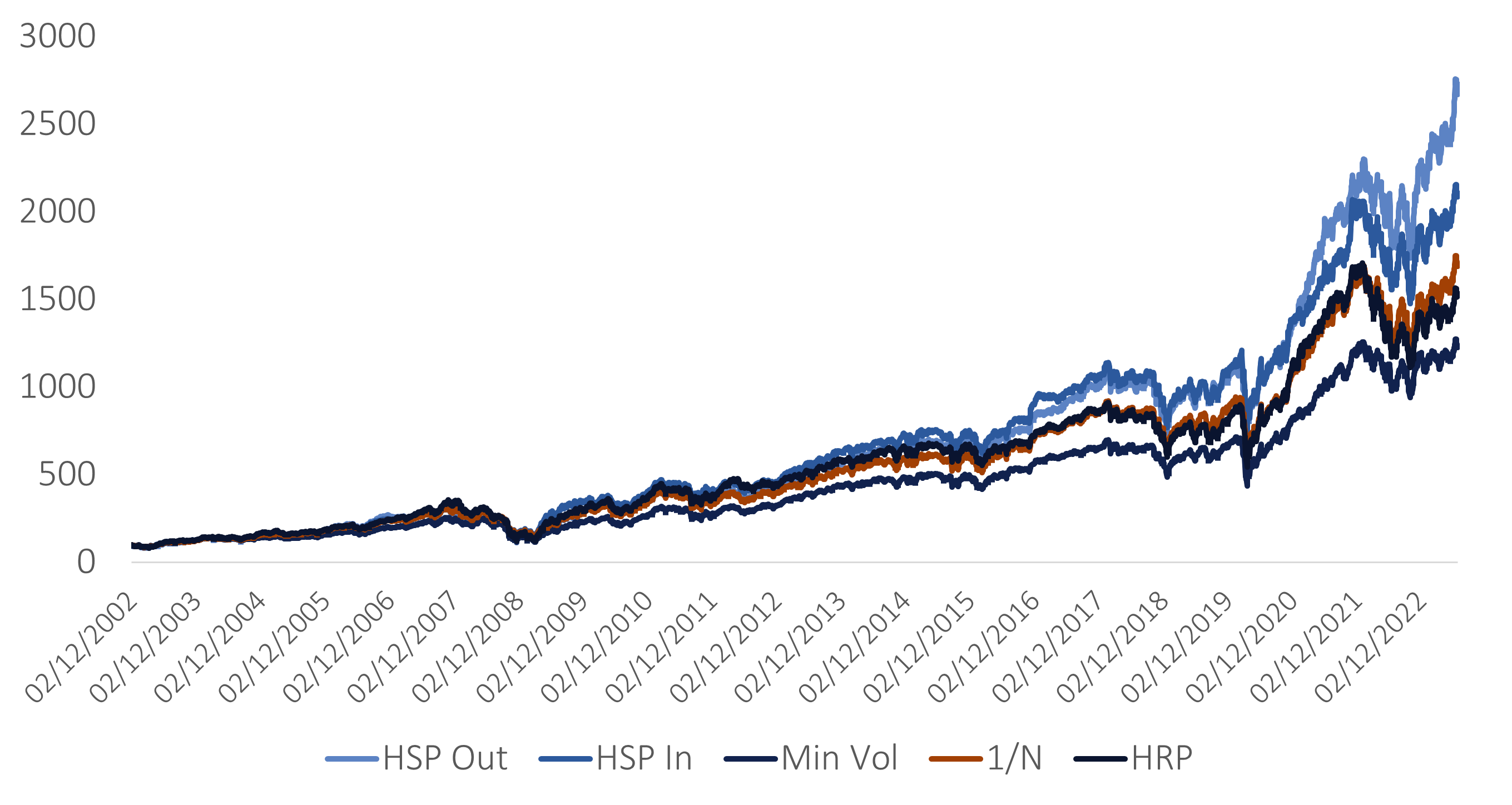}
	\caption{NAV performance over two decades (12/2002–12/2022) for the U.S. portfolio: comparison of top-performing strategies.}
	\label{figure192dec}
\end{figure}

To illustrate the performance of the HSP method as a suitable solution for retirement plans or other long-term investment strategies, Figure \ref{figure192dec} presents results spanning two decades. Table \ref{table6} extends the evaluation period and performance comparison to a full two-decade span. HSP 6m Out OPT and HSP 6m In OPT refer to solutions where the sensitivities are computed as the average values obtained from out-of-sample and in-sample sensitivity data, respectively, using information available up to each rebalancing point.

\begin{table}[!hbt]
	\centering
	\begin{tabular}{lccccc}
		\hline\noalign{\smallskip}
		& HSP 6m Out OPT  & HSP 6m In OPT & Min Vol & 1/N  & HRP \\
		\noalign{\smallskip}\hline\noalign{\smallskip}
		$NAV_0$   & 100\%  & 100\%  & 100\%  & 100\%  & 100\%  \\
		$NAV_f$   & 2279\% & 2107\% & 1236\% & 1706\% & 1533\% \\
		Return (Ann)  & 17.43\% & 15.97\% & 12.98\% & 14.77\% & 14.20\% \\
		Vol (Ann) & 23.02\% & 23.07\% & 22.89\% & 22.06\% & 21.99\% \\
		Sharpe    & 0.76    & 0.69    & 0.57    & 0.67    & 0.65    \\
		\noalign{\smallskip}\hline
	\end{tabular}
	\caption{Performance metrics for the U.S. portfolio over the period 12/2002–12/2022. Comparison of top-performing mean-variance strategies, HRP, and HSP variants. HSP 6m Out OPT is Out-of-sample data for computing the sensitivities during training and HSP 6m In OPT is the In-sample case.}
	\label{table6}
\end{table}

Not only does the HSP method outperform all other models, but the variant using out-of-sample sensitivities also demonstrates superior performance compared to the in-sample case.

\begin{figure}[t]
    \centering
    \includegraphics[width=0.75\linewidth]{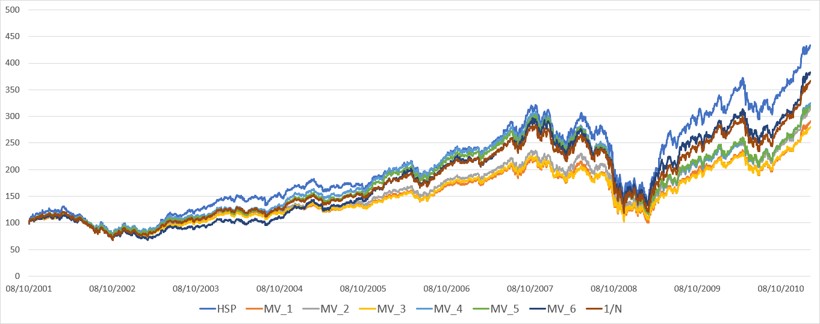}
    \caption{NAV performance during the 2008 Financial Crisis (Credit Crunch)}
    \label{CreditCrunk}
\end{figure}

The robustness of the HSP framework is further demonstrated in Figure \ref{CreditCrunk}, which depicts its behavior during the 2008 financial crisis. A novel approach to portfolio risk management—referred to as Common Causal Manifold Risk Management—is also introduced. Unlike traditional strategies that rely on monitoring random variables and absolute references such as volatility, this approach allows portfolio and risk managers to track the evolution of a persistent structure: the common causal manifold. Assets are effectively hedged through their projections onto this manifold.

\begin{figure}[t]
    \centering
    \includegraphics[width=1\linewidth]{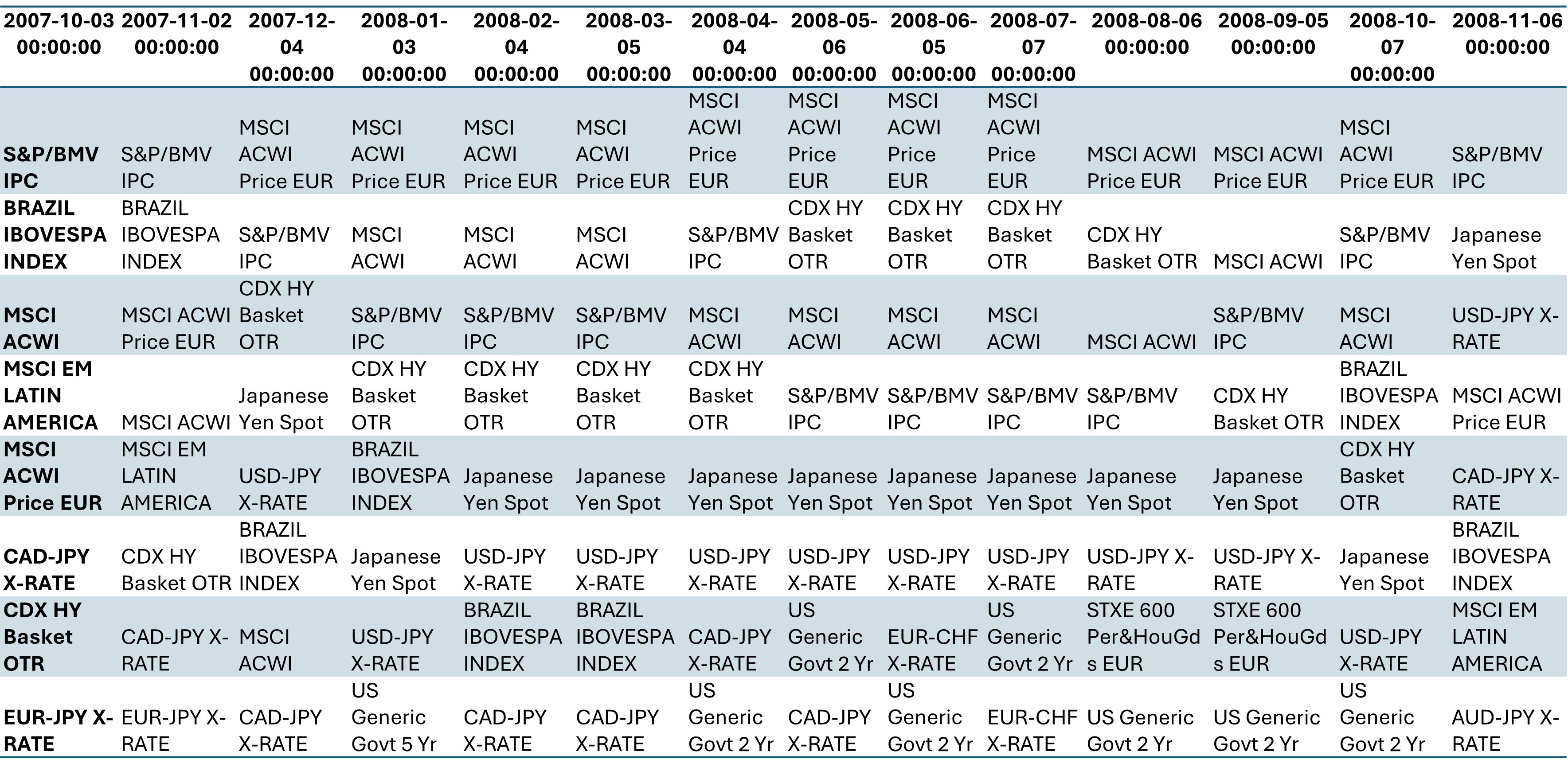}
    \caption{Monthly-selected common drivers for the U.S. portfolio during 2008.}
    \label{fig:enter-CommonCauseMonifold}
\end{figure}

The manifold, derived from common and causal drivers identified via the Commonality Principle, offers greater temporal stability and tractability than tracking the portfolio constituents directly. This makes it a more reliable reference for portfolio behavior under a wide range of market conditions. An example of such a manifold is illustrated in Figure \ref{fig:enter-CommonCauseMonifold}, representing the 2008 crisis period with monthly recalibrations of its drivers. Naturally, the shape and quality of the manifold depend on the quality of the data and the selection process employed by the user—an inherent characteristic of any investment methodology—but the figure effectively captures the core idea.

While this conceptual extension was not introduced in the original publication, it has been presented and discussed in various academic and professional settings, including the 21st WBS Quant Finance Conference in Valencia (2022), the AIFI Bootcamp Seminar Series (2024) \citep{dominguez2024portfolios}, the Santiago de Compostela Seminar (ECOBAS, 2024), Quantitative Finance Made Accessible (2024), and the 7th Machine Learning in Finance Conference by Marcus Evans in Amsterdam (2024). These presentations also preview aspects covered in the next chapter, which elaborates on manifold-based sensitivity models for portfolio optimization.

\newpage

\section{An Introduction to the Geometry of Modern Portfolio Theory and Factor Models}

The portfolio selection problem, introduced by Harry Markowitz in 1952, seeks to determine the optimal allocation of an investor’s wealth across a set of assets to achieve the best trade-off between expected return and risk \citep{10.2307/2975974}. It is formulated as follows:

Let there be $N$ assets, each with an associated random return $r_i$, for $i = 1, 2, \dots, N$. Define:
\begin{itemize}
    \item $w_i$ as the proportion of total wealth allocated to asset $i$,
    \item $\mathbf{w} = (w_1, w_2, \dots, w_N)^T$ as the portfolio weight vector,
    \item $\mu_i = \mathbb{E}[r_i]$ as the expected return of asset $i$,
    \item $\boldsymbol{\mu} = (\mu_1, \mu_2, \dots, \mu_N)^T$ as the expected return vector,
    \item $\Sigma$ as the $N \times N$ covariance matrix of asset returns, where $\sigma_{ij} = \text{Cov}(r_i, r_j)$.
\end{itemize}

The expected return of the portfolio is given by:
\begin{equation}
    R_p = \mathbf{w}^T \boldsymbol{\mu}.
\end{equation}

The risk of the portfolio, measured by variance, is given by:
\begin{equation}
    \sigma_p^2 = \mathbf{w}^T \Sigma \mathbf{w}.
\end{equation}

\subsection{The Covariance Matrix as a Metric Space of Risk}

The covariance matrix $\Sigma$ plays a crucial role in defining a metric space for risk assessment. A metric space $(X, d)$ consists of a set $X$ and a function $d: X \times X \to \mathbb{R}$ satisfying the following properties:

\begin{enumerate}
    \item \textbf{Non-negativity}: $d(x, y) \geq 0$ for all $x, y \in X$, with $d(x, y) = 0$ if and only if $x = y$.
    \item \textbf{Symmetry}: $d(x, y) = d(y, x)$ for all $x, y \in X$.
    \item \textbf{Triangle inequality}: $d(x, z) \leq d(x, y) + d(y, z)$ for all $x, y, z \in X$.
\end{enumerate}

In portfolio theory, the covariance matrix defines a pseudo-metric space where the distance between two portfolios is given by:
\begin{equation}
    d(\mathbf{w}_1, \mathbf{w}_2) = \sqrt{(\mathbf{w}_1 - \mathbf{w}_2)^T \Sigma (\mathbf{w}_1 - \mathbf{w}_2)}.
\end{equation}
This metric represents the risk difference between two portfolios, effectively quantifying diversification benefits and sensitivity to asset correlations \citep{riskmetrics}.

\subsection{Optimization Problem}

The investor seeks to optimize their portfolio by solving the following problem:
\begin{equation}
    \min_{\mathbf{w}} \quad \mathbf{w}^T \Sigma \mathbf{w},
\end{equation}
subject to:
\begin{equation}
    \mathbf{w}^T \boldsymbol{\mu} \geq R^*,
\end{equation}
\begin{equation}
    \sum_{i=1}^{N} w_i = 1,
\end{equation}
\begin{equation}
    w_i \geq 0, \quad \forall i,
\end{equation}
where $R^*$ is the minimum acceptable return level. This problem is known as the mean-variance optimization and is the foundation of modern portfolio theory \citep{markowitz1959}.

Euclidean geometry remains a cornerstone of mathematical thought, providing the basis for various applications in physics, engineering, and computer science \citep{hartshorne_geometry}. Similarly, the portfolio selection problem is a fundamental problem in financial mathematics, where optimization techniques are used to balance return and risk effectively. The covariance matrix $\Sigma$ defines a natural metric space for risk, allowing precise quantification of portfolio diversification \citep{merton_continuous}.

Financial markets exhibit varying degrees of risk and return trade-offs, often influenced by systemic shocks and investor behavior. During periods of stability, portfolio allocations remain relatively predictable, with efficient frontiers offering well-defined risk-adjusted return profiles. Investors position themselves along this frontier, balancing expected returns with their risk tolerance. However, financial crises significantly disrupt these relationships, causing extreme shifts in portfolio efficiency, risk structures, and capital market expectations.

\begin{figure}[h]
\centering
\includegraphics[width=0.8\textwidth]{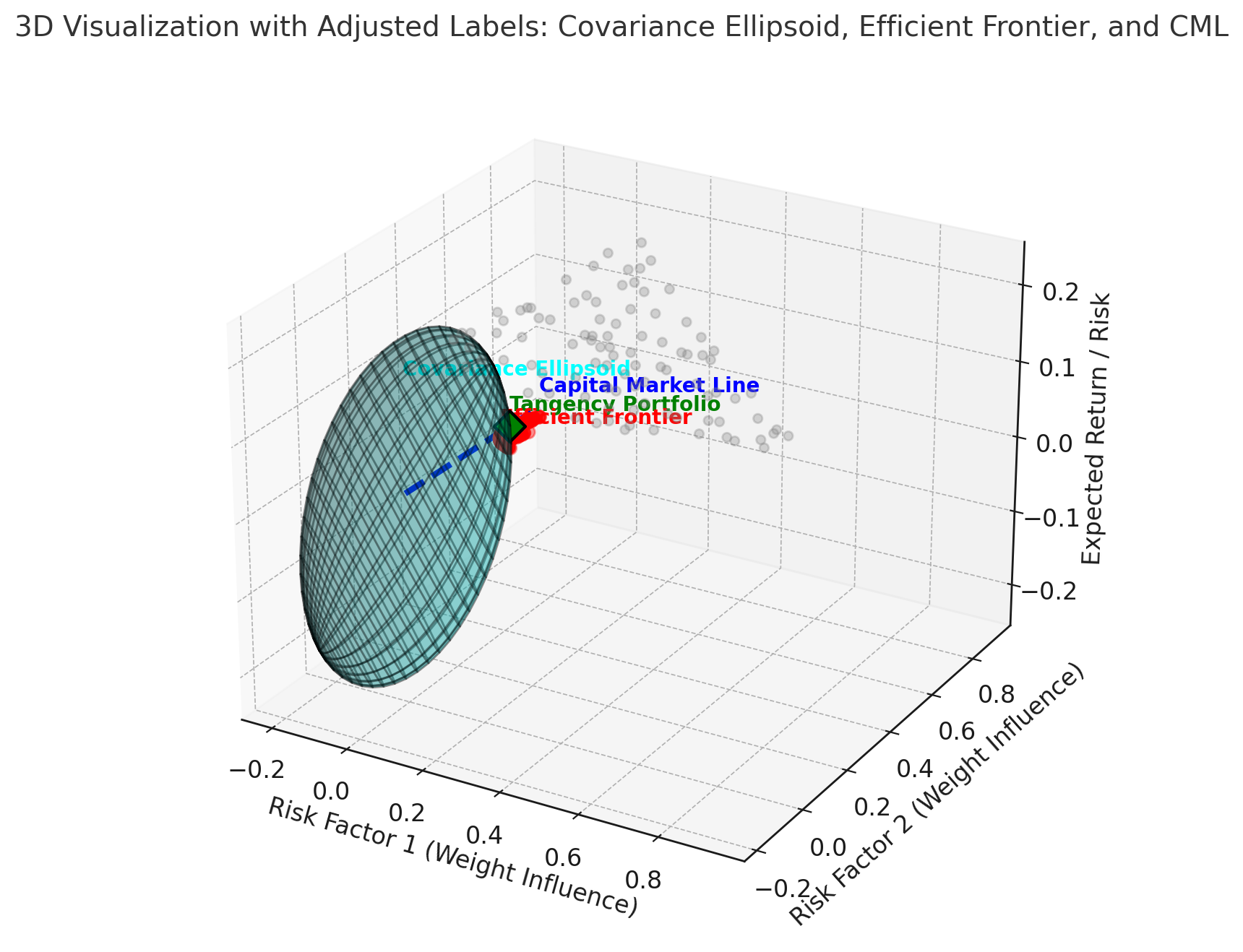}
\caption{Full covariance ellipsoid representation of risk.}
\label{fig:full_ellipsoid}
\end{figure}

In times of extreme market stress, such as the 2008 financial crisis, correlations between assets increase dramatically, reducing the benefits of diversification. This leads to a phenomenon where the efficient frontier collapses downward, reflecting deteriorating risk-adjusted returns. High-volatility assets become riskier without corresponding increases in expected return, rendering many previously optimal portfolios suboptimal. The covariance structure of the market shifts as volatility surges, often by factors of two or three, leading to a covariance ellipsoid that expands unpredictably in multiple risk dimensions \citep{10.2307/2975974}.

\begin{figure}[h]
\centering
\includegraphics[width=0.8\textwidth]{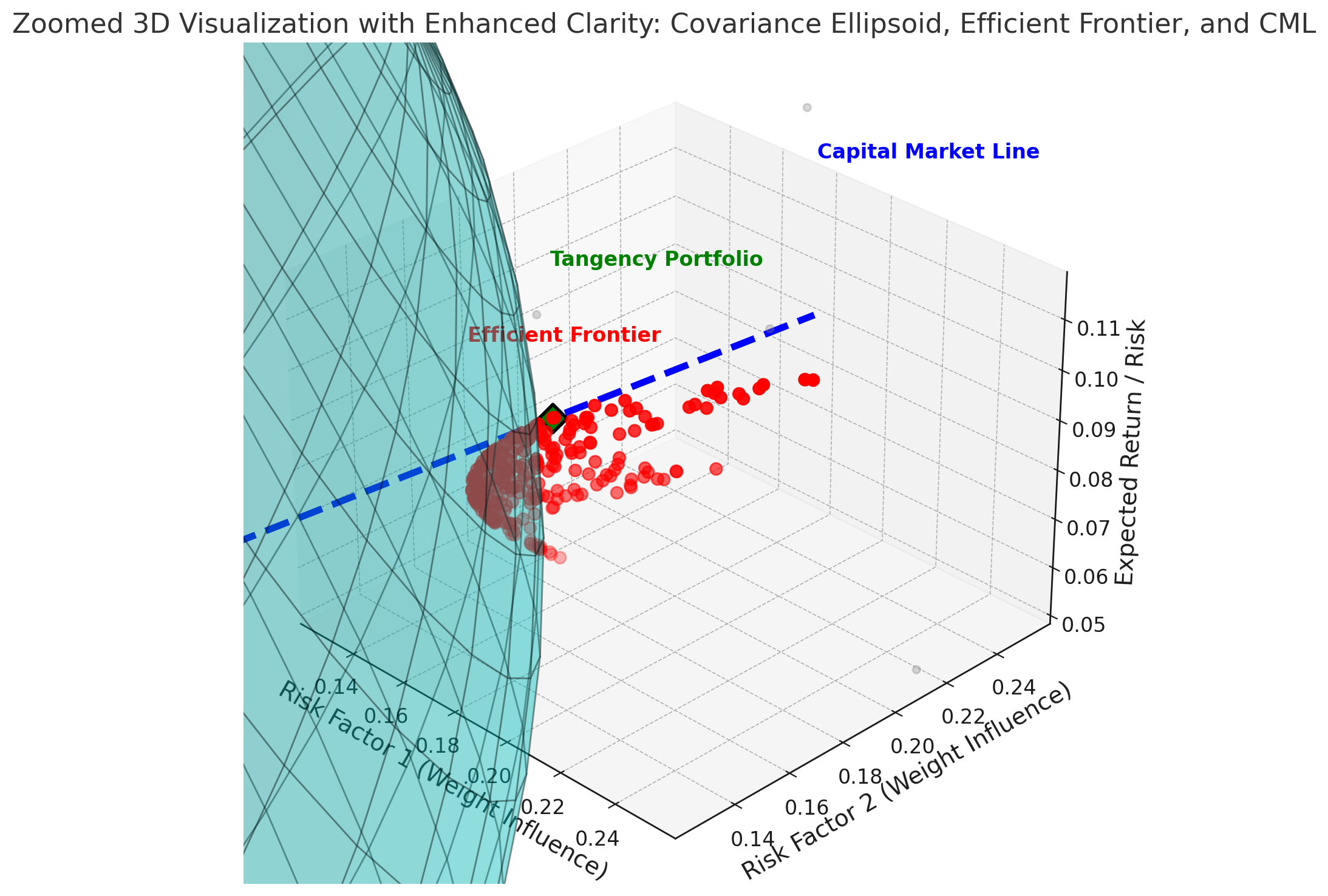}
\caption{Zoomed-in view of risk ellipsoid to emphasize localized effects.}
\label{fig:zoom_ellipsoid}
\end{figure}

The tangency portfolio, which represents the highest Sharpe ratio investment under normal conditions, migrates towards lower-risk assets during crises. Investors, fearing extreme downside risk, reallocate capital away from equities and high-yield assets, favoring government bonds and cash equivalents \citep{fama1973risk}. This shift is reflected in the Capital Market Line (CML), which steepens, favoring risk-free assets over riskier investments. As a result, aggressive investors suffer significant drawdowns, while conservative portfolios maintain relative stability.

\begin{figure}[h]
\centering
\includegraphics[width=0.8\textwidth]{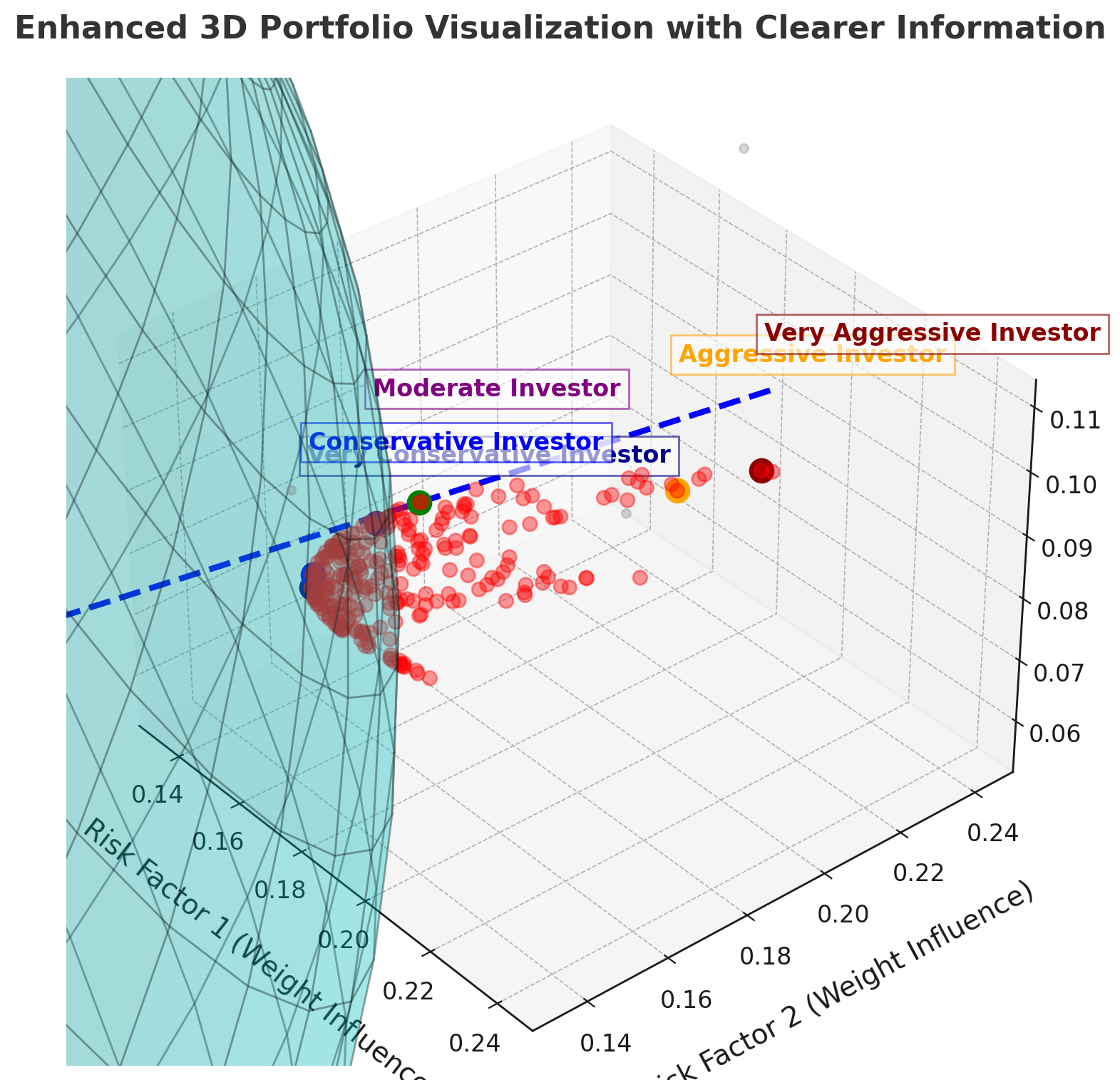}
\caption{Zoomed-in view with highlighted investor profiles.}
\label{fig:zoom_investors}
\end{figure}

During a financial crisis, capital market dynamics reinforce investor panic, exacerbating market inefficiencies. Liquidity evaporates as market participants rush for safe-haven assets, further increasing the downward pressure on returns. The role of behavioral finance becomes prominent in these periods, as fear and uncertainty drive asset mispricing \citep{shiller1981stock}. The collapse of financial institutions and regulatory interventions introduce additional distortions, creating feedback loops that prolong periods of instability.

\begin{figure}[h]
\centering
\includegraphics[width=0.8\textwidth]{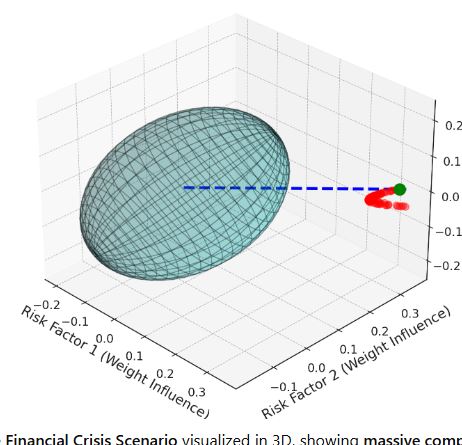}
\caption{Final extreme market crisis scenario showing collapsed efficient frontier.}
\label{fig:extreme_crisis}
\end{figure}

Following extreme downturns, markets enter recovery phases where risk premiums adjust, and asset correlations gradually revert to pre-crisis levels. The efficient frontier regains an upward trajectory, though with lingering volatility. The CML begins to flatten as risk-free rates normalize, leading investors to reconsider risk-taking strategies. Portfolio construction during recovery periods requires a reassessment of asset allocations, emphasizing resilience against future shocks \citep{merton1973intertemporal}.

Understanding these extreme cases provides insight into portfolio management strategies under market stress. Defensive allocations, such as increasing exposure to uncorrelated assets or hedging against volatility spikes, become critical. The study of financial crises highlights the importance of dynamic portfolio adjustments and robust risk assessment frameworks, ensuring that investors can navigate market cycles with greater confidence and efficiency.

\subsection{Dynamic Portfolio Movement Under Changing Risk Conditions}

Portfolio optimization typically assumes a static risk environment, where asset correlations and volatilities remain stable. However, real-world financial markets exhibit dynamic shifts in expected returns and risk levels, especially during periods of crisis. This paper explores the movement of a portfolio on the covariance ellipsoid when risk is allowed to change over time, including scenarios where covariance matrices shift due to market crises.

\subsubsection{Portfolio Movement on the Covariance Ellipsoid}

A portfolio's risk is determined by the covariance matrix  and the weight vector , with the variance given by:
\begin{equation}
\sigma_p^2 = w^T \Sigma w.
\end{equation}
When risk is fixed, portfolio movements occur along the surface of the covariance ellipsoid, meaning weight adjustments maintain a constant variance while optimizing expected return:
\begin{equation}
E[R_p] = w^T E[R].
\end{equation}
The optimal direction of movement follows the gradient of expected return, constrained by:
\begin{equation}
2 w^T \Sigma \delta w = 0,
\end{equation}
which ensures that movements occur within the tangent space of the ellipsoid.

\subsubsection{Extending the Model: Allowing Risk to Change}

In a more realistic scenario, portfolio risk evolves due to market fluctuations or investor-driven changes in risk appetite. If the covariance matrix shifts over time, portfolio variance becomes:
\begin{equation}
\sigma_p^2 (t) = w(t)^T \Sigma(t) w(t).
\end{equation}
Differentiating this equation with respect to time gives the rate of change of risk:
\begin{equation}
\frac{d}{dt} \sigma_p^2 (t) = 2 w^T \Sigma \frac{dw}{dt} + w^T \frac{d\Sigma}{dt} w.
\end{equation}
Here, $\frac{dw}{dt}$ represents portfolio re-balancing, while $\frac{d\Sigma}{dt}$ captures shifts in market risk dynamics.

\subsubsection{Crisis-Induced Covariance Shifts}

During financial crises, risk relationships between assets intensify, leading to increased correlation and volatility. This results in an expanded covariance ellipsoid, capturing the new risk landscape:
\begin{equation}
\Sigma_{crisis} = \Sigma_0 + \Gamma(t),
\end{equation}
where $\Gamma(t)$ represents the covariance shock. As a result, portfolio risk grows dynamically:
\begin{equation}
\sigma_p^2 (t) = w^T \Sigma_{crisis} w.
\end{equation}
Portfolio trajectories are no longer restricted to the original ellipsoid but can move outward as systemic risk escalates.

\begin{figure}[h]
\centering
\includegraphics[width=0.6\textwidth]{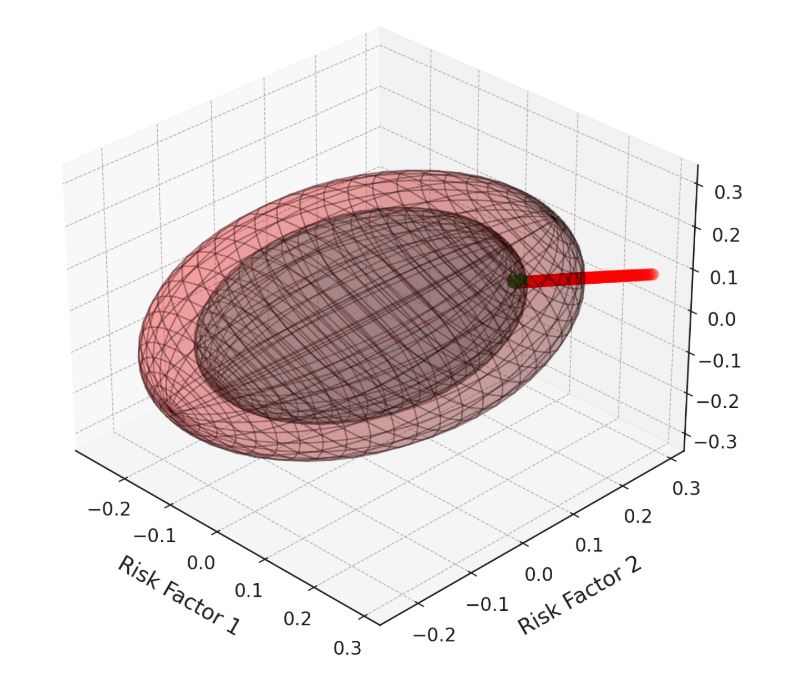}
\caption{Expansion of the covariance ellipsoid during a financial crisis, illustrating increased portfolio risk.}
\label{fig:crisis_scenario}
\end{figure}

\subsubsection{Non-Linear Crisis Dynamics}

Instead of a smooth transition from normal to crisis risk conditions, real-world market crises often unfold non-linearly. A sigmoid function can be used to model a crisis's abrupt escalation:
\begin{equation}
\lambda(t) = \frac{1}{1 + e^{-k(t - t_0)}},
\end{equation}
where $\lambda(t)$ determines the crisis transition intensity over time, with $k$ controlling the sharpness of escalation. The evolving portfolio risk follows:
\begin{equation}
\sigma_p^2 (t) = \sigma_p^2 (0) + \lambda(t) \Delta \sigma^2,
\end{equation}
where $\Delta\sigma^2$ is the additional variance induced by the crisis.

 Traditional movement along the covariance ellipsoid assumes fixed risk, but allowing risk to change introduces new dynamics that better reflect real-world market behavior. Future work should explore stochastic models where crises unfold with varying probabilities and magnitudes, incorporating investor sentiment and liquidity constraints.

\subsection{The One-Factor Model}

In the one-factor model, the return of an asset  is expressed as:
\begin{equation}
R_i = \alpha_i + \beta_i F + \epsilon_i,
\end{equation}
where:
\begin{itemize}
\item  is the asset's expected return independent of the factor,
\item  represents the sensitivity of the asset to the factor,
\item  is the common systematic factor,
\item  is the idiosyncratic error term, assumed to be uncorrelated with the factor .
\end{itemize}

The expected return of asset  is then:
\begin{equation}
E[R_i] = \alpha_i + \beta_i E[F].
\end{equation}

\subsubsection{Portfolio Return and Variance}

For a portfolio with weights , the portfolio return is given by:
\begin{equation}
R_p = \sum_{i=1}^{n} w_i R_i = \sum_{i=1}^{n} w_i (\alpha_i + \beta_i F + \epsilon_i).
\end{equation}
Taking expectations,
\begin{equation}
E[R_p] = \sum_{i=1}^{n} w_i (\alpha_i + \beta_i E[F]) = \alpha_p + \beta_p E[F],
\end{equation}
where:
\begin{equation}
\alpha_p = \sum_{i=1}^{n} w_i \alpha_i, \quad \beta_p = \sum_{i=1}^{n} w_i \beta_i.
\end{equation}

The variance of the portfolio return is given by:
\begin{equation}
\sigma_p^2 = \beta_p^2 \sigma_F^2 + \sum_{i=1}^{n} w_i^2 \sigma_{\epsilon_i}^2,
\end{equation}
where:
\begin{itemize}
\item  is the variance of the factor,
\item  is the variance of the idiosyncratic error term.
\end{itemize}

\subsubsection{Optimization Problem}

The investor seeks to minimize portfolio risk while achieving a target expected return. The optimization problem can be formulated as:
\begin{equation}
\min_w \quad \sigma_p = \sqrt{\beta_p^2 \sigma_F^2 + \sum_{i=1}^{n} w_i^2 \sigma_{\epsilon_i}^2} \quad \text{subject to} \quad \alpha_p + \beta_p E[F] \geq R_{target}, \quad \sum_{i=1}^{n} w_i = 1.
\end{equation}
This formulation ensures that the portfolio's expected return meets or exceeds a specified target while minimizing overall risk. The resulting allocation balances systematic risk exposure and return expectations.

\subsubsection{Risk Components and Their Effect on the Covariance Ellipsoid}

The shape of the covariance ellipsoid in a one-factor model is determined by systematic and idiosyncratic risk components. Systematic risk, captured by factor variance, stretches the ellipsoid along the dominant risk direction, while idiosyncratic risk contributes to a more uniform spread across different assets. This interaction defines how portfolio risk is structured and how asset correlations impact overall risk distribution.


\begin{table}[h]
\centering
\begin{tabularx}{\textwidth}{lX}
\textbf{Risk Component} & \textbf{Effect on Ellipsoid Shape} \\
\toprule
Higher factor variance ($\sigma_F^2$) & Stretches the ellipsoid along the systematic risk direction. \\
Higher idiosyncratic risk ($\sigma_{\epsilon_i}^2$) & Makes the ellipsoid more spherical, spreading risk across all assets. \\
Higher factor loadings ($\beta$) & Causes a more elongated ellipsoid along the dominant factor direction. \\
Lower factor impact, higher idiosyncratic risk & Results in a more uniform, rounded ellipsoid. \\
\bottomrule
\end{tabularx}
\caption{Effect of Risk Components on the Shape of the Covariance Ellipsoid}
\label{tab:risk_ellipsoid}
\end{table}

\begin{figure}[h]
\centering
\includegraphics[width=0.8\textwidth]{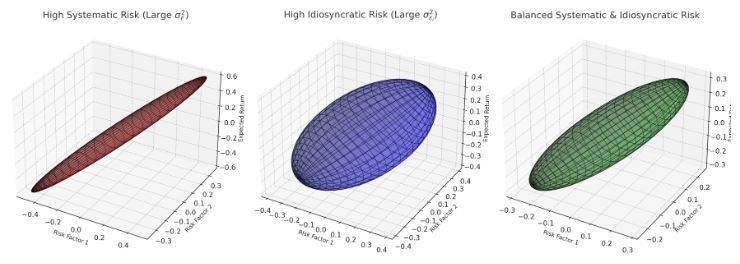}
\caption{Three cases illustrating the impact of systematic and idiosyncratic risk on the covariance ellipsoid. Left: High systematic risk (red). Middle: High idiosyncratic risk (blue). Right: Balanced systematic and idiosyncratic risk (green).}
\label{fig:risk_ellipsoid_cases}
\end{figure}

In Figure \ref{fig:risk_ellipsoid_cases}, the three cases illustrate how systematic and idiosyncratic risk shape the covariance ellipsoid. The leftmost ellipsoid (red) is stretched in a dominant direction due to high systematic risk, meaning most risk stems from market-wide factors. The middle ellipsoid (blue) is more spread out in all directions, indicating greater individual asset influence and less systematic dominance. The rightmost ellipsoid (green) represents a more realistic balance, where both common market movements and asset-specific risks contribute to portfolio risk.

\subsubsection{Why Some Portfolios Are Inside the Ellipsoid but Not on the Efficient Frontier}

Not all portfolios that fall within the covariance ellipsoid are efficient. In fact, many of them exhibit higher risk than is necessary for a given level of expected return. If a portfolio lies inside the ellipsoid but not precisely on its surface, it implies that there exists another portfolio with the same expected return but lower variance. This makes the interior portfolio inefficient. The efficient frontier, by construction, must reside on the surface of the ellipsoid. It represents the boundary of risk-return combinations that are not only feasible but optimal.

\subsubsection{Why Some Portfolios on the Ellipsoid Are Not on the Efficient Frontier}

Even when a portfolio lies on the surface of the covariance ellipsoid, it is not necessarily part of the efficient frontier. This situation arises when such a portfolio yields a lower return than another portfolio with the same level of risk. In this case, despite being on the boundary defined by the risk constraint, the portfolio is suboptimal from a return perspective. The efficient frontier is, therefore, not the entire ellipsoid surface, but rather a subset where the trade-off between risk and return is maximized. Portfolios that do not meet this optimization criterion are excluded, despite satisfying the variance condition.

\subsubsection{The Mathematical Intersection of the Efficient Frontier and the Ellipsoid}

The efficient frontier is mathematically defined as the locus of portfolios that minimize portfolio variance for a given expected return. Geometrically, this corresponds to a curve on the surface of the ellipsoid where optimal risk-return combinations lie. Portfolios inside the ellipsoid are considered inefficient, as one can always find an alternative with lower risk for the same return. Similarly, portfolios on the ellipsoid surface but not on the efficient frontier are also suboptimal—they satisfy the variance constraint but fail to maximize return given that risk level. In other words, these portfolios do not solve the mean-variance optimization problem.

\subsubsection{Mathematical Conditions for Efficiency}

A portfolio is said to be efficient if it minimizes variance subject to both a target expected return and full investment (i.e., the sum of weights equals one). This is typically solved using the method of Lagrange multipliers, where the objective function is:

\begin{equation}
\mathcal{L}(w, \lambda, \mu) = w^T \Sigma w - \lambda (w^T E[R] - R_{\text{target}}) - \mu (\sum w_i - 1),
\end{equation}

where \( w \) is the vector of portfolio weights, \( \Sigma \) is the covariance matrix of asset returns, \( E[R] \) is the expected return vector, and \( \lambda \), \( \mu \) are Lagrange multipliers for the return and budget constraints, respectively.

The first-order condition for optimality is obtained by taking the gradient of \( \mathcal{L} \) with respect to \( w \), leading to:

\begin{equation}
2 \Sigma w - \lambda E[R] - \mu \mathbf{1} = 0.
\end{equation}

Solving for \( w \), we get:

\begin{equation}
w = \frac{1}{2} \Sigma^{-1} (\lambda E[R] + \mu \mathbf{1}).
\end{equation}

To ensure this solution corresponds to an efficient portfolio, we must enforce the constraints:

\begin{equation}
w^T \Sigma w = \sigma_p^2, \quad w^T E[R] = R_{\text{target}}, \quad \sum w_i = 1.
\end{equation}

Together, these equations fully characterize the set of efficient portfolios lying on the efficient frontier.

From this analysis, we can conclude that the efficient frontier must lie on the surface of the ellipsoid, representing the optimal set of portfolios. Portfolios inside the ellipsoid are inefficient because they assume unnecessary risk for their return, and those on the ellipsoid but not on the frontier fail to optimize the return-risk trade-off. Only portfolios that simultaneously satisfy all constraints will lie both on the ellipsoid and on the efficient frontier.

\begin{figure}[h]
\centering
\includegraphics[width=0.8\textwidth]{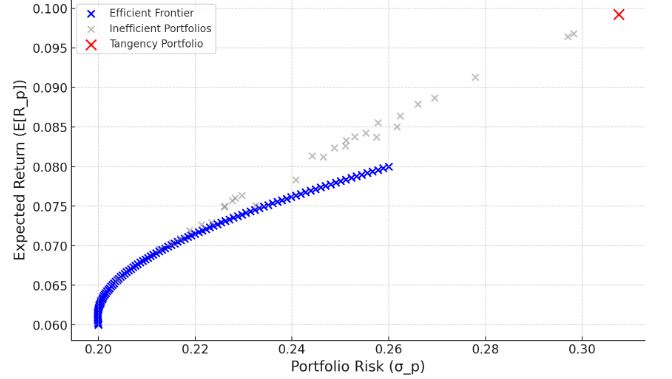}
\caption{2D Projection of Efficient and Inefficient Portfolios. Blue points represent the efficient frontier, gray points indicate inefficient portfolios, and the red point is the tangency portfolio.}
\label{fig:efficient_inefficient}
\end{figure}

The efficient frontier represents the optimal set of portfolios that provide the highest expected return for a given level of risk. Investors should allocate their portfolios along this curve to achieve the best possible risk-return trade-off. The tangency portfolio, identified by the red point in the figure, is the portfolio that maximizes the Sharpe ratio and provides the highest risk-adjusted return. Additionally, investors with different risk aversion levels will choose different points along the efficient frontier, forming utility-based portfolios. Higher risk-averse investors will prefer portfolios closer to the minimum variance portfolio, while risk-seeking investors will position themselves towards the right end of the curve, accepting higher risk in exchange for greater expected return.

\begin{figure}[h]
\centering
\includegraphics[width=0.8\textwidth]{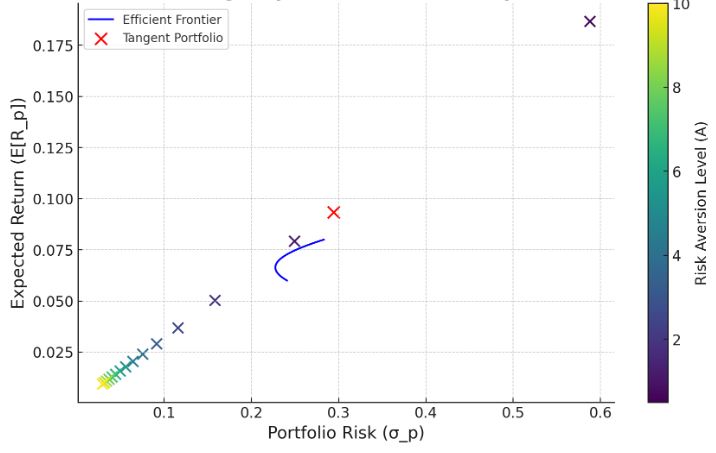}
\caption{Visualization of the Efficient Frontier, Tangency Portfolio, and Utility-Based Portfolios. The blue curve represents the efficient frontier, the red dot represents the tangency portfolio, and the color gradient of utility-based portfolios reflects varying investor risk preferences.}
\label{fig:efficient_frontier_tangency_utility}
\end{figure}

\subsection{N-Factor Model Representation}

A standard \textit{N}-factor model expresses the return of asset  as:
\begin{equation}
R_i = \alpha_i + \sum_{j=1}^{N} \beta_{ij} F_j + \epsilon_i,
\end{equation}
where:
\begin{itemize}
\item $\alpha_i$ is the expected return independent of factors.
\item $\beta_{ij}$ represents the sensitivity of asset $i$ to factor $j$.
\item $F_j$ is the systematic factor.
\item $\epsilon_i$ is the idiosyncratic risk component, assumed to be uncorrelated with factors.
\end{itemize}
Taking expectations, the expected return of asset $i$ is:
\begin{equation}
E[R_i] = \alpha_i + \sum_{j=1}^{N} \beta_{ij} E[F_j].
\end{equation}

\subsubsection{Constructing the Covariance Matrix}

The variance of asset $i$ is given by:
\begin{equation}
\text{Var}(R_i) = \sum_{j=1}^{N} \beta_{ij}^2 \sigma_{F_j}^2 + \sigma_{\epsilon_i}^2,
\end{equation}
where:
\begin{itemize}
\item $\sigma_{F_j}^2$ is the variance of factor $j$.
\item $\sigma_{F_j}^2$ is the idiosyncratic variance of asset $i$.
\end{itemize}
The covariance between two assets $i$ and $k$ is given by:
\begin{equation}
\text{Cov}(R_i, R_k) = \sum_{j=1}^{N} \beta_{ij} \beta_{kj} \sigma_{F_j}^2.
\end{equation}
Thus, the full covariance matrix is:
\begin{equation}
\Sigma = B \cdot \text{diag}(\sigma_F^2) \cdot B^T + D,
\end{equation}
where:
\begin{itemize}
\item $B$ is the $m\times N$ factor loading matrix.
\item $\text{diag}(\sigma_F^2)$ is the diagonal matrix of factor variances.
\item $D$ is the diagonal matrix of idiosyncratic variances.
\end{itemize}

\subsubsection{Impact of Factor Interactions on the Covariance Ellipsoid}

The covariance matrix shapes the risk ellipsoid based on factor interactions:

\begin{itemize}
\item \textbf{Higher factor variance} ($\sigma_F^2$) stretches the ellipsoid along systematic risk directions.
\item \textbf{Higher factor correlations} distort the ellipsoid by increasing off-diagonal covariance elements.
\item \textbf{Higher factor loadings} ($\beta$) elongate the ellipsoid along dominant risk dimensions.
\item \textbf{Higher idiosyncratic risk} ($\sigma_{\epsilon}^2$) makes the ellipsoid more spherical.
\end{itemize}

\begin{table}[h]
\centering
\begin{tabularx}{\textwidth}{lX}
\textbf{Factor Influence} & \textbf{Effect on Covariance Matrix and Ellipsoid} \\
\toprule
Higher factor variance ($\sigma_F^2$) & Increases diagonal elements, stretching the ellipsoid along systematic risk axes. \\
Higher factor correlations & Increases off-diagonal interactions, distorting ellipsoid alignment and causing directional skew. \\
Higher factor loadings ($\beta$) & Amplifies systematic risk impact, elongating the ellipsoid along dominant factor directions. \\
Higher idiosyncratic risk ($\sigma_{\epsilon}^2$) & Increases diagonal elements of $D$, making the ellipsoid more rounded and reducing systematic dominance. \\
\bottomrule
\end{tabularx}
\caption{How Factor Interactions Shape the Covariance Ellipsoid}
\label{tab:factor_ellipsoid}
\end{table}

\subsubsection{Principal Components and Factor Dominance}

Since $\Sigma$ is symmetric, its eigenvalues and eigenvectors determine its shape:
\begin{equation}
\Sigma v = \lambda v.
\end{equation}
Larger eigenvalues correspond to dominant systematic factors, while smaller ones indicate lower-impact variations. The ellipsoid aligns with the principal eigenvectors, stretching in major risk directions.

\subsubsection{Projection of Factor Model onto Lower Dimensions}

The factor model can be interpreted as a projection of the high-dimensional return space onto a lower-dimensional factor space. The projection process follows:

\begin{enumerate}
\item Compute the eigenvalues and eigenvectors of the covariance matrix :
\begin{equation}
\Sigma v_i = \lambda_i v_i.
\end{equation}
The eigenvectors  define the principal directions of risk, while the eigenvalues  measure the magnitude of variance along each eigenvector.

\item Construct the projection matrix using the first \( K \) dominant eigenvectors:
\begin{equation}
    P_K = [v_1, v_2, \dots, v_K].
\end{equation}

\item Project the original return space onto the lower \( K \)-dimensional space:
\begin{equation}
    R^*_i = P_K^T R_i.
\end{equation}
This transformation extracts the most significant risk components while reducing noise from idiosyncratic variation.

\end{enumerate}
Projecting asset returns onto a lower-dimensional factor space has significant implications for portfolio optimization:

\begin{itemize}
\item \textbf{Reduced Dimensionality:} Optimization becomes computationally efficient as only key factors are used.
\item \textbf{Better Risk Diversification:} By focusing on dominant risk sources, factor investing enables improved diversification strategies.
\item \textbf{Improved Sharpe Ratios:} Portfolios constructed using dominant factor exposures often achieve better risk-adjusted returns.
\item \textbf{Principal Component Selection:} Choosing an appropriate number of factors (i.e., eigenvectors) determines the trade-off between capturing essential risk and avoiding noise.
\end{itemize}

\newpage

\section{Common Causal Manifold Risk Management}

This chapter builds upon the foundational concept of a causal system, framing the common causal manifold as such a system. Moreover, the sensitivity space introduced in earlier chapters is now interpreted as the tangent space to that manifold. This perspective opens a novel avenue for portfolio risk management by drawing from the rich literature in differential geometry and manifold learning.

For readers less familiar with the mathematical details, consider the analogy from the previous chapter where covariance matrices were visualized as ellipsoids, and the solution lay on a tangent plane where the Capital Market Line (CML) intersected both the ellipsoid and the plane (see Figure \ref{fig:zoom_ellipsoid}). A similar geometric intuition applies here. However, instead of an ellipsoid, we now deal with a common causal manifold of arbitrary shape, and the sensitivity space serves as its tangent space, where the optimal solution resides.

A mapping between these two spaces—common causal manifold and sensitivity space—was previously explored using hierarchical methods in Chapter \ref{HSP_Section}, alongside other proposed mappings. In contrast, this chapter shifts focus from static mapping to dynamically modeling the trajectory of optimal diversification over time. Given that the common causal manifold is defined as a causal time system \citep{windeknecht1967causality}, the objective is to model both the manifold and its tangent space at time $t$, leveraging tools from differential geometry, particularly stochastic differential equation (SDE) projections on manifolds \citep{10.1007/978-3-319-25040-3_76, Armstrong2022}.

This approach extends the model’s capabilities beyond prior formulations by incorporating causal properties of diversification dynamics into a forward-looking framework. To the author's knowledge, this represents the first application in the portfolio optimization literature where causal systems and differential geometric techniques are combined in a unified methodological and theoretical framework, surpassing the scope of previous causal approaches.

A definition of Causal Time System Manifolds is given next.

\subsection{Causality and Time Systems in the Common Causal Framework}

Let \( S \) be a time system, and consider a trajectory prediction \( A \), along with its subsequent states \( A', A'', \dots \).

\subsubsection{Non-Anticipatory Systems}

The system \( S \) is said to be non-anticipatory if, for any two trajectories \( A \) and \( A' \) that are equal up to time \( t \), and under the same causal manifold \( S(t) \), it holds that \citep{windeknecht1967causality}:
\[
A(t) = A'(t) \Rightarrow S[A](t) = S[A'](t).
\]
In other words, the system's output at time \( t \) depends only on the past trajectory up to that point, and not on future values.

\textit{Remark:} A time system \( S \) admits a set of initial states if and only if it is non-anticipatory \citep{windeknecht1967causality}.

\subsubsection{Causality Conditions}

Let \( \mathcal{Q}_0 \) denote a set of input functions and let \( \tau, \theta \) be time parameters. Then, the system \( S \) is causal with respect to \( \mathcal{Q}_0(\tau, \theta) \) if the following conditions are satisfied \citep{windeknecht1967causality}:

\begin{enumerate}
    \item \textbf{Domain Causality:}
    \[
    \forall q \in \mathcal{Q}_0(\tau, \theta), \quad A \in \mathcal{D}_S(q) \Rightarrow A(t_0, q) \in \mathcal{D}_S.
    \]

    \item \textbf{Causal Consistency Over Time:}
    \[
    \forall A \in \mathcal{D}_S(q), \quad A(t_1) = A'(t_1) \text{ for } t_1 \in [t_0, t] \Rightarrow S[A](t) = S[A'](t).
    \]

    \item \textbf{Time-Relative Output Consistency:}
    \[
    A(t) = A'(t) \Rightarrow \mathcal{D}_S(A(t+1)) = \mathcal{D}_S(A'(t+1)), \quad S(A(t+1)) = z(t, A(t+1)),
    \]
    where \( z(t, A(t+1)) \) is the output point in the relative trajectory, and \( q \in \mathcal{Q}_0(t, t') \subset \mathcal{T} \) \citep{windeknecht1967causality}.
\end{enumerate}

\subsubsection{Dynamics of the Common Causal System}

For the common causal system \( S \), and at any absolute time \( t' \), a relative trajectory \( q \) for the asset dynamics \( A \) exists, parameterized by a local time variable \( t \).

If \( S \) is causal, then for each fixed \( t' \), the corresponding trajectory over \( t \) is well-defined. Once the state \( S(t') \) is determined, the evolution \( A(t+1) \) follows deterministically.

Furthermore, as time progresses through:
\[
S \rightarrow S' \rightarrow S'' \rightarrow \cdots,
\]
the following occurs:
\begin{itemize}
    \item One point in the relative trajectory, namely \( A(t+1) \), is fixed by the current state \( S \).
    \item The rest of the trajectory evolves according to the updated system states \( S', S'', \dots \).
\end{itemize}

It can be concluded that, the system \( S \) has both a static component, representing its current configuration or manifold at time \( t' \), and a dynamic component, governing its transition and trajectory evolution. This framework provides a structured basis for modeling \textit{diversification dynamics} and \textit{portfolio evolution} in the context of causal manifolds.

\begin{figure}
    \centering
    \includegraphics[width=1\linewidth]{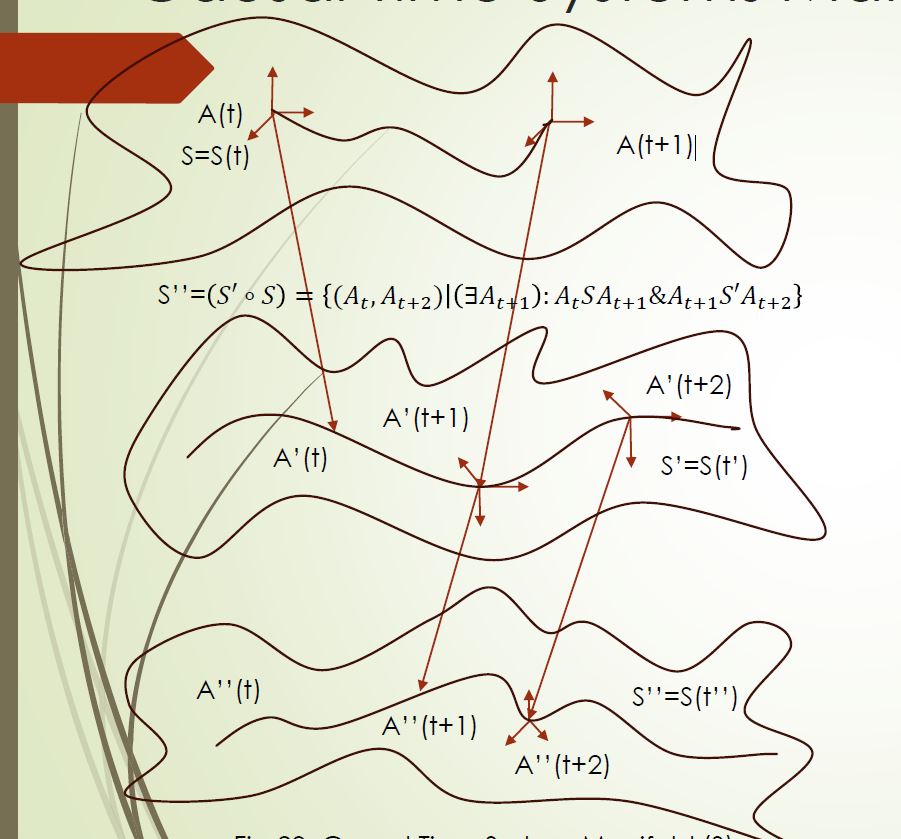}
    \caption{Illustration of time compounding in asset trajectory predictions over Causal Time System Manifolds. The diagram shows how static predictions at each time layer align with the evolution of the causal manifold, ensuring trajectory consistency across dynamic transitions. Illustration from WBS Quantitative Finance Conference in 2023 in Valencia presenting \citep{RODRIGUEZDOMINGUEZ2023100447}}
    \label{fig:enter-CommonTimeSystem}
\end{figure}

In Figure \ref{fig:enter-CommonTimeSystem}, a representation of a causal time system is shown, illustrating the two time components \( t \) and \( t' \) for a given asset \( A \), whose optimal trajectory can be forecasted at future absolute timestamps. If the drivers forming the manifold are causal, then the asset's optimal trajectory is predetermined by its current state and causal structure. The upper part of the figure represents a static prediction trajectory at fixed time \( t \) on the causal manifold. When the system evolves from \( t \) to \( t' \), the asset transitions to a new state \( A' \), and the corresponding optimal forecast updates since the manifold itself has changed.

A key consistency condition follows: if the manifold forms a causal system and the forecast is optimal, then the prediction made at \( t \), denoted \( A(t+1) \), must coincide with the realized value \( A'(t+1) \) at time \( t' \). If the process repeats and a new forecast \( A'(t+2) \) is computed at time \( t' \), the subsequent realization at time \( t'' \), denoted \( A''(t+2) \), must match it. This recurrence is guaranteed by the system's causal nature, where each trajectory segment aligns with prior predictions.

This sequential consistency reflects the notion of time composition and system interconnection in causal time systems, as formalized in~\cite{windeknecht1967causality}, and is presented next as a theorem.

\begin{theorem}[Causality and System Decomposition \cite{windeknecht1967causality}]
If $S$ is causal with respect to $(Q, Q_0, \tau, \theta)$, the map 
\[
\tau_{\{q,A\}} : T \to Q \quad \text{with} \quad t \mapsto \tau(q, A, t)
\]
is a state trajectory of $S$.

$S$ is a static system if there exists a map $c : A_t \to A_{t+1}$ such that:
\[
\forall A_t \in \mathcal{D}_S, A_{t+1} \in RS: A_t S A_{t+1} \iff (\forall t): A_{t+1}(t) = c(A_t(t))
\]

If $S$ and $S'$ are time systems and $RS \subseteq \mathcal{D}_{S'}$, the series interconnection of $S$ and $S'$ is a composition time system defined by:
\[
S'' = (S' \circ S) = \left\{ (A_t, A_{t+2}) \middle| \exists A_{t+1} : A_t S A_{t+1} \wedge A_{t+1} S' A_{t+2} \right\}
\]

Then, $S$ is causal if and only if $S$ is the series interconnection of some transition system $S'$ and some static system $S''$.

\end{theorem}

\subsection{Motivation}

The concepts of causal time systems \citep{windeknecht1967causality} and compounding time rules, together with the mathematical developments in stochastic differential equations (SDEs) on manifolds over the past decade \citep{10.1007/978-3-319-25040-3_76, Armstrong2022}, as well as prior work on portfolio optimization over common causal manifolds and sensitivity spaces \citep{RODRIGUEZDOMINGUEZ2023100447, dominguez2024portfolios, rodriguez2024aifi}, constitute the essential theoretical background for the developments presented in this chapter.

\begin{figure}[h]
\centering
\includegraphics[width=0.8\textwidth]{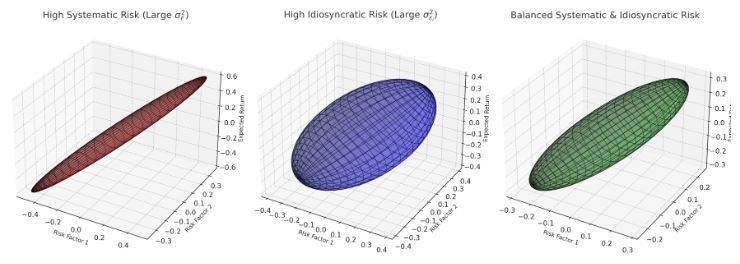}
\caption{Three cases illustrating the impact of systematic and idiosyncratic risk on the covariance ellipsoid. Left: High systematic risk (red). Middle: High idiosyncratic risk (blue). Right: Balanced systematic and idiosyncratic risk (green).}
\label{fig:risk_ellipsoid_cases2}
\end{figure}

\begin{figure}[h]
\centering
\includegraphics[width=0.6\textwidth]{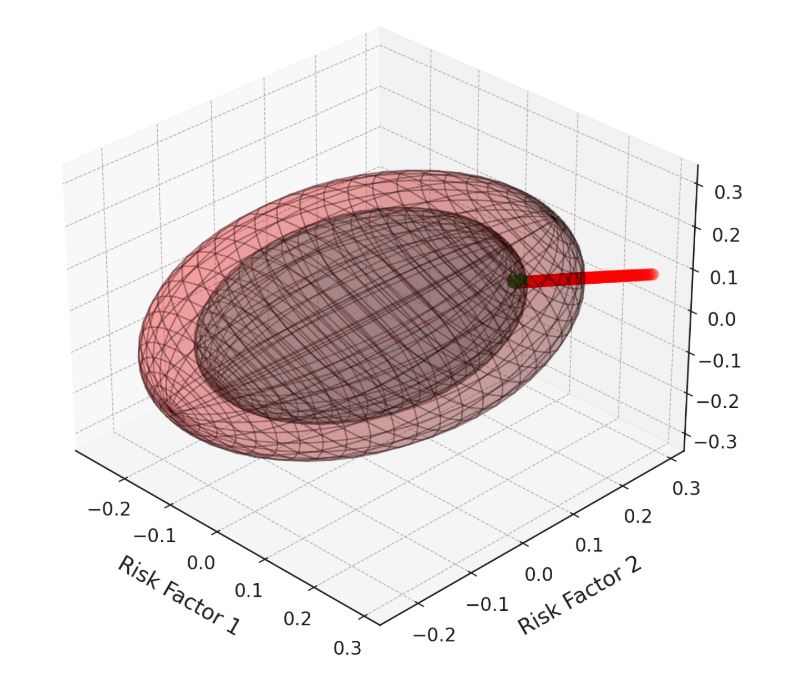}
\caption{Expansion of the covariance ellipsoid during a financial crisis, illustrating increased portfolio risk.}
\label{fig:crisis_scenario2}
\end{figure}

As further motivation—particularly for readers interested in physics and causality—several illustrations are included to highlight the natural emergence of these connections. For instance, reconsider Figures~\ref{fig:risk_ellipsoid_cases2}, which depict three scenarios showing the impact of systematic and idiosyncratic risk on the covariance ellipsoid, or Figure~\ref{fig:crisis_scenario2}, illustrating the expansion of the covariance ellipsoid during a financial crisis and the corresponding increase in portfolio risk. These visualizations make it evident that such geometric risk representations evolve over time, and this time dependence directly influences the location of the optimal solution. Crucially, although these objects deform with time, they remain constrained by structural properties—such as retaining their ellipsoidal shape—which governs the boundaries within which portfolio dynamics must evolve.

In the case of common causal manifolds, a similar behavior is observed: these structures can be conceptualized as nonlinear geometric objects whose shape evolves over time. Depending on the embedding space, their deformation exhibits different characteristics. When embedded into a sensitivity space, such manifolds can either expand or contract, adapting to the information dynamics of causal sensitivities. In contrast, when embedded into a time-embedding space—similar to approaches in theoretical physics—the manifolds typically exhibit only expansive behavior over time due to the accumulation of informational flow and system evolution.

\subsubsection{Tangent Spaces, SDEs and Manifolds}

Let \( (U, \varphi) \) be a chart around a point \( p \in \mathcal{S} \), where \( \mathcal{S} \) is a smooth manifold of dimension \( m \). The tangent space \( T_p \mathcal{S} \) at point \( p \) is the set of linear derivations \( v : C^\infty(\mathcal{S}) \rightarrow \mathbb{R} \) of the form:
\[
v(f) = \sum_{i=1}^m a_i \left.\frac{\partial (f \circ \varphi^{-1})}{\partial u_i}\right|_{\varphi(p)},
\]
for some coefficients \( a = (a_1, \dots, a_m) \in \mathbb{R}^m \), and \( u = (u_1, \dots, u_m) \) are the local coordinates given by \( \varphi \).

\begin{figure}[H]
    \centering
    \includegraphics[width=0.7\linewidth]{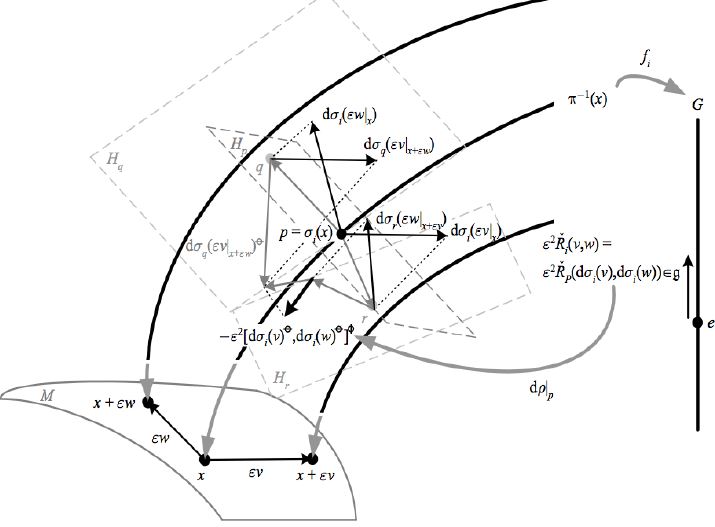}
    \caption{Illustration of a fibre bundle, with base space and associated tangent fibres}
    \label{fig:enter-FibreBundle}
\end{figure}

For any smooth manifold \( \mathcal{S} \) of dimension \( m \), the tangent bundle \( T\mathcal{S} \) is defined as:
\[
T\mathcal{S} = \bigsqcup_{p \in \mathcal{S}} T_p \mathcal{S},
\]
the disjoint union of all tangent spaces at each point \( p \in \mathcal{S} \). The bundle projection
\[
\pi : T\mathcal{S} \rightarrow \mathcal{S}, \quad \pi(v) = p \quad \text{for } v \in T_p \mathcal{S},
\]
is a smooth submersion. The fibers of this map are the individual tangent spaces. An example of a fibre bundle is shown in Figure~\ref{fig:enter-FibreBundle}, where the total space is composed of a base manifold and attached tangent spaces, forming a smooth structure that evolves with the base.

\begin{figure}[H]
    \centering
    \includegraphics[width=0.65\linewidth]{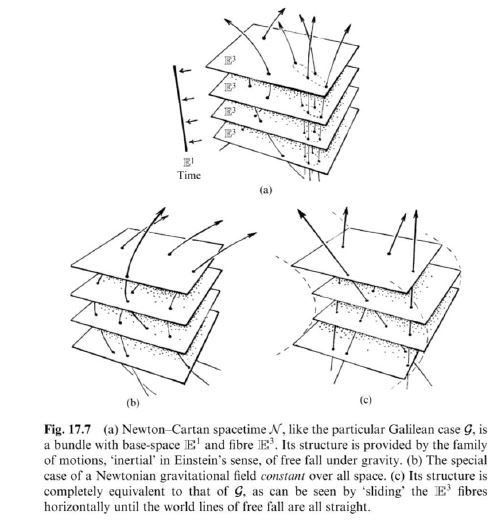}
    \caption{Schematic representation of Newton–Cartan space-time, highlighting evolving spatial slices (fibre bundles) over absolute time}
    \label{fig:enter-Newton}
\end{figure}

As mentioned, these types of fibre bundle constructions also appear in theoretical physics, for example, in Newton–Cartan space-time models, which provide a geometric formulation of Newtonian gravity. Such a setting can be visualized in Figure~\ref{fig:enter-Newton}.

\vspace{1em}
\noindent
\textbf{Application to Portfolio Optimization.}

In the context of portfolio theory, the manifold \( \mathcal{S} \) is defined by the common causal drivers of the portfolio. Portfolio constituents are projected onto the tangent bundle \( T\mathcal{S} \), allowing us to interpret asset dynamics locally in the space defined by causal structure.

Following the work of \citep{Armstrong2022}, portfolio dynamics described by stochastic differential equations (SDEs) can be projected onto the tangent bundle using \emph{Itô jet projections}. This framework enables optimal projection of the asset dynamics as a convex optimization problem on the manifold. This geometric approach provides a powerful foundation for representing portfolio risk and sensitivity in a causally consistent way.

\begin{figure}[H]
    \centering
    \includegraphics[width=0.8\textwidth]{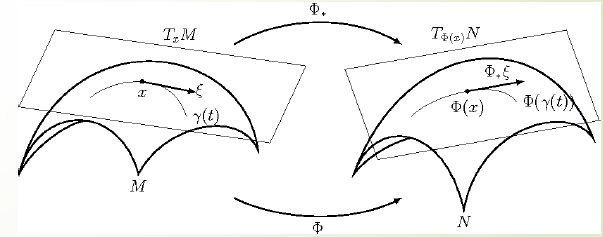}
    \caption{Causal time system interconnection step: representation of dynamics across fiber bundles, with \( M = \mathcal{S} \) and \( N = \mathcal{S}' \). Adapted from \citep{seibert2012bfgs}.}
    \label{fig:tangent_bundle}
\end{figure}

\newpage

\subsection{Research Background}

Portfolio optimization has been a fundamental area of financial research since the introduction of Modern Portfolio Theory by H. Markowitz in 1952 \citep{10.2307/2975974}. Subsequent developments, such as the efficient market hypothesis by H. Fama in 1970 \citep{fama1970efficient}, have shaped the field significantly. Despite these advancements, portfolio optimization faces several challenges, including overfitting \citep{Maggiolo2023}, parameter uncertainty \citep{Kumar2024}, transaction costs, regime switches, and changing market conditions \citep{Fabozzi2015}. Addressing these issues has led to extensive research into forecasting returns and correlations using linear and non-linear models, including econometrics, machine learning, and deep learning methods \citep{Butler2023,Yu2023}. Factor models have been incorporated to add structural knowledge, although these forecasts often remain difficult and factors can be non-causal and spurious \citep{Lopez_de_Prado_2023}.

Recent research has focused on modeling assets and portfolio dynamics using partial differential equations (PDEs) and stochastic PDEs (SPDEs) \citep{Fernholz2002}. Forward-looking techniques such as dynamic programming, optimal control, and reinforcement learning represent more advanced methods, relying on policies, rules, and additional constraints beyond traditional out-of-sample methods \citep{Kolm2019,Dixon2020}. Similarity learning approaches and methods incorporating causality, based on causal factors and manifolds, have also shown promise \citep{RODRIGUEZDOMINGUEZ2023100447,dominguez2024portfolios}.

This chapter analyzes the manifold of the conditional probability of a portfolio and its constituents, conditional on portfolio common drivers. Utilizing a sensitivity-based approach for portfolio optimization, the tangent space on this manifold is studied to obtain trajectories for the sensitivities of each constituent with respect to the same set of portfolio drivers. This involves using multi-step ahead conditional probabilities of portfolio constituents, given the set of portfolio drivers at a specific point in time.

When these conditional probabilities are approximated using predictors such as neural networks, a second type of manifold emerges, characterized by the L-norm losses of these predictors. For a given horizon $T$, a set of $T$ predictors, each with different forecasting steps from 1 to $T$, is defined. This set of predictors varies for each portfolio constituent and set of portfolio common drivers, indexed by two times: the timestamp of the common drivers, $t$, which are the inputs of the predictors, and the timestamp of the multi-step ahead forecast associated with each predictor, $t'$, from 1 to $T$.

Assuming that, starting at time $t$, future values are obtained for the inputs of these predictors (the common portfolio drivers), a term structure of predictors arises for different values of $t$ and $t'$. At each $t$, there are $T-t$ predictors with inputs at $t$ and forecasting steps $t' = 1, \ldots, T-t$. Additionally, there are predictors starting at $t$ and ending at $T$, creating a term structure that depends on the forecasting horizon and the input timestamp.

Two types of manifolds are differentiated:
\begin{enumerate}
    \item A static manifold, dependent on $t'$, represented by the predictors’ parameters for different multi-step ahead forecasting problems. Each $t'$ has an associated set of model parameters.
    \item A dynamic manifold, dependent on $t$, represented by the inputs of each predictor, the common drivers for the portfolio.
\end{enumerate}

From these manifolds, two tangent spaces emerge:
\begin{enumerate}
    \item One given by the derivatives of the loss predictors with respect to their model parameters, for all predictors in the term structure.
    \item Another given by the sensitivities of each predictor to the inputs, or the derivatives of the predictor forecasts with respect to the common drivers set used as inputs.
\end{enumerate}

A compound manifold that integrates these two types of manifolds, along with a compound tangent space, is defined. This framework can be utilized to model the trajectory of the sensitivity of the portfolio constituents with respect to the portfolio drivers.

This work extends the idea that assets and portfolio dynamics can be modeled by PDEs/SPDEs \citep{Musiela2010} and approximated efficiently with machine learning techniques \citep{Christian2023}. The focus is on optimal diversification, which allows for searching for connections between assets and portfolio dynamics to select optimal portfolio drivers for diversification in terms of probability of causality and persistence. Sensitivities of portfolio constituents with respect to optimal portfolio drivers, approximated with market data, can be used as embedding spaces in which constituents can be projected, and diversification, including trajectory information, can be optimized \citep{RODRIGUEZDOMINGUEZ2023100447}. In this work, these sensitivities are treated as path-dependent, modeled, simulated into the future, and used for forward-looking portfolio optimization \citep{Davis2021}.

\subsection{Literature Review}
Recent advancements in portfolio optimization have emphasized the importance of incorporating dynamic asset behavior and the selection of common drivers for enhanced diversification. Rodriguez Dominguez (2023) introduced a framework that models asset and portfolio dynamics using Partial Differential Equations (PDEs) and neural networks \citep{RODRIGUEZDOMINGUEZ2023100447}. This approach leverages the sensitivities of portfolio constituents with respect to common drivers, obtained through Automatic Adjoint Differentiation, to optimize portfolio diversification. The framework defines the Commonality Principle for selecting optimal portfolio drivers and utilizes a sensitivity distance matrix to measure the similarity of projections in the sensitivity space, enabling hierarchical clustering to solve the convex optimization problem. This method addresses major challenges in portfolio management, such as regimes, non-stationarity, overfitting, and selection bias, demonstrating superior performance across various markets and datasets. In a subsequent study, Rodriguez Dominguez (2024) further explored the dynamics of portfolio optimization by presenting a closed-form formula for the conditional probability of a portfolio given its optimal common drivers \citep{dominguez2024portfolios}. This study models the joint distribution of portfolio constituents and common drivers using Gaussian copulas, resulting in a conditional risk-neutral PDE. The PDE framework facilitates dynamic risk management by providing implied conditional portfolio volatilities and weights, which serve as new risk metrics. These metrics can be dynamically monitored or obtained from the solutions of the PDEs, enhancing the ability to manage portfolio risks in a dynamically changing market environment.

The authors propose a hierarchical approach to portfolio management that uses sensitivity analysis to manage risk and optimize returns. This approach involves decomposing the portfolio into sub-portfolios based on their sensitivities to different risk factors, allowing for more targeted and effective risk management \citep{Kritzman2003}. Christoffersen and Jacobs (2004) highlight the significant impact that the choice of loss function has on the performance of option pricing models. Their research underscores the need for careful selection and evaluation of loss functions to improve model accuracy, parameter stability, and overall effectiveness in financial applications \citep{Christoffersen2004}.

Windeknecht (1967) explores the mathematical systems theory with a focus on causality, emphasizing the importance of non-anticipatory systems and state transitions. Non-anticipatory systems are those where the current state and evolution depend only on past and present inputs, not future inputs, ensuring realistic simulations. Windeknecht proves that causal systems are precisely non-anticipatory systems with well-defined state transitions, providing a theoretical basis for modeling dynamic systems. This concept is crucial for simulating stochastic differential equations (SDEs) on manifolds, where it is essential to maintain the geometric constraints of the manifold. By ensuring that the simulated paths are non-anticipatory and adhere to the manifold's structure, Windeknecht's principles enhance the accuracy and realism of financial models, such as those used for simulating the term structure of forward rates \citep{windeknecht1967causality}. A distributional approach to analyzing the sensitivity of control systems, focusing on non-anticipatory systems and state transitions, is crucial for simulating stochastic differential equations (SDEs) on manifolds. These principles ensure that simulated paths adhere to geometric constraints and maintain realism. The sensitivity matrix quantifies how variations in system parameters impact behavior, which is vital for accurate simulations. This approach is particularly relevant in financial modeling, such as simulating the term structure of forward rates, where maintaining non-anticipatory properties and manifold constraints is essential for achieving accurate and arbitrage-free results \citep{Newcomb1967}.

The integration of Stochastic Differential Equations (SDEs) on manifolds into financial prediction models, particularly for portfolio optimization and the term structure of forward rates, is a burgeoning field. Armstrong, Brigo, and Ferrucci (2022) explore the projection of SDEs onto submanifolds, providing foundational methods for representing these equations in a coordinate-free manner and deriving formulae for their optimality in both weak and mean-square senses. This work is crucial for financial applications where maintaining the geometric constraints of the manifold ensures accurate and realistic simulations \citep{Armstrong2022}.

Brigo and Mercurio's research (2015) on interest rate models delves into the practical applications of SDEs in finance, particularly emphasizing the maintenance of no-arbitrage conditions and realistic market behavior through sophisticated mathematical frameworks. Their work provides essential methodologies for incorporating SDEs into financial models, ensuring that predictions and simulations adhere to market dynamics \citep{10.1007/978-3-319-25040-3_76}.

Furthermore, recent advances by Gogioso and Pinzani (2023) introduce geometric frameworks like causaltopes, which can define and simulate state transitions on manifolds while respecting probabilistic behaviors and constraints. This approach enhances the prediction accuracy and robustness of financial models by ensuring that the simulations remain within the bounds of the defined geometric structures \citep{Gogioso2023}.

The exploration of term structures in financial prediction using machine learning predictors is also significant. Masini, Medeiros, and Mendes (2023) review machine learning advances for time series forecasting, demonstrating how these techniques can be applied to financial metrics over different time horizons. Additionally, Binsbergen, Han, and Lopez-Lira (2020) highlight the term structure of earnings expectations, showcasing improvements in long-term financial forecasting accuracy through machine learning models \citep{Masini2023, Binsbergen2020}.

A simulation-based approach analyzes the impact of noise in financial correlations on portfolio and risk management, highlighting the sensitivity of financial techniques to noisy inputs and demonstrating the framework's potential for investigating various financial problems \citep{Szilard2004}. A class of multivariate models based on variance-correlation separation uses univariate GARCH models for individual asset variances and parsimonious parametric models for time-varying correlation matrices, reducing correlation parameters and simplifying constraints while offering faster estimation and favorable forecasting power \citep{Dellaportas2004}. A nonlinear common factor (NCF) method for modeling and forecasting correlation matrices in economic and business data simplifies estimation and offers greater flexibility, as shown in an application to Boston energy prices \citep{Yongli2021}. The analysis reveals how changes in asset correlations and volatilities affect the minimum variance portfolio's composition, expected return, and risk, providing guidelines for maintaining minimized risk under variable market conditions. Additionally, it is demonstrated that estimation errors in market data have a limited impact on the actual portfolio variance, ensuring the stability of the minimum variance portfolio \citep{Ji2013}. The application of risk-sensitive control to portfolio optimization in a general factor model is discussed, where economic factors and security prices influence mean returns and volatilities, and optimal strategies are constructed through the analysis of Bellman equations under a finite time horizon, including considerations for partial information \citep{Nagai2004}. A continuous-time model is proposed where investors use expert forecasts to construct a benchmark-outperforming portfolio, employing a Kalman filter for estimation and deriving the optimal investment policy in closed form, with results showing a range of strategies from passive to active, highlighting the importance of factor choice and debiasing in investment performance.

\subsection{Preliminary}
The efficient frontier can be represented as a problem in quadratic curves. Given the assets returns $\boldsymbol{r}=r_1,\dots,r_n$ in a market, each portfolio can be represented as a vector $\boldsymbol{w}=w_1,\dots,w_n$ such that $\sum_{i}{w_i=1}$, holding assets $\boldsymbol{w}^T\boldsymbol{r}=\sum_{i}{w_ir_i}$. The following quadratic optimization is solved to maximize risk-adjusted returns:
\begin{equation}
\left\{\begin{matrix}E[w^Tr]\\\min_w\sigma^2=Var[w^Tr]\\\sum_{i}{w_i=1}\\\end{matrix}\right.
\end{equation}

Portfolios are points in the Euclidean space $\mathbb{R}^n$. The third equation states that the portfolio should fall on a plane defined by $\sum _{i}w_{i}=1$. The first equation states that the portfolio should fall on a plane defined by $w^{T}E[R]=\mu$. The second condition states that the portfolio should fall on the contour surface for $\sum _{ij}w_{i}\rho _{ij}w_{j}$ that is as close to the origin as possible. Since the equation is quadratic, each such contour surface is an ellipsoid (assuming that the covariance matrix $\rho _{ij}$ is invertible). Therefore, we can solve the quadratic optimization graphically by drawing ellipsoidal contours on the plane $\sum _{i}w_{i}=1$, then intersect the contours with the plane:
\begin{equation}
    \{w:w^{T}E[R]=\mu ;\sum _{i}w_{i}=1\}
\end{equation}
As the ellipsoidal contours shrink, eventually one of them would become exactly tangent to the plane, before the contours become completely disjoint from the plane. The tangent point is the optimal portfolio at this level of expected return. As we vary $\mu$, the tangent point varies as well, but always falling on a single line (this is the two mutual funds theorem). Let the line be parameterized as $\{w+w't:t\in\mathbb{R}\}$. We find that along the line:

\begin{equation}
    {\begin{cases}\mu &=(w'^{T}E[R])t+w^{T}E[R]\\\sigma ^{2}&=(w'^{T}\rho w')t^{2}+2(w^{T}\rho w')t+(w^{T}\rho w)\end{cases}}
\end{equation}
giving a hyperbola in the $(\sigma ,\mu )$ plane. The hyperbola has two branches, symmetric with respect to the $\mu$ axis. However, only the branch with $\sigma >0$ is meaningful. By symmetry, the two asymptotes of the hyperbola intersect at a point $\mu _{MVP}$ on the $\mu$ axis. The point $\mu _{mid}$ is the height of the leftmost point of the hyperbola, and can be interpreted as the expected return of the global minimum-variance portfolio (global MVP).

\begin{figure}
	\centering
	\includegraphics[width=1\textwidth]{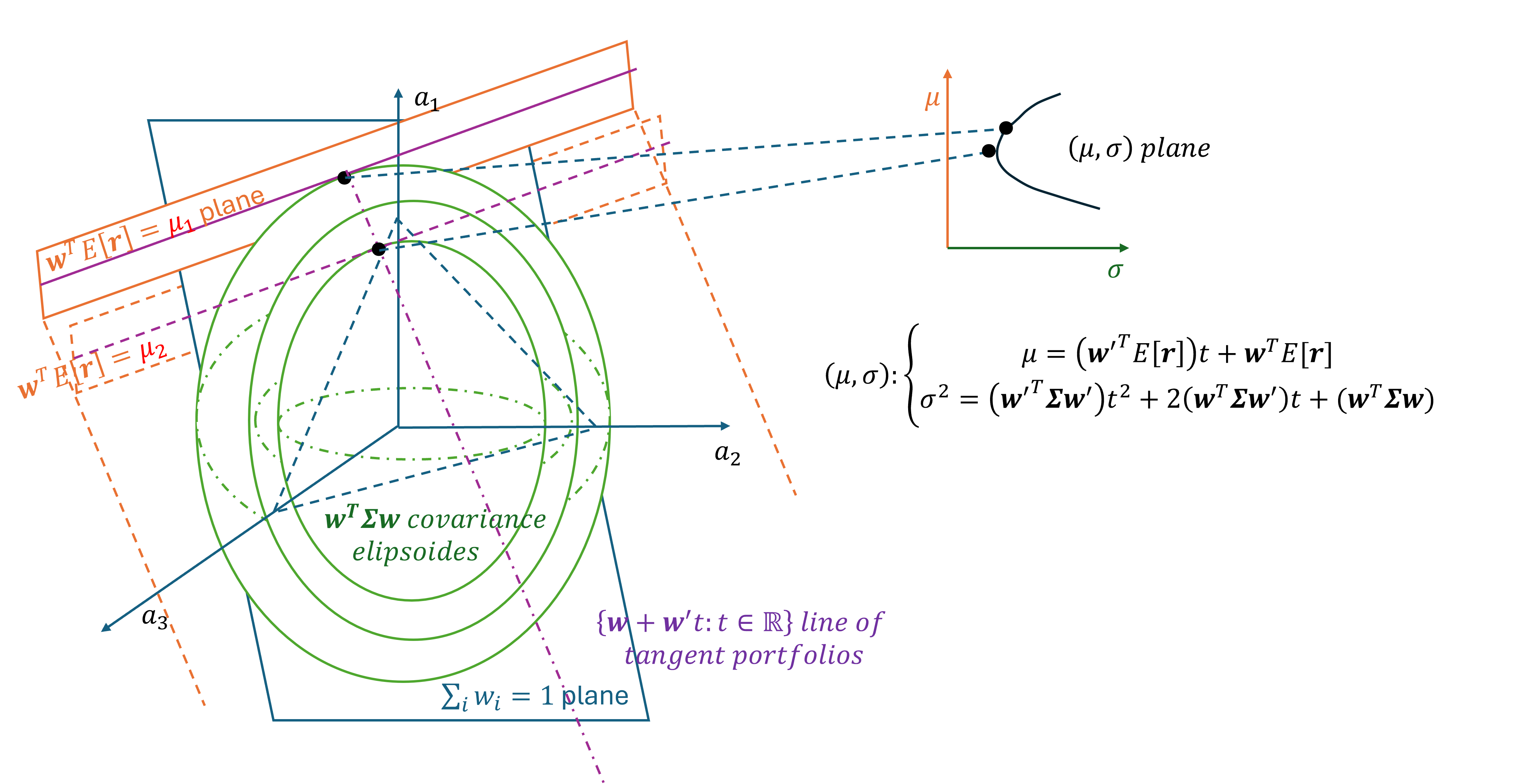}
	\caption{The ellipsoid is the contour of constant variance. The $x+y+z=1$ plane is the space of possible portfolios. The other plane is the contour of constant expected return. The ellipsoid intersects the plane to give an ellipse of portfolios of constant variance. On this ellipse, the point of maximal (or minimal) expected return is the point where it is tangent to the contour of constant expected return. All these portfolios fall on one line.}
 \label{MPTIssues}
\end{figure}

\subsection{Common Causal Manifolds}
Assume a joint probability space $(\Omega, \mathcal{F}, \boldsymbol{P})$ of a financial market. $\Omega$ is a countable set of outcomes, $\left\{\left(x_1(i_1),x_2(i_2),\dots,x_N(i_N)\right)\right\}_{i_1,\dots,i_N}$, where $x_j(i_j)$ are all possible values of $X_j$ for $j=1,\dots,N$. $\mathcal{F}$ be a $\sigma-algebra$. The index $i=(i_1,\dots,i_N)$ uniquely identifies the specific outcomes for each of the $N$ random variables in the joint probability space, representing the joint probability mass function as a countable set of random variables. The joint probability mass function $\boldsymbol{P}_{X_1,X_2,\dots,X_N}$ over the countable set of outcomes. Each point $(x_1(i_1),x_2(i_2),\dots,x_N(i_N),p_{i_1,\dots,i_N})$ represents a point on the manifold $\mathcal{M}_{(\Omega, \mathcal{F}, \boldsymbol{P})}$ in the joint probability space:
\begin{equation}
    \mathcal{M}_{(\Omega, \mathcal{F}, \boldsymbol{P})}=\left\{(x_1(i_1),x_2(i_2),\dots,x_N(i_N),p_{i_1,\dots,i_N})| x_j(i_j)\in Range(X_j)\right\}
\end{equation}
A portfolio consists of $n$ financial assets $\boldsymbol{p}=[a_{1},\dots,a_{n}]\in\Omega$ with respective weights $\boldsymbol{w}$. The optimal common causal drivers for the portfolio are selected based on the Commonality Principle \citep{RODRIGUEZDOMINGUEZ2023100447}. Specifically, the subset of $m$ optimal common causal drivers for a portfolio are selected so that the Reichembach's Common Cause Principle independent conditions have the highest probability \citep{Reichenbach1956-REITDO-2}, from a set of drivers' candidates $N$ in the financial markets $\Omega$ with $N>>m$. The subset of optimal portfolio common drivers $\boldsymbol{D}=[D_{1},\dots,D_{m}]\in\Omega$, satisfying the Reichembach Common Cause Principle (RCCP) conditions, make the portfolio constituents conditional on $\boldsymbol{D}$, independent. The Commonality principle implies that:

    Every portfolio $\boldsymbol{p}_t=[a_{1t},\dots,a_{nt}]$ under $(\Omega, \mathcal{F}_t, P)$, such that $\boldsymbol{p}_t\subset\Omega$, has associated a unique subset of optimal common causal drivers $\boldsymbol{D}_\tau\subset\Omega$ with $0\leq\tau\leq t$, $\forall t, n$, so that the RCCP conditions are maximized in probability. Also, every subset $\boldsymbol{D}_\tau\subset\Omega$ has associated a portfolio consisting of $\boldsymbol{p}=[a_{1},\dots,a_{n}]$ assets

This allows to define the tuple $(\boldsymbol{p}, \boldsymbol{D})\subset\Omega$, with joint probability $\boldsymbol{P}(\boldsymbol{p},\boldsymbol{D})$ which has associated a particular manifold, for each tuple. The conditional probability of $\boldsymbol{p}=\boldsymbol{p}_{n}^t$ given $\boldsymbol{D}=\boldsymbol{D}_\tau$ can be represented as a sub-manifold $\mathcal{M}_{\boldsymbol{D}_\tau}^{\boldsymbol{p}_{n}^t}$, of the joint probability manifold,  in the non-linear space of common causal drivers $\boldsymbol{D}_\tau\in\mathbb{R}^M$ of $\boldsymbol{p}_{n}^t$, for a given $n$ and $t$:

\begin{equation}
\mathcal{M}_{\boldsymbol{D}_\tau}^{\boldsymbol{p}_{n}^t}=\left\{P\left[\boldsymbol{p}_{n}^t\left|\boldsymbol{D}_\tau\right.\right]\left|\boldsymbol{D}_\tau\in\mathbb{R}^m,n,t\right.\ \right\}
\end{equation}

Next, the problem must be framed from a statistical perspective in order to enable its resolution using empirical data. This formulation is presented in the following section, with corresponding experiments detailed in the subsequent one.

\subsection{Statistical Approximation}
Assuming the conditional probability $P\left[\boldsymbol{p}_{n}^t\left|\boldsymbol{D}_\tau\right.\right]$ can be model with a neural network via supervised learning and market data, and the conditional expectation estimated. Let $\boldsymbol{D}_\tau$ be a random variable with $E[(\boldsymbol{D}_\tau)^2]<\infty$. Then $E[\boldsymbol{D}_\tau|\mathcal{F}]$ is the orthogonal projection of $\boldsymbol{D}_\tau$ on $\mathcal{L}^2(\Omega,\mathcal{F}, \boldsymbol{P})$. For any $\mathcal{F}-measurable$ $\boldsymbol{p}_{n}^t$ with $E[(\boldsymbol{p}_{n}^t)^2]<\infty$:
\begin{equation}
    E[((\boldsymbol{D}_\tau)^2-(\boldsymbol{p}_{n}^t)^2)]\geq E[(\boldsymbol{D}_\tau-E[\boldsymbol{D}_\tau|\mathcal{F}])^2]    
\end{equation}
with equality if and only if $\boldsymbol{p}_{n}^t=E[\boldsymbol{D}_\tau|\mathcal{F}]$ \citep{KlenkeAchim2007}. On the other hand, it is well known that $E[\boldsymbol{p}_{n}^t|\boldsymbol{D}_\tau]$ is of the form $\bar{f}(\boldsymbol{D}_\tau)$ for a regression function $\bar{f}: \mathcal{R}^m\rightarrow \mathcal{R}$ which can be characterized as a minimizer of the mean squared distance $E[(\boldsymbol{p}_{n}^t - f(\boldsymbol{D}_\tau))^2]$ over all Borel functions $f: \mathcal{R}^m\rightarrow\mathcal{R}$ \citep{Bru1985}. It can be
approximated with a least squares regression, consisting in minimizing an empirical mean squared distance:
\begin{equation}
   \frac{1}{M}\sum_{k=1}^{m}{((\boldsymbol{p})^k-f((\boldsymbol{D})^k))}^2
\end{equation}
based on realizations $(\boldsymbol{D}, \boldsymbol{p})$ of $(\boldsymbol{D}_\tau, \boldsymbol{p}_{n}^t)$ over a suitable family $S$ of Borel functions $f: \mathcal{R}^m\rightarrow \mathcal{R}$. An $L-layer$ neural network is denoted by:
\begin{align}
    &f_{\boldsymbol{\theta}}\left(\boldsymbol{D}_\tau\right)=\boldsymbol{W}^{\left[L-1\right]}\sigma\circ\left(\boldsymbol{W}^{\left[L-1\right]}\sigma\circ\left(\cdots\left(\boldsymbol{W}^{\left[1\right]}\sigma\circ\left(\boldsymbol{W}^{\left[0\right]}\sigma\boldsymbol{D}_\tau+b^{\left[0\right]}\right)+b^{\left[1\right]}\right)\cdots\right)+\right.\nonumber\\
    &+\left.b^{\left[L-2\right]}\right)+b^{\left[L-1\right]}
\end{align}
Where $\boldsymbol{W}^{\left[l\right]}\in\mathbb{R}^{m_{l+1}\times m_l}$, $\boldsymbol{b}^{\left[l\right]}=\mathbb{R}^{m_{l+1}}$, $m_0=d_{in}=d$, $m_L=d_0$, $\sigma$ is a scalar function and $"\circ"$ means entry-wise operation. The set of parameters are $\boldsymbol{\theta}=\left(\boldsymbol{W},\boldsymbol{b}\right)$.

\begin{figure}
	\centering
	\includegraphics[width=1\textwidth]{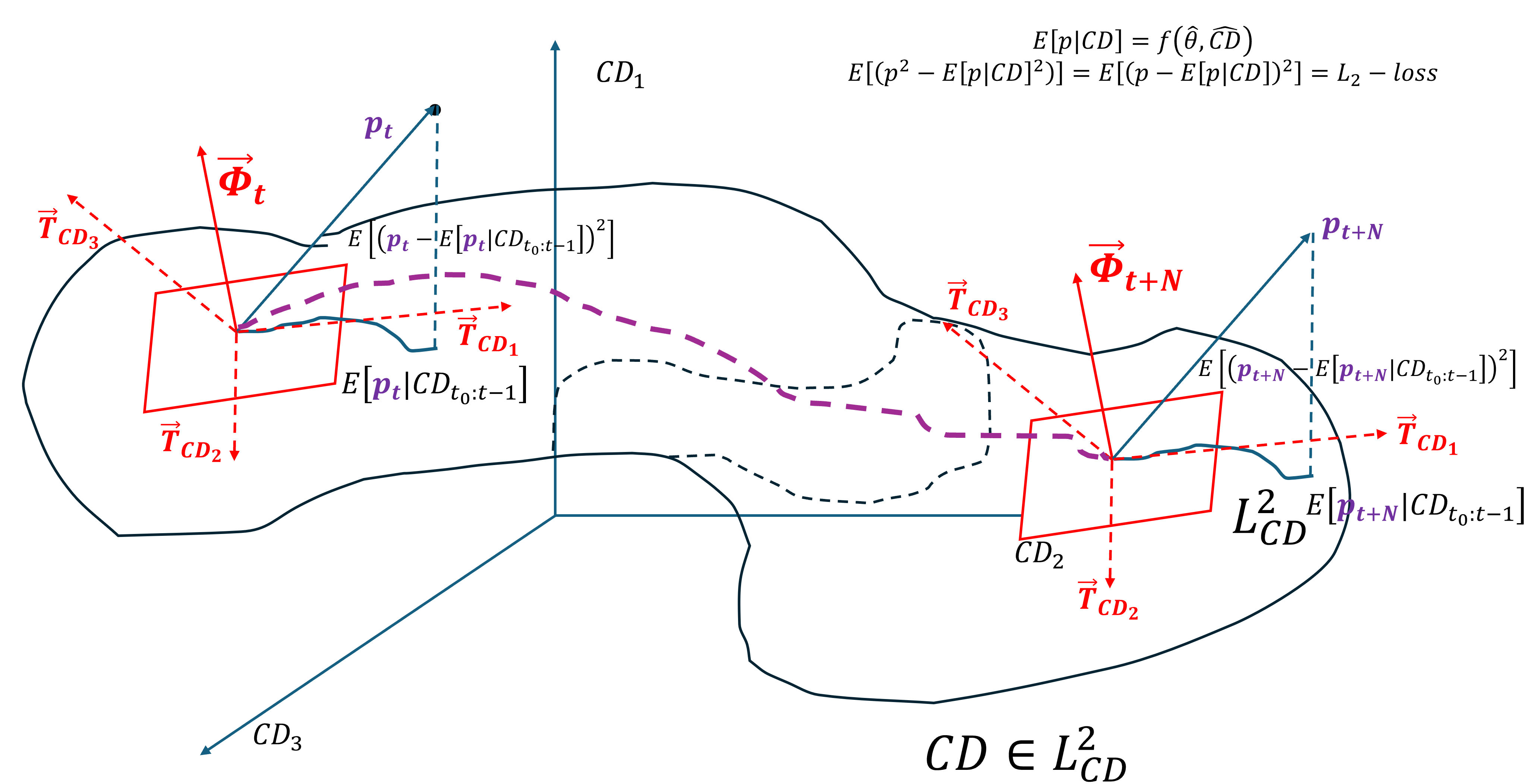}
    \caption{Common Causal Manifolds as a static system or submanifold in which the portfolio trajectory prediction evolves across different tangent spaces, also referred to as sensitivity spaces. The predictions are generated by forecasting the sensitivities and projecting the assets and the portfolio onto the predicted sensitivity (tangent) space, as will be demonstrated in Chapter~\ref{experiemntssenspath}.}

 \label{causalstatic}
\end{figure}

\begin{figure}
	\centering
	\includegraphics[width=1\textwidth]{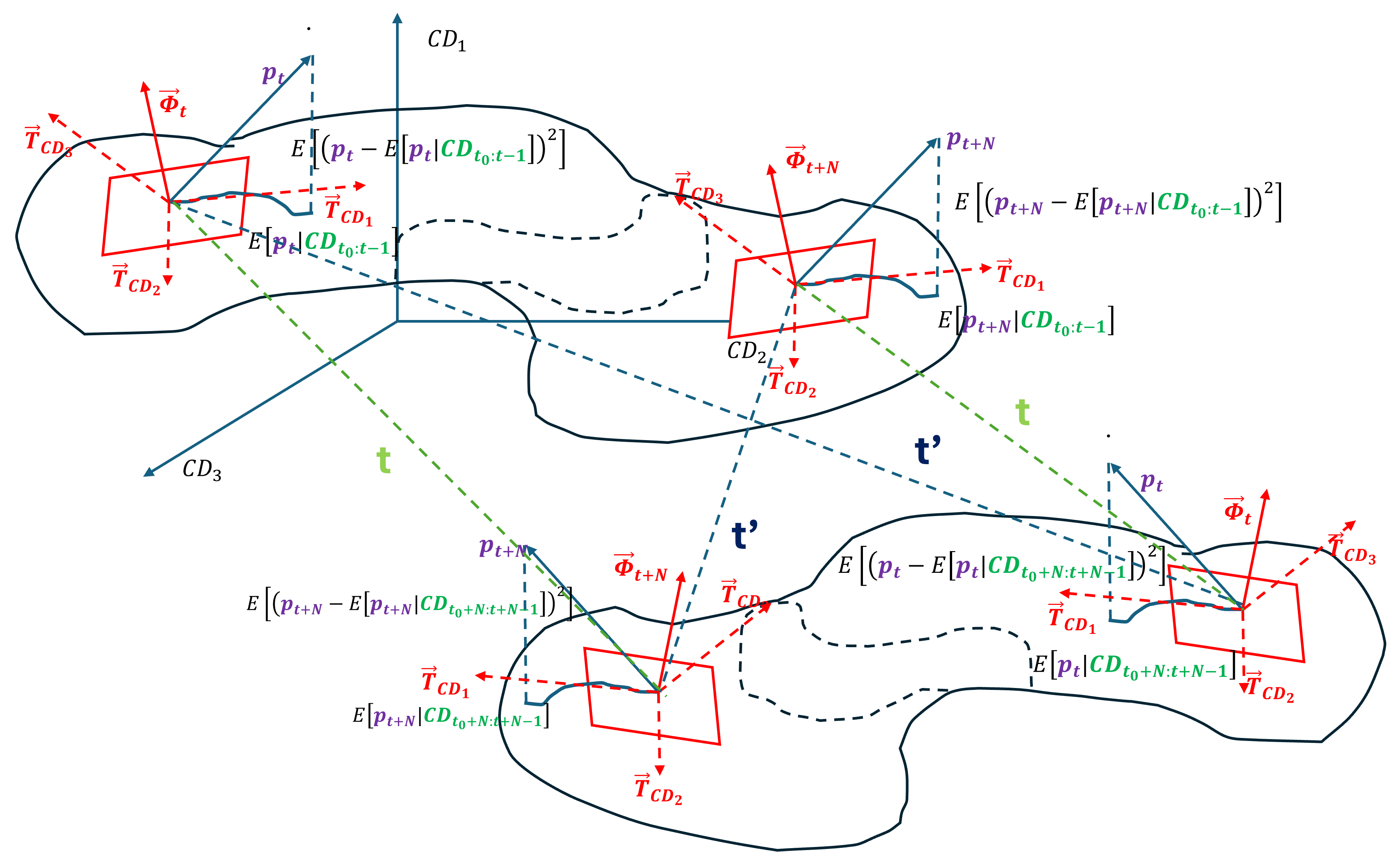}
    \caption{Illustration of both the dynamic and static components of the Common Causal Manifold, aligned with the rules of causal time systems. The embedded sensitivity space evolves as a tangent space over the manifold, representing predictions of sensitivities. The portfolio is projected into this predicted sensitivity space, which corresponds to a static submanifold extracted from a Common Causal Time Manifold, following the formulation in \citep{windeknecht1967causality}.}

 \label{causalstaticdinamic}
\end{figure}

 The conditional expectation $E[\boldsymbol{p}_{n}^t|\boldsymbol{D}_\tau]$ can be viewed a projection of the joint distribution $\boldsymbol{P}(\boldsymbol{p},\boldsymbol{D})$ onto the sub-manifold defined by the conditional distribution $\boldsymbol{P}(\boldsymbol{p}|\boldsymbol{D})$. If a neural network $f_{\boldsymbol{\theta}}(\boldsymbol{D})$ parameterized by weights and biases collected in the vector $\boldsymbol{\theta}$ is used to approximate the conditional probability, the statistical manifold in which each point represents a specific probability distribution parameterized by $\boldsymbol{\theta}$, represents the joint distribution $\boldsymbol{P}(\boldsymbol{p},\boldsymbol{D};\boldsymbol{\theta})$. The conditional distribution $\boldsymbol{P}(\boldsymbol{p}|\boldsymbol{D};\boldsymbol{\theta})$ is a function of $\boldsymbol{\theta}$ and its derivatives with respect to $\boldsymbol{D}$ can be expressed as: 
\begin{equation}
\frac{\partial P(\boldsymbol{p}|\boldsymbol{D}=\boldsymbol{D}_\tau;\boldsymbol{\theta})}{\partial \boldsymbol{D}_\tau}
\end{equation}

These derivatives can be viewed as a tangent vector in the tangent space at the point $\boldsymbol{P}(\boldsymbol{p}|\boldsymbol{D}=\boldsymbol{D}_\tau;\boldsymbol{\theta})$.

Given $n$ and $t$, the conditional probability $P\left[\boldsymbol{p}_{n}^t\left|\boldsymbol{D}_\tau\right.\right]$ represented as a sub-manifold $\mathcal{M}_{\boldsymbol{D}_\tau}^{\boldsymbol{p}_{n}^t}$ approximated by a neural network $f_{\boldsymbol{\theta}}(\boldsymbol{D})$, the variables that define the tangent vector are the inputs of the neural network $\boldsymbol{D}_\tau$, with $\tau=t-s$, for lags $s$, and the output of the neural network $f_{\boldsymbol{\theta}}(\boldsymbol{D}_\tau)$. The tangent vector components are given as a gradient representation by:
\begin{equation}
\nabla_{\boldsymbol{D}_\tau}f_{\boldsymbol{\theta}}(\boldsymbol{D}_\tau)=\left(\frac{\partial f_{\boldsymbol{\theta}}(\boldsymbol{D}_\tau)}{\partial\boldsymbol{D}_1(\tau)},\dots,\frac{\partial f_{\boldsymbol{\theta}}(\boldsymbol{D}_\tau)}{\partial\boldsymbol{D}_m(\tau)}\right)
\end{equation}
This are called sensitivities of the neural network and can be approximated with market data via automatic differentiation. The output of the neural network is influenced by the parameters $\boldsymbol{\theta}$. To express the relationship between the network parameters $\boldsymbol{\theta}$ and the input variables
$\boldsymbol{D}_\tau$ in the tangent space, we need to consider the derivatives of the output with respect to both $\boldsymbol{\theta}$ and $\boldsymbol{D}_\tau$. The tangent space at a point $(\boldsymbol{D}_\tau(0),\boldsymbol{\theta}_0)$ in the manifold of network outputs is spanned by the gradients with respect to both $\boldsymbol{D}_\tau(0)$ and $\boldsymbol{\theta}_0$. Tangent vectors are composed of partial derivatives of the network output with respect to these variables. The total differential is:
\begin{equation}
    df_{\boldsymbol{\theta}}(\boldsymbol{D}_\tau)=\sum_{j=1}^{m}{\frac{\partial f_{\boldsymbol{\theta}}(\boldsymbol{D}_\tau)}{\partial\boldsymbol{D}_j(\tau)}{d\boldsymbol{D}}_j(\tau)}+\sum_{j=1}^{p}{\frac{\partial f_{\boldsymbol{\theta}}(\boldsymbol{D}_\tau)}{\partial\theta_j}\partial\boldsymbol{\theta}_j}
\end{equation}

\subsection{Prediction Loss as a Manifold}
For the single-step ahead forecast using future inputs \(\boldsymbol{D}_{\tau_i}\), the predicted returns at \( t+i \) are:

\[
\mathbf{R}_{t+i} = f_{\boldsymbol{\theta}}(\boldsymbol{D}_{\tau_i}), \quad \text{for } i=1,\ldots,T
\]

where \( \boldsymbol{\theta} \) represents the parameters of the single-step ahead forecast network. The loss is given by:

\[
\mathcal{L} = \left( \mathbb{E}[R_{t+1} \mid \boldsymbol{D}_{\tau_0}] - f_{\boldsymbol{\theta}}(\boldsymbol{D}_{\tau_0}) \right)^2, \quad 
\]

where \( f_{\boldsymbol{\theta}}(\boldsymbol{D}_{\tau_i}) \) is the \(i\)-th predicted return by the neural network. \( \mathbb{E}[R_{t+i} \mid \boldsymbol{D}_{\tau_i}] \) is the true conditional expectation of the portfolio returns at \( t+i \) given the common drivers \(\boldsymbol{D}_{\tau_i}\). For the multi-step ahead forecast using the same input \(\boldsymbol{D}_{\tau_0}\) and different networks and losses for each step ahead:
\[
\mathbf{R}_{t+i} = f_{\boldsymbol{\theta}^{(\tau_0 \rightarrow t+i)}}(\boldsymbol{D}_{\tau_0}), \quad \text{for } i=1,\ldots,T
\]
where \( \boldsymbol{\theta}^{(\tau_0 \rightarrow t+i)} \) represents the parameters of the neural network predicting at \( t+i \) with input $\boldsymbol{D}_{\tau_0}$. The loss for the \(i\)-th neural network is given by:
\[
\mathcal{L}_i = \left( \mathbb{E}[R_{t+i} \mid \boldsymbol{D}_{\tau_0}] - f_{\boldsymbol{\theta}^{(\tau_0 \rightarrow t+i)}}(\boldsymbol{D}_{\tau_0}) \right)^2, \quad \text{for } i=1,\ldots,T
\]
where:
\begin{itemize}
    \item \( \mathcal{L}_i \) is the loss for the \(i\)-th network.
    \item \( \mathbb{E}[R_{t+i} \mid \boldsymbol{D}_{\tau_0}] \) is the true conditional expectation of the portfolio returns at \( t+i \) given the common drivers \(\boldsymbol{D}_{\tau_0}\).
    \item \( f_{\boldsymbol{\theta}^{(\tau_0 \rightarrow t+i)}}(\boldsymbol{D}_{\tau_0}) \) is the predicted by the \(i\)-th neural network.
\end{itemize}

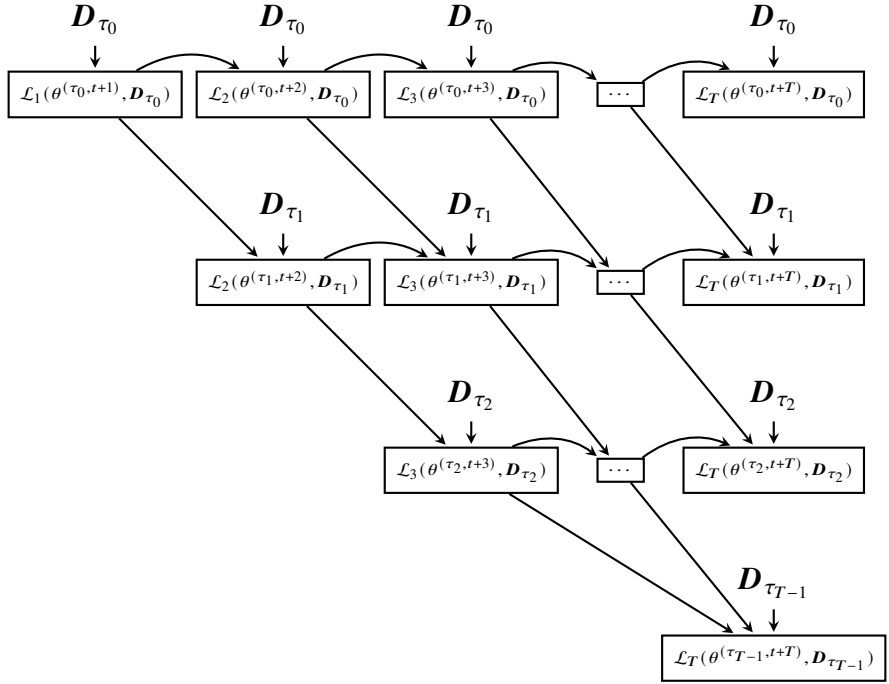
\begin{figure}[h!]
\centering
\begin{tikzpicture}[->,>=stealth,shorten >=1pt,auto,node distance=2.5cm,
                    thick,main node/.style={rectangle,draw,font=\sffamily\tiny}]

  \node[main node] (L1) {$\mathcal{L}_1 (\theta^{(\tau_0,t+1)}, \boldsymbol{D}_{\tau_0})$};
  \node[main node] (L2) [right of=L1] {$\mathcal{L}_2 (\theta^{(\tau_0,t+2)}, \boldsymbol{D}_{\tau_0})$};
  \node[main node] (L3) [right of=L2] {$\mathcal{L}_3 (\theta^{(\tau_0,t+3)}, \boldsymbol{D}_{\tau_0})$};
  \node[main node] (dots1) [right=0.5cm of L3] {$\dots$};
  \node[main node] (LT) [right=0.5cm of dots1] {$\mathcal{L}_T (\theta^{(\tau_0,t+T)}, \boldsymbol{D}_{\tau_0})$};

  \node[main node] (L2_2) [below of=L2, node distance=2.5cm] {$\mathcal{L}_2 (\theta^{(\tau_1,t+2)}, \boldsymbol{D}_{\tau_1})$};
  \node[main node] (L3_2) [right of=L2_2] {$\mathcal{L}_3 (\theta^{(\tau_1,t+3)}, \boldsymbol{D}_{\tau_1})$};
  \node[main node] (dots2) [right=0.5cm of L3_2] {$\dots$};
  \node[main node] (LT_2) [right=0.5cm of dots2] {$\mathcal{L}_T (\theta^{(\tau_1,t+T)}, \boldsymbol{D}_{\tau_1})$};

  \node[main node] (L3_3) [below of=L3_2, node distance=2.5cm] {$\mathcal{L}_3 (\theta^{(\tau_2,t+3)}, \boldsymbol{D}_{\tau_2})$};
  \node[main node] (dots3) [right=0.5cm of L3_3] {$\dots$};
  \node[main node] (LT_3) [right=0.5cm of dots3] {$\mathcal{L}_T (\theta^{(\tau_2,t+T)}, \boldsymbol{D}_{\tau_2})$};

  \node[main node] (LT_4) [below of=LT_3, node distance=2.5cm] {$\mathcal{L}_T (\theta^{(\tau_{T-1},t+T)}, \boldsymbol{D}_{\tau_{T-1}})$};

  \path[every node/.style={font=\sffamily\tiny}]
    (L1) edge [bend left] node [below] {} (L2)
    (L2) edge [bend left] node [below] {} (L3)
    (L3) edge [bend left] node [below] {} (dots1)
    (dots1) edge [bend left] node [below] {} (LT);

  \path[every node/.style={font=\sffamily\tiny}]
    (L2_2) edge [bend left] node [below] {} (L3_2)
    (L3_2) edge [bend left] node [below] {} (dots2)
    (dots2) edge [bend left] node [below] {} (LT_2);

  \path[every node/.style={font=\sffamily\tiny}]
    (L3_3) edge [bend left] node [below] {} (dots3)
    (dots3) edge [bend left] node [below] {} (LT_3);

  \node (D1) [above of=L1, node distance=1cm] {$\boldsymbol{D}_{\tau_0}$};
  \node (D2) [above of=L2, node distance=1cm] {$\boldsymbol{D}_{\tau_0}$};
  \node (D3) [above of=L3, node distance=1cm] {$\boldsymbol{D}_{\tau_0}$};
  \node (D4) [above of=LT, node distance=1cm] {$\boldsymbol{D}_{\tau_0}$};

  \node (D2_2) [above of=L2_2, node distance=1cm] {$\boldsymbol{D}_{\tau_1}$};
  \node (D3_2) [above of=L3_2, node distance=1cm] {$\boldsymbol{D}_{\tau_1}$};
  \node (D4_2) [above of=LT_2, node distance=1cm] {$\boldsymbol{D}_{\tau_1}$};

  \node (D3_3) [above of=L3_3, node distance=1cm] {$\boldsymbol{D}_{\tau_2}$};
  \node (D4_3) [above of=LT_3, node distance=1cm] {$\boldsymbol{D}_{\tau_2}$};

  \node (D4_4) [above of=LT_4, node distance=1cm] {$\boldsymbol{D}_{\tau_{T-1}}$};

  \path[every node/.style={font=\sffamily\tiny}]
    (D1) edge node [left] {} (L1)
    (D2) edge node [left] {} (L2)
    (D3) edge node [left] {} (L3)
    (D4) edge node [left] {} (LT);                                                                                                                                                                                                                                                                                                                                                                                                                                                                                                                                                                      

  \path[every node/.style={font=\sffamily\tiny}]
    (D2_2) edge node [left] {} (L2_2)
    (D3_2) edge node [left] {} (L3_2)
    (D4_2) edge node [left] {} (LT_2);

  \path[every node/.style={font=\sffamily\tiny}]
    (D3_3) edge node [left] {} (L3_3)
    (D4_3) edge node [left] {} (LT_3);

  \path[every node/.style={font=\sffamily\tiny}]
    (D4_4) edge node [left] {} (LT_4);

  \path[every node/.style={font=\sffamily\tiny}]
    (L1) edge node [below] {} (L2_2)
    (L2) edge node [below] {} (L3_2)
    (L3) edge node [below] {} (dots2)
    (dots1) edge node [below] {} (LT_2)
    (L2_2) edge node [below] {} (L3_3)
    (L3_2) edge node [below] {} (dots3)
    (dots2) edge node [below] {} (LT_3)
    (L3_3) edge node [below] {} (LT_4)
    (dots3) edge node [below] {} (LT_4);

\end{tikzpicture}
\caption{Grid of Networks as Predictors and Losses Predicting Different Horizons}
\label{fig:losses_grid}
\end{figure}

The set of multi-step ahead forecast neural networks $f_{\boldsymbol{\theta}^{(\tau_0 \rightarrow t+i)}}(\boldsymbol{D}_{\tau_0})$, with associated losses \( \mathcal{L}_i \) for the \(i\)-th network, $i=1,\dots T$, are combined with the iterative single-step ahead forecast network in a grid. Each multi-step ahead forecast network can be applied in a iterative form by using different input data. Also, the single-step and multi-step ahead forecast networks are equivalent for the case $i=1$. Figure \ref{fig:losses_grid}, shows a diagram for the grid of sequential losses and losses with forecasts at different horizon lengths. This visualization shows the interaction between sequential predictions and predictions at different horizons, with connections indicating corresponding prediction timestamps and parameters, and includes the common drivers for the upper set of nodes.
\begin{itemize}
    \item Nodes: Each node represents a loss for a neural network predicting returns at \( t+i \) with inputs \(\boldsymbol{D}_{\tau_j}\) and parameters \(\boldsymbol{\theta}^{(\tau_j \rightarrow t+i)}\), $i,j = 1,\dots T$. The uppermost array predicts from \( t+1 \) to \( t+T \), the next array from \( t+2 \) to \( t+T \), and so on. Each array predicts from \( t+j \) to \( t+T \), $j=1,\dots T$.
    \item Edges: Arrows within each array connect sequential predictions. Each array shows the predictions made by a set of neural networks, indexed by the parameters \(\boldsymbol{\theta}^{(\tau_j \rightarrow t+i)}\).
    \item Each set of predictions uses common drivers \(\boldsymbol{D}_{\tau_j}\) for the respective nodes.
\end{itemize}

This visualization illustrates the prediction process for different horizons, showing how each network in the array contributes to predicting returns at successive time steps.

\subsection{A Term Structure of Predictors and their Sensitivities}
Let $a_{i,t+k}$ be the predicted value for asset $i$ at future time $t+k$, $\mathbf{D}_{\tau}$ represents the common drivers set at $t=\tau$ with $\tau<t$, and $f_{\boldsymbol{\theta}}^{\tau \rightarrow t+k}$ is the the model, parameterized by $\boldsymbol{\theta}^{\tau \rightarrow t+k}$ that predicts the asset return at time $t+k$ with input, the common drivers at time $t=\tau$ so that:
\begin{equation}
 a_{i,t+k} = f_{\boldsymbol{\theta}}^{\tau \rightarrow t+k}(\mathbf{D}_{\tau})   
\end{equation}

The following grid represents the predictors for different horizons \(k=1,\ldots,T\), with inputs \(\mathbf{D}_{\tau_j}\) at \(t=\tau_j\) for \(j=1,\ldots,T\), ensuring that \(\tau_j < t\). In this grid, \(a_{i,t+k}\) is the predicted value for entity \(i\) at future time \(t+k\). The inputs \(\mathbf{D}_{\tau_j}\) represent the data or features up to time \(\tau_j\). The function \(f_{\boldsymbol{\theta}}^{\tau_j \rightarrow t+k}\) defines the model, parameterized by \(\boldsymbol{\theta}^{\tau_j \rightarrow t+k}\), indicating the prediction from time \(\tau_j\) to \(t+k\).

\[
\begin{array}{c|cccc}
 & k=1 & k=2 & \cdots & k=T \\
\hline
t=1 & f_{\boldsymbol{\theta}}^{\tau_1 \rightarrow 2}(\mathbf{D}_{\tau_1}) & f_{\boldsymbol{\theta}}^{\tau_1 \rightarrow 3}(\mathbf{D}_{\tau_1}) & \cdots & f_{\boldsymbol{\theta}}^{\tau_1 \rightarrow T}(\mathbf{D}_{\tau_1}) \\
t=2 & f_{\boldsymbol{\theta}}^{\tau_2 \rightarrow 3}(\mathbf{D}_{\tau_2}) & f_{\boldsymbol{\theta}}^{\tau_2 \rightarrow 4}(\mathbf{D}_{\tau_2}) & \cdots & f_{\boldsymbol{\theta}}^{\tau_2 \rightarrow T}(\mathbf{D}_{\tau_2}) \\
t=3 & f_{\boldsymbol{\theta}}^{\tau_3 \rightarrow 4}(\mathbf{D}_{\tau_3}) & f_{\boldsymbol{\theta}}^{\tau_3 \rightarrow 5}(\mathbf{D}_{\tau_3}) & \cdots & f_{\boldsymbol{\theta}}^{\tau_3 \rightarrow T}(\mathbf{D}_{\tau_3}) \\
\vdots & \vdots & \vdots & \ddots & \vdots \\
t=T & f_{\boldsymbol{\theta}}^{\tau_T \rightarrow T}(\mathbf{D}_{\tau_T}) & f_{\boldsymbol{\theta}}^{\tau_T \rightarrow T}(\mathbf{D}_{\tau_T}) & \cdots & f_{\boldsymbol{\theta}}^{\tau_T \rightarrow T}(\mathbf{D}_{\tau_T}) \\
\end{array}
\]

\subsection{Derivation of Sensitivities Term Structure}
For each time step \(t\) and prediction horizon \(i\), the sensitivities \(\mathbf{S}_{t+i}\) with respect to the inputs \(\boldsymbol{D}_{\tau_j}\) are given by:

\begin{itemize}
    \item Sensitivities for Single-Step Ahead Predictions
\[
\mathbf{S}_{t+i} = \frac{\partial f_{\boldsymbol{\theta}}(\boldsymbol{D}_{\tau_i})}{\partial \boldsymbol{D}_{\tau_i}}, \quad \text{for } i=1,\ldots,T
\]
    \item Sensitivities for Multi-Step Ahead Predictions
\[
\mathbf{S}_{t+i} = \frac{\partial f_{\boldsymbol{\theta}^{(\tau_j \rightarrow t+i)}}(\boldsymbol{D}_{\tau_j})}{\partial \boldsymbol{D}_{\tau_j}}, \quad \text{for } j=0,\ldots,T, \quad i=1,\ldots,T
\] 
\end{itemize}

For each time step \(t\) and prediction horizon \(i\), the sensitivities \(\mathbf{S}_{t+i}\) are structured in a grid. The first array of sensitivities (Single-Step and Multi-Step coincide at \(i=j-1\)), $i,j=1,\dots,T$:
\[
\begin{aligned}
&\mathbf{S}_{t+1} = \frac{\partial f_{\boldsymbol{\theta}}(\boldsymbol{D}_{\tau_0})}{\partial \boldsymbol{D}_{\tau_0}} = \frac{\partial f_{\boldsymbol{\theta}^{(\tau_0 \rightarrow t+1)}}(\boldsymbol{D}_{\tau_0})}{\partial \boldsymbol{D}_{\tau_0}}, \\
&\mathbf{S}_{t+2} = \frac{\partial f_{\boldsymbol{\theta}}(\boldsymbol{D}_{\tau_1})}{\partial \boldsymbol{D}_{\tau_1}} =  \frac{\partial f_{\boldsymbol{\theta}^{(\tau_1 \rightarrow t+2)}}(\boldsymbol{D}_{\tau_1})}{\partial \boldsymbol{D}_{\tau_1}}, \\
&\mathbf{S}_{t+3} = \frac{\partial f_{\boldsymbol{\theta}}(\boldsymbol{D}_{\tau_2})}{\partial \boldsymbol{D}_{\tau_2}} =  \frac{\partial f_{\boldsymbol{\theta}^{(\tau_2 \rightarrow t+3)}}(\boldsymbol{D}_{\tau_2})}{\partial \boldsymbol{D}_{\tau_2}}, \\
&\dots \\
&\mathbf{S}_{t+T} = \frac{\partial f_{\boldsymbol{\theta}}(\boldsymbol{D}_{\tau_{T-1}})}{\partial \boldsymbol{D}_{\tau_{T-1}}} = 
 \frac{\partial f_{\boldsymbol{\theta}^{(\tau_{T-1} \rightarrow t+T)}}(\boldsymbol{D}_{\tau_{T-1}})}{\partial \boldsymbol{D}_{\tau_{T-1}}}
\end{aligned}
\]

so that $\boldsymbol{\theta}\equiv\boldsymbol{\theta}^{(\tau_{j} \rightarrow t+i)}$. The overall sensitivities across all horizons and input times can be described by combining the arrays into a structured grid. For \(j=0\), \(i=1,\ldots,T\):
\[
\mathbf{S}_{t+1}^1 = \frac{\partial f_{\boldsymbol{\theta}}(\boldsymbol{D}_{\tau_0})}{\partial \boldsymbol{D}_{\tau_0}} = \frac{\partial f_{\boldsymbol{\theta}^{(\tau_0 \rightarrow t+1)}}(\boldsymbol{D}_{\tau_0})}{\partial \boldsymbol{D}_{\tau_0}},\mathbf{S}_{t+2}^2 = \frac{\partial f_{\boldsymbol{\theta}^{(\tau_0 \rightarrow t+2)}}(\boldsymbol{D}_{\tau_0})}{\partial \boldsymbol{D}_{\tau_0}},\dots,\mathbf{S}_{t+T}^T = \frac{\partial f_{\boldsymbol{\theta}^{(\tau_0 \rightarrow t+T)}}(\boldsymbol{D}_{\tau_0})}{\partial \boldsymbol{D}_{\tau_0}}
\]

For \(j=1\), \(i=1,\ldots,T\):
\[
\mathbf{S}_{t+2}^1 = \frac{\partial f_{\boldsymbol{\theta}}(\boldsymbol{D}_{\tau_1})}{\partial \boldsymbol{D}_{\tau_1}} = \frac{\partial f_{\boldsymbol{\theta}^{(\tau_1 \rightarrow t+2)}}(\boldsymbol{D}_{\tau_1})}{\partial \boldsymbol{D}_{\tau_1}},\mathbf{S}_{t+3}^2 = \frac{\partial f_{\boldsymbol{\theta}^{(\tau_1 \rightarrow t+3)}}(\boldsymbol{D}_{\tau_1})}{\partial \boldsymbol{D}_{\tau_1}},\dots,\mathbf{S}_{t+T}^{T-1} = \frac{\partial f_{\boldsymbol{\theta}^{(\tau_1 \rightarrow t+T)}}(\boldsymbol{D}_{\tau_1})}{\partial \boldsymbol{D}_{\tau_1}}
\]

For \(j=2\), \(i=1,\ldots,T\):
\[
\mathbf{S}_{t+3}^1 = \frac{\partial f_{\boldsymbol{\theta}}(\boldsymbol{D}_{\tau_2})}{\partial \boldsymbol{D}_{\tau_2}} = \frac{\partial f_{\boldsymbol{\theta}^{(\tau_2 \rightarrow t+3)}}(\boldsymbol{D}_{\tau_2})}{\partial \boldsymbol{D}_{\tau_2}},\mathbf{S}_{t+4}^2 = \frac{\partial f_{\boldsymbol{\theta}^{(\tau_2 \rightarrow t+4)}}(\boldsymbol{D}_{\tau_2})}{\partial \boldsymbol{D}_{\tau_2}},\dots,\mathbf{S}_{t+T}^{T-2} = \frac{\partial f_{\boldsymbol{\theta}^{(\tau_2 \rightarrow t+T)}}(\boldsymbol{D}_{\tau_2})}{\partial \boldsymbol{D}_{\tau_2}}
\]

\[
\vdots
\]

For \(j=T-2\), \(i=1,\ldots,T\):
\[
\mathbf{S}_{t+T-1}^1 = \frac{\partial f_{\boldsymbol{\theta}}(\boldsymbol{D}_{\tau_{T-2}})}{\partial \boldsymbol{D}_{\tau_{T-2}}} = \frac{\partial f_{\boldsymbol{\theta}^{({\tau_{T-2}} \rightarrow t+T-1)}}(\boldsymbol{D}_{\tau_{T-2}})}{\partial \boldsymbol{D}_{\tau_{T-2}}},\mathbf{S}_{t+T}^2 = \frac{\partial f_{\boldsymbol{\theta}^{(\tau_{T-2} \rightarrow t+T)}}(\boldsymbol{D}_{\tau_{T-2}})}{\partial \boldsymbol{D}_{\tau_{T-2}}}
\]

For \(j=T-1\), \(i=1,\ldots,T\):
\[
\mathbf{S}_{t+T}^1 = \frac{\partial f_{\boldsymbol{\theta}}(\boldsymbol{D}_{\tau_{T-1}})}{\partial \boldsymbol{D}_{\tau_{T-1}}} = \frac{\partial f_{\boldsymbol{\theta}^{({\tau_{T-1}} \rightarrow t+T)}}(\boldsymbol{D}_{\tau_{T-1}})}{\partial \boldsymbol{D}_{\tau_{T-1}}}
\]

This formulation allows for the calculation of sensitivities at each prediction horizon \(t+i\) using the corresponding neural network parameters and input drivers.

\[
\mathbf{S}_{t+i}^{i-j} = \begin{cases}
\frac{\partial f_{\boldsymbol{\theta}}(\boldsymbol{D}_{\tau_i})}{\partial \boldsymbol{D}_{\tau_i}}, & \text{for single-step ahead predictions} \quad \text{for }  i=0,\ldots,T,\\
\frac{\partial f_{\boldsymbol{\theta}^{(\tau_j \rightarrow t+i)}}(\boldsymbol{D}_{\tau_j})}{\partial \boldsymbol{D}_{\tau_j}}, & \text{for multi-step ahead predictions}, \quad \text{for } i,j=0,\ldots,T,
\end{cases}
\]


\subsection{Combined Manifold Representation}
Let $T$ neural networks $f_{\boldsymbol{\theta}^{(\tau_0 \rightarrow t+i)}}(\boldsymbol{D}_{\tau_0})$ for $k=1,\dots,T$ be the $k-th$ step ahead predictors of the portfolio contituent $a_{i}^{t+k}$ given the input $\boldsymbol{D}_{\tau_0}$ as common causal drivers. The combined $L_2$ loss is:
\begin{equation}
\mathcal{L}=\frac{1}{2}\sum_{k=1}^{T}\left(a_{i}^{t+k}-f_{\boldsymbol{\theta}^{(\tau_0 \rightarrow t+k)}}(\boldsymbol{D}_{\tau_0})\right)^2
\end{equation}
For each network, the gradient of the loss with respect to the parameters \(\boldsymbol{\theta}^{(\tau_0 \rightarrow t+k)}\) is:
\begin{equation}
\nabla_{\boldsymbol{\theta}^{(\tau_0 \rightarrow t+k)}} \mathcal{L}_i = \sum_{k=1}^{T} \left( f_{\boldsymbol{\theta}^{(\tau_0 \rightarrow t+k)}}(\boldsymbol{D}_{\tau_0}) - a_{i}^{t+k} \right) \nabla_{\boldsymbol{\theta}^{(\tau_0 \rightarrow t+k)}} f_{\boldsymbol{\theta}^{(\tau_0 \rightarrow t+k)}}(\boldsymbol{D}_{\tau_0})
\end{equation}
The combined gradient in the parameter space is:
\begin{equation}
\nabla_{\boldsymbol{\theta}} \mathcal{L} = \left( \begin{array}{c}
\nabla_{\boldsymbol{\theta}^{(\tau_0 \rightarrow t+1)}} \mathcal{L} \\
\nabla_{\boldsymbol{\theta}^{(\tau_0 \rightarrow t+2)}} \mathcal{L} \\
\vdots \\
\nabla_{\boldsymbol{\theta}^{(\tau_0 \rightarrow t+T)}} \mathcal{L}
\end{array} \right) = \left( \begin{array}{c}
\mathbf{J}_1^T (\mathbf{F}_1 - a_{i}^{t+1}) \\
\mathbf{J}_2^T (\mathbf{F}_2 - a_{i}^{t+2}) \\
\vdots \\
\mathbf{J}_T^T (\mathbf{F}_T - a_{i}^{t+T})
\end{array} \right)
\end{equation}
with $\mathbf{J}_k^T=\nabla_{\boldsymbol{\theta}^{(\tau_0 \rightarrow t+k)}} f_{\boldsymbol{\theta}^{(\tau_0 \rightarrow t+k)}}(\boldsymbol{D}_{\tau_0})$ and $\mathbf{F}_k=f_{\boldsymbol{\theta}^{(\tau_0 \rightarrow t+k)}}(\boldsymbol{D}_{\tau_0})$. For each network, the gradient of the loss with respect to the inputs $(\boldsymbol{D}_{\tau_j})$, $j=1,\dots,T$ is:
\begin{equation}
\nabla_{\boldsymbol{D}_{\tau_j}} \mathcal{L}_k = \sum_{j=1}^{T}\sum_{t=1}^{T} \left( f_{\boldsymbol{\theta}^{(\tau_0 \rightarrow t+k)}}(\boldsymbol{D}_{\tau_j}) - a_{i}^{t+k}\right) \nabla_{\boldsymbol{D}_{\tau_j}} f_{\boldsymbol{\theta}^{(\tau_0 \rightarrow t+k)}}(\boldsymbol{D}_{\tau_j})
\end{equation}

The combined gradient in the input space is:
\begin{equation}
\begin{split}
 \nabla_{\boldsymbol{D}_{\tau_j}} \mathcal{L} = \left( \begin{array}{c}
\nabla_{\boldsymbol{D}_{\tau_1}} \mathcal{L}_1\quad\dots\quad \nabla_{\boldsymbol{D}_{\tau_1}} \mathcal{L}_T\\
\nabla_{\boldsymbol{D}_{\tau_2}} \mathcal{L}_1\quad\dots\quad\nabla_{\boldsymbol{D}_{\tau_2}} \mathcal{L}_T\\
\vdots \\
\nabla_{\boldsymbol{D}_{\tau_T}} \mathcal{L}_1\quad\dots\quad \nabla_{\boldsymbol{D}_{\tau_T}} \mathcal{L}_T
\end{array} \right) =\\
\left( \begin{array}{c}
\mathbf{J}_1^T(\boldsymbol{D}_{\tau_1}) (\mathbf{F}_1(\boldsymbol{D}_{\tau_1}) - a_{i}^{t+1}) \quad \dots\quad\mathbf{J}_T^T(\boldsymbol{D}_{\tau_2}) (\mathbf{F}_T(\boldsymbol{D}_{\tau_1}) - a_{i}^{t+1}) \\
\mathbf{J}_1^T(\boldsymbol{D}_{\tau_2}) (\mathbf{F}_1(\boldsymbol{D}_{\tau_2}) - a_{i}^{t+2})\quad \dots\quad\mathbf{J}_T^T(\boldsymbol{D}_{\tau_1}) (\mathbf{F}_T(\boldsymbol{D}_{\tau_2}) - a_{i}^{t+2}) \\
\vdots \\
\mathbf{J}_1^T(\boldsymbol{D}_{\tau_T}) (\mathbf{F}_1(\boldsymbol{D}_{\tau_T}) - a_{i}^{t+T})\quad\dots\quad\mathbf{J}_T^T(\boldsymbol{D}_{\tau_T}) (\mathbf{F}_T(\boldsymbol{D}_{\tau_T}) - a_{i}^{t+T})
\end{array} \right)   
\end{split}
\end{equation}

Combining the gradients in both the parameter and input space, the tangent vector at each point $((\boldsymbol{\theta}^{(\tau_0 \rightarrow t+k)}, \boldsymbol{D}_{\tau_j})$ can be written as:
\begin{equation}
\mathbf{T}_{(\boldsymbol{\theta}^{(\tau_0 \rightarrow t+k)}, \boldsymbol{D_{\tau_j}})} \mathcal{M} = \left\{ \left( \nabla_{\boldsymbol{\theta}^{(\tau_0 \rightarrow t+k)}} \mathcal{L}, \nabla_{\boldsymbol{D}_{\tau_j}} \mathcal{L} \right) \bigg| (\boldsymbol{\boldsymbol{\theta}}^{(\tau_0 \rightarrow t+k)}, \boldsymbol{D}_{\tau_j}) \in \mathcal{M} \right\}
\end{equation}

\subsection{Combined Causal Time System Manifold}
The combined manifold is composed of a static and dynamic components. The static manifold \(\mathcal{M}_{\text{static}}\) depends on the parameters of the $T$ multi-step ahead neural networks  $\boldsymbol{\boldsymbol{\theta}}^{(\tau_0 \rightarrow t+k)}$, $k=1,\dots,T$. It is static because it does not depend on a time-dependent input but a set of parametric models for different forecasting horizons from \(\tau_o\) to \(t+T\). The static manifold is defined as:
\[ \mathcal{M}_{\text{static}} = \left\{ \left( \theta^{(\tau_0 \rightarrow t+1)}, \theta^{(\tau_0 \rightarrow t+2)}, \ldots, \theta^{(\tau_0 \rightarrow t+T)} \right) \right\} \]
The dynamic manifold \(\mathcal{M}_{\text{dynamic}}\) depends on the time-dependent input variables \(\boldsymbol{D}_{\tau_j}\), $j=1,\dots,T$, and is defined as:
\[ \mathcal{M}_{\text{dynamic}} = \left\{ \boldsymbol{D}_{\tau_1}, \boldsymbol{D}_{\tau_2}, \ldots, \boldsymbol{D}_{\tau_T} \right\} \]
The combined manifold \(\mathcal{M}_{\text{combined}}\) integrates both static and dynamic components in a causal time system manifold:
\[ \mathcal{M}_{\text{combined}} = \mathcal{M}_{\text{static}} \times \mathcal{M}_{\text{dynamic}} \]
The static manifold tangent vectors:
\[ \mathbf{T}_{(\boldsymbol{\theta}^{(\tau_0 \rightarrow t+1)}, \boldsymbol{\theta}^{(\tau_0 \rightarrow t+2)}, \ldots, \boldsymbol{\theta}^{(\tau_0 \rightarrow t+T)})} \mathcal{M}_{\text{static}} = \left( \nabla_{\boldsymbol{\theta}^{(\tau_0 \rightarrow t+1)}} \mathcal{L}, \nabla_{\boldsymbol{\theta}^{(\tau_0 \rightarrow t+2)}} \mathcal{L}, \ldots, \nabla_{\boldsymbol{\theta}^{(\tau_0 \rightarrow t+T)}} \mathcal{L} \right) \]
Dynamic manifold tangent vectors:
\[ \mathbf{T}_{(\boldsymbol{D}_{\tau_0}, \boldsymbol{D}_{\tau_1}, \ldots, \boldsymbol{D}_{\tau_T})} \mathcal{M}_{\text{dynamic}} = \left( \nabla_{\boldsymbol{D}_{\tau_0}} \mathcal{L}, \nabla_{\boldsymbol{D}_{\tau_1}} \mathcal{L}, \ldots, \nabla_{\boldsymbol{D}_{\tau_T}} \mathcal{L} \right) \]

The combined tangent vectors for the causal time system manifold are:
\begin{equation}
    \begin{split}
\mathcal{T}_{\text{combined}} = \mathcal{T}_{\text{static}} \times \mathcal{T}_{\text{dynamic}}=\\
\left(\begin{matrix}
\nabla_{\boldsymbol{\theta}^{(\tau_0 \rightarrow t+1)}}\nabla_{\boldsymbol{D}_{\tau_0}} & \nabla_{\boldsymbol{\theta}^{(\tau_0 \rightarrow t+2)}}\nabla_{\boldsymbol{D}_{\tau_0}} & \begin{matrix}\dots & \nabla_{\boldsymbol{\theta}^{(\tau_0 \rightarrow t+T)}}\nabla_{\boldsymbol{D}_{\tau_0}} \\ \end{matrix} \\
\nabla_{\boldsymbol{\theta}^{(\tau_0 \rightarrow t+1)}}\nabla_{\boldsymbol{D}_{\tau_1}} & \nabla_{\boldsymbol{\theta}^{(\tau_0 \rightarrow t+2)}}\nabla_{\boldsymbol{D}_{\tau_1}} & \begin{matrix}\dots & \nabla_{\boldsymbol{\theta}^{(\tau_0 \rightarrow t+T)}}\nabla_{\boldsymbol{D}_{\tau_1}} \\ \end{matrix} \\
\vdots & \vdots & \begin{matrix}\ddots & \vdots \\ \end{matrix} \\
\nabla_{\boldsymbol{\theta}^{(\tau_0 \rightarrow t+1)}}\nabla_{\boldsymbol{D}_{\tau_T}} & \nabla_{\boldsymbol{\theta}^{(\tau_0 \rightarrow t+2)}}\nabla_{\boldsymbol{D}_{\tau_T}} & \begin{matrix}\dots & \nabla_{\boldsymbol{\theta}^{(\tau_0 \rightarrow t+T)}}\nabla_{\boldsymbol{D}_{\tau_T}} \\ \end{matrix} \\
\end{matrix}\right)
    \end{split}
\end{equation}

\subsection{Sensitivities as Functions of the Tangent Space to the Combined Manifold}
\label{experiemntssenspath}

The tangent vectors to the combined manifold can be represented as:

\[
\mathbf{T}_{(\boldsymbol{\theta}^{(\tau_0 \rightarrow t+k)}, \boldsymbol{D}_{\tau_j})} \mathcal{M} = \left( \nabla_{\boldsymbol{\theta}^{(\tau_0 \rightarrow t+k)}} \mathcal{L}, \nabla_{\boldsymbol{D}_{\tau_j}} \mathcal{L} \right)
\]

where \(\nabla_{\boldsymbol{\theta}^{(\tau_0 \rightarrow t+k)}} \mathcal{L}\) are the gradients with respect to the model parameters and \(\nabla_{\boldsymbol{D}_{\tau_j}} \mathcal{L}\) are the gradients with respect to the input data.

The sensitivities \(\frac{\partial f_{\boldsymbol{\theta}}^{\tau_j \rightarrow t+k}(\mathbf{D}_{\tau_j})}{\partial \mathbf{D}_{\tau_j}}\) can be expressed as components of the tangent vectors to the combined manifold. Specifically, for each prediction step \(k\), the sensitivity can be written as:

\[
S_{\boldsymbol{\theta}}^{\tau_j \rightarrow t+k} = \frac{\partial f_{\boldsymbol{\theta}}^{\tau_j \rightarrow t+k}(\mathbf{D}_{\tau_j})}{\partial \mathbf{D}_{\tau_j}} = \nabla_{\boldsymbol{D}_{\tau_j}} f_{\boldsymbol{\theta}}^{\tau_j \rightarrow t+k}(\mathbf{D}_{\tau_j})
\]

Therefore, the sensitivity at each point \(((\boldsymbol{\theta}^{(\tau_0 \rightarrow t+k)}, \mathbf{D}_{\tau_j}))\) in the combined manifold can be represented as:

\[
S_{\boldsymbol{\theta}}^{\tau_j \rightarrow t+k} = \mathbf{T}_{(\boldsymbol{\theta}^{(\tau_0 \rightarrow t+k)}, \mathbf{D}_{\tau_j})} \mathcal{M} \cdot \mathbf{e}_{\mathbf{D}_{\tau_j}}
\]

where \(\mathbf{e}_{\mathbf{D}_{\tau_j}}\) is the unit vector in the direction of the input data \(\mathbf{D}_{\tau_j}\).

The sensitivities \(\frac{\partial f_{\boldsymbol{\theta}}^{\tau_j \rightarrow t+k}(\mathbf{D}_{\tau_j})}{\partial \mathbf{D}_{\tau_j}}\) are projections of the tangent vectors to the combined manifold onto the directions of the input data \(\mathbf{D}_{\tau_j}\). These projections capture how small changes in the input data affect the outputs of the prediction models, thus reflecting the sensitivity of the predictors with respect to their inputs.

By expressing the sensitivities as functions of the tangent space to the combined manifold, we can leverage the geometric properties of the manifold to analyze and interpret the behavior of the prediction models and their dependencies on the input data.

The trajectory of the function \( f \) in the manifold as a function of time \( t \) describes how the function value evolves over time due to changes in both the model parameters \(\boldsymbol{\theta}\) and the input data \(\mathbf{D}_{\tau}\). The changes in the function value are influenced by the sensitivities (gradients) of the function with respect to the parameters and the inputs.

The trajectory is given by:

\begin{equation}
f_{\boldsymbol{\theta}(t)}(\mathbf{D}_{\tau}(t)) = f_{\boldsymbol{\theta}(0)}(\mathbf{D}_{\tau}(0)) + \int_0^t \left( \nabla_{\boldsymbol{\theta}} f_{\boldsymbol{\theta}(\tau)}(\mathbf{D}_{\tau}(\tau)) \cdot \frac{d\boldsymbol{\theta}(\tau)}{d\tau} + \nabla_{\mathbf{D}_{\tau}} f_{\boldsymbol{\theta}(\tau)}(\mathbf{D}_{\tau}(\tau)) \cdot \frac{d\mathbf{D}_{\tau}(\tau)}{d\tau} \right) d\tau
\end{equation}

This equation shows that the function \( f \) at time \( t \) is equal to its initial value at time \( 0 \) plus the integral of the sum of the products of the sensitivities with respect to the parameters and inputs and their respective rates of change over time.

To derive the expression for the trajectory of the function \( f \) with minimal loss as a function of the sensitivities of the predictors with respect to the inputs \(\mathbf{D}_{\tau}\), we start with the simplified assumption that the gradients \(\nabla_{\mathbf{D}_{\tau}} f\) are constant over time. Let \(\bar{e}_i\) represent the average error between the predicted and true values over the integration interval. Additionally, \(\eta_{\mathbf{D}_{\tau}}\) denotes the learning rate for the inputs.

Given these assumptions, the trajectory of \( f \) up to time \( t \) can be expressed by integrating the sensitivities over time. The initial value of the function is \( f_{\boldsymbol{\theta}(0)}(\mathbf{D}_{\tau}(0)) \), and it evolves based on the average error, the learning rate, and the constant sensitivity.

The function's trajectory is given by:

\[
f_{\boldsymbol{\theta}(t)}(\mathbf{D}_{\tau}(t)) \approx f_{\boldsymbol{\theta}(0)}(\mathbf{D}_{\tau}(0)) + \int_0^t \left( \eta_{\mathbf{D}_{\tau}} \sum_{i} \bar{e}_i \nabla_{\mathbf{D}_{\tau}} f \right) d\tau
\]

Since the integral of a constant over time is simply the product of the constant and the integration limit, this simplifies to:

\[
f_{\boldsymbol{\theta}(t)}(\mathbf{D}_{\tau}(t)) \approx f_{\boldsymbol{\theta}(0)}(\mathbf{D}_{\tau}(0)) + t \left( \eta_{\mathbf{D}_{\tau}} \sum_{i} \bar{e}_i \nabla_{\mathbf{D}_{\tau}} f \right)
\]

Therefore, the trajectory of the function \( f \) with minimal loss, as a function of the sensitivities of the predictors with respect to the input \(\mathbf{D}_{\tau}\), is:

\[
f_{\boldsymbol{\theta}(t)}(\mathbf{D}_{\tau}(t)) \approx f_{\boldsymbol{\theta}(0)}(\mathbf{D}_{\tau}(0)) + t \left( \sum_{i} \bar{e}_i \right) \eta_{\mathbf{D}_{\tau}} \nabla_{\mathbf{D}_{\tau}} f
\]

This expression illustrates that the function \( f \) evolves over time \( t \), starting from its initial value and increasing linearly based on the average error, the learning rate for the inputs, and the constant sensitivity of the predictor with respect to the inputs.

This new framework, which conceptualizes common causal manifolds as causal time systems with tangent spaces, and which is approximated using manifold learning theory and sensitivity modeling, constitutes a natural and logical extension of the framework introduced in \citep{RODRIGUEZDOMINGUEZ2023100447} and developed throughout the previous chapters of this monograph. It enables the integration of a broad range of methodologies, including various predictive models, causal driver identification techniques, sensitivity estimation tools, and convex optimization strategies, offering a highly flexible and robust foundation for advanced portfolio optimization.

\newpage

\subsection{Methodology}
\label{HSP_sens_fut}

In this section, we propose a methodology within the presented framework based on numerical experiments. These experiments use neural networks as predictors and Automatic Adjoint Differentiation (AAD) for sensitivity approximation, consistent with the approach in \citep{RODRIGUEZDOMINGUEZ2023100447}.

Before proceeding, sensitivities are examined from a statistical perspective in order to identify appropriate modeling approaches.

\subsubsection{Statistical Analysis of Sensitivities to Common Drivers}
\label{statanaly}

A comprehensive statistical analysis is performed on sensitivities across thousands of combinations of common causal drivers and portfolio constituents. Sensitivities are approximated, and their distributional properties are analyzed. This analysis is conducted both at the individual driver-asset level and in aggregated buckets by grouping portfolios and drivers, in order to determine whether the behavior of sensitivities varies across categories.

\begin{figure}
    \centering
    \includegraphics[width=0.65\linewidth]{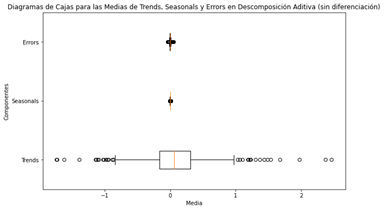}
    \caption{Distribution of sensitivities for all drivers and portfolio constituents in box plot format. As noted by Hugo Valle Varcarcel, the distribution closely resembles a Cauchy distribution.}
    \label{fig:enter-HugoImagen3m}
\end{figure}

\begin{figure}
    \centering
    \includegraphics[width=1\linewidth]{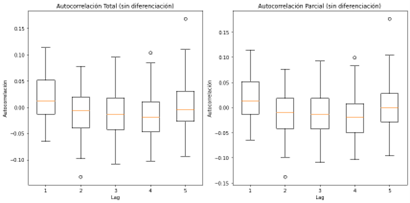}
    \caption{Autocorrelation analysis of the sensitivity time series, indicating the presence of significant autocorrelation. Graph by Hugo Valle Varcarcel.}
    \label{fig:enter-HugoImagen5m}
\end{figure}

\begin{figure}
    \centering
    \includegraphics[width=1\linewidth]{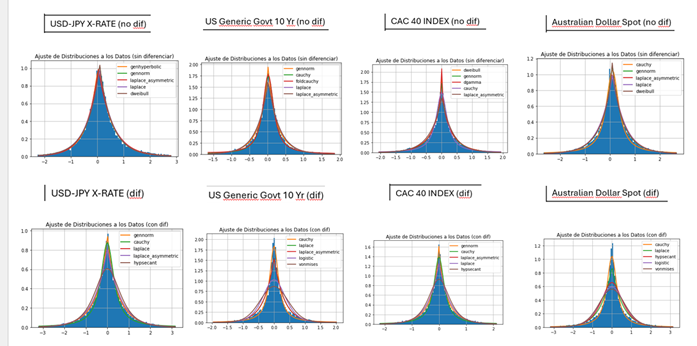}
    \caption{Distribution of sensitivities across multiple portfolio constituents for a common driver. The results show the distributions are not driver-dependent, supporting the idea that sensitivities behave as a general financial process. Graph by Hugo Valle Varcarcel.}
    \label{fig:enter-HugoImagen9m}
\end{figure}

One of the key findings is that, in nearly all cases, the distribution of sensitivities remains consistent between in-sample and out-of-sample data. As shown in Figure \ref{fig:enter-HugoImagen3m}, the distribution of sensitivities across all drivers and constituents resembles a Cauchy distribution. Figure \ref{fig:enter-HugoImagen5m} presents evidence of autocorrelation in the time series of sensitivities, suggesting temporal dependencies that should be accounted for in modeling. Most notably, Figure \ref{fig:enter-HugoImagen9m} demonstrates that the distribution of sensitivities does not vary significantly by driver or asset-driver pair, suggesting that sensitivities exhibit general statistical properties across financial processes. This insight substantially simplifies the modeling of sensitivities for portfolio optimization.

With appropriate models in place, sensitivities can now be forecasted, enabling the construction of predicted tangent spaces and embedded sensitivity spaces. This facilitates the computation of predicted sensitivity matrices, allowing for the tracking of diversification dynamics further into the future. The approach builds upon the foundational framework introduced in~\citep{RODRIGUEZDOMINGUEZ2023100447}, leveraging geometric principles from differential geometry applied to causality and dynamical systems.

\subsubsection{Trajectory-Wide Sensitivity Distance Matrix from Predicted Tangent Spaces}
\label{sec:trajectory_sens_matrix}

Let \( \mathcal{M} \) denote the \textit{Common Causal Manifold}, formed by a set of common causal drivers \( \boldsymbol{CD} = \{D_1, \dots, D_m\} \), and let \( \{a_1, \dots, a_n\} \) represent the portfolio constituents.

At each future time step \( t+\tau \), for \( \tau = 1, \dots, T \), a predicted \textit{tangent space} \( T_{t+\tau} \mathcal{M} \) is constructed from the estimated \textit{sensitivity matrix} \( \hat{\boldsymbol{S}}_{t+\tau} \in \mathbb{R}^{n \times m} \), where each row \( \hat{\boldsymbol{s}}_{t+\tau}^i \) corresponds to the sensitivities of constituent \( a_i \) with respect to the common drivers \( \boldsymbol{CD} \).

The pairwise \textit{sensitivity distance matrix} \( \boldsymbol{\Sigma}_{t+\tau} \in \mathbb{R}^{n \times n} \) is then defined using the Euclidean distance between sensitivity vectors (any distance can be applied thought):
\[
\boldsymbol{\Sigma}_{t+\tau}(i, j) = \left\| \hat{\boldsymbol{s}}_{t+\tau}^i - \hat{\boldsymbol{s}}_{t+\tau}^j \right\|_2.
\]

This matrix captures the geometric structure of diversification across the predicted tangent space at time \( t+\tau \), based on causal sensitivity relationships.

To aggregate information across the entire trajectory, the following summary matrices are defined:

\paragraph{Mean Sensitivity Distance Matrix:}
\[
\bar{\boldsymbol{\Sigma}}_t = \frac{1}{T} \sum_{\tau=1}^{T} \boldsymbol{\Sigma}_{t+\tau}
\]

\paragraph{Cumulative Sensitivity Distance Matrix:}
\[
\tilde{\boldsymbol{\Sigma}}_t = \sum_{\tau=1}^{T} \boldsymbol{\Sigma}_{t+\tau}
\]

Due to the additive property of Euclidean distances, both the mean and cumulative matrices preserve the intrinsic geometry of the sensitivity relationships across the horizon. In practice, these matrices are found to be equivalent for the Euclidean case, as verified empirically in the experiments discussed in subsequent sections.

\textbf{Portfolio Optimization via Hierarchical Sensitivity Parity:}

In the experiments presented in this monograph, portfolio optimization is carried out using the \textit{Hierarchical Sensitivity Parity (HSP)} method applied directly to the cumulative sensitivity distance matrix \( \tilde{\boldsymbol{\Sigma}}_t \). This technique leverages the entire predicted trajectory of the manifold and provides more robust allocation outcomes by capturing future causal dynamics in the tangent space framework. 

\begin{figure}[t]
    \centering
    \includegraphics[width=0.8\textwidth]{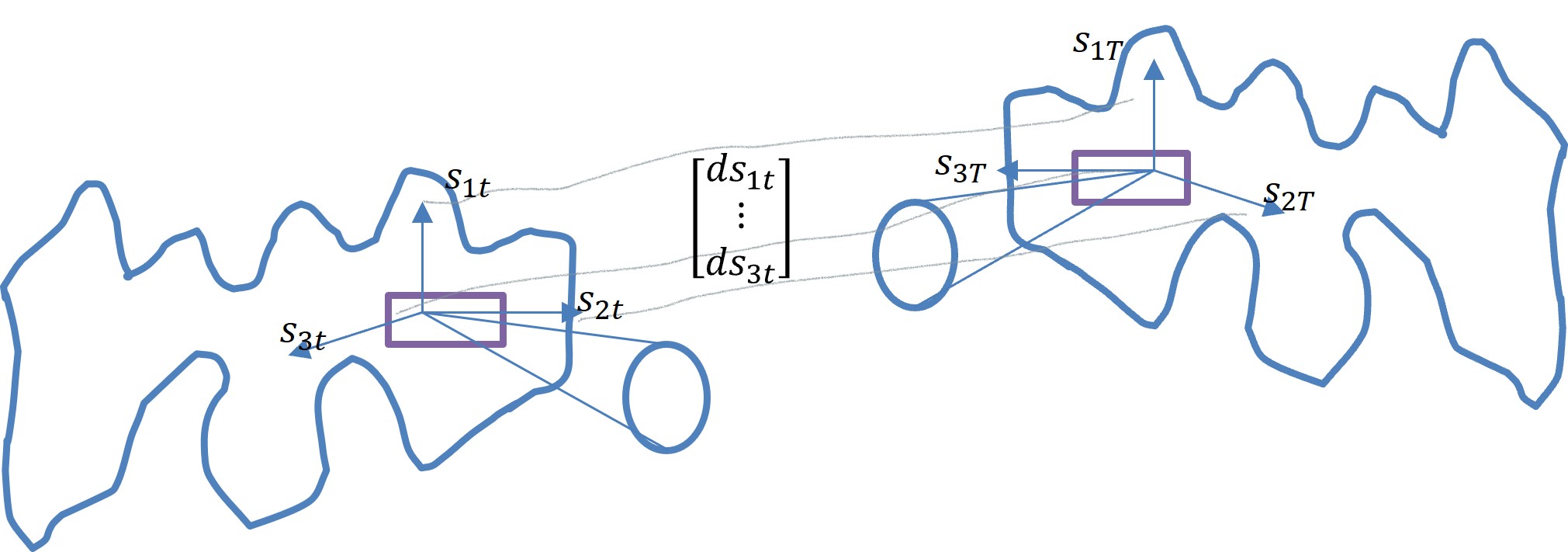} 
    \caption{Two sensitivity space at different absolute times $t$ and $T$. The sensitivity space forecast of time $T$, from time $t$.}
    \label{fig:VasicekSens}
\end{figure}

In Figure \ref{fig:VasicekSens} it can be shown the sensitivity space forecast of time $T$, from time $t$.  

\subsubsection{Modeling Choice for the Sensitivity Process}
Sensitivities are modeled in future states of the causal manifold using Stochastic Differential Equations (SDEs). In this section some examples that are a good fit are discussed.

Consider the state variable \( \mathbf{D}_{t-k} \) following an SDE with a mean-reverting drift term:
\[
d\mathbf{D}_{t-k} = \kappa(\theta - \mathbf{D}_{t-k}) dt + \sqrt{f_{t-k}} d\mathbf{Z}_t
\]
where \( f_{t-k} \) is a function of the state variable \( \mathbf{D}_{t-k} \). The asset price \( a_t \) follows an SDE with stochastic volatility:
\[
da_t = \mu_{t-k, t} dt + \sqrt{g_{t-k, t}} dW_{a, t}
\]
where \( g_{t-k, t} \) is a function of both the state variable \( \mathbf{D}_{t-k} \) and the asset price \( a_t \). We are interested in the sensitivity \( S_t \) defined as:
\[
S_t = \frac{\partial a_t}{\partial \mathbf{D}_{t-k}}
\]

Using Itô's Lemma, the dynamics for the sensitivity \( S_t \) are derived:

\begin{equation}
    \begin{split}
dS_t = \left( \frac{\partial \mu_{t-k, t}}{\partial \mathbf{D}_{t-k}} + S_t \frac{\partial (\kappa(\theta - \mathbf{D}_{t-k}))}{\partial \mathbf{D}_{t-k}} +\right.\\
+\left.\frac{1}{2} \frac{\partial^2 a_t}{\partial \mathbf{D}_{t-k}^2} f_{t-k} \frac{\partial f_{t-k}}{\partial \mathbf{D}_{t-k}} \right) dt + S_t \frac{\partial f_{t-k}}{\partial \mathbf{D}_{t-k}} d\mathbf{Z}_t        
    \end{split}
\end{equation}
The discretize expression of the SDE using the Euler-Maruyama method is:
\begin{equation}
    \begin{split}
S_{t_{n+1}} = S_{t_n} + \left( \frac{\partial \mu_{t-k, t_n}}{\partial \mathbf{D}_{t-k}} + S_{t_n} \frac{\partial (\kappa(\theta - \mathbf{D}_{t-k, t_n}))}{\partial \mathbf{D}_{t-k}} +\right.\\
+\left.\frac{1}{2} \frac{\partial^2 a_{t_n}}{\partial \mathbf{D}_{t-k}^2} f_{t-k, t_n} \frac{\partial f_{t-k, t_n}}{\partial \mathbf{D}_{t-k}} \right) \Delta t + S_{t_n} \frac{\partial f_{t-k, t_n}}{\partial \mathbf{D}_{t-k}} \sqrt{\Delta t} \zeta_n   
    \end{split}
\end{equation}

The partial derivative of \( \mu_{t-k, t} \) with respect to \( \mathbf{D}_{t-k} \), with $\mu_{t-k, t} = \kappa_a (\theta_a - a_t)$, is $\frac{\partial \mu_{t-k, t}}{\partial \mathbf{D}_{t-k}} = 0$ as \( \mu_{t-k, t} \) does not explicitly depend on \( \mathbf{D}_{t-k} \). The partial derivative of the mean-reverting term \( \kappa(\theta - \mathbf{D}_{t-k}) \) is $\frac{\partial (\kappa(\theta - \mathbf{D}_{t-k}))}{\partial \mathbf{D}_{t-k}} = -\kappa$. The second-order partial derivative of \( a_t \) with respect to \( \mathbf{D}_{t-k} \) is $\frac{\partial^2 a_t}{\partial \mathbf{D}_{t-k}^2} \approx 0$. The partial derivative of the local stochastic volatility \( f_{t-k} \) with respect to \( \mathbf{D}_{t-k} \) is $\frac{\partial f_{t-k}}{\partial \mathbf{D}_{t-k}}$. The partial derivative of the local stochastic volatility \( g_{t-k, t} \) with respect to \( \mathbf{D}_{t-k} \) and \( a_t \) are $\frac{\partial g_{t-k, t}}{\partial \mathbf{D}_{t-k}}$ and $\frac{\partial g_{t-k, t}}{\partial a_t}$ respectively. Substituting these derivatives into the discretized update rule:
\[
S_{t_{n+1}} = S_{t_n} + \left( 0 + S_{t_n} (-\kappa) + \frac{1}{2} \cdot 0 \cdot f_{t-k, t_n} \cdot \frac{\partial f_{t-k, t_n}}{\partial \mathbf{D}_{t-k}} \right) \Delta t + S_{t_n} \frac{\partial f_{t-k, t_n}}{\partial \mathbf{D}_{t-k}}\sqrt{\Delta t} \zeta_n
\]

Simplifying, the discretized update for the sensitivity \( S_t \) is:
\[
S_{t_{n+1}} = S_{t_n} (1 - \kappa \Delta t) + S_{t_n} \frac{\partial f_{t-k, t_n}}{\partial \mathbf{D}_{t-k}} \sqrt{\Delta t} \zeta_n
\]

\subsubsection{Example with Specific SDEs for \( f \) and \( g \)}

Consider the state variable \( \mathbf{D}_{t-k} \) following the SDE:
\[
d\mathbf{D}_{t-k} = \kappa(\theta - \mathbf{D}_{t-k}) dt + \sqrt{\nu_{t-k}} d\mathbf{Z}_t
\]
where \( \nu_{t-k} \) is given by:
\[
\nu_{t-k} = \alpha \mathbf{D}_{t-k} + \beta \int_0^{t} \mathbf{D}_s ds
\]

The asset price \( a_t \) follows the SDE:
\[
da_t = \mu_{t-k, t} dt + \sqrt{\sigma_{t-k, t}} dW_{a, t}
\]
where \( \sigma_{t-k, t} \) is given by:
\[
\sigma_{t-k, t} = \gamma a_t + \delta \int_0^{t} a_s ds + \rho \mathbf{D}_{t-k}
\]

We define the sensitivity \( S_t \) as:
\[
S_t = \frac{\partial a_t}{\partial \mathbf{D}_{t-k}}
\]

Using Itô's Lemma, the dynamics for the sensitivity \( S_t \) are:
\[
dS_t = \left( \frac{\partial \mu_{t-k, t}}{\partial \mathbf{D}_{t-k}} + S_t \frac{\partial (\kappa(\theta - \mathbf{D}_{t-k}))}{\partial \mathbf{D}_{t-k}} + \frac{1}{2} \frac{\partial^2 a_t}{\partial \mathbf{D}_{t-k}^2} \nu_{t-k} \frac{\partial \nu_{t-k}}{\partial \mathbf{D}_{t-k}} \right) dt + S_t \frac{\partial \nu_{t-k}}{\partial \mathbf{D}_{t-k}} d\mathbf{Z}_t
\]

Using the Euler-Maruyama method to discretize the SDE:

\begin{equation}
    \begin{split}
    S_{t_{n+1}} = S_{t_n} + \left( \frac{\partial \mu_{t-k, t_n}}{\partial \mathbf{D}_{t-k}} + S_{t_n} \frac{\partial (\kappa(\theta - \mathbf{D}_{t-k, t_n}))}{\partial \mathbf{D}_{t-k}} +\right.\\
    +\left.\frac{1}{2} \frac{\partial^2 a_{t_n}}{\partial \mathbf{D}_{t-k}^2} \nu_{t-k, t_n} \frac{\partial \nu_{t-k, t_n}}{\partial \mathbf{D}_{t-k}} \right) \Delta t + S_{t_n} \frac{\partial \nu_{t-k, t_n}}{\partial \mathbf{D}_{t-k}} \sqrt{\Delta t} \zeta_n    
    \end{split}
\end{equation}

The partial derivative of \( \mu_{t-k, t} = \kappa_a (\theta_a - a_t)\) with respect to \( \mathbf{D}_{t-k} \) is $\frac{\partial \mu_{t-k, t}}{\partial \mathbf{D}_{t-k}} = 0$ since Since \( \mu_{t-k, t} \) does not explicitly depend on \( \mathbf{D}_{t-k} \). The partial derivative of the mean-reverting term \( \kappa(\theta - \mathbf{D}_{t-k}) \) is $\frac{\partial (\kappa(\theta - \mathbf{D}_{t-k}))}{\partial \mathbf{D}_{t-k}} = -\kappa$. The second-order partial derivative of \( a_t \) with respect to \( \mathbf{D}_{t-k} \) is $\frac{\partial^2 a_t}{\partial \mathbf{D}_{t-k}^2} \approx 0$. The partial derivative of the local stochastic volatility \( \nu_{t-k} \) with respect to \( \mathbf{D}_{t-k} \):
\[
\nu_{t-k} = \alpha \mathbf{D}_{t-k} + \beta \int_0^{t} \mathbf{D}_s ds
\]
\[
\frac{\partial \nu_{t-k}}{\partial \mathbf{D}_{t-k}} = \alpha + \beta t
\]

The partial derivative of the local stochastic volatility \( \sigma_{t-k, t} \) with respect to \( \mathbf{D}_{t-k} \) and \( a_t \):
\[
\sigma_{t-k, t} = \gamma a_t + \delta \int_0^{t} a_s ds + \rho \mathbf{D}_{t-k}
\]
\[
\frac{\partial \sigma_{t-k, t}}{\partial \mathbf{D}_{t-k}} = \rho
\]
\[
\frac{\partial \sigma_{t-k, t}}{\partial a_t} = \gamma + \delta t
\]

Substituting these derivatives into the discretized update rule:
\[
S_{t_{n+1}} = S_{t_n} + \left( 0 + S_{t_n} (-\kappa) + \frac{1}{2} \cdot 0 \cdot \nu_{t-k, t_n}(\alpha + \beta t) \right) \Delta t + S_{t_n} (\alpha + \beta t) \sqrt{\Delta t} \zeta_n
\]

Simplifying, the discretized update for the sensitivity \( S_t \) is:
\[
S_{t_{n+1}} = S_{t_n} (1 - \kappa \Delta t) + S_{t_n} (\alpha + \beta t) \sqrt{\Delta t} \zeta_n
\]

\subsubsection{Vasicek}

Consider the state variable \( \mathbf{D}_{t-k} \) following an SDE with a mean-reverting drift term:
\[
d\mathbf{D}_{t-k} = \kappa(\theta - \mathbf{D}_{t-k}) dt + \sigma_D d\mathbf{Z}_t
\]

The asset price \( a_t \) follows an SDE with stochastic volatility:
\[
da_t = \mu_a dt + \sigma_a dW_{a, t}
\]

To match the Vasicek sensitivity formula, we choose \( f \) and \( g \) as functions of the asset's and state's drifts and volatilities. Let:
\[
f_{t-k} = \left( \frac{\sigma_D^2}{2 \kappa} \right) \left(1 - e^{-2\kappa (t-k)} \right)
\]
\[
g_{t-k, t} = \left( \frac{\sigma_a^2}{2 \kappa_a} \right) \left(1 - e^{-2\kappa_a t} \right)
\]
Using Itô's Lemma, the dynamics for the sensitivity \( S_t = \frac{\partial a_t}{\partial \mathbf{D}_{t-k}} \) are:
\[
dS_t = \left( \frac{\partial \mu_a}{\partial \mathbf{D}_{t-k}} + S_t \frac{\partial (\kappa(\theta - \mathbf{D}_{t-k}))}{\partial \mathbf{D}_{t-k}} + \frac{1}{2} \frac{\partial^2 a_t}{\partial \mathbf{D}_{t-k}^2} f_{t-k} \frac{\partial f_{t-k}}{\partial \mathbf{D}_{t-k}} \right) dt + S_t \frac{\partial f_{t-k}}{\partial \mathbf{D}_{t-k}} d\mathbf{Z}_t
\]
with $\frac{\partial \mu_a}{\partial \mathbf{D}_{t-k}} = 0$, $\frac{\partial (\kappa(\theta - \mathbf{D}_{t-k}))}{\partial \mathbf{D}_{t-k}} = -\kappa$ and $\frac{\partial^2 a_t}{\partial \mathbf{D}_{t-k}^2} \approx 0$.  The partial derivative of the local stochastic volatility \( f_{t-k} = \left( \frac{\sigma_D^2}{2 \kappa} \right) \left(1 - e^{-2\kappa (t-k)} \right) \) with respect to \( \mathbf{D}_{t-k} \) is $\frac{\partial f_{t-k}}{\partial \mathbf{D}_{t-k}} = 0$. The partial derivative of the local stochastic volatility \( g_{t-k, t} = \left( \frac{\sigma_a^2}{2 \kappa_a} \right) \left(1 - e^{-2\kappa_a t} \right) \) with respect to \( \mathbf{D}_{t-k} \) and \( a_t \) are $\frac{\partial g_{t-k, t}}{\partial \mathbf{D}_{t-k}} = 0$ and $\frac{\partial g_{t-k, t}}{\partial a_t} = 0$ respectively. Substituting these derivatives into the dynamics of the sensitivity \( S_t \):
\[
dS_t = \kappa (\theta - S_t) dt + \sigma_S dW_t
\]
where \( \sigma_S \) is the combined volatility term. Using the Euler-Maruyama method to discretize the SDE, the discretized form is:
\[
S_{t_{n+1}} = S_{t_n} + \kappa (\theta - S_{t_n}) \Delta t + \sigma_S \sqrt{\Delta t} \zeta_n
\]
where \( \zeta_n \) is a standard normal random variable.

\begin{table}[h!]
\centering
\begin{tabular}{|c|c|c|c|}
\hline
\textbf{Model} & \textbf{Formula (Discrete Form)} & \textbf{Parameters} & \textbf{Letter} \\
\hline
ARIMA & 
$ S_t = c + \sum_{i=1}^{p} \phi_i S_{t-i} + \sum_{j=1}^{q} \theta_j \epsilon_{t-j} + \epsilon_t $ & 
$c, \phi_i, \theta_j, \epsilon_t$ & 
$\alpha$ \\
\hline
Vasicek  & 
$ S_{t+1} = \kappa (\mu - S_t) + S_t + \sigma \epsilon_t $ & 
$\kappa, \mu, \sigma, \epsilon_t$ & 
$\beta$ \\
\hline
Hull-White  & 
$ S_{t+1} = S_t + \kappa (\theta_t - S_t) + \sigma \epsilon_t $ & 
$\kappa, \theta_t, \sigma, \epsilon_t$ & 
$\gamma$ \\
\hline
HJM & 
$ S_{t+1} = S_t + \mu_t \Delta t + \sigma_t \epsilon_t \sqrt{\Delta t} $ & 
$\mu_t, \sigma_t, \epsilon_t$ & 
$\eta$ \\
\hline
N. N.  & 
$ S_t = f_{\theta}(\boldsymbol{D}_\tau) $ & 
$\theta$ & 
$\delta$ \\
\hline
Recurrent N.N. & 
$ S_t = f_{\theta}(S_{t-1}, \boldsymbol{D}_\tau) $ & 
$\theta$ & 
$\epsilon$ \\
\hline
Non-Linear & 
$ S_t = \sum_{i=1}^{p} \phi_i S_{t-i} + \sum_{j=1}^{q} \theta_j g(S_{t-j}) + \epsilon_t $ & 
$\phi_i, \theta_j, \epsilon_t$ & 
$\zeta$ \\
\hline
\end{tabular}
\caption{Various Time Series Models with Formulas, Parameters, and Greek Letter Notation}
\label{tab:models}
\end{table}

\newpage

\subsection{Results}
Finally, the methodology is shown for the case in which sensitivities are modeled as stochastic processes. In this setting, stochastic models are calibrated to sensitivity data obtained from neural networks using Automatic Adjoint Differentiation (AAD) as in \citep{RODRIGUEZDOMINGUEZ2023100447} and Chapter \ref{PDEsAAD}. Once calibrated, these models enable the simulation of future sensitivity paths for a given investment horizon, such as the upcoming month. Portfolio constituents can then be projected into the corresponding sensitivity space at each simulated time step. For each step, a sensitivity-based distance matrix is computed, and these are subsequently aggregated into a single matrix that captures both the magnitude and temporal trajectory of diversification dynamics.

\begin{figure}[t]
    \centering
    \includegraphics[width=0.8\textwidth]{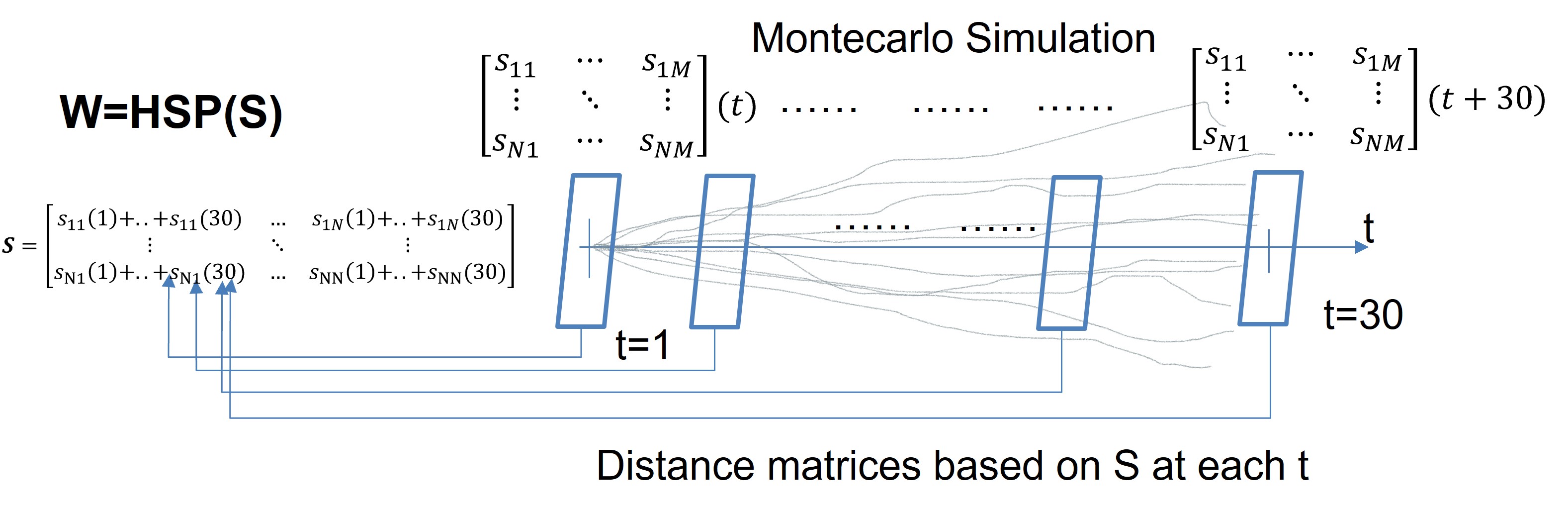} 
    \caption{Full method using SDEs calibrated from data obtained with ADD and Neural Nets as introduced in \citep{RODRIGUEZDOMINGUEZ2023100447} and further developed in Chapter~\ref{PDEsAAD}, the sensitivities are simulated using Monte Carlo techniques. A sensitivity space is constructed for each predicted timestamp, and a corresponding distance (sensitivity) matrix is computed. These matrices are aggregated into a single cumulative matrix (euclidean case), which serves as the input to the Hierarchical Sensitivity Parity (HSP) method. This approach enables portfolio optimization by incorporating the entire forecasted trajectory of optimal diversification dynamics over the future month.}

    \label{fig:VasicekSens2}
\end{figure}

In Figure \ref{fig:VasicekSens2} the full method is shown for the case of the Vasicek model~\citep{VASICEK1977177} and the Hull–White model~\citep{White1990} for the sensitivity calibration and simulated paths. 

From an approximation standpoint, the idea is that, once the sensitivity function is discretely fitted to the data, one can calibrate a model and perform future simulations. This process can be applied to all sensitivities. At each future time step, a sensitivity space can be formed with properties consistent with the modeling approach used to obtain it. Within this space, a sensitivity distance matrix can be computed. If this process is repeated sequentially over time—using appropriate metrics, such as Euclidean distance for the sensitivity matrices—it becomes straightforward to aggregate all the matrices into one. This aggregate can then be used in a Hierarchical Sensitivity Parity (HSP) framework or another sensitivity-based method from Chapter \ref{RiskMap} to incorporate the trajectory information into the manifold of the common causal system over time in the portfolio optimization process.

\begin{figure}[H]
    \centering
    \includegraphics[width=\textwidth]{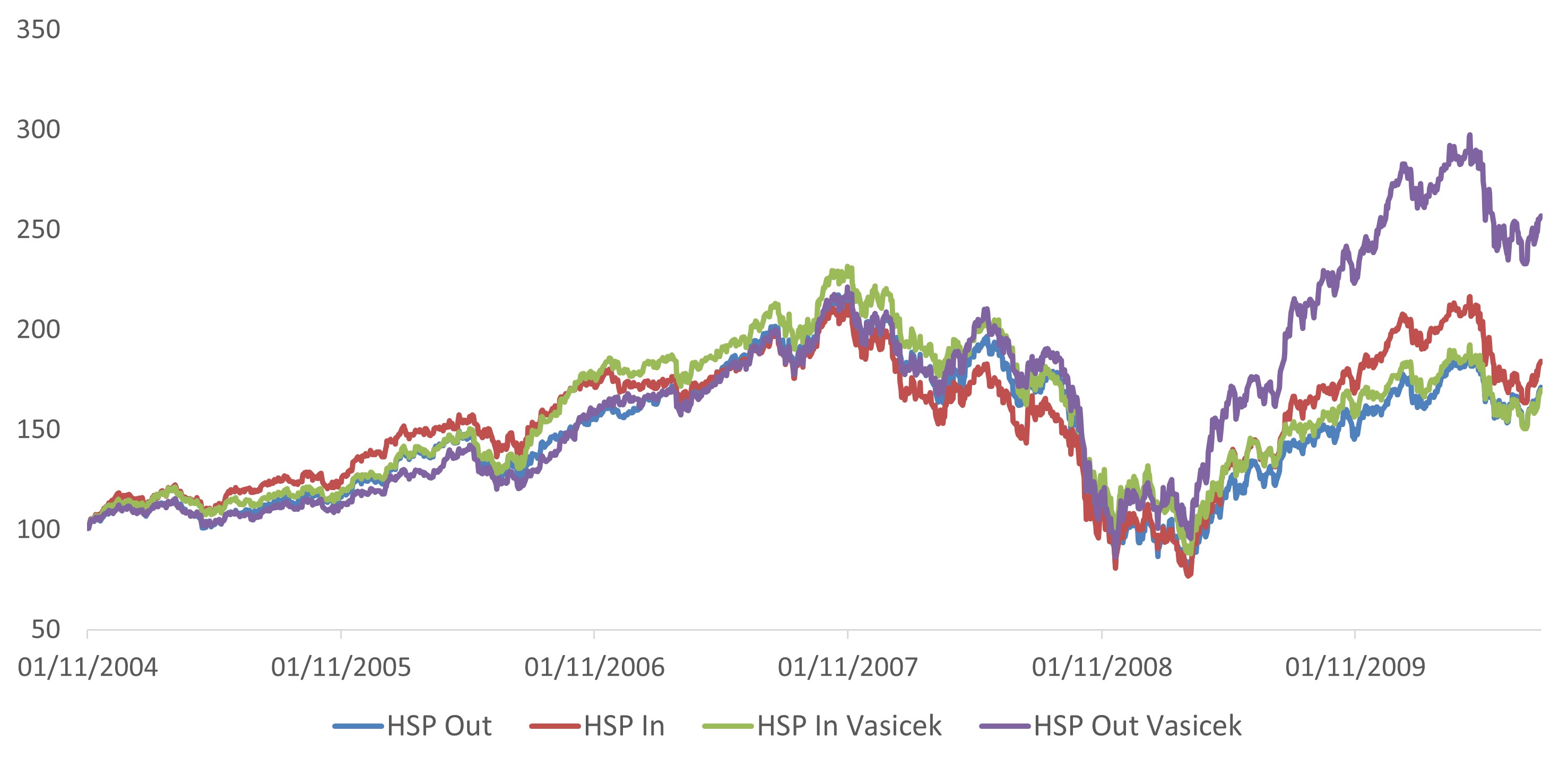} 
    \caption{Net Asset Values (NAVs) of the Path-dependent Hierarchical Sensitivity Parity (HSP) method with Vasicek-modeled sensitivities compared to the original HSP approach during the 2008 credit crisis period. Euclidean distance is used for the sensitivity distance matrix. Sensitivities are estimated using neural networks (non-deep architectures) and Monte Carlo simulation. “Out” refers to models calibrated with out-of-sample sensitivity data, while “In” refers to those calibrated with in-sample data, both obtained prior to the rebalancing date.}

    \label{fig:VasicekSens3}
\end{figure}

This approach enables portfolio optimization by incorporating the entire forecasted trajectory of optimal diversification dynamics over the future month. In Figure~\ref{fig:VasicekSens3}, the results of the proposed \textit{Path-dependent Hierarchical Sensitivity Parity} (HSP) method are presented for the 2008 Credit Crunk period, alongside the standard HSP implementation described in earlier chapters. The comparison demonstrates that the path-dependent extension outperforms other HSP variants.

Although other competing portfolio optimization models are not included in this figure—as this constitutes an ablation study focusing exclusively on HSP variants—it is worth noting that in prior chapters, the standard HSP methodology already outperformed all other methods across all the shown experimental scenarios.

\newpage

\appendix

\section{Examples of Common Drivers' Selections and Re-balancing Fittings Results for the Experiments}
\label{sec:sample:appendix}


\begin{table}[H]
\label{tablenumbers}
\centering
\begin{tabularx}{\columnwidth} {@{} l*{3}{X} @{}}
\toprule
  & 6m periods              & 1y periods              \\
\toprule
\midrule
1 & 01/12/2019 - 01/06/2020 & 01/06/2019 - 01/06/2020 \\
2 & 01/06/2020 - 01/01/2021 & 01/01/2020 - 01/01/2021 \\
3 & 01/01/2021 - 01/07/2021 & 01/07/2021 - 01/07/2021 \\
\bottomrule
\end{tabularx}
\caption{Start and end dates of past windows on the three dates for the common drivers’ selection: In the first column, we assign numbers (1,2,3) to each window for the subsequent tables. 6 and 12-month widows lengths on 01/06/2020, 01/01/2021 and 01/07/2021}
\end{table}

\begin{table}[H]
\fontsize{8}{8}\selectfont
\centering
\begin{tabularx}{\columnwidth} {@{} l*{3}{X} @{}}
\toprule
\textbf{SPX 1 6m   OPT}   & \textbf{SPX 2 6m OPT}        & \textbf{SPX 3 6m OPT}        \\
\toprule
\midrule
MSCI INDIA                & S\&P 500 HEALTH CARE IDX     & DOW JONES INDUS. AVG         \\
USD-NOK RR 25D 3M         & S\&P 500 CONS STAPLES IDX    & S\&P 500 INDEX               \\
USD-SEK RR 25D 3M         & ISHARES MSCI USA QUALITY FAC & MSCI WORLD                   \\
IBEX 35 INDEX             & ISHARES MSCI USA SIZE FACTOR & MSCI Daily TR Net World      \\
S\&P 500 HEALTH CARE IDX  & ISHARES MSCI USA MIN VOL FAC & S\&P 500 HEALTH CARE IDX     \\
S\&P 500 CONS STAPLES IDX & MSCI World Quality Pr \$     & ISHARES MSCI USA QUALITY FAC \\
STXE 600 Utilities EUR    & World Size Tilt              & ISHARES MSCI USA SIZE FACTOR \\
STXE 600 Telcomm EUR      & MSCI WORLD Min Vol PR        & ISHARES MSCI USA MIN VOL FAC \\
MSCI EM LATIN AMERICA     & MSCI World ESG               & MSCI WORLD Min Vol PR        \\
MSCI World High Dividend  & MSCI WORLD/REAL EST          & MSCI World ESG               \\
\multicolumn{2}{l}{MSCI   WORLD/HLTH CARE}               & MSCI Daily Net TR World      \\
\multicolumn{2}{l}{MSCI   WORLD/CON STPL}                &                              \\
\bottomrule
\end{tabularx}
\caption{For the three dates of common drivers' selection, for USA portfolio, and 6-month past window length, we have OPT selection of common drivers winners, which means the full algorithmic selection based on thresholds for correlation values as in  Section \ref{Subsection41App} is used.}
\end{table}

\begin{table}[H]
\fontsize{8}{8}\selectfont
\centering
\begin{tabularx}{\columnwidth} {@{} l*{3}{X} @{}}
\toprule
\textbf{SPX 1 6m SELECT}          & \textbf{SPX 2 6m SELECT}           & \textbf{SPX 3 6m SELECT}\\   
\toprule
\midrule
USD SWAP SEMI 30/360 10Y & Generic 1st 'FV' Future   & DOW JONES INDUS. AVG         \\
EUR-CZK X-RATE           & BONOS Y OBLIG DEL ESTADO  & Generic 1st 'S ' Future      \\
CHF-JPY X-RATE           & NASDAQ COMPOSITE          & S\&P 500 INDUSTRIALS IDX     \\
BUONI POLIENNALI DEL TES & EUR SWAP ANN (VS 6M) 10Y  & SOYBEAN FUTURE    Nov21      \\
UK Gilts 30 Year         & U.S. TIPS                 & ISHARES MSCI USA VALUE FACTO \\
US Generic Govt 10 Yr    & MSCI World Momentum Pri\$ & ISHARES MSCI USA SIZE FACTOR \\
MSCI INDIA               & MSCI WORLD/REAL EST       & MSCI WORLD VALUE INDEX       \\
\multicolumn{2}{l}{U.S.   Treasury}                  & World Size Tilt              \\
\multicolumn{2}{l}{USD-NOK RR   25D 3M}              & MSCI WORLD/INDUSTRL          \\
\multicolumn{2}{l}{USD-SEK RR   25D 3M}              &                              \\
\multicolumn{2}{l}{NASDAQ   COMPOSITE}               &                              \\
\multicolumn{2}{l}{MSCI US REIT   INDEX}             &                              \\
\multicolumn{2}{l}{Japanese Yen   Spot}              &                              \\
\multicolumn{2}{l}{Indian Rupee   Spot}              &                              \\
U.S. TIPS                &                           &                              \\
1-3 Year EU              &                           &                              \\
\bottomrule
\end{tabularx}
\caption{Common drivers' selection with the SELECT method for final selection among the common drivers set for three different selection dates (6-months past window length). The algorithm in Section \ref{Subsection41App} has been tuned based on spurious correlation, stock as part of an index driver, or multicollinearity.}
\end{table}

\begin{table}[H]
\caption{Common drivers selection with the SELECT method for final selection among the common drivers set for three different selection dates (12 months past window length). The selection was performed by a Junior PM}
\centering
\begin{tabularx}{\columnwidth} {@{} l*{3}{X} @{}}
\toprule
\textbf{SPX 1 1y SELECT}                 & \textbf{SPX 2 1y SELECT}           & \textbf{SPX 3 1y SELECT}               \\       \\
\toprule
\midrule
USD SWAP SEMI 30/360 10Y       & CAC 40 INDEX             & DOW JONES INDUS. AVG         \\
USD SWAP SEMI 30/360 30Y       & BONOS Y OBLIG DEL ESTADO & S\&P 500 FINANCIALS INDEX    \\
EUR-CZK X-RATE                 & BUONI POLIENNALI DEL TES & S\&P 500 INDUSTRIALS IDX     \\
UK Gilts 10 Yr                 & US Generic Govt 10 Yr    & S\&P 500 MATERIALS INDEX     \\
UK Gilts 30 Year               & NASDAQ 100 STOCK INDX    & ISHARES MSCI USA VALUE FACTO \\
US Generic Govt 5 Yr           & MSCI INDIA               & ISHARES MSCI USA QUALITY FAC \\
USD-JPY X-RATE                 & U.S. Aggregate           & ISHARES MSCI USA SIZE FACTOR \\
NASDAQ 100 STOCK INDX          & USD-NOK RR 25D 3M        & ISHARES MSCI USA MIN VOL FAC \\
MSCI INDIA                     & USD-SEK RR 25D 3M        & MSCI WORLD VALUE INDEX       \\
U.S. Treasury                  & NASDAQ COMPOSITE         & World Size Tilt              \\
J.P. Morgan EMBI Global Spread & BRAZIL IBOVESPA INDEX    & MSCI WORLD Min Vol PR        \\
USD-NOK RR 25D 3M              & IBEX 35 INDEX            & MSCI World High Dividend     \\
USD-SEK RR 25D 3M              & FTSE MIB INDEX           & MSCI World ESG               \\
BRAZIL IBOVESPA INDEX          & S\&P 500 REAL ESTATE IDX & MSCI WORLD/FINANCE           \\
Japanese Yen Spot              & MSCI EM LATIN AMERICA    & MSCI WORLD/INDUSTRL          \\
\multicolumn{2}{l}{MSCI EM   LATIN AMERICA}               &                              \\
U.S. TIPS                      &                          &                              \\  
\bottomrule
\end{tabularx}
\end{table}

\newpage

\section{Proof of Probabilistic Causality for the Commonality Principle in \citep{RODRIGUEZDOMINGUEZ2023100447}}

\begin{proof}Probabilistic common causality and optimal portfolio drivers :\\

\begin{itemize}
\item Probability of causality for an asset or a portfolio, given a set of drivers, is defined as the probability that their dynamics are caused by this set of drivers.

\item Given portfolio constituents $a_1,\dots, a_N$ of a portfolio $p$, each constituent has associated $M_1,\dots,M_N$ number of specific drivers. Constituent $a_1$ has $SD_{11},\dots,SD_{1M_1}$ specific drivers, $a_2$ has $SD_{21},\dots,SD_{1M_2}$, and the same up to $a_N$ with  $SD_{N1},\dots,SD_{NM_N}$. There exists a probability of causality, probability of the drivers causing the constituents' dynamics at $k$ time steps in the future. Given by the following vector X, using Judea Pearl notation \citep{neuberg_2003}:  

\begin{equation}
\begin{split}
P\left({a}_{1,t+k}\middle|do(\left[SD_{11},\dots SD_{1M_1}\right]_t\right))\le X_1,\dots,\\ P\left({a}_{N,t+k}\middle|do(\left[SD_{N1},\dots SD_{NM_N}\right]_t\right))\le X_N    
\end{split}
\end{equation}

 \item	At a portfolio level, with $\boldsymbol{D}_t=[D_1,\dots D_M]_t$ the common drivers for the portfolio as per the commonality principle:
\begin{equation}P\left({p}_{t+k}\middle|\left[D_1,\dots D_M\right]_t\right)\le Y\end{equation}    
  \item	To prove the principle (portfolio drivers’ optimality) for the probability of causality, the following proposition must be verified almost surely. It is shown now that it is, but only for the special case that the focus is at a portfolio level, which is coincidentally the only interest in portfolio optimization. $\forall \ p=[a_1,\dots, a_N]$ and specific drivers $SD=[SD_{11},\dots,SD_{1M_1},SD_{21},\dots,SD_{1M_2}\dots,SD_{N1},\dots,SD_{NM_N}]$ and common drivers $\boldsymbol{D}=[D_1,\dots D_M]$:

\begin{equation}
\begin{split}
P\left({p}_{t+k}\middle|\left[D_1,\dots D_M\right]_t\right)=\\
P\left({a}_{1,t+k}\cap\dots\cap{a}_{N,t+k}\middle|\left[D_1,\dots D_M\right]_t\right)>\\
P\left({a}_{1,t+k}\middle|\left[SD_{11},\dots SD_{1M_1}\right]_t\right)*\dots*P\left({a}_{N,t+p}\middle|\left[SD_{N1},\dots SD_{NM_N}\right]_t\right)=\\P({a}_{1,t+k})*\dots*P({a}_{N,t+k})
\end{split}
\end{equation}

\item For the proof, the Common Cause Principle (CCP)  \citep{Reichenbach1956-REITDO-2} is used:\\
Suppose that events A and B are positively probabilistically correlated: \begin{equation} p(A \cap B) > p(A)p(B) 
\label{equation9}
\end{equation} Reichenbach’s Common Cause Principle states that when such a probabilistic correlation between A and B exists, this is because one of the following causal relations exists: A is a cause of B; B is a cause of A; or A and B are both caused by a third factor, C. In the last case, the common cause C occurs prior to A and B, and must satisfy the following four independent conditions:
\begin{equation} p(A \cap B|C) = p(A|C)p(B|C)
\label{equation10}
\end{equation}
\begin{equation} p(A \cap B|\overline{\rm C}) = p(A|\overline{\rm C})p(B|\overline{\rm C}) 
\label{equation11}
\end{equation}
\begin{equation} p(A |C) > p(A |\overline{\rm C}))
\label{equation12}\end{equation}
\begin{equation} p(B|C) > p(B|\overline{\rm C})
\label{equation13}\end{equation}  
$\overline{\rm C}$ denotes the absence of event C (the negation of the proposition that C happens) and it is assumed that neither C nor $\overline{\rm C}$  has probability zero. Condition (\ref{equation10}) states that A and B are conditionally independent, given C. In Reichenbach’s terminology, C screens A off from B. Condition (\ref{equation11}) states that $\overline{\rm C}$  also screens A off from B. Conditions (\ref{equation12}) and (\ref{equation13}) state that A and B are more probable, conditional on C, than conditional on the absence of C. These inequalities are natural consequences of C being a cause of A and of B. Together, conditions (\ref{equation10}) through (\ref{equation13}) mathematically entail (\ref{equation9}). The common cause can thus be understood to explain the correlation in (\ref{equation9}) \citep{Reichenbach1956-REITDO-2}.

\item For the general (CCP) case that the correlated effects are random variables like in the portfolio case: Suppose X and Y are random variables that are correlated, ie, there exist  $x_{i}$ and $y_{j}$ such that \begin{equation} p(X = x_{i} \cap Y = y_{j} ) \neq p(X = x_{i}) p(Y = y_{j})\end{equation} 
Then there exists a set of variables $Z_{1},\dots,Z_{M}$ so that each variable is the cause of X and Y, and 
\begin{equation}
\begin{split}
p(X = x_{i} \cap Y = y_{j}| Z_{1} = z_{k_{1}},\dots, Z_{m} = z_{k_{m}}) =\\
p(X = x_{i}| Z_{1} = z_{k_{1}},\dots, Z_{m} = z_{k_{m}})p(Y = y_{j}| Z_{1} = z_{k_{1}},\dots, Z_{m} = z_{k_{m}}) \end{split}
\end{equation}

How the independent conditions are met for the portfolio case it is verified $\forall \ \boldsymbol{p}=[a_1,\dots, a_N]$, $\forall i,j=1,\dots,N$, $i\neq j$, and common drivers $\boldsymbol{D}=[D_1,\dots D_M]$:

\begin{equation} p(a_i \cap a_j|\boldsymbol{D}) = p(a_i|\boldsymbol{D})p(a_j|\boldsymbol{D})
\label{equation20}
\end{equation}
\begin{equation} p(a_i \cap a_j|\overline{\rm \boldsymbol{D}}) = p(a_i|\overline{\rm \boldsymbol{D}})p(a_j|\overline{\rm \boldsymbol{D}}) 
\label{equation21}
\end{equation}
\begin{equation} p(a_i |\boldsymbol{D}) > p(a_i |\overline{\rm \boldsymbol{D}}))
\label{equation22}
\end{equation}
\begin{equation} p(a_j|\boldsymbol{D}) > p(a_j|\overline{\rm \boldsymbol{D}})
\label{equation23}
\end{equation}

This mathematically entails:

\begin{equation} p(a_i \cap a_j|\boldsymbol{D}) > p(a_i)p(a_j)
\label{Equation24}
\end{equation}

The common drivers (common cause) can thus be understood to explain the correlation between assets in the portfolio. The common cause must occur prior, which is the case for the common portfolio drivers and asset dynamics. The generalization of CCP is given by the Causal Markov Condition (CMC): A variable $a_i$ is independent of every other variable (except $a_i$’s effects) and conditional on all its direct causes. CMC can be applied to all pairs of portfolio constituents as a generalization of CCP given the subset of common drivers such that (\ref{equation20}-\ref{equation23}) holds. For that, it is necessary that the common driver’s subset is a direct cause of portfolio constituent dynamics, which can be probabilistic approximated with correlation, by making use of the CCP for the particular case that the common cause (common drivers) is, at most, the same for all portfolio constituents. If $A$ is a subset of $\boldsymbol{p}=[a_1,\dots, a_N]$, $\boldsymbol{S_{i}}$ is the set of specific drivers for asset $a_i$, $\boldsymbol{SD}$ is the set of all specific drivers for the portfolio constituents:
\begin{equation}
\forall a_i\in \boldsymbol{A}, \forall \boldsymbol{S_{i}}\in \boldsymbol{SD}, P(a_i|do(\boldsymbol{S_{i}})) > P(a_i|do(\sim \boldsymbol{S_{i}}))
\label{Equation25}
\end{equation}
Then, for the case of two assets: $\boldsymbol{D}=[D_1,\dots D_M]$:
\begin{equation}
\begin{split}
 \forall\ a_i,a_j\in A, i,j=1,\dots,N,i\neq j\ , \forall\ \boldsymbol{S_{i}}, \boldsymbol{S_{j}}\in \boldsymbol{SD}:\\ 
 \left[P(a_i\ |\ do(\boldsymbol{S_{i}}))\ >\ P(a_i\ |\ do(\sim \boldsymbol{S_{i}}))\right]\\
 \land\ \left[P(a_j\ |\ do(\boldsymbol{S_{j}}))\ >\ P(a_j\ |\ do\left(\sim \boldsymbol{S_{j}}\right))\right]\\
 \rightarrow\ \left[P\left(a_i\cap a_j\middle| S_X,S_Y\right)=
 P\left(a_i\middle| S_X\right)P\left(a_j\middle| S_Y\right)\right]=\\
 \left[P\left(a_i\cap Y\middle| S\right)= P(a_i)P(a_j)\right], \boldsymbol{S} \equiv \boldsymbol{S_{i}}\equiv \boldsymbol{S_{j}}
 \label{Equation26}
\end{split}
\end{equation}
with $\boldsymbol{S}$ a set of common drivers for constituents $a_i$ and $a_j$. Finally, for any $N$-assets portfolio: 
\begin{equation}
\begin{split}
\left\{\forall p=[a_1,\dots, a_N]\equiv A,\forall\ \{\boldsymbol{S_1},\boldsymbol{S_2},\dots,\boldsymbol{S_N}\}\equiv\boldsymbol{SD},\right.\\
\left.\left[P(p|do(\boldsymbol{SD}))>P(p|do(\sim \boldsymbol{SD}))\right]\right\}\\
\longleftrightarrow\\
[P\left(a_1\cap a_2,\cap \dots \middle| \boldsymbol{S_{1}},\boldsymbol{S_{2}}, \dots,\boldsymbol{S_N}\right)=
P\left(a_1\middle| \boldsymbol{S_{1}}\right)P\left(a_2\middle| \boldsymbol{S_{2}}\right), \dots]\equiv\\
[P\left(a_1\cap a_2,\cap \dots \middle|\boldsymbol{S} \right)=P{(a}_1)P{(a}_2)\dots]\\
\forall a_1,\ a_2, \dots a_N \in A,\forall \boldsymbol{S_{1}}\equiv \boldsymbol{S_{2}}\equiv,\dots \equiv\boldsymbol{S_{N}}\equiv\boldsymbol{S}
\label{Equation27}
\end{split}
\end{equation}

\item In (\ref{Equation25}), it is shown that for any $a_i$, there exists a set of specific drivers that cause its dynamics optimally in probability, using Judea Pearl notation \citep{neuberg_2003}. In (\ref{Equation26}), it is shown how for a portfolio of two assets, and its specific drivers’ selection, how, if they have the highest probability of causality for the assets’ dynamics, the CCP conditions and (\ref{Equation24}) are met only if both $\boldsymbol{S_i}$ and $\boldsymbol{S_j}$ are equivalent, as per the commonality principle, and equal to their optimal common drivers in terms of probabilistic causality. This means that the common drivers are the common source of causality of portfolio constituents’ dynamics, they are the greatest source in the probability of causality for portfolio dynamics, and they explain the correlation between portfolio constituents, for the two assets case, by applying \citep{Reichenbach1956-REITDO-2}. 
\item In (\ref{Equation27}), the generalization for any combination of assets (portfolio). Here, the implication goes two ways in that, for any portfolio of assets, their common drivers being the source of the highest probability of causality of portfolio dynamics (not their constituents), imply CCP conditions and (\ref{Equation24}) are met. But, if CCP and (\ref{Equation24}) are met, which occurs only in the case that the common drivers are selected based on the commonality principle, which in turn makes the equivalence in sets S possible, CCP conditions and (\ref{Equation24}) imply that they are  the greatest source in the probability of causality for portfolio dynamics. This is true for any combinations of assets or portfolio $(\boldsymbol{p})$ and common drivers set $(\boldsymbol{D})$ chosen as in the commonality principle. But also, like in (\ref{Equation26}), this means that common drivers explain the correlation between portfolio constituents by applying \citep{Reichenbach1956-REITDO-2}.
\end{itemize}
\end{proof}
The fact that common drivers explain the correlation between portfolio constituents, as outlined by \citep{Reichenbach1956-REITDO-2}, justifies the possible selection of common drivers (common causes) as those most correlated with the greatest number of portfolio constituents. This is the simplest driver selection method and has been shown to work effectively in experiments. Other selection methods may also be applicable, but they must adhere to the commonality principle, such as using Bayesian networks to check the CCP independence conditions.

\newpage

\section{Proof of the Commonality Principle as a Necessary and Sufficient Condition for Optimal Diversification and Portfolio Optimization Efficiency from \citep{RODRIGUEZDOMINGUEZ2023100447}}
\label{appselectdrov}

In the mean-variance framework from MPT, the portfolio’s expected returns lie on a hyperplane of the constituents' expected returns, and portfolio risk lies on a hypersurface, as seen in Figure \ref{fig:enter-Markowitztimeembedding}. The hyperplane is given by: 
\begin{equation}
E\left[r_p\right]=\sum_{i=1}^{n}w_iE[r_{a_i}]    
\end{equation} 

where, $E\left[r_p\right]$ are the portfolio’s expected returns, Tp is the Tangency Portfolio, $E\left[r_{a_i}\right]=\mu_i$ are the constituents' expected returns, as in Figure 2. $E\left[r_p\right]$ is linear in $E[r_{a_i}]$, portfolio risk $\sigma_p$ is non-linear in constituents' risk ${\sigma_{a_i}}$, and w are the weights, solution to the quadratic optimization in: 
\begin{equation}
\label{equationwww}
w = \min_w{w^{T}\mathrm{\Sigma}\ w}
\end{equation}
with the tangency portfolio as the optimal solution. In Figure \ref{figure2}, we show the representation of portfolio constituents' expected returns for a period, in the time-dimensional space. Axis are points in time, $\theta$ are angles between expected returns, and the cosines are the correlations, ie, $\rho_{23}=\cos{\theta_2}$ is the correlation between $r_{a_2}$ and $r_{a_3}$ and $\theta_2=\hat{E[r_{a_2}]E[r_{a_3}]}$.
The expected portfolio returns conditional on the common drivers is a linear combination of constituents’ expected returns conditional on the same drivers: 
\begin{equation}
E\left[r_p|CD\right]=\sum_{i=1}^{n}w_iE[r_{a_i}|CD]    
\end{equation}
is a hypersurface. In Figure \ref{fig:enter-Markowitztimeembedding}, we show the hypersurface and the tangency portfolio when solving the portfolio optimization. In Figure \ref{fig:enter-1stconformalmap.jpg} (right graph), we can see portfolio constituents' expected returns conditional on common drivers in a time-dimensional space with points in time as an axis. We demonstrate the existence of a conformal map from this space to another in which the conditional expectations are embedded in the space of sensitivities of constituents with respect to the common drivers (see the right side of Figure \ref{fig:enter-SecondConformalMap}). In the embedded space, angles between conditional expected returns are a sum of two components, a systematic component, and an idiosyncratic component from the unconditional expectation case from MPT: 
\begin{equation}
\theta_{XY}^\prime={\alpha_1\theta}_{XY}+\alpha_2\gamma_{XY}\end{equation}
where $\cos{\theta_{XY}}$\ is $\rho_{XY}$ from $\mathrm{\Sigma}$\ \ in (\ref{equationwww}), with ${\ \theta}_{XY}$ the idiosyncratic and $\gamma_{XY}$  the systematic component. The conformal map is such that the idiosyncratic component maintains the same proportion between angles in both spaces, and the idiosyncratic risk representation is kept at most in the new space. We will show next that the embedding of the time-dimensional space into a space of sensitivities with respect to common drivers is a conformal map.
\begin{lemma}
A Conformal map implies: 
\begin{equation}
\begin{split}
\theta_{XY}=f\left(t,\ E\left[r_{a_X}\right],E\left[r_{a_Y}\right]\right)=f\left(t,\ E\left[r_{a_X}\right](t),E\left[r_{a_Y}\right](t)\right)=\\
f(t)\Longrightarrow f\left(t,\ E\left[r_{a_X}|CD\right],E\left[r_{a_Y}\middle| C D\right]\right)=f\left(t\right)  
\end{split}
\end{equation}
\end{lemma}
If the angle in the mean-variance framework between unconditional expected returns is a function of both expected returns, which are also a function of time, the angle can be reduced to a function of time alone. A conformal map implies that the angle in the embedded sensitivity space, between the expected returns conditional on common drivers, which is a function of these conditional expectations and time, can be reduced to a function of time alone too. The map then is conformal at a given point in time.

\begin{proof}For the proof we analyze all possible cases.
\begin{itemize}
    \item Case I: Non-Common Drivers \begin{equation}
        \theta_{XY}=f(t,\ E\left[r_{a_X}|D_X\right],E[r_{a_Y}|D_Y])
        \label{eqapend1}
    \end{equation}
    
\begin{equation}
\begin{split}
 \forall\ D_X,D_Y,D_X\neq D_Y:E\left[r_{a_X}|D_X\right]=\widehat{\beta_X}D_X, \widehat{\beta_X}=[\beta_{X_1},\ldots,\beta_{X_M}],\\ E\left[r_{a_Y}|D_Y\right]=\widehat{\beta_Y}D_Y, \widehat{\beta_Y}=\left[\beta_{Y_1},\ldots,\beta_{Y_M}\right]\\
 \Longrightarrow E\left[r_{a_X}|D_X\right]\bot E\left[r_{a_Y}|D_Y\right]
\end{split}
\label{eqapend2}
\end{equation}

\begin{equation}
 \left[\begin{matrix}E\left[r_{a_X}|D_X\right]\\E\left[r_{a_Y}|D_Y\right]\\\end{matrix}\right]=\left[\begin{matrix}\beta_{X_1},\ldots,\beta_{X_M},000,\ldots,0\\000,\ldots0,\beta_{Y_1},\ldots,\beta_{Y_M}\\\end{matrix}\right]\left[\begin{matrix}D_X\\D_Y\\\end{matrix}\right]
 \label{eqapend3}
\end{equation}
Case I focuses on examples where drivers are not common. In this case, conditional expected returns are orthogonal in the embedded space of sensitivities with respect to drivers’ constituents, as seen in (\ref{eqapend1})-(\ref{eqapend3}), which is not a valid case.
    \item 	Case II: Common Non-Casual Drivers \begin{equation}
        \theta_{XY}=f(t,\ E\left[r_{a_X}|D\right],E[r_{a_Y}|D])
    \end{equation} 
    \begin{equation}
    \begin{split}
      	E\left[r_{a_X}|D\right]=\widehat{\beta_X}D,\widehat{\beta_X}=\left[\beta_{X_1},\ldots,\beta_{X_M}\right],E\left[r_{a_Y}|D\right]=\widehat{\beta_Y}D, \\ \widehat{\beta_Y}=\left[\beta_{Y_1},\ldots,\beta_{Y_M}\right];\left[\begin{matrix}E\left[r_{a_X}|D\right]\\E\left[r_{a_Y}|D\right]\\\end{matrix}\right]=\left[\begin{matrix}\beta_{X_1},\ldots,\beta_{X_M}\\\beta_{Y_1},\ldots,\beta_{Y_M}\\\end{matrix}\right]\left[\begin{matrix}D_1\\\ldots\\D_M\\\end{matrix}\right]        
    \end{split}
    \end{equation}
\begin{equation}
	\left[\begin{matrix}E\left[r_{a_X}|D\right](t)\\E\left[r_{a_Y}|D\right](t)\\\end{matrix}\right]=\left[\begin{matrix}\frac{\partial r_{a_X}}{\partial D}\left(t\right)\ast D\\\frac{\partial r_{a_Y}}{\partial D}\left(t\right)\ast D\\\end{matrix}\right]
	\end{equation}
	if D is not cause of $r_{a_X}$ and \begin{equation}
	r_{a_Y}\Rightarrow r_{a_X}\left(t\right)\neq f\left(D\left(t\right)\right),r_{a_Y}\left(t\right) \neq f(D\left(t\right))
	\end{equation}
	Hence 
	\begin{equation}f\left(t,E\left[r_{a_X}|D\right](t),E\left[r_{a_Y}|D\right](t)\right)\neq f(t) \end{equation}
Case II focuses on common but non-casual drivers, in this case, if drivers are not the source of causality of constituents’ returns, these are not a function of the drivers, which is also not valid.	
	\item 	Case III: Common Causal Persistent Drivers (Commonality Principle):
	If D causes $r_{a_X}$ and $r_{a_Y}$:
	\begin{equation}\Rightarrow r_{a_X}\left(t\right)=f\left(CD\left(t\right)\right),r_{a_Y}\left(t\right) =f(CD\left(t\right))\end{equation}
	And we have: 
	\begin{equation}\theta_{XY}^\prime=f\left(t,\ E\left[r_{a_X}|CD\right],E\left[r_{a_Y}\middle| C D\right]\right)=f\left(t\right)
	\end{equation}
	embedding of t.
	\begin{equation}
	\begin{split}
	f\left(D_X\left(t\right)\right)=NN(D_X(t)\cong E[r_{a_X}|D_X(t)]\neq r_{a_X}(t);\\ f\left(D_Y\left(t\right)\right)=NN(D_Y(t)\cong E[r_{a_Y}|D_Y(t)]\neq r_{a_Y}\left(t\right);\\
	D_X\left(t\right)=D_Y(t) \iff D_X\left(t\right)=D_Y\left(t\right)=CD(t) 	    
	\end{split}
	\end{equation}
	Case III focuses on the commonality principle examples in which drivers are commonly casual and persistent. In this case, if common drivers (CD) cause constituents’ returns, these are functions of CD. If and only if those specific drivers are equal (commonality principle selection), the relationship between conditional expected returns is a function of time alone and the angle in the embedded space of sensitivities is also a function of time alone. Hence proving the conformal map.
\end{itemize}
\end{proof}

\newpage

\section{Selection of Common Drivers via RCCP Reverse Engineering for the Hierarchical Sensitivity Parity Experiments in \citep{RODRIGUEZDOMINGUEZ2023100447}}

\label{Subsection41App}
From a set of M drivers with  $M >> N$, N being the number of constituents of the portfolio. For each constituent, we rank correlations with respect to all drivers for different lags and time horizons, with a threshold that depends on the lag. We select drivers that have passed the thresholds the greatest number of times among all portfolio constituents. We now show the algorithm: \par

\begin{equation}\forall\ {Asset}_i,{SD\ }_i\ \ i=1,\dots N,\ {\forall\ Driver}_j\ j=1,\dots M
\label{equation31}
\end{equation}

\begin{equation}
\begin{split}
 \ 
{(Driver}_j\in\ {SD\ }_i)\Rightarrow\\ (corr\left(\ {Asset}_i\left(t\right),\ {Driver}_j\left(t\right)\right)>T0\\
\wedge \ corr\left(\ {Asset}_i\left(t\right),\ {Driver}_j\left(t-1\right)\right)>T1)   
\label{equation32}
\end{split}
\end{equation}

\begin{equation}
A=A_0,\ B=\ \left\{b\in A_i:b\geq a\ \forall a\ \in A_0\ \right\},\ A_1=\ A_0\setminus B_1, 
\label{equation33}
\end{equation}

\begin{equation}
B_{i+1}\ =\ \left\{b\in A_i:b\geq a\ \forall a\ \in A_i\ \right\},\ A_{i+1}=\ A_i\setminus B_{i+1}
\label{equation34}
\end{equation}

\begin{equation}
\begin{split}
 (A={{Driver}_j}| \max( \#({Driver}_j\ \in \ {SD\ }_i)\\ 
 \forall\ {Driver}_j\ , {SD\ }_i \ i=1,\dots N,\ \ j=1,\dots M)   
 \label{equation35}
\end{split}
\end{equation}
 
\begin{equation} B_k={CD}_i,\ k=\max{\left(\#CD\right)},\ i=1,\dots N
\label{equation36}
\end{equation}

Equation (\ref{equation32}), states that, for a driver to be a specific driver for a particular constituent, it must have correlations above thresholds T1 and T0 for respective lags 1 and 0 (hyperparameters 4 and 5). Equations (\ref{equation33}) and (\ref{equation34}) are the formulations for the problem of finding the set of (i+1) elements that have a greater value than a threshold from other sets of elements. Equation (\ref{equation35}) is adapting Set A to our problem because we want, from all drivers of the drivers set, those that are simultaneously specific for the greatest number of portfolio constituents. K is the hyperparameter 1 of choice that indicates the maximum number of common drivers to select for the model implementation. $B_k$ will be the k common drivers optimally chosen. Optimal in terms of passing the thresholds and being repeated the maximum number of times among portfolio constituents.


\newpage

\bibliographystyle{elsarticle-harv}
\bibliography{cas-refs}





\end{document}